\documentclass[reqno,a4paper,11pt,final,oneside]{book}
\usepackage{amsmath,amssymb,amstext,amsthm,amscd,mathrsfs,eucal}
\pdfoutput=1

\usepackage[Rejne]{fncychap}
\usepackage[british]{babel}
\usepackage{a4wide}
\usepackage[utf8x]{inputenc}
\usepackage{graphicx,color}
\usepackage[nohbox]{pdfsync}
\usepackage{array}
\usepackage{cite}
\usepackage{hyperref}
\usepackage[left=4.0cm,right=2.5cm,top=3cm,bottom=4cm]{geometry}
\usepackage{setspace}
\usepackage[all]{xy} 
\usepackage[vcentermath]{youngtab}
\Yboxdim{4pt}

\hypersetup{%
 pdftitle   = {Metric 3-Leibniz algebras and M2-branes},
 pdfkeywords = {M2, M Theory, Bagger-Lambert, Lie algebra, Filippov, 3-algebra, Lie triple system},
 pdfauthor  = {Elena Méndez-Escobar},
 pdfcreator = {\LaTeX\ with package \flqq hyperref\frqq}
}

\let\into\hookrightarrow
\newcommand{\half}{\tfrac12}
\newcommand{\fg}{\mathfrak{g}}

\newcommand{\fa}{\mathfrak{a}}
\newcommand{\fB}{\mathfrak{B}}

\normalfont
\newcommand{\fd}{\mathfrak{d}}

\newcommand{\fF}{\mathfrak{F}}

\newcommand{\fM}{\mathfrak{M}}
\newcommand{\fV}{\mathfrak{V}}
\newcommand{\fW}{\mathfrak{W}}
\newcommand{\fgl}{\mathfrak{gl}}
\newcommand{\fh}{\mathfrak{h}}

\newcommand{\fk}{\mathfrak{k}}
\newcommand{\fl}{\mathfrak{l}}

\newcommand{\fr}{\mathfrak{r}}
\newcommand{\fs}{\mathfrak{s}}

\newcommand{\fw}{\mathfrak{w}}
\newcommand{\fz}{\mathfrak{z}}

\newcommand{\fso}{\mathfrak{so}}
\newcommand{\fosp}{\mathfrak{osp}}
\newcommand{\fspin}{\mathfrak{spin}}
\newcommand{\fusp}{\mathfrak{usp}}
\newcommand{\fsl}{\mathfrak{sl}}
\newcommand{\fsp}{\mathfrak{sp}}
\newcommand{\fsu}{\mathfrak{su}}
\newcommand{\fu}{\mathfrak{u}}

\newcommand{\Cl}{\mathrm{C}\ell}

\newcommand{\TT}{\mathbb{T}}

\newcommand{\RR}{\mathbb{R}}
\newcommand{\OO}{\mathbb{O}}
\newcommand{\CC}{\mathbb{C}}
\newcommand{\HH}{\mathbb{H}}
\newcommand{\KK}{\mathbb{K}}
\newcommand{\VV}{\mathbb{V}}
\newcommand{\ZZ}{\mathbb{Z}}
\newcommand{\XX}{\mathbb{X}}
\newcommand{\YY}{\mathbb{Y}}
\newcommand{\XXbar}{\overline{\mathbb{X}}}
\newcommand{\eC}{\mathscr{C}}

\newcommand{\eL}{\mathscr{L}}

\newcommand{\eR}{\mathscr{R}}

\newcommand{\be}{\boldsymbol{e}}

\newcommand{\sW}{\mathsf{W}}

\newcommand{\Ubar}{\overline{U}}
\newcommand{\Wbar}{\overline{W}}
\newcommand{\Vbar}{\overline{V}}
\renewcommand{\Im}{\mathrm{Im}}
\renewcommand{\Re}{\mathrm{Re}}
\newcommand{\AdS}{\mathrm{AdS}}

\DeclareMathOperator{\Dar}{Rep}
\DeclareMathOperator{\Irr}{Irr}

\DeclareMathOperator{\End}{End}
\DeclareMathOperator{\Der}{Der}
\DeclareMathOperator{\Mat}{Mat}

\DeclareMathOperator{\ad}{ad}

\DeclareMathOperator{\im}{im}
\DeclareMathOperator{\re}{Re}
\DeclareMathOperator{\tr}{tr}

\newcommand{\rf}[1]{[\![#1]\!]}
\newcommand{\rh}[1]{(\!(#1)\!)}

\theoremstyle{plain}
\newtheorem{lemma}{Lemma}
\newtheorem{proposition}[lemma]{Proposition}
\newtheorem{theorem}[lemma]{Theorem}
\newtheorem{corollary}[lemma]{Corollary}
\theoremstyle{definition}
\newtheorem{definition}[lemma]{Definition}
\newtheorem{remark}[lemma]{Remark}
\newtheorem{example}[lemma]{Example}
\allowdisplaybreaks
\begin{document}
\title{Metric 3-Leibniz algebras and M2-branes}
\author{Elena Méndez-Escobar}

\pagenumbering{Roman}

\begin{titlepage}

\vspace{1mm}
\begin{center}
{\LARGE{\bf Metric 3-Leibniz algebras and M2-branes}}\\
\vspace{2mm}
\end{center}
\vspace{17mm}
\par
\noindent

\vspace{1mm}
\begin{center}
{\Large{\bf Elena Méndez-Escobar}}\\
\vspace{2mm}
\end{center}
\vspace{87mm}
\par
\noindent

\vspace{1mm}
\begin{center}
{\Large{\bf Doctor of Philosophy}}\\
\vspace{2mm}
\end{center}
\begin{center}
{{\bf The University of Edinburgh}}\\
\vspace{1mm}
\end{center}
\begin{center}
{{\bf 2010}}\\
\vspace{1mm}
\end{center}
\vspace{1mm}
\par
\noindent

\end{titlepage}
\newpage
\pagestyle{plain}
\begin{center}
\section*{Declaration}
\end{center}
\vspace{2in}
I hereby declare that this thesis is of my own composition, and that
it contains no material previously submitted for the award of any
other degree.  The work reported in this thesis has been executed by
myself, except where due acknowledgement is made in the text. 
\vspace{2in}
\begin{flushright}
Elena Méndez Escobar
\end{flushright}
\vspace{2in}
On August 16th 2010 the examiners Professor Neil Lambert (external) and Professor Agata Smoktunowicz (internal) recommended to award the degree of Doctor of Philosophy to Elena Méndez-Escobar with no corrections.

\newpage

\begin{center} \section*{Abstract} \end{center}
{\small 
\singlespacing
Since the early days of the AdS/CFT correspondence, a Lagrangian formulation for a superconformal gauge theory in three dimensions with $N=8$ supersymmetry has been sought, as the dual theory to a stack of M2-branes in an $AdS_4 \times S^7$ background. This search was fruitless until late 2007 when Bagger and Lambert proposed a candidate Lagrangian for such theory using a 3-Lie algebra. In the past three years generalisations of this theory have been found and candidate Lagrangians for dual theories to M2-branes in other backgrounds have been suggested. 

The interest in these theories is two-fold. On the one hand, it is a new family of theories in which to test the AdS/CFT correspondence and on the other, they are important to study one of the main objects of M-theory (M2-branes) and will perhaps shed some light into a microscopical interpretation for them. All these theories have something in common: they can be written in terms of 3-Leibniz algebras. In this thesis we study the structure theory of such algebras, paying special attention to a subclass of them that gives rise to maximal supersymmetry and that was the first to appear in this context: 3-Lie algebras.

In chapter 2, we review the structure theory of metric Lie algebras and their unitary representations. The purpose of this review is to set a uniform notation for the rest of the thesis and pave the way to some of the results which can be seen as a generalisation of certain known theorems of metric Lie algebras. In chapter 3, we study metric 3-Leibniz algebras and show, by specialising a construction originally due to Faulkner, that they are in one to one correspondence with pairs of real metric Lie algebras and unitary representations of them. We also show a third characterisation for six extreme cases of 3-Leibniz algebras as graded Lie (super)algebras. In chapter 4, we study metric 3-Lie algebras in detail. We prove a structural result and also classify those with a maximally isotropic centre, which is the requirement that ensures unitarity of the corresponding conformal field theory. Finally, in chapter 5, we study the universal structure of superpotentials in this class of superconformal Chern-Simons theories with matter in three dimensions. We provide a uniform formulation for all these theories and establish the connection between the amount of supersymmetry preserved and the gauge Lie algebra and the appropriate unitary representation to be used to write down the Lagrangian. The conditions for supersymmetry enhancement are then expressed equivalently in the language of representation theory of Lie algebras or the language of 3-Leibniz algebras.}

\newpage
\begin{center} \section*{Acknowledgements}  \end{center}

First and foremost I would like to thank my supervisor José Figueroa-O'Farrill, who has not only been a mentor for me but also a great role model of excellence in academia. By treating me as a collaborator from day one he offered me the best learning experience of all: witnessing first hand how ideas are generated and developed. I will always be grateful for the inspiring ideas he shared with me and the insightful and unique explanations he patiently gave to all of my questions over the years. 

The other person without whom this thesis would not exist is Paul de Medeiros, whose ability to give you those key facts to understand something that nobody else told you before I admire. He is always devoted to finding the most interesting problems and I feel privileged that he shared those problems with me and provided the tools I needed to work with him in trying to solve them. I am in debt with him for the patience with which he went through very detailed calculations showing me useful shortcuts.

From Paul and José I did not only learn physics and mathematics but also what team work really means and the pleasure of a true scientific collaboration. Moreover, José and Atsuko have always been the best hosts one could imagine, giving a whole new meaning to the word \textit{transkills} to include such things as learning about Japanese culture and cooking sushi and \textit{tostones}.

I am also grateful to the rest of the EMPG group, especially to my second supervisor Joan Simon, whose guidance has been extremely valuable for me and has given me more food for thought than he is probably aware of. Most of all I am in debt with the members of the EMPJ. Very especially with Patricia and Antonella for sharing their knowledge and experience with me since my very first day. Without their lead, the student activities that I have enjoyed so deeply would have never happened. Moreover, they have become great friends over the years and I will forever treasure our conversations on life and academia which made everything make sense even when it did not seem it would.

I also want to thank the rest of my colleagues in the department, especially the head of the graduate school Andrew Ranicki whose dedication improves the lives of the \textsl{PGs} in the department every day and whose confidence in us when we were PG representatives made me learn valuable things about myself that I did not know. My deepest gratitude to Gina, Gill and the rest of the administrative staff who are truly at the core of the department and without whose experience and knowledge I would have probably not made it through the bureaucratic wall more than once.

Life as a PhD student would have not been the same without the great support of the rest of the PGs. Special thanks to my office mates Pamela and Daniele (I'm still convinced we are doing the same thing!) and lunch mates - present and past - Eric, Chris, Andrew, George, Rosemary, Richard, Marina, Jesús, Lisa, Achim, Nairn, Steven, Val (by extension) and many others without whom I am sure days would have seem much longer.

There are many people who, even in the distance, have accompanied me through the years. I am thankful to all the incredible friends that I have. I treasure every single one of you and hope we will continue to visit each other everywhere we go. My colleagues from undergraduate days at UAM, that unforgettable summer in DESY (Hamburg), Part III in Cambridge and also those whom I met in Edinburgh and left before I did. Thank you for your support despite of the kilometres and for constantly surprising me with how well you know me.

My deepest gratitude to my friend Paola who in the last 10 years has come to be more than a sister for me and whose unconditional friendship has made me a better person. Thank you for always having the right words to comfort me when I need it the most.

I am grateful to my very extended family, including the newly acquired Italian branch, for being there always and especially to Patxi and María who have been very inspiring role models and showed me a preview of what life in academia looked like. Most of all I am thankful to my brother Ricardo and my parents José and Merche who taught me to love science from a very young age, always believed in me giving me the confidence that I could do anything I wanted and supported me in every possible way to reach my dreams. 

Very special thanks to someone who fits in all of the categories above: a colleague, a friend and since two months ago also officially my family. My husband Dario has been my driving engine for the past five years, encouraging me to pursue each one of my goals and offering patiently his endless support when I needed it the most. Thank you for making easy what seemed impossible, I would have not made it without you.

\onehalfspacing

\tableofcontents
\listoftables 
\clearpage
\pagenumbering{arabic}
\pagestyle{headings}

\chapter{Introduction}
\label{sec:introduction}

In this introduction we motivate and formulate the research questions to be addressed in this thesis. First, we introduce M2-branes in the context of M-theory and review the potential benefits of applying the AdS/CFT correspondence to them. Then, we explain the challenges of this approach since the early days of Maldacena's duality and the breakthrough that came with the work of Bagger and Lambert in 2007. We then go on to explain how the considerable amount of papers that followed their work raised many questions on a mathematical structure that seemed to be at the core of M2-branes dynamics: 3-algebras. This will allow us to formulate the precise questions we want to address here on the structure of 3-algebras, their relation to the better known Lie algebras and their role in the context of modelling M2-branes. Finally we go through the organisation of the thesis by chapters. 

\section{What is an M2-brane?}

The history of membranes in theoretical physics dates back to the times before string theory. In 1962, Paul Dirac suggested for the first time that some elementary particles (in particular the electron and the muon) could correspond to excited states of a two-dimensional object, in that case a sphere\cite{dirac1962extensible}. This idea was abandoned amongst other reasons for the lack of spin in that theory, and it was not until the supergravity days that membranes came back to theoretical physics.

With the development of string theory in the 1970s, a natural question arose: if one can have one-dimensional strings, why not 2-dimensional membranes or, in general, p-dimensional objects? Theories for (closed and open) branes were explored in parallel with string theory as generalisations of the Green-Schwarz action. Some analogous results followed, for example what was called \textit{brane scan} determined the restrictions supersymmetry imposed on p-branes propagating in D dimensions; just like in the early eighties Green and Schwarz \cite{GreenSchwarzString} had shown that supersymmetry allows classical superstrings to live in spacetime dimensions 3, 4, 6 and 10. 

After some failed attempts to quantise these theories for branes, a first breakthrough came in 1986 when Hughes, Liu and Polchinski \cite{Hughes:1986fa} showed that it was possible to incorporate supersymmetry to a theory of 3-branes. Their argument was later generalised by Bergshoeff, Sezgin and Townsend in \cite{Bergshoeff:1987cm} to p-branes propagating in D dimensions. They had written down for the first time a Lagrangian for a \textit{single} supersymmetric membrane, but it would not be until twenty years later that the first hint of a Lagrangian describing the dynamics of \textit{multiple} membranes would appear in the literature. In that paper they also argued that p-branes could only consistently propagate in supergravities containing a closed (p+1)-form potential they could couple to.

Membranes in particular regained a certain popularity due to the fact that 11D supergravity contains a 3-form, which strongly suggested a relation with a supermembrane theory. Indeed, this supermembrane was later identified with the solitonic membrane of 11-dimensional supergravity. The same happened with other p-branes that propagated in supergravity backgrounds containing (p+1)-forms in them, thus melding \textit{super-brane} theories and supergravity theories.

Interest in 11-dimensional supergravity revived as the web of string dualities was unveiled in the mid-90s. As it turned out, two of the string theories (type IIA and the $E_8 \times E_8$ heterotic string theory) develop an eleventh dimension in the strong coupling limit, giving rise to a theory that Witten baptised as M-theory at the \textit{Strings '95} conference in the University of Southern California. Witten proposed M-theory to be an extension of string theory that would contain all known five string theories via dualities and have 11-dimensional gravity as its low energy limit. At the same time, another reason for string theorists to reconcile with supergravity theories and their branes was given by Polchinski when he identified the extended objects in which strings can end (D-branes) with certain supergravity p-branes\cite{Polchinski:1995mt}.

Even nowadays not much is known about M-theory. The spectrum of 11D supergravity contains a graviton, a Majorana gravitino and a third rank anti-symmetric tensor (also known as C-field) which can couple electrically to 2-branes and magnetically to 5-branes. In particular it contains no strings. Since the spectrum should be protected by supersymmetry, M-theory is expected to contain the same objects. This means that it is not a theory of strings and its constituents are instead 2 and 5 branes also known as \textit{M2-branes} and \textit{M5-branes}. This motivates the study of the properties and dynamics of M2-branes, as we hope to learn more about the mysterious M-theory via one of its constituents.

\section{M2-branes and AdS/CFT}

The mid 90s were an exciting time for string theory. On top of the identification between p-branes and boundary conditions for open strings and the the web of dualities discovered during the \textit{second superstring revolution}, there was Maldacena's AdS/CFT correspondence \cite{AdSCFT_Mald}. Since 1997, many things have been learnt of the numerous instances of this duality. In particular its application to systems of coincident D-branes has led to a deeper understanding of the dynamics of these nonperturbative objects. It was only natural to try a similar approach with stacks of M2 and M5 branes. In that case one would start from the gravity side of the correspondence, where these branes are defined, and hope to learn new things from the corresponding dual field theory. For example such construction could shed some light on a microscopical interpretation for them, just like the dual description to supergravity D-branes provides their microscopical interpretation in string theory as ending points for open strings.

One of the most exploited instances of this conjecture is that known as $AdS_5/CFT_4$, obtained by considering a stack of D3-branes. The conjecture says that type IIB superstring theory in an $AdS_5 \times S^5$ background is dual to $N=4$ $SU(n)$ super Yang-Mills theory. The supergravity background corresponds to the ground state of the gauge theory (which is known to be conformal) and excitations and interactions in one description correspond to excitations and interactions in the other. The rank of the gauge group $n$ coincides with the flux units going through the five-sphere and the isometry group for the string theory background, $PSU(2,2|4)$, is also the conformal supergroup of the $N=4$ gauge theory. The Yang-Mills coupling constant is related to the string coupling constant and the fact that it does not depend on the energy scale corresponds to the fact that the dilaton is constant for the black D3-brane solution. Finally, the fourth power of the common radius $R$ of $AdS_5$ and $S^5$ is proportional to the 't Hooft parameter of the gauge theory.

The corresponding situation involving M2-branes, often referred to as $AdS_4/CFT_3$, predicts the following. A stack of M2-branes in 11-dimensional supergravity has an $AdS_4 \times S^7$ near-horizon geometry and its dual is a conformal field theory (CFT) in three dimensions with superalgebra $\fosp(8|4)$. Since the number of fermionic and bosonic degrees of freedom coincide when considering only the matter fields, gauge fields added to the theory should not propagate in order to preserve supersymmetry. This suggests that the gauge theory should be of Chern-Simons type. We would be looking for a 3-dimensional superconformal Chern-Simons theory coupled to matter (3D SCCSM) with $N=8$ supersymmetry and even part of the superalgebra isomorphic to $\fso(2,3) \oplus \fso(8)$, which is the isometry algebra of $AdS_4 \times S^7$.

Although  many CFTs in two and four dimensions were know, until 2007 very few examples in three dimensions had arisen, let alone one with such specific requirements. Even as recently as 2004 it was claimed\cite{SchwarzCS} that a Lagrangian description for such theories might not exist at all. After all, this was by no means guaranteed given the absence of a dilaton field in the M-theory spectrum. This implies in particular that the conformal field theory would not have an adjustable coupling constant and would necessarily have to be strongly coupled, and only weakly coupled theories are presumed to have a Lagrangian description.

We want to point out that this adds extra motivation for looking for such theories. Their interest lies not only in the information on M2-branes and M-theory they might encode, but also as highly supersymmetric 3-dimensional CFTs they are of particular importance on their own. Also, their discovery would mean a new broad test field for the AdS/CFT correspondence that has brought so many unexpected results and applications to string theory over the past years.

\section{Recent history of M2-branes}

What has been called \textit{M2-branes mini-revolution} started in 2007 with a series of three papers \cite{BL1,BL2,BL3} by Jonathan Bagger and Neil Lambert. In those papers, inspired by the generalised Nahm's equation for a system of M2 and M5 branes of Basu and Harvey \cite{BasuHarvey}, they proposed for the first time a Lagrangian possessing all the symmetries expected for the dual theory to a stack of coincident M2-branes: a maximally supersymmetric ($N=8$) conformal field theory in three dimensions with an explicit $\fso(8)$ R-symmetry. In these papers, and also in the independent work of Gustavsson \cite{GustavssonAlgM2}, they showed that for the supersymmetry transformations to close the matter fields have to live in a \textit{3-Lie algebra}. Such a structure is a generalisation of a Lie algebra: it is a vector space equipped with a \textit{3-bracket} $[-,-,-]$, which is totally antisymmetric and satisfies an identity called the fundamental identity that generalises the Jacobi identity for Lie algebras.

A key ingredient in order to write a Lagrangian was the existence of an inner product in the vector space where the matter fields live compatible with the 3-bracket. For unitarity of the resulting theory, one requires this inner product to be positive-definite. It was soon noticed that there is a single simple 3-Lie algebra admitting a positive-definite inner product \cite{NagykLie,GP3Lie,GG3Lie}.  The corresponding Bagger-Lambert (BL) theory, upon quantisation, gives rise to a family of theories labelled by the integer Chern-Simons level $k$. This theory was argued to describe at most two M2-branes \cite{LambertTong, DistlerBL} on a certain M-theory orbifold, at least for $k=1,2$. For other values of $k$ its spacetime interpretation is still unclear.

It seemed like a no-go for these theories as candidates to describe the dynamics of multiple M2-branes. Two ways of overcoming this were suggested. The first one was to drop the requirement of positive-definiteness of the metric. In \cite{GMRBL,BRGTV,HIM-M2toD2rev} 3-Lie algebras equipped with a Lorentzian metric were considered. In principle one could run into trouble with unitarity of the theory, but it turned out that the ghost-like fields, that might give rise to negative-norm states, decoupled from the physical Hilbert space. This opened the door to further study of these theories using 3-Lie algebras of indefinite signature.

The second option was to drop the requirement of maximal supersymmetry. After some $N=4$ superconformal Chern-Simons theories in three dimensions were constructed \cite{GaiottoWitten,MasahitoBL} a big break through came with the work of Aharony, Bergman, Jafferis and Maldacena \cite{MaldacenaBL}. In that paper not only did they write a 3D SCCSM Lagrangian with explicit N=6 supersymmetry, but also they were able to argue that their Lagrangian was actually describing the low energy limit of an arbitrary number of M2-branes probing a $\RR^8/\ZZ_k$ singularity. The matter fields in this theory (known as ABJM) are valued in the bifundamental representation of $\fsu(n) \oplus \fsu(n)$ and for $n,k=2$ it coincides with the original Bagger-Lambert model, see also \cite{Lambert:2010ji}. This paper opened a wide range of possibilities that generated many 3D SCCSM theories with different amounts of supersymmetry preserved by using a variety of techniques. 

\section{Ternary algebras and M2-branes}

When the first Bagger-Lambert theory appeared, it became clear that a better understanding of the structure theory of 3-Lie algebras was needed in order to understand the potential different theories that could arise from this description. Moreover, the limitations of the Euclidean model led to the problem of classifying Lorentzian 3-Lie algebras.

It was also natural to ask why not go to higher indices. Once the requirement of positive-definiteness had been dropped, why not explore all other options? A better understanding of metric 3-Lie algebras of any signature was needed. This would allow to explore the possibility of applying the same techniques used in the Lorentzian case to recover unitarity. 

On the other hand, the fact that the original BL theory was recast in terms of a standard Lie algebra, $\fsu(2) \oplus \fsu(2)$\cite{MukhiBL,BandresCS,VanRaamsdonkBL}, together with the fact that the ABJM model was constructed also using a Lie algebra, raised concerns on whether 3-algebras were necessary at all in order to build these theories.

However, Bagger and Lambert were then able to re-write the ABJM Lagrangian in terms of a 3-algebra \cite{BL4}, only not of Lie type (that is with the 3-bracket not being totally antisymmetric). Also, new 3D SCCSM theories were constructed based on other ternary algebras by relaxing the condition of total antisymmetry of the 3-bracket \cite{CherSaem}. 

It seemed that at least some of these theories could be written both in terms of 3-algebras or in terms of Lie algebras. This raised several algebraic questions. If total antisymmetry of the bracket was not required, what were the minimal conditions for a 3-algebra to give rise to a 3D SCCSM? Was there a one-to-one correspondence between this family of 3-algebras and standard Lie algebras? Why was it that only very particular Lie algebras could be used for highly supersymmetric theories? These are the questions we have addressed with our research in the past two years.

\section{Ternary algebras in mathematics}

Ternary algebras (or 3-algebras) were known to mathematicians well before the times of string theory. By ternary algebra we mean a vector space together with an map $V \times V \times V \to V$. Different classes of 3-algebras are defined by imposing conditions that this map must satisfy.

The earliest examples of  3-algebras in the mathematical literature are \emph{Lie triple systems}, dating back to the work of Élie Cartan (see for example \cite{cartan1952oeuvres}), which are to symmetric spaces what Lie algebras are to Lie groups.  They were studied algebraically by Jacobson in \cite{JacobsonLTS}, who also developed the structure theory for the closely related \emph{Jordan triple systems} in \cite{JacobsonJordan}. The structure theory of Lie triple systems was studied by Lister \cite{Lister}.  The modern definition stems from the work of Yamaguchi (also romanized as Yamaguti) in \cite{LTSreducedaxioms}, who has a vast body of work encompassing many aspects of triple systems. 

In the 80s, Filippov introduced one more family of ternary algebras: \textit{3-Lie algebras}. He defined them  as a straight-forward generalisation of Lie algebras: a vector space together with a totally antisymmetric ternary bracket satisfying an identity that generalised the Jacobi identity. Actually, he did more than that and defined \textit{n-Lie algebras} \cite{Filippov} (where the antisymmetric bracket has now n entries) and studied their structure theory extensively. In subsequent years Kasymov \cite{kasymov1987theory} and Ling built on his work. In particular, Ling classified simple n-Lie algebras in his PhD thesis \cite{LingSimple}. A variety of other triple systems have been considered: anti-Lie, anti-Jordan, Kantor, etc. some of which play a role in theoretical physics. 

Of special relevance to this thesis is the work of Faulkner \cite{faulkner1971construction,FaulknerIdeals} on the relation of some of these ternary algebras to Lie algebras. 

Despite the long lived interest in 3-algebras amongst mathematicians, there is no literature on the metric case, which is of special interest to physicists. This is perhaps not surprising given that even metric Lie algebras have not attracted much attention in mathematics. Part of the work in this thesis is aimed at covering this gap by specialising some known results on 3-algebras to the metric case. 

The interplay between physics and mathematics in this area has been constant throughout the years. Already the early mathematical literature on the topic refers to the work of Petiau \cite{Petiau}, Duffin \cite{0020.09006} and Kemmer \cite{MR0000571,MR0009939} on covariant field equations, which features what Jacobson \cite{JacobsonLTS} would later call \emph{meson algebras}, which are algebras defined on an inner product space satisfying a cubic relation $ X Y X = \left<X,Y\right> X$. 

From that time until their present resurgence in M-theory, ternary (and more generally $n$-ary) algebras have appeared occasionally in the mathematical physics landscape.  One of the best known examples is the Nambu bracket: a ternary generalisation of the Poisson bracket introduced by Nambu \cite{Nambu:1973qe} in an attempt to generalise Hamiltonian mechanics using the Liouville theorem as a guiding principle.  Nambu was apparently motivated by the subtleties of the quark model at the time, and the notoriously difficult quantisation of the ternary Nambu bracket and its generalisations of order $n$ has witnessed a considerable activity since then. 

Ternary algebras have also proved useful in finding solutions of the Yang-Baxter equation \cite{MR1224212,MR1224213}. More recently, $n$-algebras have appeared, although tangentially, in connection with the classification \cite{FOPMax,FOPPluecker} of maximally supersymmetric backgrounds of ten-dimensional IIB supergravity. Indeed, maximal supersymmetry implies the flatness of a connection on the spinor bundle of a ten-dimensional Lorentzian spin manifold, and this in turn makes the tangent bundle into a bundle of metric $4$-Lie algebras. In particular the concept of a \textit{metric} n-Lie algebra was introduced in \cite{FOPPluecker}.

\section{Research questions}

This is a thesis on algebra, motivated by and applied to M-theory. In particular, these are the problems we want to confront:

\begin{itemize}
	\item To classify Lorentzian 3-Lie algebras and study their relation to Lie algebras.
	\item To classify metric 3-Lie algebras of any signature and study \textsl{unitarisability} of the resulting maximally supersymmetric Bagger-Lambert theories.
	\item To study the ternary algebras appearing in superconformal Chern-Simons-matter theories in three dimensions and their relation to Lie algebras.
	\item To study the mechanism for supersymmetry enhancement in such theories and how it is encoded in algebraic terms.
\end{itemize}

\section{Thesis roadmap}

In the next three chapters we discuss how we approached these problems mathematically and in the final chapter we integrate our findings and their consequences in the context of modelling M2-branes. This thesis is organised as follows. 

In chapter 2, we review the structure theory of metric Lie algebras and their unitary representations. We believe that it is useful to review metric Lie algebras here due to a lack of specific literature in the topic of metricity. Also, as it turns out, several of the results to be discussed in this thesis can be seen as generalisations of the theory of metric Lie algebras. With this in mind this chapter serves to set a uniform notation and pave the way to our results in following chapters.

In chapter 3, we study metric 3-Leibniz algebras and show, by specialising a construction originally due to Faulkner, that they are in one-to-one correspondence with pairs of real metric Lie algebras and unitary representations of them. We also show a third characterisation for six extreme cases of 3-Leibniz algebras as graded Lie (super)algebras. 

In chapter 4, we study metric 3-Lie algebras in detail. We prove a structural result and also classify those with a maximally isotropic centre, which is the requirement that ensures unitarity of the corresponding maximally supersymmetric conformal field theory. 

Finally, in chapter 5, we use the results found in chapters 3 and 4 to study in a systematic way classical superconformal Chern-Simons theories with matter in three dimensions. We provide a uniform formulation for these theories in $N=1$ superspace formalism and review the universal structure of the $N=1$ superpotentials that give rise to them. The conditions for supersymmetry enhancement are then expressed equivalently in the language of representation theory of Lie algebras or the language of 3-Leibniz algebras. We close the chapter with a section dedicated to theories with maximal supersymmetry. We discuss in it our findings on unitary theories constructed from non positive-definite 3-Lie algebras in an attempt to overcome the shortage of positive-definite ones. 

The contents of this thesis are based on the already published papers \cite{Lor3Lie,2p3Lie,Lie3Algs,2pBL,SCCS3Algs}. For a complete list of publications of the author see appendix \ref{publications}.

\chapter[Review of Lie algebras]{Review of Lie algebras and their representations}
\label{ch:liealgebras}

The core of this thesis are two extensions of the theory of Lie algebras that have found recent applications in the context of M-theory. On the one hand, we will present structural results on metric 3-Leibniz algebras, which can be seen as a generalisation of Lie algebras; on the other, we will study the relation of these 3-algebras with the representation theory of Lie algebras. 

In fact, what is being generalised here is the theory of \textbf{metric} Lie algebras, that is those admitting a (not necessarily positive-definite) non-degenerate and ad-invariant inner product. The lack of specific literature on this topic is already a good reason to review metric Lie algebras here, but we also hope that this summary will set a uniform notation and provide a solid starting point for rest of this document.

This chapter does not intend to be an exhaustive review, instead we just go through the facts that we will need in what follows. For more comprehensive reviews see for example \cite{Humphreys, IntroLA, FultonHarris}. 

In section \ref{sec:metricLA} we introduce metric Lie algebras. This definition will allow a transparent generalization to 3-algebras in chapter \ref{sec:3algebras}. In section \ref{sec:structureLA} we present some known structural results and in section \ref{sec:metric-lie-algebras} a classification theorem for those Lie algebras with a maximally isotropic centre. These results will be generalized to 3-Lie algebras in chapter \ref{sec:3LieAlg}. Finally, in section \ref{sec:RepsLA} we review unitary representations of real Lie algebras. In chapter \ref{sec:3algebras} we will see how one can build 3-algebras from each kind of representation and vice-versa.

\section{Real Lie algebras}
\label{sec:rLA}
Like many other mathematical objects, a Lie algebra is defined as a set in which one can do some operations that satisfy certain properties. Essentially, a real Lie algebra is a real vector space with an extra bilinear operation that assigns to every pair of vectors a third one. This extra operation is usually called multiplication and, in the case of Lie algebras, Lie bracket. More precisely:

\begin{definition}
\label{def:LieAlg}
A \textbf{real Lie algebra} is a real vector space $\fg$ together with a binary operation $\left[-,-\right]: \fg \times \fg \rightarrow \fg$ called \textbf{Lie bracket} which satisfies: 

\begin{enumerate}
	\item \textbf{Bilinearity}: 
	\begin{equation}
	[a x + b y, z] = a [x, z] + b [y, z], \quad [z, a x + b y] = a[z, x] + b [z, y] ,
	\end{equation}
  \item anticommutativity, or \textbf{skew-symmetry}: 
		\begin{equation}
	[x,y]=-[y,x]
	\end{equation}
  \item and the \textbf{Jacobi identity}: 
	\begin{equation}
	\label{eq:Jacobi}
	[x,[y,z]] + [z,[x,y]] + [[x,z],y]  = 0 ;
	\end{equation}	
\end{enumerate}
for all $x, y, z \in \fg$ and $a,b \in \RR$.
\end{definition}

Some remarks about this definition:
\begin{itemize}
	\item Although strictly speaking a Lie algebra is a pair $\left(\fg, [-,-]\right)$, we will often denote Lie algebras, as is commonly done in the literature, by $\fg$, $\fh$, $\fk$, etc. where the existence of the Lie bracket is implicit.
	\item We will only consider finite-dimensional, real Lie algebras. For the remaining of this chapter, the underlying vector space is therefore always real and finite-dimensional.
	\item There is one way to define the Jacobi identity that will be useful for us later on. Consider the adjoint maps $ad_x(y) := [x,y]$. For every $x \in \fg$ this defines a linear map from the Lie algebra to itself, $ad_x(-): \fg \rightarrow \fg$. Requiring that this map acts as a \textbf{derivation} on the Lie bracket is equivalent to the Jacobi identity \eqref{eq:Jacobi}.
	
Acting as a derivation means satisfying the \textbf{Leibniz rule} or product rule that derivatives satisfy $D(fg) = (Df)g + f (Dg)$. In this case:

\begin{equation}
ad_x([y,z]) = [ad_x(y),z] + [y,ad_x(z)],
\label{eq:adjDer}
\end{equation}
which plugging back $ad_x(-) := [x,-]$ becomes the Jacobi identity \eqref{eq:Jacobi}. Hence, by definition, if $\fg$ is a Lie algebra, the adjoint maps are derivations of it. In general, not all derivations of a Lie algebra are of the form $[x,-]$ for some $x \in \fg$. Those that are of this form are called \textbf{inner}.
	\item Given a basis $\left\{e_A\right\}$ for the vector space $\fg$, one can fully define the Lie algebra by stating how the bracket acts on elements of the basis, which can also be done by specifying the \textbf{structure constants} $f_{AB}{}^{C}$:

\begin{equation}
\label{def:LieStrConst}
    [e_A,e_B] = f_{AB}{}^{C}e_C.
\end{equation}
\end{itemize}
\subsection{Metric Lie algebras}
\label{sec:metricLA}
In physics, one is usually interested in constructing scalars out of vectors in order to write down Lagrangians. For this reason, one typically requires Lie algebras to be \textbf{metric}, this assigns a scalar to every pair of vectors in a way that is compatible with the Lie bracket.

\begin{definition}
\label{def:MetLieAlg}
A real Lie algebra $(\fg, [-,-])$ is a said to be \textbf{metric} if it is equipped with a non-degenerate, symmetric bilinear form $( -,-)$ which is \textbf{ad-invariant}, that is:

  \begin{align}
		\left( ad_x(y), z \right) + \left( y,ad_x(z) \right) & = 0 
\intertext{or, equivalently}
		\left( [x,y], z \right) + \left( y,[x,z] \right) &= 0,
	\end{align}
for all $x, y, z \in \fg$.	
\end{definition}

The components of this inner product $( -,-)$ in terms of the basis $\left\{e_A\right\}$ are

\begin{equation}
\label{def:MetComp}
    \left( e_A,e_B \right) = b_{AB}.
\end{equation}
  
One can then use $b_{AB}$ to raise and lower indices and define, for example, the \textbf{canonical 3-form}:
  
\begin{equation}
\label{def:3form}
    f_{ABC} =  f_{AB}{}^{D}b_{DC} = \left( [e_A,e_B],e_C \right) .
\end{equation}

Real symmetric inner products have a notion of \textbf{signature} $(n,p)$ which describes the number of negative $n$ and positive $p$ eigenvalues. Signature $(0,p)$ is also known as \textbf{Euclidean} or \textbf{positive-definite} and $(1,p)$ as \textbf{Lorentzian}. The \textbf{index} of the inner product is the number of negative eigenvalues $n$. By extension, a metric Lie algebra $\fg$ is said to have index $n$ if the ad-invariant inner product has index $n$, which is the same as saying that the maximally negative-definite subspace of $\fg$ is $n$-dimensional.

\subsection{Structure of metric Lie algebras}
\label{sec:structureLA}

In this section we investigate the building blocks of metric Lie algebras and how they are constructed. Although, to the best of our knowledge, there is no classification theorem for metric Lie algebras, a number of structural results are known and classifications do exist for index less than 3. First, we need to review some more definitions and classic results. 

\subsubsection{Some definitions and results}

A metric Lie algebra is said to be \textbf{indecomposable} if it cannot be written as an orthogonal direct sum of metric Lie algebras $(\fg_1\oplus \fg_2, [-,-]_1 \oplus [-,-]_2, b_1\oplus b_2)$ with $\dim \fg_i > 0$. In order to classify metric Lie algebras, it is enough to classify indecomposable ones. 

One dimensional Lie algebras are clearly indecomposable. We will see now that another class of indecomposable Lie algebras are simple ones. However, the converse statement is not true in general. That is, indecomposable algebras need not be simple. To define simple Lie algebras, we need to define first a Lie subalgebra and an ideal.

\begin{definition}
\label{def:Lsubalgebra}
A subspace, $\fh$, of elements of a Lie algebra, $\fg$, is called a \textbf{Lie sublagebra} if it is closed under the Lie bracket:
\begin{equation}
[h_1, h_2] \in \fh,
\label{eq:subalg}
\end{equation}
for all $h_1, h_2 \in \fh \subset \fg$. Schematically, this condition is written:
\begin{equation}
[\fh,\fh] \subset \fh ,
\end{equation}
where we use $[\fw_1,\fw_2]$ as short notation for the span of $\left\{[w_1,w_2]  | w_1 \in \fw_1, w_2 \in \fw_2 \right\}$ for any two sets $\fw_1,\fw_2$. It is then guaranteed that $\fh$ satisfies on its own all the axioms of a Lie algebra. 
\end{definition}
An example of Lie subalgebra, given a subspace $\fs \subset \fg$, is the \textbf{centraliser} $Z(\fs)$ which is defined as those elements of $\fg$ that commute with all elements in $\fs$: $Z(\fs) = \left\{z \in \fg \middle | [z,s] = 0, \forall s \in \fs \right\}$. In other words, $[Z(\fs),\fs] = 0$.
 
\begin{definition}
\label{def:LAIdeal}
A subset, $I$, of elements of a Lie algebra $\fg$ is called an \textbf{ideal} and denoted $I \lhd \fg$, if it is \textit{absorbent} with respect to the Lie bracket, that is:
\begin{equation}
[I,\fg] \subset I.
\label{eq:Ideal}
\end{equation}
\end{definition}

Notice that in particular this implies $[I,I] \subset I$, therefore $I$ is a Lie subalgebra of $\fg$. The total space $\fg$ and the element 0 are always ideals of $\fg$. All other ideals of $\fg$ are called \textbf{proper}.

An important ideal is the \textbf{derived ideal} $\fg' $, which is the image of of the Lie bracket. In other words, it is the subalgebra generated by the elements of $\fg$ that appear on the right hand side of each Lie bracket or $\fg' := [\fg,\fg] \subset \fg$.

Another ideal of especial relevance is the \textbf{centre} $Z$ which is defined as the elements of $\fg$ that commute with all other elements: $Z := \left\{z \in \fg \middle | [z,g] = 0, \forall g \in \fg \right\}$. In other words, $[Z,\fg] = 0$. Notice that this is also the centralizer of all $\fg$, $Z = Z(\fg)$.

An ideal is said to be \textbf{minimal} if it does not contain any other proper ideals and \textbf{maximal} if it is not contained in any ideal other than the total space $\fg$.

The reason why ideals are so important in structure theory of Lie algebras is because they have very special properties that are summarized in the isomorphism theorems. The most relevant for us being that when quotienting a Lie algebra by one of its ideals the resulting space also has a Lie algebra structure. It is then useful to distinguish those Lie algebras that posses proper ideals from those that don't. 

\begin{definition}
\label{def:LASimple}
A Lie algebra $\fg$ is said to be \textbf{simple} if it is not one-dimensional and it does not contain any proper ideals. 
\end{definition}

\begin{definition}
\label{def:LASSimple}
A Lie algebra $\fg$ is said to be \textbf{semisimple} if it is isomorphic to a direct sum of simple Lie algebras. 
\end{definition}

Semisimple Lie algebras have two important properties:

\begin{itemize}
	\item They coincide with their first derived ideal. That is, if  $\fg' = [\fg,\fg] = \fg$. 
	\item All derivations are inner. That is, if $D: \fg \to \fg$ is a derivation, then there exists a unique element $x \in \fg$ such that $D = [x,-]$.
\end{itemize}

It can be shown that any simple Lie algebra is indecomposable, but the converse is not true in general. Cartan classified \textsl{complex} simple Lie algebras in his PhD thesis in the 1890s. There are three infinite series ($\fsl(n), \fso(n), \fsp(n)$) called \textit{classical} simple Lie algebras, and five exceptional ones ($E_6, E_7, E_8, F_4, G_2$). Classifying \textsl{real} Lie algebras is a harder job and it took Cartan a few more years to achieve this result. The main idea is that any real simple Lie algebra $\fg_0$  is the \textbf{real form} of a complex Lie algebra $\fg$. That is, complexifying $\fg_0$ one obtains a complex Lie algebra $\fg$:

\begin{equation}
    \fg \simeq \CC \otimes_{\RR} \fg_0.
\end{equation}

Moreover, it can be proved that the resulting algebra is semisimple, hence a direct sum of complex simple Lie algebras that had previously been classified. The trouble is that a given complex Lie algebra has often more than one real form and finding them all can be tricky. The real simple Lie algebras which are real forms of the classical ones are:  $\fsl_n(\RR), \fsl_n(\CC), \fsl_n(\HH), \fso_{p,q}(\RR), \fsp_{2n}(\RR), \fsp_{2n}(\CC), \fsu_{p,q}, \fu_{p,q}(\HH), \fu_n^*(\HH)$. Then there are 17 ones which are real forms of the five exceptional complex Lie algebras. For more details see, for instance, \cite{FultonHarris} pg. 433.

Lets focus now on metric Lie algebras. In that case, for any subspace (which needs not be a subalgebra) $\fs \subset \fg$ one can define its \textbf{orthogonal complement} $\fs^\perp$:

\begin{equation}
\fs^\perp = \left\{x \in \fg \middle | (x,y) = 0, \forall y \in \fs\right\}.
\label{eq:orthcompl}
\end{equation}

We say that $\fs$ is:

\begin{itemize}
	\item \textbf{nondegenerate} if $\fs \cap \fs^\perp = 0$, hence $\fg = \fs \oplus
\fs^\perp$,
	\item \textbf{isotropic} if $\fs \subset \fs^\perp$,
	\item \textbf{coisotropic} if $\fs \supset \fs^\perp$.
\end{itemize}

\subsubsection{Classification of low index metric Lie algebras}

In positive-definite signature all subspaces are nondegenerate, hence there are no isotropic or coisotropic subspaces. It follows that the notion of indecomposability coincides with that of simplicity because, if a positive-definite Lie algebra is not simple, there exists at least one proper ideal $I\lhd \fg$ that is nondegenerate and $\fg = I \oplus I^\perp$, hence it is decomposable. We conclude that any Euclidean Lie algebra is an orthogonal direct sum of simple and abelian (or equivalently one-dimensional) Lie algebras. Such Lie algebras are called \textbf{reductive}.

However, if the metric is not positive-definite, there could be a degenerate ideal $I$ in $\fg$ so that $\fg \neq I \oplus I^\perp$. and the Lie algebra could still be indecomposable. Hence, in signature $> 0$, there are Lie algebras which are not simple but still are indecomposable. The main idea behind structural results and classifications of low index metric Lie algebras is to exploit the properties of such ideal. 

In Lorentzian signature there is a classification due to Medina \cite{MedinaLorentzian}. He proved that indecomposable Lorentzian Lie algebras are constructed out of simple and one-dimensional Lie algebras by iterating two constructions: orthogonal direct sum and \emph{double extension}, to be defined later on.

This was later extended by Medina and Revoy \cite{MedinaRevoy} (see also work of Stanciu and Figueroa-O'Farrill \cite{FSalgebra}), who showed that indecomposable metric Lie algebras of \textit{any} signature are constructed by again iterating the operations of direct sum and the (generalised) double extension, using again as ingredients the simple and one-dimensional Lie algebras.  This result was used in \cite{MedinaLorentzian} to construct all possible indecomposable metric Lie algebras of index 2.  Contrary to the Lorentzian case, there is a certain ambiguity in this construction, which prompted Kath and Olbrich \cite{KathOlbrich2p} to approach the classification problem for metric Lie algebras from a cohomological perspective.  In particular they classified indecomposable metric Lie algebras with index 2, a result which had been announced in \cite{baumkath2p}.  

For more indefinite signatures, the classification problem is still largely open. However, there is a particular class of indefinite signature Lie algebras for which there exists a classification theorem due to Kath and Olbrich \cite{KathOlbrich2p}: those with a maximally isotropic centre. We dedicate next section to this result because, although we were not aware of its existence at the time, the approach is similar to the one we use to prove an analogous theorem for 3-Lie algebras in section \ref{sec:max-iso-cent}. The latter case is more involved and new ideas are needed, but they both share the same spirit. The study of this particular class of 3-Lie algebras will be later motivated in order to construct maximally supersymmetric unitary Bagger-Lambert theories.

\subsection{Metric Lie algebras with maximally isotropic centre.}
\label{sec:metric-lie-algebras}

Although, as we said, there is no classification theorem for metric Lie algebras in general, there is a subset of them that can be classified: those admitting a maximally isotropic centre. In this section we review this result, originally due to Kath and Olbrich \cite{KathOlbrich2p}. 

Lets first define the class of metric Lie algebras that we want to classify. Recall that the centre $Z$ of a Lie algebra $\fg$ is the set of all elements in $\fg$ that commute with all other elements in $\fg$. Since the Lie algebra is metric, this centre is \textit{isotropic} if $Z \subset Z^\perp$ as a subspace implying $\left\langle Z,Z\right\rangle=0$. By maximally isotropic we mean that it has the maximal dimension allowed for an isotropic subspace of $\fg$. This size is limited by the index of the metric, $r$, hence a maximally isotropic subspace must have dimension $r$. In rigour, a maximally isotropic ideal is one of dimension equal to the minimum of $(p,q)$, the signature of the inner product. In summary, a \textbf{Lie algebra with a maximally isotropic centre} is an algebra such that its centre $Z$ ($[Z,\fg] = 0$) has dimension equal to its index $r$ and satisfies $\left\langle Z,Z\right\rangle=0$.

The idea is to first find a useful basis of the underlying vector space by expanding a basis for the centre and imposing the fact that the Lie algebra is metric. Then, write the most general form of the Lie bracket in that basis satisfying the condition of admitting a maximally isotropic centre. Imposing ad-invariance of the metric and the Jacobi identity provides a series of equations that the structure constants must satisfy. Finally, one solves those equations and ensures that the resulting Lie algebra is indecomposable. 

\subsubsection{Preliminary form of the Lie algebra}
\label{sec:preliminaries}

Let $\fg$ be a finite-dimensional indecomposable metric Lie algebra of index $r>0$ admitting a maximally isotropic centre.   Let $v_i$, $i=1,\dots,r$, denote a basis for the centre $Z$, which as we said must have dimension $r$. 

By definition of the orthogonal complement, we have that $\left<v_i,x\right>=0, \forall x \in Z^\perp$. Since $Z$ is isotropic, all elements in $Z$ are also in its orthogonal complement, so we have that $\left<v_i,v_j\right>=0, \forall i,j=1,\dots,r$. In particular all the $v_i$ vectors are null (that is they have zero norm).

Since the inner product on $\fg$ is nondegenerate, there exist $u_i$, $i=1,\dots,r$, which obey $\left<u_i,v_j\right> = \delta_{ij}, \forall i,j=1,\dots,r$. Notice that the space spanned by the $v_i$ and that spanned by the $u_i$ are by definition dual to each other. 

 It is always possible to choose the $u_i$ such that $\left<u_i,u_j\right>=0$ and so they also span a maximally isotropic subspace. Indeed, if the $u_i$ do not, then redefine them
by $u_i \mapsto u_i - \half \sum_{j=1}^r \left<u_i,u_j\right> v_j$ so that they do.  The perpendicular complement to the $2r$-dimensional
subspace spanned by the $u_i$ and the $v_j$ is then positive-definite.  In summary, $\fg$ admits the following vector
space decomposition

\begin{equation}
  \label{eq:decomp-2Lie}
  \fg = \bigoplus_{i=1}^r \left(\RR u_i \oplus \RR v_i\right) \oplus
  \fr,
\end{equation}
where $\fr$ is the positive-definite subspace of $\fg$ perpendicular
to all the $u_i$ and $v_j$.

By definition of the centre spanned by the $v$'s, any Lie bracket containing them vanishes. All other brackets could be non-vanishing. Since the image of the bracket has to be in $\fg$, in principle, each bracket could have a component in $\bigoplus_{i=1}^r \RR u_i$, another one in $\bigoplus_{i=1}^r \RR v_i$ and another one in $\fr$. However, the facts that the $v$'s are in the centre and that the metric is ad-invariant imply that the brackets cannot have a $u$ component on the right hand side. Indeed, if we had that $[x,y] = a u_i + ...$ for some $x,y \in \fg$ and for some $u_i$, we would have:

\begin{equation}
\left\langle [x,y], v_i \right\rangle = a \left\langle u_i, v_i \right\rangle  = a.
\end{equation}

Recalling that the $v$'s are orthogonal to $\fr$ and to the $v$'s themselves, but not to the $u$'s. Now, ad-invariance tells us that:

\begin{equation}
a=\left\langle [x,y], v_i \right\rangle = a \left\langle y, [v_i,x] \right\rangle  = 0,
\end{equation}
hence, we conclude that $a=0$ and $[x,y]$ cannot have a component along the $u_i$ directions.

If $\left\{e_a\right\}$ is a basis for $\fr$, the most general brackets we can have are:

\begin{equation}
  \label{eq:pre-2-brackets}
  \begin{aligned}[m]
    [u_i,u_j] &= f_{ij}{}^a e_a + \sum_{k=1}^r f_{ij}{}^k v_k\\
    [u_i,e_a] &=  f_{ia}{}^b e_b + \sum_{k=1}^r f_{ia}{}^j v_j\\
    [e_a,e_b] &= f_{ab}{}^c e_c + \sum_{k=1}^r f_{ab}{}^i v_i,
  \end{aligned}
\end{equation}
where brackets not listed vanish and indices $a, b, c$ indicate a component in $\fr$, downstairs $i, j, k$ refer to $u_i, u_j, u_k$ components respectively whereas upstairs ones refer to $v_i, v_j, v_k$ components. This can be done because, as we explained, the $v$'s don't appear inside the brackets and the $u$'s don't appear on the right-hand side.

Ad-invariance of the inner product relates some of the structure constants, for instance:

\begin{equation}
\left\langle  [u_i,e_a] , u_j\right\rangle = f_{iaj} = - \left\langle  e_a , [u_i,u_j]\right\rangle = - \left\langle  e_a , f_{ij}{}^b e_b\right\rangle
\end{equation}
hence, $f_{iaj} = - f_{ij}{}^b \left\langle  e_a ,  e_b\right\rangle = - f_{ija}$. In fact, ad-invariance implies that, if $\left\{e_A\right\}$ is a basis for $\fg$, $f_{ABC}$ is totally antisymmetric. After imposing this property, the most general Lie brackets on $\fg$ are of the form:

\begin{equation}
  \label{eq:2-brackets}
  \begin{aligned}[m]
    [u_i,u_j] &= K_{ij} + \sum_{k=1}^r L_{ijk} v_k\\
    [u_i,x] &=  J_ix - \sum_{j=1}^r \left<K_{ij},x\right> v_j\\
    [x,y] &= [x,y]_{\fr} - \sum_{i=1}^r \left<x, J_i y\right> v_i,
  \end{aligned}
\end{equation}
where we have changed slightly the notation by using $x,y \in \fr$ rather than the basis on $\fr$ and we have given special names to the non-vanishing structure constants. In particular, we have replaced $f_{ij}{}^a$ with $K_{ij} = - K_{ji} \in \fr$ and $f_{ij}{}^k$ with $L_{ijk} \in \RR$ which is totally skewsymmetric in the indices. $f_{ia}{}^b$ has been replaced with a family of  endomorphisms $J_i : \fr \to \fr$ which, by ad-invariance of the metric, preserve the metric itself. That is, they satisfy: $\left<J_i  x,  y\right> + \left<x, J_i y\right> =0$ which is summarised by saying that $J_i \in \fso(\fr)$ . Finally, $f_{ab}{}^c$ has been replaced by a bracket $[-,-]_\fr:\fr \times \fr \to \fr$ which is bilinear and skewsymmetric, the reason being that the Jacobi identity \eqref{eq:Jacobi} in $\fg$ will imply that $f_{ab}{}^c$ (only in $\fr$) also satisfy the Jacobi identity.

  Finally, metricity also implies that
\begin{equation}
  \label{eq:metric-Lie}
   \left<[x,y]_\fr, z \right> = \left<x, [y,z]_\fr \right>,
\end{equation}
for all $x,y,z \in \fr$. 

Computing the Jacobi identity \eqref{eq:Jacobi} for $\fg$ is
equivalent to the following identities on $[-,-]_{\fr}$, $J_i$ and
$K_{ij}$, whereas $L_{ijk}$ is unconstrained:
\begin{subequations}
  \begin{align}
    [x,[y,z]_\fr]_\fr - [[x,y]_\fr,z]_\fr - [y,[x,z]_\fr]_\fr &= 0    \label{eq:h-is-Lie}\\
    J_i [x,y]_{\fr} - [J_ix,y]_{\fr} -  [x, J_i y]_{\fr} &=  0 \label{eq:J-is-deriv}\\
    J_i J_j x - J_j J_i x - [K_{ij},x]_\fr &= 0 \label{eq:JJK}\\
    J_i K_{jk} + J_j K_{ki} +     J_k K_{ij}  &= 0 \label{eq:JK}\\
    \left<K_{\ell i},K_{jk}\right> + \left<K_{\ell j},K_{ki}\right> +
    \left<K_{\ell k},K_{ij}\right> &=0, \label{eq:KK}
  \end{align}
\end{subequations}
for all $x,y,z \in \fr$. Notice that in particular $[-,-]_\fr$ is skewsymmetric, satisfies the Jacobi identity \eqref{eq:h-is-Lie} and the ad-invariance condition \eqref{eq:metric-Lie}. Thus, it defines a metric Lie algebra on $\fr$.

In what follows, we will show how solving these equations imposes further restrictions on each of the structure constants.

\subsubsection{$\fr$ is abelian}
\label{sec:fr-abelian}

We prove next that $\fr$ is not only a metric Lie algebra, but it is an abelian one. Equivalently, $f_{ab}{}^c = 0$ or $[x,y]_{\fr} \equiv 0$.

Being positive-definite, it is reductive, hence an orthogonal direct sum $\fr = \fs \oplus \fa$, where $\fs$ is semisimple and $\fa$
is abelian.  We will show by contradiction that $\fg$ indecomposable, forces $\fs=0$. 

If $\fs$ did not vanish, then we will show that we could write $\fg = \fs \oplus \fs^\perp$
as a metric Lie algebra\footnote{Notice that $\fs^\perp= \bigoplus_{i=1}^r \left(\RR u_i \oplus \RR v_i\right) \oplus \fa$ which implies that $\fs \cap \fs^\perp =0$. In order to show that $\fg = \fs \oplus \fs^\perp$ as a \textsl{Lie algebra} it only remains to show that there are no mixed terms of the Lie brackets,  $[\fs , \fs^\perp]=0$.} hence $\fg$ would be decomposable.

Equation \eqref{eq:J-is-deriv} says that $J_i$ is a derivation of $\fr$, which we know to be skewsymmetric. The Lie algebra of
skewsymmetric derivations of $\fr$ is given by the sum of that of $\fs$ plus that of $\fa$. All derivations of a semisimple Lie algebra are inner, or equivalently of the form of an adjoint map, hence the inner derivations of $\fs$ form a Lie algebra called $\ad \fs$. Those of $\fa$ are generically just skew-symmetric endomorphisms. Hence the  Lie algebra of
skewsymmetric derivations of $\fr$ is $\ad \fs \oplus \fso(\fa)$ and we may write $J_i = \ad z_i + J^\fa_i$, for some unique $z_i \in \fs$ and $J^\fa_i \in \fso(\fa)$. 

Decompose $K_{ij} = K^{\fs}_{ij} + K^{\fa}_{ij}$, with $K^{\fs}_{ij} \in \fs$ and $K^{\fa}_{ij} \in \fa$.  Then equation \eqref{eq:JJK} becomes the following two conditions
\begin{align}
  [z_i,z_j]_\fr &= K^{\fs}_{ij}   \label{eq:zzKs}\\
  \intertext{and}
  [J^\fa_i, J^\fa_j] &= 0.  \label{eq:Js-commute}
\end{align}

One can now check that the $\fs$-component of the Jacobi identity for
$\fg$ is automatically satisfied, whereas the $\fa$-component gives
rise to the two equations
\begin{align}
  J^\fa_i K^{\fa}_{jk} +   J^\fa_j K^{\fa}_{ki} +   J^\fa_k K^{\fa}_{ij} &= 0 \label{eq:JKa}\\
  \intertext{and}
    \left<K^{\fa}_{\ell i},K^{\fa}_{jk}\right> + \left<K^{\fa}_{\ell j},K^{\fa}_{ki}\right> +
    \left<K^{\fa}_{\ell k},K^{\fa}_{ij}\right> &=0. \label{eq:KKa}
\end{align}

We can now show that $\fg \cong \fs \oplus \fs^\perp$, which violates the indecomposability of $\fg$ unless $\fs = 0$.  Consider the isometry $\varphi$ of the vector space $\fg$ defined by
\begin{equation}
  \label{eq:isometry-2}
  \begin{aligned}[m]
    \varphi(u_i) &= u_i - z_i - \half \sum_{j=1}^r \left<z_i,z_j\right> v_j\\
    \varphi(v_i) &= v_i\\
    \varphi(x) &= x + \sum_{i=1}^r \left<z_i, x\right> v_i,
  \end{aligned}
\end{equation}
for all $x \in \fr$.  Notice that if $x \in \fa$, then $\varphi(x) =
x$.  It is a simple calculation to see that for all $x,y\in \fs$,
\begin{equation}
  [\varphi(u_i), \varphi(x)] = 0 \qquad\text{and}\qquad [\varphi(x),
  \varphi(y)] = \varphi ([x,y]_\fr).
\end{equation}

Hence, the image of $\fs$ under $\varphi$ is a Lie subalgebra
of $\fg$ isomorphic to $\fs$ and commuting with its perpendicular
complement in $\fg$.  In other words, as a metric Lie algebra $\fg
\cong \fs \oplus \fs^\perp$, violating the decomposability of $\fg$
unless $\fs = 0$.

It turns out then that our initial Lie algebra $\fg$ has become an abelian, positive-definite one, hence just a Euclidean vector space that we rename $E$.

In summary, we have proved the following

\begin{lemma}\label{lem:h-is-abelian}
  Let $\fg$ be a finite-dimensional indecomposable metric Lie algebra
  with index $r>0$ and admitting a maximally isotropic centre.  Then
  as a vector space
  \begin{equation}
    \fg = \bigoplus_{i=1}^r \left(\RR u_i \oplus \RR v_i\right) \oplus E,
  \end{equation}
  where $E$ is a euclidean space, $u_i, v_i \perp E$ and
  $\left<u_i,v_j\right>= \delta_{ij}$, $\left<u_i,u_j\right> =
  \left<v_i,v_j\right> = 0$.  Moreover the Lie bracket is given by
  \begin{equation}
    \label{eq:Lie-brackets}
    \begin{aligned}[m]
      [u_i,u_j] &= K_{ij} + \sum_{k=1}^r L_{ijk} v_k\\
      [u_i,x] &=  J_ix - \sum_{j=1}^r \left<K_{ij},x\right> v_j\\
      [x,y] &= - \sum_{i=1}^r \left<x, J_i y\right> v_i,
    \end{aligned}
  \end{equation}
  where $K_{ij} = - K_{ji} \in E$, $L_{ijk} \in \RR$ is totally
  skewsymmetric in its indices, $J_i \in \fso(E)$ and in addition obey
  the following conditions:
  \begin{subequations}
    \begin{align}
      J_i J_j - J_j J_i &= 0 \label{eq:Jscommute}\\
      J_i K_{jk} + J_j K_{ki} +     J_k K_{ij}  &= 0 \label{eq:JK2}\\
      \left<K_{\ell i},K_{jk}\right> + \left<K_{\ell j},K_{ki}\right> +
      \left<K_{\ell k},K_{ij}\right> &=0. \label{eq:KK2}
    \end{align}
  \end{subequations}
\end{lemma}
The analysis of the above equations will take the rest of this
section.


\subsubsection{Solving for the $J_i$}
\label{sec:solving-js}

Equation \eqref{eq:Jscommute} says that the $J_i \in \fso(E)$ are
mutually commuting, hence they span an abelian subalgebra $\fh
\subset \fso(E)$.  Since $E$ is positive-definite, $E$ decomposes as the
following orthogonal direct sum as a representation of $\fh$:
\begin{equation}
  \label{eq:E-split}
  E = \bigoplus_{\pi=1}^s E_\pi \oplus E_0,
\end{equation}
where
\begin{equation}
  \label{eq:e0}
  E_0 = \left\{ x\in E \middle | J_i x = 0\forall i\right\}
\end{equation}
and each $E_\pi$ is a two-dimensional real irreducible representation of
$\fh$ with certain nonzero weight.  Let $(H_\pi)$ denote the basis for
$\fh$ where
\begin{equation}
  \label{eq:Hs}
  H_\pi H_\varrho =
  \begin{cases}
    0 & \text{if $\pi\neq \varrho$,}\\
    - \Pi_\pi & \text{if $\pi = \varrho$,}
  \end{cases}
\end{equation}
where $\Pi_\pi \in \End(E)$ is the orthogonal projector onto $E_\pi$.
Relative to this basis we can then write $J_i =\sum_\pi J_i^\pi H_\pi$, for
some real numbers $J_i^\pi$.

\subsubsection{Solving for the $K_{ij}$}
\label{sec:solving-ks}

Since $K_{ij} \in E$, we may decompose according to \eqref{eq:E-split} as
\begin{equation}
  K_{ij} = \sum_{\pi=1}^s K^\pi_{ij} + K^0_{ij}.
\end{equation}
We may identify each $E_\pi$ with a complex line where $H_\pi$ acts by
multiplication by $i$.  This turns the complex number $K^\pi_{ij}$ into
one component of a complex bivector $K^\pi \in \Lambda^2\CC^r$.
Equation \eqref{eq:JK2} splits into one equation for each $K^\pi$ and
that equation says that
\begin{equation}
  J^\pi_i K^\pi_{jk} + J^\pi_j K^\pi_{ki} + J^\pi_k K^\pi_{ij} = 0,
\end{equation}
or equivalently that $J^\pi \wedge K^\pi = 0$, which has as unique
solution $K^\pi = J^\pi \wedge t^\pi$, for some $t^\pi \in \RR^r$.  In other
words,
\begin{equation}
  K^\pi_{ij} = J^\pi_i t^\pi_j - J^\pi_j t^\pi_i.
\end{equation}

Now consider the following vector space isometry $\varphi: \fg \to
\fg$, defined by
\begin{equation}
  \label{eq:isometry-3}
  \begin{aligned}[m]
    \varphi(u_i) &= u_i - t_i - \half \sum_{j=1}^r \left<t_i,t_j\right> v_j\\
    \varphi(v_i) &= v_i\\
    \varphi(x) &= x + \sum_{i=1}^r \left<t_i, x\right> v_i,
  \end{aligned}
\end{equation}
for all $x \in E$, where $t_i \in E$ and hence $t_i = \sum_{\pi=1}^s
t_i^\pi + t_i^0$.  Under this isometry the form of the Lie algebra
remains invariant, but $K_{ij}$ changes as
\begin{equation}
  K_{ij} \mapsto K_{ij} - J_i t_j + J_j t_i
\end{equation}
and $L_{ijk}$ changes in a manner which need not concern us here.
Therefore we see that $K^\pi_{ij}$ has been put to zero via this
transformation, whereas $K^0_{ij}$ remains unchanged.  In other words,
we can assume without loss of generality that $K_{ij} \in E_0$, so
that $J_i K_{kl} = 0$, while still being subject to the quadratic
equation \eqref{eq:KK2}.

In summary, we have proved the following theorem, originally due to
Kath and Olbrich \cite{KathOlbrich2p}:

\begin{theorem}\label{thm:main-Lie}
  Let $\fg$ be a finite-dimensional indecomposable metric Lie algebra
  of index $r>0$ admitting a maximally isotropic centre.  Then as a
  vector space
  \begin{equation}
    \fg = \bigoplus_{i=1}^r \left(\RR u_i \oplus \RR v_i\right) \oplus
    \bigoplus_{\pi=1}^s E_\pi \oplus E_0,
  \end{equation}
  where all direct sums but the one between $\RR u_i$ and $\RR v_i$
  are orthogonal and the inner product is as in lemma
  \ref{lem:h-is-abelian}.  Let $0 \neq J^\pi \in \RR^r$, $K_{ij} \in
  E_0$ and $L_{ijk} \in \RR$ and assume that the $K_{ij}$ obey the
  following quadratic relation
  \begin{equation}
    \label{eq:KK3}
    \left<K_{\ell i},K_{jk}\right> + \left<K_{\ell j},K_{ki}\right> +
    \left<K_{\ell k},K_{ij}\right> =0.
  \end{equation}
  Then the Lie bracket of $\fg$ is given by
  \begin{equation}
    \label{eq:Lie-brackets-final}
    \begin{aligned}[m]
      [u_i,u_j] &= K_{ij} + \sum_{k=1}^r L_{ijk} v_k\\
      [u_i,x] &=  J^\pi_iH_\pi x\\
      [u_i,z] &= - \sum_{j=1}^r \left<K_{ij},z\right> v_j\\
      [x,y] &= - \sum_{i=1}^r \left<x, J^\pi_i H_\pi y\right> v_i,
    \end{aligned}
  \end{equation}
  where $x,y \in E_\pi$ and $z \in E_0$.  Furthermore, indecomposability
  forces the $K_{ij}$ to span all of $E_0$, hence $\dim E_0 \leq \binom{r}{2}$.
\end{theorem}

It should be remarked that the $L_{ijk}$ are only defined up to the
following transformation
\begin{equation}
  L_{ijk} \mapsto L_{ijk} + \left<K_{ij},t_k\right> +
  \left<K_{ki},t_j\right> + \left<K_{jk},t_i\right>,
\end{equation}
for some $t_i \in E_0$.

It should also be remarked that the quadratic relation \eqref{eq:KK3}
is automatically satisfied for index $r\leq 3$, whereas for index
$r\geq 4$ it defines an algebraic variety.  In that sense, the
classification problem for indecomposable metric Lie algebras
admitting a maximally isotropic centre is not tame for index $r>3$.

\section{Representations of real Lie algebras}
\label{sec:RepsLA}

In this section we briefly review the theory of finite-dimensional representations of real Lie algebras. We are interested in a special kind of representations called unitary. These arise when the underlying vector space for the representation has an inner product that is compatible with the Lie algebra and are useful in physics because, again, this allows the construction of scalar Lagrangians out of fields living in a representation of a Lie algebra. 

We consider only those kinds of inner products which have a notion of signature since, in order to build Lagrangians for manifestly unitary theories, we want to look at positive-definite ones. There are three such inner products: real symmetric, complex hermitian and quaternionic hermitian. This prompts us to consider three kinds of representations according to the underlying field being real, complex or quaternionic. Notice that this is independent of the vector field underlying the Lie algebra itself. For example, a real Lie algebra can have real, complex and quaternionic representations.

We also define a map, that we call Faulkner map, which is associated to unitary representations which will play a central role in following chapters. 

We follow here a pattern that will be repeated in other chapters: first we define generic representations, then we focus on those which admit a metric with a signature and finally we discuss each metric case, corresponding to each underlying field, separately. Lets start by defining a representation.

\begin{definition}
\label{Def:LieAlgRep}
A \textbf{representation} of a Lie algebra $\fg$ is a linear map $\rho :\fg \to \fgl(V)$, where $V$ is a vector space, that preserves the Lie bracket:

\begin{equation}
\rho\left([x,y]\right) = \left[\rho(x), \rho(y)\right]
\label{eq:LieBraRep}
\end{equation}
for all $x,y \in \fg$.
\end{definition}

Such map is called a Lie algebra \textbf{homomorphism}. 

If the map is injective, the representation is said to be \textbf{faithful}, this means that every element in the Lie algebra is represented by a \textit{different} element in $\fgl(V)$ or, equivalently, that the map $\rho$ has trivial kernel.

By a slight abuse of notation, the vector space $V$ is often called the representation. In that case the map $\rho: \fg \rightarrow \fgl(V)$ is implicit and not specified. It is then said that the Lie algebra $\fg$ \textit{acts on} the vector space $V$ and it is written:

\begin{equation}
x \cdot u = \rho(x) u, \ \forall x \in \fg, \ \forall u \in V.
\label{eq:LieAlgAction}
\end{equation} 

\subsection{Real, complex and quaternionic representations}

We consider now real, complex and quaternionic representations of a \textit{real} Lie algebra $\fg$, which arise when $V$ is real/complex/quaternionic\footnote{There is a small caveat with quaternions: since they do not commute, they are not a field. Hence, by quaternionic vector space we actually mean a right or left vector space.}. We will denote real representations  $V \in \Dar(\fg,\RR)$,  complex representations $\VV \in \Dar(\fg,\CC)$ and quaternionic representations $\VV_H \in \Dar(\fg,\HH)$. 
We will show now that any complex or quaternionic representation can be seen as a special case of a real representation $V$. 

\subsubsection{Complex representations as special cases of real ones}

\begin{proposition}
Every complex representation $\VV$ can be seen as a special case of a real representation $V$ equipped with a complex structure $I$.
\end{proposition}

\begin{proof}

We will show first that every real representation equipped with a complex structure defines a complex representation. 

Let $V$ be a real representation equipped with a \textit{complex structure}, $I$, which is just an element of $\fgl(V)$ satisfying $I^2 = -1$, that commutes with the action of $\fg$ on $V$. Notice that this implies that $V$ has even real dimension.

To prove that this defines a complex representation of $\fg$ we need to prove that the pair $(V,I)$ has the structure of a complex vector space that we will call $\VV$ and provide a map $\rho_{\CC}: \fg \rightarrow \fgl(\VV)$ preserving the Lie bracket.

Since $V$ already has the structure of a real vector space, we know that the elements there can be added and that this addition satisfies all desired properties. All we need to do is to define how one can multiply these vectors with complex numbers and prove that this external operation is distributive with respect to addition (both in the vectors and in the complex numbers) and compatible with multiplication in the complex numbers.

We define the multiplication of a complex number $z = a +ib$ with a vector $v \in \VV$ using the complex structure $I$:

\begin{equation}
\label{eq:complexmult}
z v = (a + i b) v = a v + b I v,
\end{equation}
then it is not difficult to check that it satisfies the desired properties:

\begin{enumerate}
	\item $z (v + w) = zv + zw$
	\item $(z_1 + z_2)v = (a_1 + a_2) v + I (b_1 + b_2) v = (a_1 + Ib_1) v + (a_2 + Ib_2) v= z_1 v + z_2v.$
	\item $z_1 \left( z_2 v \right)  =  (a_1 + i b_1)  \left( (a_2 + Ib_2) v \right)  = \left( a_1 a_2 + I b_1 b_2 + I b_1 a_2 - b_1b_2 \right) v \\ $
	$=  \left( a_1 a_2 - b_1b_2 \right) v +  \left( I b_1 b_2 + I b_1 a_2\right) v=\left( z_1 z_2 \right) v$.
\end{enumerate}

Therefore $\VV = (V,I)$ is indeed a complex vector space. Now, since $I$ commutes with $\rho$, it is enough to take $\rho_{\CC} = \rho$ to define the complex representation.

We have shown that every real representation with a complex structure defines a complex representation. Now we just need to show that every complex representation can be written in that way (as a real representation $V$ equipped with a complex structure $I$). This is straight forward since any complex vector space defines a real one by taking the real and imaginary parts and the complex structure $I$ is defined by the multiplication by $i$. 
\end{proof}

\begin{example}
$\left(\RR^2, I = \left(\begin{array}{rr} 0 & -1  \\ 1 & 0 \end{array}\right) \right)$ is a complex vector space and in fact it is isomorphic to $\CC$.

In this case the real vector space $V = \RR^2$ and the complex structure $I \in \fgl(\RR^2)$. We define multiplication of a vector by a complex number as in \eqref{eq:complexmult}, more explicitly:

\begin{equation}
z v = (a + i b)  \left(\begin{array}{r} v_1 \\ v_2 \end{array}\right) = a  \left(\begin{array}{r} v_1 \\ v_2 \end{array}\right) + b \left(\begin{array}{rr} 0 & -1  \\ 1 & 0 \end{array}\right) \left(\begin{array}{r} v_1 \\ v_2 \end{array}\right) = %
\left(\begin{array}{r} a v_1 - b v_2  \\  a v_2 + b v_1\end{array}\right).
\end{equation}

To see that this is isomorphic to $\CC$, lets see how multiplication by complex numbers works in $\CC$. Let $w = w_1 + i w_2 \in \CC$:

\begin{equation}
  \begin{aligned}[m]
z w &= (a + i b)  (w_1 + i w_2) = a w_1 + i a w_2 + i b w_1 - b w_2 
&= a w_1 - b w_2 + i (b w_1 + a w_2).
  \end{aligned}
\end{equation}

Hence the map that assigns to every $w \in \CC$ a vector in $\RR^2$ by $v = (v_1, v_2) = \left(Re(w), Im(w) \right)$ is an isomorphism between the two spaces. 
\end{example}

\subsubsection{Quaternionic representations as special cases of real ones}

\begin{proposition}
Every quaternionic representation $\VV_H$ can be seen as a special case of a real representation $V$ equipped with a complex structure $I$ and a \textit{quaternionic structure} $J$, which is a linear map $J: V \rightarrow V$ that squares to minus the identity, $J^2 = -1$, and anticommutes with $I$, $IJ = -JI$. Both $I$ and $J$ commute with the action of $\fg$ on $V$.
\end{proposition}

\begin{proof}

We show first that every real representation $V$ equipped with a complex structure $I$ and a quaternionic structure $J$ defines a quaternionic representation. Notice that this implies that $(V,I)$ has even complex dimension or, equivalently, $V$ has real dimension which is a multiple of four.

To prove this, we need to prove that $(V,I, J)$ has the structure of a right/left quaternionic vector space that we will call $\VV_H$ and provide a map $\rho_{\HH}: \fg \rightarrow \fgl(\VV_H)$ preserving the Lie bracket. For example, for a left quaternionic vector space we define

\begin{equation}
q v = (a + b i + c j + d k) v= a v + b Iv + c J v + d IJ v ,
\label{eq:LQuatAct}
\end{equation}
for every quaternion $q = a + b i + c j + d k$. It is not difficult to check that this satisfies all the desired properties of a left vector space. Again, since $I$ and $J$ commute with $\rho$, it is enough to take $\rho_{\HH} = \rho$ to define the quaternionic representation.

To prove the converse, we just need to show that every quaternionic representation can be written in that way (as a real representation $V$ equipped with a complex structure $I$ and a quaternionic structure $J$), which is again is straight forward since any quaternionic vector space defines a real one by forgetting the quaternionic structure and the complex and quaternionic structures $I, J$ are defined by the multiplication by $i$ and $j$ respectively. 
\end{proof}

In table \ref{tab:LAreps} we summarise the representations of real Lie algebras that we have defined in this section.

\begin{table}[ht!]
  \centering
  \begin{tabular}{|c| >{$}c<{$}|>{$}c<{$}|}
    \hline
    Representation & \text{Notation} & \text{Vector space}\\\hline
    Real &  \Dar(\fg,\RR) & \text{Real,} V\\
    Complex &  \Dar(\fg,\CC) & \text{Complex,} \VV = (V,I)\\
    Quaternionic &  \Dar(\fg,\HH) & \text{L/R Quaternionic,} \VV_\HH=(V,I,J)\\\hline
  \end{tabular}
  \vspace{8pt}
  \caption{Representations of real Lie algebras}
  \label{tab:LAreps}
\end{table}

\subsubsection{Relations between real, complex and quaternionic representations}
\label{sec:relat-betw-real}

Actually, there are many natural maps between representations obtained by altering the ground field $\KK$. Using the more compact notation of denoting $U, V$ and $W$, real, complex and quaternionic representations respectively, these maps correspond to

\begin{itemize}
\item \emph{extending from $\RR$ to $\CC$}: $U \mapsto V = U_\CC$ (complexification)
\item \emph{extending from $\CC$ to $\HH$}: $V \mapsto W = V_\HH$ (quaternionification)
\item \emph{conjugation}: $V \mapsto \Vbar$
\item \emph{restricting from $\HH$ to $\CC$}: $W \mapsto V = \rh{W}$
\item \emph{restricting from $\CC$ to $\RR$}: $V \mapsto U = \rf{V}$. 
\end{itemize}

These maps are summarised succinctly in the following (noncommutative!) diagram, borrowed from \cite{Adams} via \cite{MR781344}:
\begin{equation}
  \label{eq:adams}
  \xymatrix{ & \Dar(\fg,\CC) & \\
    \Dar(\fg,\RR) \ar[ur]^c & & \Dar(\fg,\HH) \ar[ul]_{r'} \\
    & \Dar(\fg,\CC) \ar[uu]^t \ar[ul]^r \ar[ur]_q }
\end{equation}
where $\Dar(\fg,\KK)$ means the category of representations of the Lie algebra $\fg$ over the field $\KK$, and the arrows denote the following functors:
\begin{itemize}
\item[$t$:] if $V \in \Dar(\fg,\CC)$, then $t(V) = \Vbar$ denote the complex conjugate representation;
\item[$q$:] if $V \in \Dar(\fg,\CC)$, then $q(V) = V_\HH = \HH \otimes_\CC V \in \Dar(\fg,\HH)$, where $\HH$ is a right $\CC$-module;
\item[$c$:] if $U\in \Dar(\fg,\RR)$, then $c(U) = U_\CC = \CC \otimes_\RR U$ is its complexification;
\item[$r$:] if $V \in \Dar(\fg,\CC)$, then $r(V) = \rf{V} \in \Dar(\fg,\RR)$ is obtained by restricting scalars; and
\item[$r'$:] if $W \in \Dar(\fg,\HH)$, then $r'(W) = \rh{W} \in \Dar(\fg,\CC)$ is obtained by restricting scalars.
\end{itemize}

The map $t$ does not change the dimension, and neither do $c$ or $q$ in the sense that $\dim_\CC U_\CC = \dim_\RR U$ and $\dim_\HH V_\HH = \dim_\CC V$.  However $r$ and $r'$ double the dimension: $\dim_\RR \rf{V} = 2 \dim_\CC V$ and $\dim_\CC \rh{W} = 2 \dim_\HH W$.  In this thesis we are not working with quaternionic representations themselves but with their image under $r'$; although this will not always be reflected in our notation.  In other words, we will often write simply $W$ for $\rh{W}$ if in so doing the possibility of confusion is minimal.

The above maps obey some relations that we list below. For proofs see \cite[proposition 3.6]{Adams} or \cite[proposition(6.1)]{MR781344}.

\begin{proposition}
\label{prop:diamond}
  The following relations hold:
  \begin{enumerate}
  \item $t^2 = 1$ or $\overline{\overline{V}} \cong V$ for all $V \in \Dar(\fg,\CC)$;
  \item $tc = c$ or $\overline{U_\CC} \cong U_\CC$ for all $U \in \Dar(\fg,\RR)$;
  \item $qt = q$ or ${\overline V}_\HH \cong V_\HH$ for all $V \in \Dar(\fg,\CC)$;
  \item $rc = 2$ or $\rf{U_\CC} \cong U \oplus U$ for all $U \in \Dar(\fg,\RR)$;
  \item $cr = 1 + t$ or $\rf{V}_\CC \cong V \oplus \Vbar$ for all $V \in \Dar(\fg,\CC)$;
  \item $rt = r$ or $\rf{\Vbar} \cong \rf{V}$ for all $V \in \Dar(\fg,\CC)$;
  \item $t r' = r'$ or $\overline{\rh{W}} \cong \rh{W}$ for all $W \in \Dar(\fg,\HH)$;
  \item $q r' = 2$ or $\rh{W}_\HH \cong W \oplus W$ for all $W \in \Dar(\fg,\HH)$; and
  \item $r' q = 1 + t$ or $\rh{V_\HH} \cong V \oplus \Vbar$ for all $V \in \Dar(\fg,\CC)$.
\end{enumerate}
\end{proposition}

Recall that a real (resp. complex, quaternionic) representation is \textbf{irreducible} if it admits no proper real (resp. complex, quaternionic) subrepresentations.  We denote by $\Irr(\fg,\KK)$ the irreducible representations of $\fg$ of type $\KK$.  The following proposition states what happens to irreducible representations under the above maps.  It is not always the case that irreducibles go to irreducibles, but their images are under control in any case.

\begin{proposition}
\label{prop:irreducible}
  The following hold:
  \begin{enumerate}
  \item $V \in \Irr(\fg,\CC) \iff \Vbar \in \Irr(\fg,\CC)$;
  \item if $U \in \Irr(\fg,\RR)$ then $U_\CC \in \Irr(\fg,\CC)$, unless $U = \rf{V}$ for some $V\in\Irr(\fg,\CC)$, in which case $U_\CC \cong V \oplus \Vbar$;
  \item if $V\in \Irr(\fg,\CC)$ then $\rf{V} \in \Irr(\fg,\RR)$, unless $V = U_\CC$ for some $U\in\Irr(\fg,\RR)$, in which case $\rf{V} \cong U \oplus U$;
  \item if $W \in \Irr(\fg,\HH)$ then $\rh{W} \in \Irr(\fg,\CC)$, unless $W = V_\HH$ for some $V\in\Irr(\fg,\CC)$, in which case $\rh{W} \cong V \oplus \Vbar$; and
  \item if $V\in \Irr(\fg,\CC)$ then $V_\HH \in \Irr(\fg,\HH)$, unless $V = \rh{W}$ for some $W \in \Irr(\fg,\HH)$, in which case $V_\HH \cong V \oplus V$.
  \end{enumerate}
\end{proposition}

See \cite[proposition(6.6)]{MR781344} for a partial proof.

\subsection{Unitary representations}
\label{sec:unitrep}

We consider now representations admitting an inner product which has a notion of signature. There are three such inner products: real symmetric, complex hermitian and quaternionic hermitian. The corresponding representations are called orthogonal, complex unitary and quaternionic unitary respectively. Collectively, we call them \textit{unitary representations}.


\subsubsection{Real orthogonal representations}

\begin{definition}
\label{def:orthogRep}
Let $V$ be a real representation of the Lie algebra $\fg$ with homomorphism $\rho: \fg \rightarrow \fgl(V)$. This representation is \textbf{orthogonal} if there exists a non-degenerate, symmetric bilinear 2-form in $V$, $\left\langle -,-\right\rangle$, that satisfies:

\begin{equation}
\label{eq:orthRep}
\left\langle x \cdot u,v\right\rangle  + \left\langle u,x \cdot v\right\rangle  =0
\end{equation}
$\forall x \in \fg, \forall u, v \in V$.
\end{definition}

In terms of the maps $\rho(x) \in \fgl(V)$ this condition is equivalent to:

\begin{equation}
\label{eq:orthLA}
\left\langle \rho(x) u,v\right\rangle  + \left\langle u,\rho(x) v\right\rangle  =0, 
\end{equation}
which is the definition of  $\rho(x)$ belonging to the Lie algebra $\fso(V)$. Therefore, alternatively, an orthogonal representation of $\fg$ can be defined as a Lie algebra homomorphism $\rho: \fg \rightarrow \fso(V)$.

The name \textit{orthogonal} is more natural when referred to Lie groups. Remember that exponentiating a Lie algebra, one obtains the corresponding Lie group. The condition for a Lie group representation to be orthogonal is to admit an inner product that the representation preserves in the following way. Let $V$ be now a representation of a Lie group $G$. If we choose a basis in $V$, then in terms of that basis the elements of the representation take the form of a matrix  $M$. Let $u, v \in V$, the condition is: 

\begin{equation}
\label{eq:orthLG}
\left\langle u,v\right\rangle  = \left\langle M u,M v\right\rangle  =\left\langle  u,M^TM v\right\rangle
\end{equation}
where $M^T$ is the transpose of $M$ with respect to the inner product. Hence, the inner product is preserved if and only if $M M^T = Id$, or equivalently $M = M^T$, that is every matrix $M$ in the representation of $G$ must be orthogonal. In particular, this ensures that the angles between vectors are preserved and orthogonal vectors remain orthogonal. 

The condition \eqref{eq:orthLG} translated to the corresponding Lie algebra via differentiation and evaluation at the identity, becomes precisely \eqref{eq:orthLA}. It is for this reason that a representation of a Lie algebra that satisfies this condition is called orthogonal. 

Orthogonal representations come with a map which will play a crucial role in this thesis:

\begin{definition}
\label{def:RFaulkMap}
The \textbf{real Faulkner map} is an application $T: V \times V \rightarrow \fg$ defined by the relation:

\begin{equation}
\left(T(u,v),x\right) = \left\langle \rho(x) u, v\right\rangle = \left\langle x \cdot u, v\right\rangle
\label{eq:T-map-R}
\end{equation}
$\forall  u, v \in V$ and $x \in \fg$. 
\end{definition}

Where $(-,-)$ is the inner product in $\fg$ and $\left\langle -,-\right\rangle$ in $V$. Notice that, from \eqref{eq:orthRep} and the symmetry of $\left\langle -,-\right\rangle$, it follows that $T$ is alternating:

\begin{equation}
\label{eq:T-skew}
T(u,v) = - T(v,u)
\end{equation}

Of especial interest will be the fourth-rank tensor:

\begin{equation}
  \label{eq:R4tensor}
  \eR_R (u,v,w,s) := \left(T(u, v), T(w,s)\right),
\end{equation} 
which is antisymmetric in the first and last two slots and symmetric under pair-interchange of the first two entries with the last two, this is $\eR$ belongs to:

\begin{equation}
  \label{eq:real-rank4-rep}
  S^2 \Lambda^2 V \cong \Lambda^4 V \oplus V^{\yng(2,2)}.
\end{equation}


\subsubsection{Complex unitary representations}
\label{sec:CUrep}
The next inner product with a notion of signature is defined on a complex vector space $\VV$ by a \textbf{Hermitian form} $h(-,-)$, which is an inner product on $\VV$ that is \textit{sesquilinear}, this is, complex linear on the first entry but complex antilinear on the second: 

\begin{equation}
  \begin{aligned}[m]
	h(\lambda u,v)=  \lambda h(u,v) \\
		h(u,\lambda v) = \bar{\lambda}  h(u,v),
	\end{aligned}
\end{equation}
$\forall \lambda \in \CC$. It satisfies $h(u,v) = \overline{h(v,u)} $. An example of this is the standard inner product on $\CC^n$ is given by:

\begin{equation}
h(u,v) = u v^\dag 
\label{eq:HermCn}
\end{equation}
where $v^\dag $ stands for transpose and complex conjugate. We then define a unitary representation as follows:

\begin{definition}
\label{def:CunitRep}
Let $\VV$ be a complex representation of the Lie algebra $\fg$ with homomorphism $\rho_{\CC}: \fg \rightarrow \fgl(\VV)$. This representation is \textbf{unitary} if there exists a Hermitian form $h: V \times V \rightarrow \CC$ which the Lie algebra preserves:

\begin{equation}
\label{eq:unitRep}
h \left( x \cdot u,v \right)  +h \left( u,x \cdot v \right)  =0
\end{equation}
$\forall x \in \fg, \forall u, v \in \VV$.
\end{definition}

In terms of the maps $\rho(x) \in \fgl(\VV)$ with the standard inner product \eqref{eq:HermCn}, this condition is equivalent to:

\begin{equation}
h \left( \rho(x) u,v \right)  +h \left( u, \rho(x) v \right)   =0,
\end{equation}
which is the definition of a $\rho(x)$ belonging to $\fu(\VV)$. Therefore, alternatively, a complex unitary of $\fg$ can be defined as a Lie algebra homomorphism $\rho: \fg \rightarrow \fu(\VV)$.

A unitary representation can be defined from an orthogonal one equipped with a complex structure. For this to be possible, the complex structure $I$ must be orthogonal with respect to the inner product:

\begin{equation}
\left\langle u,v\right\rangle = \left\langle Iu,Iv\right\rangle 
\end{equation}

Then, the hermitian form is related to the real inner product $\left\langle -,-\right\rangle$ by:

\begin{equation}
h(u,v) = \left\langle u,v\right\rangle + i \left\langle u, Iv\right\rangle
\end{equation}

Unitary representations also come with an extra map which, again, will play a crucial role. However to define this map we need to use the complexification $\fg_\CC = \CC \otimes_\RR \fg$ of the Lie algebra $\fg$.  As a vector space this is just: $\fg_\CC = \fg \oplus i \fg $. We denote elements of $\fg_\CC$ by $\XX = x + i y$, with $x, y \in \fg$. We then extend the bracket and the inner product on $\fg$ complex bilinearly in such a way as to make $\fg_\CC$ into a complex metric Lie algebra.

There is a small caveat as a complex Lie algebra cannot leave a hermitian inner product invariant in the strict sense, due to sesquilinearity of the inner product. Instead we have that

\begin{equation}
  \label{eq:pre-unitarity-C}
  h(\XX \cdot v , w) + h(v, \XXbar\cdot w) = 0,\qquad\text{for all $\XX \in \fg_\CC$ and $v,w\in \VV$.}
\end{equation}

Indeed, letting $\XX = x + i y$,

\begin{align}
\label{eq:had-invariance}
  h(\XX\cdot v, w) &= h((x + iy) \cdot v, w)\\
  &= h(x\cdot v, w) + i h(y\cdot v, w)\\
  &= - h(v, x \cdot w) - i h(v, y \cdot w) && \text{by equation\eqref{eq:unitRep}}\\
  &= - h(v, x \cdot w) + h(v, i y \cdot w) && \text{by sesquilinearity of $h$}\\
  &= - h(v, (x-iy)\cdot w)\\
  &= - h(v,\XXbar\cdot w).
\end{align}

Then, we can also define a Faulkner map like in the real case that we now denote $\TT$ and is defined by:

\begin{definition}
\label{def:CFaulkMap}
The \textbf{complex Faulkner map} is an application $\TT: \VV \times \VV \rightarrow \fg_\CC$ defined by the relation:
\begin{equation}
  \label{eq:T-map-C}
  \left(\TT(u,v),\XX\right) = h(\XX \cdot u, v),
\end{equation}
for all $u,v \in V$ and $\XX \in \fg_\CC$.  
\end{definition}

If $\VV = (V,I)$, the following lemma shows how $\TT$ can be written in terms of the real Faulkner map:

\begin{lemma}\label{le:T-map-C}
  For all $u,v\in U$, we have $\TT(u,v) = T(u,v) + i T(u, Iv)$.
\end{lemma}

\begin{proof}
  Write $\TT(u,v) = A(u,v) + i B(u,v)$ and expand equation \eqref{eq:T-map-C} for $\XX = X + i Y$ complex bilinearly.  The left-hand side becomes
  \begin{align*}
    \left(\TT(u,v),\XX\right) &= \left(A(u,v) + i B(u,v),X + i Y\right) \\
    &= \left(A(u,v),X \right) + i \left(A(u,v), Y\right)  + i \left(B(u,v),X \right)  - \left(B(u,v), Y\right),
  \end{align*}
  whereas the right-hand side becomes
  \begin{align*}
    h(\XX \cdot u, v) &= \left<(X + i Y) \cdot u,v\right> + i \left<(X + i Y) \cdot u,Iv\right>\\
    &= \left<X \cdot u,v\right> + \left<I (Y \cdot u),v\right> + i \left<X \cdot u,Iv\right> +  i \left<I (Y \cdot u),Iv\right>\\
    &= \left<X \cdot u,v\right> - \left<Y \cdot u, I v\right> + i \left<X \cdot u, Iv \right> +  i \left<Y \cdot u,v\right>.
  \end{align*}
  Comparing real and imaginary parts and the terms depending on $X$ and $Y$, we arrive at the following two equations:
  \begin{equation*}
    \left(A(u,v),X \right)  = \left<X \cdot u,v\right> \qquad\text{and}\qquad \left(B(u,v),X \right)  = \left<X \cdot u, Iv \right>.
  \end{equation*}
  Now comparing the first of the above equations with the definition of $T$ \eqref{eq:T-map-R}, we obtain $A(u,v) = T(u,v)$, whereas the second equation says that $B(u,v) = T(u, I v)$.
\end{proof}

In the next proposition we prove two symmetries Faulkner map: a symmetry property and that it is a sesquilinear map.

\begin{proposition}\label{pr:T-map-C}
  The map $\TT: V \times V \to \fg_\CC$ satisfies the following properties
  \begin{equation*}
   \overline{\TT(u,v)} = - \TT(v,u)\qquad  \TT(I u,v) = i \TT(u,v) \qquad \TT(u, I v) = -i \TT(u,v).
  \end{equation*}
\end{proposition}

\begin{proof}
  Notice first some properties of the real Faulkner map under the action of $I$:
	
  \begin{align*}
    \left(T(u,Iv),X\right) &= \left<X \cdot u , I v\right>  && \text{by equation \eqref{eq:T-map-R}}\\
    &= - \left<I (X \cdot u) , v\right> && \text{by skewsymmetry of $I$}\\
    &= - \left<X \cdot I u , v\right> && \text{by $\fg$-invariance of $I$}\\
    &= - \left(T(Iu,v),X\right) && \text{again by equation \eqref{eq:T-map-R},}
  \end{align*}
  hence
  \begin{equation}
    \label{eq:T-I}
    T(u,Iv) = - T(Iu, v).
  \end{equation}
  Together with equation \eqref{eq:T-skew}, we have in addition that
  \begin{equation}
    \label{eq:T-I-too}
    T(u,Iv) = T(v,Iu).
  \end{equation}
  To show the first identity, we calculate
  \begin{align*}
    \overline{\TT(u,v)} &= \overline{T(u,v) + i T(u,Iv)}\\
    &= T(u,v) - i T(u,Iv)\\
    &= - T(v,u) -i T(v,Iu) && \text{using equations \eqref{eq:T-skew} and \eqref{eq:T-I-too}}\\
    &= - \TT(v,u);
  \end{align*}
  and to show the second, we calculate
  \begin{align*}
    \TT(I u,v) &= T(Iu,v) + i T(Iu,Iv)\\
    &= - T(u,Iv) + i T(u,v) && \text{using equation \eqref{eq:T-I}}\\
    &= i ( T(u,v) + i T(u,Iv) ) \\
    &= i \TT(u,v).
  \end{align*}
	
	The third property then follows from the first two. We have proved that $\TT$ is complex linear in the first entry but complex antilinear in the second, hence it is a sesquilinear map.
\end{proof}

Again, we can define the forth-rank tensor:

\begin{equation}
  \label{eq:complex-rank4-rep}
  \eR_C (u,v,w,s) := \left(\TT(u, v), \TT(w,s)\right),
\end{equation}
which is symmetric under pair interchange and hence belongs to:

\begin{equation}
  S^2 (\VV \otimes \bar{\VV}) \cong \left(S^2 \VV \otimes S^2\bar{\VV}\right) \oplus \left(\Lambda^2 \VV \otimes \Lambda^2 \bar{\VV} \right)
\end{equation}

\subsubsection{Quaternionic unitary representations}
\label{sec:QUrep}

Let $\VV_H = (\VV,J)$ be a quaternionic representation that comes from a unitary complex one, that is, it has an hermitian form $h(-,-)$ compatible with the Lie bracket. Then, one can define a complex skew-symmetric or \textbf{symplectic form} on top of the hermitian form:

\begin{equation}
  \label{eq:complex-symplectic}
  \omega(u,v) := h(u,Jv) = - \omega(v,u)
\end{equation}

Notice that this implies that $\VV = (V,I)$ must have even complex dimension. A quaternionic unitary representation will be one consisting of endomorphisms of $\VV$ which commute with $J$ and preserve $\omega$. The first condition means that the endomorphisms must be in $\fu(\VV)$ and the second implies they are in $\fsp(\VV)$. Therefore, we define a \textbf{quaternionic unitary} representation of $\fg$ as a Lie algebra homomorphism $\rho$ from $\fg$ to the intersection of both, $\fusp(\VV) = \fu(\VV) \cap \fsp(\VV)$.

The complex symplectic structure $\omega$ can be written in terms of the real inner product on $V$:

\begin{equation}
  \label{eq:omega-from-I-and-J}
  \omega(u,v) = \left<u, J v\right> + i \left<u, I J v\right>.
\end{equation}

Using this new form, we can define one last map that we call \textbf{quaternionic Faulkner map}:

\begin{equation}
  \label{eq:T-map-H}
  \left(\TT_H(u,v), \XX\right) = \omega(\XX\cdot u, v).
\end{equation}

Using the expression of $\omega$ in terms of $h$ \eqref{eq:complex-symplectic} we have that:

\begin{equation}
  \TT_H(u,v) = \TT(u,Jv)
\end{equation}

This implies that $\TT_H$ is now complex bilinear. Also, since $\omega$ is symplectic, it follows that $\TT_H(u,v)$ is symmetric:

\begin{equation}
  \label{eq:TTh-symmetric}
  \TT_H(u,v) = \TT_H(v,u),
\end{equation}

Finally, we can define the fourth-rank tensor $\eR_H(u,v,w,s) := \left(\TT_H(u,v), \TT_H(w,s)\right)=\omega(\TT(u,Jv)\cdot w, s)$ belongs to
\begin{equation}
  \label{eq:quat-rank4-rep}
  S^2 S^2 \VV_H \cong \VV_H^{\yng(2,2)} \oplus S^4 \VV_H.
\end{equation}

In table \ref{tab:LAUreps} we summarise the unitary representations defined in this section. Associated to them we have defined the Faulkner maps summarised in table \ref{tab:Faulkner maps} and the fourth-rank tensors in table \ref{tab:wrtReal}.  Finally, the relation of Faulkner maps and inner products to the real ones are recalled in table \ref{tab:wrtReal}.

\begin{table}[ht!]
  \centering
  \begin{tabular}{|c| >{$}c<{$}| >{$}c<{$}|>{$}c<{$}|}
    \hline
    Representation    & \text{Vector space}& \rho : \fg \to & \text{Invariant form}\\\hline
    Orthogonal  & \text{Real } V & \fso(V) & \left\langle u,v\right\rangle\\
    $\CC$-unitary  & \text{Complex } (V,I) & \fu(\VV) & h(u,v) \\
    $\HH$-unitary & \text{L/R Quaternionic } (V,I,J)& \fusp(\VV)  & h(u,v) ,\omega(u,v)  \\\hline
  \end{tabular}
  \vspace{8pt}
  \caption{Unitary representations of real Lie algebras}
  \label{tab:LAUreps}
\end{table}

\begin{table}[ht!]\footnotesize
  \centering
  \begin{tabular}{| >{$}c<{$}| >{$}c<{$} >{$}c<{$}|c|}
    \hline
    \text{Rep}   &  \multicolumn{2}{c|}{Faulkner map}   & Properties  \\ \hline
    \RR  & T(u,v), & \left(T(u,v),x\right) =  \left\langle x \cdot u, v\right\rangle%
		& $\RR$-Bilinear, antisymmetric \\
    \CC & \TT(u,v),  & \left(\TT(u,v),\XX\right) = h(\XX \cdot u, v) & $\CC$-Sesquilinear, antihermitian\\
    \HH & \TT_H(u,v), & \left(\TT_H(u,v), \XX\right) = \omega(\XX\cdot u, v)  &   $\CC$-Bilinear, symmetric \\\hline
  \end{tabular}
  \vspace{8pt}
  \caption{Faulkner maps in unitary representations of real Lie algebras}
  \label{tab:Faulkner maps}
\end{table}

\begin{table}[ht!]
  \centering
  \begin{tabular}{| >{$}c<{$}|  >{$}c<{$}|>{$}c<{$}|}
    \hline
    \text{Rep}    & \text{Faulkner map}&   \eR(u,v,w,s)  \\ \hline
    \RR  & T(u,v) & \left(T(u, v), T(w,s)\right) = \left\langle T(u,v) \cdot w, s\right\rangle\\
    \CC  & \TT(u,v) & \left(\TT(u, v), \TT(w,s)\right) = h(\TT(u,v)\cdot w, s)\\
    \HH & \TT_H(u,v) &  \left(\TT_H(u,v), \TT_H(w,s)\right) =\omega(\TT_H(u,v)\cdot w, s)\\ \hline
  \end{tabular}
  \vspace{8pt}
  \caption{Forth-rank tensors in unitary representations of real Lie algebras}
  \label{tab:4-rank tensors}
\end{table}

\begin{table}[ht!]\footnotesize
  \centering
  \begin{tabular}{|>{$}c<{$}| >{$}c<{$}| >{$}c<{$}|}
    \hline
    \text{Rep}    & \text{Faulkner maps}&  \text{Invariant form}\\ \hline
    \RR  &  T(u,v) & \left\langle u,v\right\rangle\\
    \CC  &  \TT(u,v) = T(u,v) + i T(u, Iv)& h(u,v) = \left\langle u,v\right\rangle + i\left\langle u,Iv\right\rangle\\
    \HH &   \TT_H(u,v) = \TT(u,Jv) = T(u,Jv) + i T(u, IJv) & \omega(u,v) = \left\langle u,Jv\right\rangle + i\left\langle u,IJv\right\rangle \\\hline
  \end{tabular}
  \vspace{8pt}
  \caption{Faulkner maps and invariant forms in terms of the real ones. }
  \label{tab:wrtReal}
\end{table}

\chapter{Metric 3-Leibniz algebras}
\label{sec:3algebras}

In this chapter we define an algebraic structure that we call 3-Leibniz algebra. Recent interest in it originates in a particular class of them called 3-Lie algebras that appeared in the work of Bagger and Lambert \cite{BL1, BL2, BL3} modelling M2-branes. In subsequent works other kinds of these ternary algebras started to crop up and a relation to more common Lie algebras was suspected in specific examples. Here we present a homogeneous framework that includes all the 3-algebras that have appeared in different M2-branes theories and adapt a construction originally due to Faulkner \cite{FaulknerIdeals} that establishes a one-to-one correspondence between metric 3-Leibniz algebras and unitary representations of real Lie algebras. 

As a field which has experienced great development in a short period of time there is a bit of confusion in the literature with respect to the nomenclature. For example, on \cite{Lor3Lie, 2p3Lie} we called \textit{Lie 3-algebras}, what we now call \textit{3-Lie algebras} (also known as \textit{Filippov} algebras), after being alerted by mathematicians that the name \textit{Lie 3-algebras} was already in use for something else. Also on \cite{Lie3Algs}, following \cite{CherSaem}, we called \textit{generalized 3-Lie algebras} what we now call \textit{real orthogonal 3-Leibniz algebras}. Other changes in nomenclature with respect to recent literature will be indicated. We hope that every concept is clearly defined avoiding any confusion. 

We follow here the same pattern that was adopted in section \ref{sec:RepsLA} with representations of Lie algebras: first we discuss metric 3-Leibniz algebras generically and then we move on to discuss each of the three inner products which admit the notion of a signature separately. As it happened in that case, these three instances are far from  disjoint classes of 3-algebras, but rather they can all be seen as special cases of the real one.

This chapter is organised as follows. In section \ref{sec:3algdef} we define metric 3-Leibniz algebras. In section \ref{sec:metric3Leib} we discuss the three classes that posses an inner product with a notion of signature. In \ref{sec:Faulkner} we explain the parallelism between unitary representations of Lie algebras and 3-Leibniz algebras via the Faulkner construction that we specialized to these cases. This construction is very technical so, to clarify ideas, we discuss first the generalities of how it works in \ref{sec:FaulknerIdea} and then we move on to give all the details and proofs in each of the three cases of interest: real, complex and quaternionic. In \ref{sec:LSAembed} we focus on the six extreme cases of 3-Leibniz algebras and show how they can be embedded in Lie (super)algebras. We discuss in further detail the three extreme cases that give rise to extended supersymmetry in Bagger-Lambert theories. Finally, in section \ref{sec:restrictLA} we discuss how the possible Lie algebras associated to a given class of 3-Leibniz algebras is highly constrained by the Faulkner construction. The content of this chapter is based on \cite{Lie3Algs,SCCS3Algs}.

\section{Definitions}
\label{sec:3algdef}
We define a 3-Leibniz algebra as a straight forward generalization of a Lie algebra. Essentially it is a vector space equipped now with a \textit{triple} product that satisfies an identity known as the fundamental identity that generalises the Jacobi identity \eqref{eq:Jacobi}. 

\begin{definition}
\label{def:3alg}
We define a \textbf{metric 3-Leibniz algebra} as a vector space $V$ equipped with an ad-invariant inner product $b(-,-)$ and a 3-bracket $[-,-, -] : V \times V \times V 	\to V$ which satisfy:

\begin{enumerate}
	\item The \textit{fundamental identity}
		\begin{equation}
		\label{def:FI}
    [x,y,[z,s,t]] = [ [x,y ,z],s,t] + [ z,[x,y ,s],t] + [ z,s,[x,y ,t]]
		\end{equation}

	\item The \textit{unitarity condition} or ad-invariance of $b(-,-)$
	
		\begin{equation}
		\label{def:unitarity}
    b\left([x,y,z],s\right) + b\left(z,[x,y,s]\right) = 0
		\end{equation}
	
	\item The \textit{symmetry condition}
	
			\begin{equation}
			\label{def:symmetry}
    b\left([x,y,z],s\right) = b\left([z,s,x],y\right)
			\end{equation}
\end{enumerate}
for all $x,y,z,s,t \in V$.
\end{definition}

\subsection*{Some remarks}

\begin{itemize}
\item The fundamental identity can be understood as a generalisation of the Jacobi identity \eqref{eq:Jacobi} in the following way. Consider the maps $ad_{(x,y)} (-)= [x,y,-]$ that for each $x,y \in V$ define a linear map from $V$ to itself (or endomorphism). A derivation or \textit{Leibniz rule} (hence the name) on this triple product is defined as an endomorphism of $V$, $D$, that satisfies:
	\begin{equation}
			D\left([x,y,z] \right)= [ D(x),y,z] + [ x,D(y),z] + [ x,y,D(z)].
	\end{equation}
Requiring that the maps $ad_{(x,y)} (-)= [x,y,-]$ be derivations over the 3-bracket results in the fundamental identity \eqref{def:FI}; just like requiring that the adjoint maps on a Lie algebra $\fg$ act as derivation results in the Jacobi identity \eqref{eq:adjDer}.
\item The reason why the unitarity condition is also called ad-invariance of $b(-,-)$ is because it can be re-written as:
	\begin{equation}
	\label{def:ad-invf}
					b\left(ad_{(x,y)}(z),s\right) + b\left(z,ad_{(x,y)}(s)\right) = 0.
	\end{equation}
\item We focused on the metric case because that is the one we need to build Lagrangians. The definition of a \textit{non-metric} 3-Leibniz algebra is the same but requiring only the fundamental identity and not the unitarity and symmetry conditions.
\item We will consider only finite-dimensional 3-Leibniz algebras, so for us $V$ is always finite-dimensional.
\item Some of these definitions might seem ad-hoc at this point, but we will see that for example the symmetry condition is precisely the one that guarantees that a real \textit{metric} Lie algebra can be obtained from a 3-algebra.
\end{itemize}


Given a basis $\left\{e_A\right\}$ for the vector space $V$, it is enough to determine the 3-bracket to specify how it acts on the elements of the basis. This can be done by giving the \textbf{structure constants} $F_{ABC}{}^{D}$ which are defined as:

\begin{equation}
\label{def:3StrConst}
    [e_A,e_B,e_C] = F_{ABC}{}^{D} e_D.
\end{equation}

Again, one can use the components of this inner product $b_{AB}$ in terms of the basis $\left\{e_A\right\}$ to raise and lower indices and define now the \textbf{canonical 4-form}:
  
\begin{equation}
\label{def:4form}
    F_{ABCD} = \left\langle [e_A,e_B,e_C],e_D \right\rangle =  F_{ABC}{}^{G}b_{GD}.
\end{equation}


\section{Three classes of metric 3-Leibniz algebras}
\label{sec:metric3Leib}

As we did with representations of Lie algebras in section \ref{sec:unitrep}, we want to consider only inner products with a notion of signature. This defines three classes of metric 3-Leibniz algebras that we will be interested in and that we discuss in turn. As we will see in section \ref{sec:Faulkner}, the parallelism between the two is no coincidence. 

\subsection{Real orthogonal}
\label{sec:ro3leib}

\textbf{Real orthogonal} 3-Leibniz algebras arise when the vector space $V$ is real and the inner product that we denote $\left\langle -,-\right\rangle$ is symmetric bilinear. Then the 3-bracket $[-,-,-]$ has some further properties:

\begin{enumerate}
 \item It is trilinear.
	\item The unitarity condition and the symmetry condition together imply that it is anti-symmetric in the first 2 slots:
	
	\begin{equation}
	[x,y,z] = -[y,x,z], \ \forall x,y,z \in V
	\label{eq:R3branitsym}
	\end{equation}
\end{enumerate}

We call this property \textit{(1,2)-antisymmetry}. In \cite{CherSaem} Cherkis and Sämann introduced these algebras and used them to construct three-dimensional $N=2$ superconformal Chern-Simons theories \cite[§4.1]{CherSaem}.

There are two classes of such 3-algebras that deserve their own name. They are defined by imposing extra conditions on the 3-bracket. 

\begin{itemize}
	\item \textbf{3-Lie Algebras} (3LA). The bracket is totally skewsymmetric, therefore this corresponds to the straightforward generalization of a Lie algebra. As we will see, this is the class of 3-algebras which is relevant for maximal supersymmetry in theories dual to configurations of $M2$-branes. They were the first to appear in the recent history of ternary algebras in 3-dimensional CFTs in the $N=8$ theory of \cite{BL1,BL2,BL3,GustavssonAlgM2}.
	
	\item \textbf{Lie Triple Systems} (LTS). The bracket satisfies a cyclicity condition:
	\begin{equation}
	[x,y,z] + [y,z,x] + [z,x,y] =0
	\label{eq:LTScycl}
	\end{equation}
	
\end{itemize}

\subsection{Complex unitary}
\label{sec:cu3leib}

The vector space $\VV$ is now complex and the inner product, that we denote $h(-,-)$, is a hermitian form which is complex linear in the first entry but complex anti-linear in the second. We say that the bracket is \textit{sesquibilinear}\footnote{This name is inspired by the word \textit{sesquilinear} that is applied to maps which are linear in one entry and antilinear in the other. The prefix \textit{sesqui-} comes from Latin and means "one and a half".}, which means complex linear in two of its entries (in this case the first and last) and complex anti-linear in the remaining one.

A small caveat interpreting the equations that define the complex 3-bracket, that we write $ [\![-,-,-]\!]$,  is that the adjoint maps act differently in the antilinear slots. This is because, as we showed in \eqref{eq:had-invariance} in section \ref{sec:CUrep}, a complex Lie algebra cannot leave a hermitian inner product invariant and the image of the adjoint maps lives, as we will see, in a complex Lie algebra. In particular the ad-invariance of $h$ becomes: 

\begin{equation}
  h(ad_{(x,y)}( u), v) = - h(u, \overline{ad_{(x,y)}}(v)),
\end{equation}
and the fundamental identity:
\begin{equation}
   ad_{(x,y)}\left( [\![v,w,z]\!]\right) = [\![ad_{(x,y)}\left( v\right),w,z]\!] + [\![v,\overline{ad_{(x,y)}}\left( w\right),z]\!] + [\![v,w,ad_{(x,y)}\left( z\right)]\!] .
\end{equation}
where $\overline{ad_{(x,y)}} = - ad_{(y,x)}$. Explicitly, the conditions on the 3-bracket are then:
\begin{enumerate}
\item the \emph{unitarity} condition
  \begin{equation}
    \label{eq:unitarity-C}
    h([\![v,w,x]\!],y) = h(x,[\![w,v,y]\!]);
  \end{equation}
\item the \emph{symmetry} condition
  \begin{equation}
    \label{eq:symmetry-C}
    h([\![v,w,x]\!],y) = h([\![x,y,v]\!],w);
  \end{equation}
\item and the \emph{fundamental identity}
  \begin{equation}
    \label{eq:FI-C}
    [\![x,y,[\![v,w,z]\!]]\!] = [\![[\![x,y,v]\!],w,z]\!] - [\![v,[\![y,x,w]\!],z]\!] + [\![v,w,[\![x,y,z]\!]]\!].
  \end{equation}
\end{enumerate}
For all $x,y,z,v,w \in \VV$.

In section \ref{sec:FaulkC} we will prove that, when the complex vector space $\VV$ is thought of as a real vector space $V$ with a complex structure $I$, this 3-bracket can be written in terms of the real 3-bracket:

\begin{equation}
  \label{eq:3-brackets-R-C}
  [\![u,v,w]\!] = [u,v,w] + [u,Iv,Iw],
\end{equation}

Some special classes of these algebras are:

\begin{itemize}
	\item \textbf{Jordan Triple System} (JTS). The bracket satisfies:
	\begin{equation}
  \label{eq:JTS}
  [\![u,v,w]\!] = [\![w,v,u]\!],
\end{equation}
	\item \textbf{anti-Jordan Triple System} (aJTS). The bracket satisfies:
	\begin{equation}
  \label{eq:aJTS}
  [\![u,v,w]\!] = -[\![w,v,u]\!],
\end{equation}	
These are the relevant 3-algebras for $N=6$ supersymmetry and the skewsymmetry condition identifies this class of triple systems with those used by Bagger and Lambert in \cite{BL4} to recover the $N=6$ theories discovered by Aharony, Bergman, Jafferis and Maldacena in \cite{MaldacenaBL}.
\end{itemize}

\subsection{Quaternionic unitary}
\label{sec:qu3leib}

Finally we consider the case of a complex vector $\VV$ space equipped with a quaternionic structure map $J$ and a complex symplectic structure $\omega$. As we have mentioned before, quaternionic vector spaces as such do not exist, but one can use the quaternionic structure $J$ to equip $\VV$ with the structure of a right/left-quaternionic vector space. 

The 3-bracket is defined as: $ [\![u,v,w]\!]_H =  [\![u,Jv,w]\!] $ and satisfies the unitarity (also called symplectic) and symmetry conditions as well as the fundamental identity. Explicitly:

\begin{enumerate}
\item the \emph{symplectic} condition
  \begin{equation}
    \label{eq:symplectic-H}
    \omega([\![x,y,z]\!]_H,w) = \omega([\![x,y,w]\!]_H,z);
  \end{equation}
\item the \emph{symmetry} condition
  \begin{equation}
    \label{eq:symmetry-H}
    \omega([\![x,y,z]\!]_H,w) = \omega([\![z,w,x]\!]_H,y);
  \end{equation}
\item the \emph{fundamental identity}
  \begin{equation}
    \label{eq:FI-H}
    [\![x,y,[\![v,w,z]\!]_H ]\!]_H = [\![[\![x,y,v]\!]_H,w,z]\!]_H + [\![v,[\![x,y,w]\!]_H,z]\!]_H + [\![v,w,[\![x,y,z]\!]_H]\!]_H ,
  \end{equation}

\end{enumerate}

Some other properties of the quaternionic 3-bracket are:

\begin{itemize}
	\item It is complex trilinear. Follows from $[\![-,-,-]\!]$ being sesquibilinear and $J$ being anti-linear.
	\item symmetry under exchange of first two slots, that we call \textit{(1,2)-symmetry}, (follows from the symplectic and symmetry conditions)
	\begin{equation}
  \label{eq:symmetry-3bracket}
  [\![x,y,z]\!]_H = [\![y,x,z]\!]_H,
\end{equation}
 \item the \emph{quaternionic} condition
  \begin{equation}
    \label{eq:bracket-J}
    J[\![x,y,z]\!]_H = [\![Jx, Jy, Jz]\!]_H.
  \end{equation}
\end{itemize}

Special classes of quaternionic 3-algebras are:

\begin{itemize}
	\item \textbf{Quaternionic Triple System} (QTS). The bracket is totally symmetric. 
	\item \textbf{anti-Lie Triple System} (aLTS). The bracket satisfies a cyclicity condition:
	\begin{equation}
  \label{eq:aLTS}
  [\![u,v,w]\!]_H + [\![v,w,u]\!]_H + [\![w,u,v]\!]_H = 0,
\end{equation}	
 This is the relevant class of 3-algebras for the $N=4,5$ superconformal field theories dual to $M2$-branes configurations.
\end{itemize}

We summarise the metric 3-Leibniz algebras defined in this section in table \ref{tab:3LeibAl}.

\begin{table}[ht!]\footnotesize
  \centering
  \begin{tabular}{| >{$}c<{$}| >{$}c<{$}|>{$}c<{$}| c|c|}
    \hline
    V    & \text{3-bracket} & \text{Inner product} & Properties & Special cases  \\ \hline
    \RR  & [-,-,-] 	& \left\langle -,-\right\rangle & Trilinear & 3-Lie algebra \\ 
		&&&(1,2)-antisymmetric& Lie Triple System \\ \hline 
    \CC  & [\![-,-,-]\!] & h(-,-) & Sesquibilinear& Jordan Triple System\\
		&&& & anti-Jordan Triple System \\ \hline
    \HH &  [\![-,-,-]\!]_H & \omega(-,-) & Trilinear& Quaternionic Triple System   \\
    &&&(1,2)-symmetric& anti-Lie Triple System \\ \hline		
  \end{tabular}
  \vspace{8pt}
  \caption{Metric 3-Leibniz algebras}
  \label{tab:3LeibAl}
\end{table}

\newpage
\section{The unitary Faulkner construction}
\label{sec:Faulkner}

In this section we explain the parallelism between unitary representations of real metric Lie algebras and metric 3-Leibniz algebras. Indeed, by specialising a construction originally due to Faulkner \cite{FaulknerIdeals}, we found in \cite{Lie3Algs} that there is a one-to-one correspondence between the two. The map is highly non-trivial and explains, at least in part, why superconformal Chern-Simon theories in three dimensions have been so elusive and why they have been found first in their 3-algebraic formulation rather than their Lie algebraic one.

Like with other dualities, having these two equivalent ways to describe the same mathematical objects, is very enlightening in the sense that some properties that are straightforward in one of the formulations are not so in the other. 

\subsection{The key idea of the unitary Faulkner construction}
\label{sec:FaulknerIdea}

In this section we want to give a flavour of how the unitary Faulkner construction works, leaving the proper statements and proofs for the following sections.

The motivation we had to study this construction in the string theory context, comes from the fact that Lie algebras are well known objects that have been around for a long time whereas 3-algebras are more recent in this context and less well-known. In \cite{MaldacenaBL}, an $N=6$ theory based on a Lie algebra was introduced. Then, Bagger and Lambert were able to re-write this theory in terms of an anti-Jordan Triple System\cite{BL4}. It seemed that, at least in some cases, one could assign a standard Lie algebra to a given 3-algebra and the question of whether this could always be done arose. 

First, notice that the adjoint maps $ad_{(u,v)} = [u,v,-]: V \to V$ are endomorphisms of $V$ and endomorphisms of $V$ form a Lie algebra called $\fgl(V)$. The question is then: does the set of all adjoint maps of a 3-Leibniz algebra form a Lie subalgebra of $\fgl(V)$? In other words, is the image of $ad$, that we call $\fg := \im ad$, a Lie subalgebra of  $\fgl(V)$? As we will see, this question has a positive answer. 

Let $T(u,v) := ad_{(u,v)}$ to simplify the notation. Then, the key observation to prove that $\fg := \im T$ is a Lie algebra is that the fundamental identity \eqref{def:FI} of the 3-bracket is equivalent to what is called \textit{$\fg$-equivariance} of the maps $T(u,v)$. This condition is similar to a Leibniz condition. That is, $T$ is $\fg$-equivariant if any element $x \in \fg$ acts on it in the following way:

\begin{equation}
x \cdot T(u,v) = T(x \cdot u,v) + T(u,x \cdot v)
\label{eq:g-equiv}
\end{equation}

Since the image of $T$ is in $\fgl(V)$, $\fg$ acts on it with the Lie bracket $x \cdot T(u,v) = [x , T(u,v)]$:

\begin{equation}
T(w,s) \cdot T(u,v) = [T(w,s),T(u,v) ] =T(T(w,s) \cdot u,v) + T(u,T(w,s) \cdot v)
\label{eq:g-equiv2}
\end{equation}

However, it acts on $u$ and $v$ via some other action of $\fg$ on $V$. Remembering that in this case $\fg = \im T$, an element of $\fg$ takes the form $x = T(w,s)$ and, by definition of the adjoint map, $T(w,s) \cdot v =[w,s,v]$. Applying this equation to an element $t \in V$ and using $T(u,v) \cdot w =[u,v,w]$  we obtain precisely the fundamental identity of the 3-bracket \eqref{def:FI}. Moreover we will see how the resulting 3-algebra is metric and therefore it is a metric 3-Leibniz algebra.

Conversely, we will see how given a real Lie algebra $\left(\fg, [-,-], (-,-)\right)$ and a unitary representation $\left(V, b( -,-) \right)$ which can be real, complex or quaternionic; one can construct a metric 3-Leibniz algebra. The key now is to use the Faulkner map $T(u,v): V \times V \to \fg$ defined in \ref{sec:RepsLA} via the equation:

\begin{equation}
\label{def:FaulknerMap}
\left(T(u,v),x\right) =  b\left( x \cdot u, v \right)
\end{equation}
$\forall u, v \in V$ and $x \in \fg$\footnote{In Faulkner's initial construction \cite{FaulknerIdeals}, the map went from $V^* \times V \to \fg$ where $V^*$ is the dual space of $V$. Instead of an inner product on $V$, he then used the dual pairing between $V$ and $V^*$ to define the map.}. 

Since $V$ is a representation of $\fg$, the image of $T$ on $\fg$ acts on $V$ which allows us to define the 3-bracket:

\begin{equation}
T(u,v) \cdot w := [u,v,w],
\end{equation}
$\forall u, v, w \in V$.

 One remark is that we assume for simplicity that the Lie algebra representation is \textit{faithful}. We will argue in section \ref{sec:N1supermult} of chapter \ref{sec:3dscfts} that this can be done without loss of generality for the corresponding physical theories.

Of course it is no coincidence that we called $T$ the adjoint maps above. As we will see, this map is $\fg$-equivariant and hence the resulting 3-bracket satisfies the fundamental identity.

\begin{table}[ht!]
  \centering
  \begin{tabular}{|ccccc|}
    \hline
    && Key observation& &\\ \hline
		The fundamental &  &$\fg$-equivariance &  &$\fg = \im T$ is a \\
		identity of the		&$\Leftrightarrow$& of the adjoint &$\Leftrightarrow$& Lie subalgebra \\
		3-bracket $[-,-,-]$& &maps $T(-,-)$& & of $\fgl(V)$\\ \hline
  \end{tabular}
  \vspace{8pt}
  \caption{Key observation of the Faulkner construction}
  \label{tab:key-obs}
\end{table}

In the following sections we give proofs of these statements in the real, complex and quaternionic cases. We do this in two parts for each case. First, we show how one can define a metric 3-Leibniz algebra from a Lie algebra and a unitary representation of it. In passing, we show how this Lie algebraic origin of 3-Leibniz algebras explains some of their properties in a very natural way. Then,  we prove the converse statement: that one can extract a Lie algebra and a unitary representation from any metric 3-Leibniz algebra.

\subsection{The real case}
\label{sec:FaulkR}
\subsubsection{From the Lie algebra to the 3-algebra}

Let $\left(\fg, [-,-], (-,-)\right)$ be a metric real Lie algebra and $\left(V, \left\langle -,-\right\rangle\right)$ a faithful real orthogonal representation of $\fg$ as defined in \eqref{def:orthogRep}. 

Then we use the corresponding real Faulkner map \eqref{eq:T-map-R} $T(u,v)$ to define the 3-bracket:

\begin{equation}
\label{def:real3br}
T(u,v) \cdot w := [u,v,w],
\end{equation}
$\forall u, v, w \in V$. We claim that this defines a real metric 3-Leibniz algebra on $V$ with inner product $\left\langle -,-\right\rangle$. To prove this, we need to show that it satisfies the fundamental identity, ad-invariance of $\left\langle -,-\right\rangle$ and the symmetry condition. Before doing that, we will need two auxiliary results: one that says that $T$ is $\fg$-equivariant and hence the image of $T$ is an ideal in $\fg$ and another one that says that $T$ is onto $\fg$.

\begin{lemma}\label{le:imIdeal}
  The Faulkner map $T$ is $\fg$-equivariant, that is:
	  \begin{equation}
    \label{eq:T-g-equiv}
    [x,T(v,w)] = T(x\cdot v, w) + T(v, x\cdot w)~.
  \end{equation}
\end{lemma}

\begin{proof}

  Let $x \in \fg$ and $v,w\in V$. Then for all $y \in \fg$ we have
  \begin{align*}
    \left([T(v,w),x],y\right) &= \left(T(v,w),[x,y]\right) &&\text{since $\fg$ is metric}\\
    &= \left\langle [x,y]\cdot v, w \right\rangle &&\text{by \eqref{def:FaulknerMap}}~,
  \end{align*}
	
Now, recall that the Lie bracket in $\fso(V)$ is really just $[M,N] = MN - NM$, hence:

\begin{equation}
[\rho(x),\rho(y)] v = \rho(x)\rho(y) v - \rho(y)\rho(x) v
\end{equation} 
or equivalently:

\begin{equation}
[x,y] \cdot v =x \cdot y \cdot v - y \cdot x \cdot v.
\label{eq:auxBr}
\end{equation} 

Then, we have:
	
 \begin{align*}
    \left([T(v,w),x],y\right) &= \left\langle [x,y]\cdot v, w \right\rangle &&\text{by \eqref{def:FaulknerMap}}\\
    &= \left\langle x \cdot y \cdot v, w \right\rangle - \left\langle y \cdot x \cdot v, w \right\rangle && \text{by \eqref{eq:auxBr}}\\
    &= - \left\langle y \cdot v, x \cdot w \right\rangle -  \left\langle y \cdot x \cdot v, w \right\rangle && \text{by \eqref{eq:orthRep}}\\
    &= - \left(T(v,x \cdot w), y \right) - \left(T(x\cdot v, w),y\right) && \text{again by \eqref{def:FaulknerMap}}~,
  \end{align*}
  hence abstracting $y$,
  \begin{equation}
    \label{eq:ideal-CS}
    [x,T(v,w)] = T(x\cdot v, w) + T(v, x\cdot w) \subset \im T~.
  \end{equation}
	Notice that in particular this also means that $\im T$ is an ideal in $\fg$, that is:
	\begin{equation}
	[T(v,w),x] \in \im T,
	\label{eq:imT-Ideal}
	\end{equation}
	$\forall u,v \in V$ and $x \in \fg$.
\end{proof}

\begin{lemma}\label{le:D-onto-CS}
  $T$ is surjective onto $\fg$.
\end{lemma}

\begin{proof}

To prove this, as opposed to the previous lemma, we need to use that $V$ is a faithful representation of $\fg$.

Since $\im T$ is an ideal, it is either all of $\fg$ or else there exists a subspace of $\fg$ that we will call $\fh$ such that $\fg = \im T \oplus_{\perp} \fh$. We will show, using the fact that the inner product is not degenerate, that this complement $\fh$ must vanish.

  Let $x \in \fg$ be perpendicular to the image of $T$. This means that $\left(x,T(v,w)\right) = 0$ for all $v,w\in V$, or equivalently that  $ \left\langle x\cdot v, w \right\rangle = 0$ for all $v,w\in V$. Nondegeneracy of the inner product on $V$ implies that $x\cdot v = 0$ for all $v\in V$, which in turn implies that $x=0$ since the action of $\fg$ on $V$ is faithful. Finally, nondegeneracy of the inner product on $\fg$ says that $\fh = 0$, hence $\im T = \fg$ and $T$ is surjective.
\end{proof}

Finally, we can prove the main proposition of this section.

\begin{proposition}
  The 3-bracket $[u,v,w] = T(u,v) \cdot w$ on the real metric vector space $\left(V,\left\langle -,-\right\rangle \right)$  defines a metric 3-Leibniz algebra with the same symmetric inner product $\left\langle -,-\right\rangle$.
\end{proposition}

\begin{proof}

We need to prove that the bracket satisfies the fundamental identity and that the inner product is ad-invariant and satisfies the symmetry condition. 

\textit{Fundamental identity}

As we mentioned in the introduction, the fundamental identity \eqref{def:FI} is equivalent to $\fg$-equivariance of the map $T$ \eqref{eq:ideal-CS} by substituting $x = T(u,s)$ and applying both sides of the equation to $t \in V$.

\textit{Ad-invariance of the inner product}

Follows from the fact that $V$ is a unitary representation, hence $\fg$ preserves the inner product on $V$, simply setting $x=T(s,t)$ on \eqref{eq:orthRep} we have:

\begin{equation}
\left\langle T(s,t) \cdot v, w \right\rangle + \left\langle v, T(s,t) \cdot w \right\rangle = 0
\end{equation}

\textit{Symmetry condition}

Since $T$ is surjective, for every element in the lie algebra $x \in \fg$ there exist two elements in $V$ such that $x = T(w,s)$. Plugging this back in the definition of $T$ \eqref{def:FaulknerMap} we have:

\begin{equation}
\left(T(u,v),T(w,s)\right) = \left\langle  T(w,s) \cdot u, v \right\rangle
\end{equation}

On the other hand, using the symmetry of the inner product on $\fg$:

\begin{equation}
\left(T(u,v),T(w,s)\right) = \left(T(w,s),T(u,v)\right) = \left\langle  T(u,v) \cdot w, s \right\rangle = \left\langle  T(w,s) \cdot u, v \right\rangle
\end{equation}
which is precisely the symmetry condition. 
\end{proof}

\textbf{Some remarks}

We want to recover here some properties of the real Faulkner map 3-bracket that follow naturally from this construction. 

\begin{itemize}
\item The 3-bracket $[-,-,-]$ is trilinear. This follows from the fact that $T(u,v)$ is bilinear.
	\item The Faulkner map is alternating $T(u,v) = - T(v,u)$ which follows from the symmetry and ad-invariance of $\left\langle -,-\right\rangle$:
		\begin{equation}
\left(T(u,v),x\right) = \left\langle x \cdot u, v  \right\rangle = -\left\langle  u, x \cdot v  \right\rangle %
= -\left\langle   x \cdot v, u  \right\rangle = - \left(T(v,u),x\right)
\end{equation}
	\item From this property, the skewsymmetry of the first two entries of the 3-bracket follows immediately:
	
	\begin{equation}
	[u,v,w] = T(u,v) \cdot w = - T(v,u) \cdot w = -[v,u,w]
	\end{equation}
	\item The special cases 3-Lie Algebras and Lie Triple Systems have a natural interpretation in terms of the fourth-rank tensor $\eR_R (u,v,w,x) := \left(T(u, v), T(w,x)\right)$ which belongs to $S^2 \Lambda^2 V \cong \Lambda^4 V \oplus V^{\yng(2,2)}$. In general $\eR_R$ has components in both representations, but for these special cases one of them vanishes. When $\eR_R \in \Lambda^4 V$, the 3-bracket is totally skewsymmetric, which defines 3-Lie Algebras. If $\eR_R \in V^{\yng(2,2)}$, this is equivalent to the Bianchi-like identity
\begin{equation}
  T(u,v)\cdot w + T(v,w)\cdot u + T(w,u)\cdot v = 0,
\end{equation}
or, in terms of the 3-bracket,
\begin{equation}
\label{eq:real-LTS}
  [u,v,w] + [v,w,u] + [w,u,v] = 0.
\end{equation}

Which defines Lie Triple Systems. These are linear approximations to Riemannian symmetric spaces and indeed the tensor $\eR_R$ is nothing but the Riemann curvature tensor.
	
\end{itemize}

\subsubsection{From the 3-algebra to the Lie algebra}

In this section we show how given a real metric 3-Leibniz algebra $\left(V, [-,-,-], \left\langle -,-\right\rangle\right)$ satisfying the fundamental identity and the unitarity and symmetry conditions, we can construct a metric Lie algebra from it and then $V$ is a unitary representation of it. 

The key idea is to identify the adjoint map $ad_{(u,v)} = [u,v,-]: V \to V$ with the map $T(u,v) = ad_{(u,v)}$ and then notice that the image of $T$, which lives in $\fgl(V)$, is indeed a Lie algebra because it is a Lie subalgebra of $\fgl(V)$ thanks to the fundamental identity. We call this Lie algebra $\fg = \im T$. 

Also, we show how we can define an inner product on $\fg$ which is indeed symmetric, non degenerate and ad-invariant.

\begin{proposition}
  Let $\fg = \im T$. Then $\fg < \fgl(V)$ is a Lie subalgebra.
\end{proposition}

\begin{proof}
  This is a consequence of the fundamental identity~\eqref{def:FI}. Indeed, written in terms of $T$, the fundamental identity reads
  \begin{equation*}
   T(u,v) T(v,w) t - T(w,s) T(u,v) t = T(T(u,v) w, s) t + T(w, T(u,v) s) t~,
  \end{equation*}
  which abstracting $t$ can be rewritten as
  \begin{equation}
    \label{eq:DD-CS}
    [T(u,v), T(w,s)] = T(T(u,v) w,s) + T(w, T(u,v) s)~,
  \end{equation}
  which is the $\fg$-equivariance of $T$ by just taking $T(u,v) = x$, any element in $\fg$, and shows that $[\fg,\fg] \subset \fg$.
\end{proof}

We define an inner product on $\fg$ denoted by $\left(-,-\right)$ by extending

\begin{equation}
  \label{eq:ip-CS}
  \left(T(u,v), T(w,s)\right) = \left<T(u,v)w, s\right>
\end{equation}
bilinearly to all of $\fg$.

\begin{proposition}
  The bilinear form on $\fg$ defined by \eqref{eq:ip-CS} is symmetric, nondegenerate and ad-invariant.
\end{proposition}

\begin{proof}
  The symmetry of the bilinear form \eqref{eq:ip-CS} (which is the property that guarantees that there is a notion of signature for this inner product) is precisely the symmetry condition which says that: 
  \begin{equation}
    \left\langle [u,v,w], s \right\rangle  = \left\langle  T(u,v)w,s \right\rangle = \left(T(u,v),T(w,s)\right)
  \end{equation}
  is equal to
  \begin{equation}
    \left\langle [w,s,u], v \right\rangle= \left\langle T(w,s)u,v \right\rangle = \left(T(w,s),T(u,v)\right)~.
  \end{equation}

  To prove nondegeneracy, let $\delta = \sum_i T(w_i,s_i)$ be such that $\left(\delta, T(u,v)\right) =0$ for all $u,v\in V$. This means that
  \begin{equation}
    \left\langle \delta u, v \right\rangle = 0\quad\forall u,v\in V~.
  \end{equation}
  Since the inner product on $V$ is nondegenerate, this means that $\delta u = 0$ for all $u\in V$, hence the endomorphism $\delta = 0$. 
		
		Finally, we prove the ad-invariance of the inner product:
 \begin{align*}
    \left(T(u,v),[T(w,s),T(t,r)]\right) &= \left(T(u,v),T(T(w,s) t,r) + T(r,T(w,s)t)\right) && \text{by \eqref{eq:DD-CS}}\\
    &= \left\langle T(u,v) T(w,s) t, r\right\rangle + \left\langle T(u,v) t, T(w,s)r\right\rangle && \text{by \eqref{eq:ip-CS}}\\
    &= \left\langle T(u,v) T(w,s) t, r\right\rangle - \left\langle T(w,s) T(u,v) t, r\right\rangle && \text{by \eqref{eq:unitarity-C}}\\
    &= \left\langle [T(u,v), T(w,s)] t,r\right\rangle\\
    &= \left([T(u,v), T(w,s)], T(t,r)\right) && \text{again by \eqref{eq:ip-CS}}~.
  \end{align*}
\end{proof}

It is worth stressing that the bilinear form $(-,-)$ in \eqref{eq:ip-CS} depends on the inner product $\left\langle  -,- \right\rangle$ on $V$ and is distinct from the Killing form on $\fg$. For the case of metric 3-Lie algebras, the distinction between these two objects has been noted already by Gustavsson in \cite{Gustavsson1loop}. It is the bilinear form in \eqref{eq:ip-CS} which appears in the Chern-Simons term in the superconformal field theories associated with these 3-algebras, rather than the Killing on $\fg$ that is sometimes assumed. 

Combining the results of these two sections we have proved the following:

\begin{theorem}
  There is a one-to-one correspondence between isomorphism classes of real metric 3-Leibniz algebras $V$ and isomorphic classes of pairs $(\fg, V)$
  where $\fg$ is a metric real Lie algebra and $V$ is a faithful orthogonal representation.
\end{theorem}

Where by \textbf{isomorphism} we just mean a linear map $\varphi: V \to W$ of real metric 3-Leibniz algebras such that for all $u,v,w\in V$,
\begin{equation*}
  [\varphi u, \varphi v, \varphi w]_W = \varphi[u,v,w]_V \qquad\text{and}\qquad \left<\varphi u, \varphi v\right>_W = \left<u,v\right>_V.
\end{equation*}

Note that this last condition induces a Lie algebra isomorphism $\fso(V) \to \fso(W)$ by $x \mapsto \varphi \circ x \circ \varphi^{-1}$ which relates the Lie algebras $\fg_V$ and $\fg_W$ by $\fg_W = \varphi \circ \fg_V \circ \varphi^{-1}$. We then say that pairs $(\fg_V,V)$ and $(\fg_W,W)$ are \textbf{isomorphic} if there is an isometry $\varphi: V \to W$ which relates $\fg_V < \fso(V)$ and $\fg_W < \fso(W)$ by $\fg_W = \varphi \circ \fg_V \circ \varphi^{-1}$ and such that this is an isometry of metric Lie algebras. Hence, isomorphic 3-algebras give rise to isomorphic pairs, and conversely.

\subsubsection{Some examples}

\begin{example}\label{eg:A4}
  First of all, we consider the original Euclidean 3-Lie algebra in \cite{BL2}, denoted $S_{0,4}$. The underlying vector space is $\RR^4$ with the standard
  Euclidean structure. Relative to an orthonormal basis $(e_1,\dots,e_4)$, the 3-bracket is given by
  \begin{equation}
    [e_i,e_j,e_k] =  \sum^{4}_{\ell=1} \epsilon_{ijk\ell} e_\ell~.
  \end{equation}
	
  The Faulkner Lie algebra $\fg = \im T$ is spanned by generators $T_{ij} := T(e_i, e_j)$, for $i<j$. It is not hard to see that these six generators are
  linearly independent, hence they must span all of $\fso(4)$, which is the gauge algebra of the original Bagger-Lambert model.
  Unlike the Killing form on $\fso(4)$, the inner product $(-,-)$ induced by the 3-algebra structure
  \begin{equation}
    \left( T_{ij}, T_{k\ell} \right) =  \left< [e_i, e_j, e_k], e_\ell \right>
  \end{equation}
  has indefinite signature. Indeed, for the six generators $T_{ij}$, the only nonzero inner products are
  \begin{equation}
    \left( T_{12}, T_{34} \right) = \left( T_{14}, T_{23} \right) = \left( T_{13}, T_{42} \right) = 1~,
  \end{equation}
  hence the signature is split. Recalling that $\fso(4) \cong \fsu(2) \oplus \fsu(2)$, this manifests itself in the fact that the levels of the Chern-Simons terms coming from the different $\fsu(2)$
  factors in $\fso(4)$ have opposite signs, as seen, for example, in \cite{VanRaamsdonkBL}.
\end{example}

Lets consider now another example of a 3-Lie algebra, but this time not positive-definite. 

\begin{example}\label{eg:lor3lie}
  Let $\fs$ be a semisimple Lie algebra with a choice of ad-invariant inner product. We let $W(\fs)$ denote the Lorentzian 3-Lie algebra with underlying
  vector space $V = \fs \oplus \RR u \oplus \RR v$ with inner product extending the one on $\fs$ by declaring $u,v \perp \fs$ and $\left<u,v\right> = 1$
  and $\left<v,v\right>=0=\left<u,u\right>$. The nonzero 3-brackets are given by
  \begin{equation*}
    [u,x,y] = [x,y] \qquad\text{and}\qquad [x,y,z] = - \left<[x,y],z\right> v~,
  \end{equation*}
  for all $x,y,z\in\fs$. 
	
	The Faulkner Lie algebra $\fg$ is the Lie algebra of inner derivations of the 3-Lie algebra, which was termed $\ad V$ in
  \cite{Lor3Lie}. As shown for example in that paper, this algebra is isomorphic to $\fs \ltimes \fs_{\text{ab}}$ with generators $A_x=T(x_1,x_2)$ and
  $B_x=T(u,x)$ for $x\in\fs$ and where in the definition of $A_x$, we have $x=[x_1,x_2]$. Since $\fs$ is semisimple, $[\fs,\fs]=\fs$ and hence such
  $x_1,x_2$ can always be found and moreover $A_x$ is independent of the precise choice of $x_1,x_2$ solving $x=[x_1,x_2]$. The nonzero Lie brackets of
  $\fg$ are given by
  \begin{equation*}
    [A_x,A_y] = A_{[x,y]} \qquad\text{and}\qquad [A_x, B_y] = B_{[x,y]}~.
  \end{equation*}
	
  The inner product is then given by
  \begin{align*}
    \left(A_x,A_y\right) &= \left<[x_1,x_2,y_1],y_2\right> = - \left<x,y_1\right> \left<v,y_2\right> = 0\\
    \left(A_x,B_y\right) &= \left<[x_1,x_2,u],y\right> = \left<x,y\right>\\
    \left(B_x,B_y\right) &= \left<[u,x,u],y\right> = 0~,
  \end{align*}
  hence we see that the inner product is again split, as expected.
\end{example}

Lets consider now an example which is not of 3-Lie type: the $\eC_{2d}$ 3-algebras in \cite{CherSaem}. 

\begin{example}\label{eg:CS}
  The underlying vector space $V$ of $\eC_{2d}$ is the real vector space of off-diagonal hermitian $2d \times 2d$ matrices:
  \begin{equation*}
    V = \left\{\begin{pmatrix} 0 & A \\ A^* & 0 \end{pmatrix} \middle | A \in \Mat_d(\CC)  \right\}~,
  \end{equation*}
  with $A^*$ the hermitian adjoint and with scalar product given by the trace of the product, which agrees with (twice) the real part of the natural
  hermitian inner product on the space of complex $d\times d$ matrices:
  \begin{equation*}
    \left<\begin{pmatrix} 0 & A \\ A^* & 0 \end{pmatrix}, \begin{pmatrix} 0 & B \\ B^* & 0 \end{pmatrix}\right> = \tr (AB^* + A^*B) = 2 \Re \tr A^*B~.
  \end{equation*}
	
  The 3-bracket is defined by $[x,y,z] := [[x,y]\tau,z]$, where $\tau = \begin{pmatrix} 1 & 0 \\ 0 & -1\end{pmatrix}$, for all $x,y,z\in V$ and where
  $[-,-]$ is the matrix commutator. Letting $x = \begin{pmatrix} 0 & A \\ A^* & 0\end{pmatrix}$, $y = \begin{pmatrix} 0 & B \\ B^* & 0\end{pmatrix}$
  and $z = \begin{pmatrix} 0 & C \\ C^* & 0\end{pmatrix}$, a calculation reveals that
  \begin{equation*}
    [x,y,z] = 
    \begin{pmatrix}
      0 & (A B^* - B A^*) C + C (A^* B - B^* A) \\
      (B^* A - A^* B) C^* + C^* (B A^* - A B^*) & 0
    \end{pmatrix}~,
  \end{equation*}
  hence the action of $T(x,y)$ on $z$ is induced from the action of $T(x,y)$ on $C$, which is a linear combination of
  left multiplication by the skewhermitian $d\times d$ matrix $A B^* - B A^*$ and right multiplication by
  the skewhermitian $d\times d$ matrix $A^* B - B^* A$. Let us decompose them into traceless plus scalar matrices as follows
  \begin{equation*}
    A B^* - B A^* = S(A,B) + \tfrac{i}{d}\alpha(A,B) 1_d \qquad\text{and}\qquad
    A^* B - B^* A = S(A^*,B^*) - \tfrac{i}{d}\alpha(A,B) 1_d~, 
  \end{equation*}
  where $S(A,B)$ and $S(A^*,B^*)$ are traceless, hence in $\fsu(d)$, and $\alpha(A,B) = 2 \Im\tr A B^* \in \RR$. Notice as well that
  \begin{equation*}
    (A B^* - B A^*) C + C (A^* B - B^* A)  = S(A,B) C + C S(A^*,B^*)~,
  \end{equation*}
  hence the Lie algebra $\fg$ is isomorphic to $\fsu(d) \oplus \fsu(d)$, acting on $V$ as the underlying real representation of
  $(\boldsymbol{d},\overline{\boldsymbol{d}}) \oplus (\overline{\boldsymbol{d}},\boldsymbol{d})$, which we denote
  $\rf{(\boldsymbol{d},\overline{\boldsymbol{d}})}$. In other words, to the generalised metric 3-Lie algebra $\eC_{2d}$ one can associate the pair
  $(\fsu(d) \oplus \fsu(d), \rf{(\boldsymbol{d},\overline{\boldsymbol{d}})})$. Notice finally that the invariant inner product on $\fsu(d) \oplus
  \fsu(d)$ has split signature, being given by $\left(X_L\oplus X_R, Y_L\oplus Y_R\right) = \tr (X_L Y_L - X_RY_R)$, for $X_L,Y_L,X_R,Y_R \in \fsu(d)$.

  It is possible to modify this example in order to construct 3-algebras denoted $\eC_{m+n}$, where the underlying vector space is the space of
  hermitian $(m+n) \times (m+n)$ matrices of the form
  \begin{equation*}
    \begin{pmatrix}
      0 & A \\ A^* & 0
    \end{pmatrix}~,
  \end{equation*}
  where $A$ is a complex $m\times n$ matrix and its hermitian adjoint $A^*$ is therefore a complex $n \times m$ matrix. Then, the
  same construction as above gives rise to a generalised metric 3-Lie algebra to which one may associate the pair $(\fsu(m) \oplus \fsu(n) \oplus
  \fu(1), \rf{(\boldsymbol{m},\overline{\boldsymbol{n}})})$. The inner product on $\fsu(m) \oplus \fsu(n) \oplus \fu(1)$ is again indefinite, given by
  $\left(X_L\oplus X_R, Y_L\oplus Y_R\right) = \tr (X_L Y_L - X_RY_R)$, for $X_L,Y_L \in \fsu(m)$ and $X_R,Y_R \in \fsu(n)$, whereas the inner product
  on the $\fu(1)$ factor is either positive-definite or negative-definite depending on whether $m<n$ or $m>n$, respectively. If $m=n$ we are
  back in the original case, in which the $\fu(1)$ factor is absent since it acts trivially on the space of matrices.
\end{example}

\begin{example}
Finally, a further explicit example of real 3-Leibniz algebra appears in \cite{Nambu:1973qe} and is built out of the octonions. It was shown
in \cite{Yamazaki:2008gg} that it satisfies the axioms (CS1)-(CS3). This example may be deconstructed into the pair $(\fg_2, \OO)$, where $\OO$ denotes
the octonions, which is a reducible, faithful, orthogonal representation of $\fg_2$. The resulting 3-algebra has a nondegenerate centre spanned by
$1\in\OO$. Quotienting by the centre gives another generalised metric 3-Lie algebra associated with the pair $(\fg_2, \Im\OO)$, with $\Im\OO$ the
7-dimensional representation of imaginary octonions.
\end{example}

\subsection{The complex case}
\label{sec:FaulkC}
\subsubsection{From the Lie algebra to the 3-algebra}

Let $\left(\fg, [-,-], (-,-)\right)$ be a metric real Lie algebra and $\left(\VV, h( -,-) \right)$ a faithful complex unitary representation of $\fg$ as defined in \eqref{def:CunitRep}. 

Recall that to define the complex Faulkner map $\TT$ \eqref{eq:T-map-C} we need to compute first the complexification of $\fg$. After doing that, one can define a complex Faulkner map and 3-bracket which will prove to define a complex metric 3-Leibniz algebra on $\VV$ with inner product the hermitian form $h( -,-)$ compatible with this 3-bracket. 

Let $\fg_\CC = \CC \otimes_{\RR}\fg \cong \fg \oplus i \fg$ denote the complexification of $\fg$, turned into a complex Lie algebra by extending the Lie bracket on $\fg$ complex bilinearly. We also extend the inner product $(-,-)$ complex bilinearly. As it remains nondegenerate, it turns $\fg_\CC$ into a complex metric Lie algebra. Furthermore we extend the action of $\fg$ on $\VV$ to $\fg_\CC$ by $(x+ i y)\cdot v = x\cdot v + i y\cdot v$, for all $x,y \in \fg$ and $v\in \VV$. Recall now that a complex Lie algebra cannot leave a hermitian inner product invariant, as explained in section \ref{sec:CUrep} we have instead:

\begin{equation}
  h(\XX \cdot v , w) + h(v, \XXbar\cdot w) = 0~,\qquad\text{for all $\XX \in \fg_\CC$ and $v,w\in \VV$.}
\end{equation}

Then we have: 

\begin{lemma}
  $\VV$ remains a faithful representation of $\fg_\CC$.
\end{lemma}

\begin{proof}
   We want to show that the action of $\fg_\CC$ on $\VV$ is injective or, equivalently, if $\XX = x + i y \in \fg_\CC$ is such that $\XX \cdot v = 0$ for all $v\in \VV$, then  $\XX = 0$. 
	
Taking the complex conjugate of that equation says that $\XXbar \cdot v = 0$ for all $v\in \VV$. Therefore the real and imaginary parts of $\XX$ satisfy the same equation: $x \cdot v
  = 0$ and $u \cdot v = 0$ for all $v\in \VV$. Since $x,y\in \fg$ and $\fg$ acts faithfully on $\VV$, $x=y=0$ and hence $\XX = 0$.
\end{proof}

We can now define the complex Faulkner map and the 3-bracket:

\begin{equation}
\label{eq:T-C}
  \left(\TT(u,v),\XX\right) = h(\XX \cdot u, v),
\end{equation}

\begin{equation}
\label{def:compl3br}
\TT(u,v) \cdot w := [\![u,v,w]\!],
\end{equation}
$\forall u, v, w \in \VV$. 

In order to prove that this defines a metric 3-Leibniz algebra we need, as before two auxiliary lemmas.

\begin{lemma}\label{le:imIdeal-C}
  The Faulkner map $\TT$ is $\fg_\CC$-equivariant, that is:
	  \begin{equation}
    \label{eq:TT-g-equiv}
    [\XX,\TT(v,w)] = \TT(\XX\cdot v, w) + \TT(v, \XXbar\cdot w)~
  \end{equation}
\end{lemma}

\begin{proof}
The proof follows along the same lines as that of Lemma~\ref{le:imIdeal}. 
  Let $\XX\in \fg_\CC$ and $v,w\in \VV$. Then for all $\YY \in \fg_\CC$ we have
  \begin{align*}
    \left([\TT(v,w),\XX],\YY\right) &= \left(\TT(v,w),[\XX,\YY]\right) &&\text{since $\fg_\CC$ is metric}\\
    &= h([\XX,\YY]\cdot v, w) &&\text{by \eqref{eq:T-C}}\\
    &= h(\XX \cdot \YY \cdot v, w) - h(\YY \cdot \XX \cdot v, w) && \text{since $V$ is a $\fg_\CC$-module}\\
    &= - h(\YY \cdot v, \XXbar \cdot w) - h(\YY \cdot \XX \cdot v, w) && \text{by \eqref{eq:pre-unitarity-C}}\\
    &= - \left(\TT(v,\XXbar \cdot w), \YY \right) - \left(\TT(\XX\cdot v, w),\YY\right) && \text{again by \eqref{eq:T-C}}~,
  \end{align*}
  hence abstracting $\YY$,
  \begin{equation}
    \label{eq:ideal-C}
    [\XX,\TT(v,w)] = \TT(\XX\cdot v, w) + \TT(v, \XXbar\cdot w)~.
  \end{equation}
	
	This proves that $\TT$ is $\fg_\CC$-equivariant and also that its image is an ideal of $\fg_\CC$.
	
\end{proof}
	
\begin{lemma}\label{le:T-onto}
  $\TT$ is surjective onto $\fg_\CC$.
\end{lemma}

\begin{proof}

  Again, it is enough to prove that the orthogonal complement of the image of $\TT$ is empty. Let $\XX \in \fg_\CC$ be perpendicular to the image of $\TT$. This means that $\left(\XX,\TT(v,w)\right) = 0$ for all $v,w\in \VV$, or equivalently   that $h(\XX\cdot v, w) = 0$ for all $v,w\in \VV$. Nondegeneracy of $h$ implies that $\XX\cdot v = 0$ for all $v\in \VV$, which in turn   implies that $\XX=0$ since the action of $\fg_\CC$ on $V$ is still faithful. Nondegeneracy of the inner product on $\fg_\CC$ says that $(\im D)^\perp   = 0$, which implies surjectivity.
\end{proof}

The main proposition now follows. 

\begin{proposition}
  The 3-bracket $[\![u,v,w]\!]: = \TT(u,v) \cdot w$ on the complex vector space $\VV$  defines a complex metric 3-Leibniz algebra with hermitian inner product $h( -,-)$.
\end{proposition}

\begin{proof}

We need to prove that the bracket satisfies the fundamental identity \eqref{eq:FI-C} and the unitarity \eqref{eq:unitarity-C} and symmetry \eqref{eq:symmetry-C} conditions of a complex 3-Leibniz algebra. 

Again the fundamental identity is basically the $\fg_\CC$-equivariance of $\TT$, whereas the other two conditions follow as before from the fact that $\overline{\TT(u,v)} = - \TT(v,u)$ and that for all $\XX \in \fg_\CC$, we have that
\begin{equation}
\label{eq:C-ad-inv}
  h(\XX\cdot u, v) = - h(u, \overline{\XX} \cdot v),
\end{equation}
and from the fact that the inner product $(-,-)$ on $\fg_\CC$ that has been obtained by extending that on $\fg$ complex bi-linearly is also symmetric:

\begin{equation}
  \label{eq:complex-symmetry}
  \left(\TT(u, v), \TT(w,x)\right) = h(\TT(w,x)\cdot u, v) = h(\TT(u,v)\cdot w, x),
\end{equation}

\end{proof}

\textit{Some remarks}

In section \ref{sec:CUrep} we had proved some properties of the complex Faulkner map that now imply the following properties of the 3-bracket:

\begin{itemize}

	\item Seeing $\VV$ as a real vector space $V$ with complex structure $I$, the complex Faulkner map can be written in terms of the real one as $\TT(u,v) = T(u,v) + i T(u, Iv)$.
	\item $\overline{\TT(u,v)} = - \TT(v,u)$
	\item $\TT$ is sesquilinear. 
	\item $[\![-,-,-]\!]$ is sesquibilinear. Follows immediately from $\TT$ sesquilinear.
	\item Compatibility with the quaternionic structure says that $J \circ \TT(u,Jv) = \overline{\TT(u,Jv)} \circ J$, where $\overline{\TT(u,Jv)} = -\TT(Jv,u)$, which in turn implies for the 3-bracket that $J[\![u,Jv,w]\!] = - [\![Jv,u,Jw]\!]$.
	\item The special cases Jordan Triple Systems and anti-Jordan Triple Systems have again a natural interpretation in terms of the fourth-rank tensor $\eR_C (u,v,w,x) := \left(\TT(u, v), \TT(w,x)\right) \in S^2 (\VV \otimes \bar{\VV}) \cong \left(S^2 \VV \otimes S^2\bar{\VV}\right) \oplus \left(\Lambda^2 \VV \otimes \Lambda^2 \bar{\VV} \right)$. The first extreme case is $\eR_C (u,v,w,x) \in S^2\VV \otimes S^2\bar{\VV}$, which corresponds to those $\VV$ where
\begin{equation}
  \label{eq:T-HSS}
  \TT(u,v)\cdot w = \TT(w,v)\cdot u 
\end{equation}
or equivalently where $[\![u,v,w]\!] = [\![w,v,u]\!]$. Such a bracket defines on $\VV$ the structure of a \textbf{Jordan triple system} (JTS) \cite{JacobsonLTS} and we will say such a representation $\VV$ is \textbf{JTS}. 

The other special class is $\eR_C \in \Lambda^2 \VV \otimes \Lambda^2\bar{\VV}$, which corresponds to those $\VV$ where
\begin{equation}
  \label{eq:T-BL4}
  \TT(u,v)\cdot w = - \TT(w,v)\cdot u
\end{equation}
or, equivalently, where $[\![u,v,w]\!] = - [\![w,v,u]\!]$. Such a 3-bracket defines precisely \textbf{anti-Jordan triple system} (see, e.g., \cite[Remark~4.3]{FaulknerFerrarAJP}). They are the relevant representations for $N=6$ supersymmetry and the skewsymmetry condition \eqref{eq:T-BL4} identifies this class of triple systems with those used by Bagger and Lambert in \cite{BL4} to recover the $N=6$ theories discovered by Aharony, Bergman, Jafferis and Maldacena in \cite{MaldacenaBL}.

\end{itemize}

\subsubsection{From the 3-algebra to the Lie algebra}

We show now how given a complex metric 3-Leibniz algebra $\left(\VV, [\![-,-,-]\!], h(-,-)\right)$ satisfying the fundamental identity and the unitarity and symmetry conditions, we can construct a real metric Lie algebra from it. We will do this in two steps. First, the Lie algebra that corresponds naturally to a complex 3-Leibniz algebra is a complex one. Again, it will be just the image of the complex Faulkner map $\fg_\CC = \im \TT$. However this algebra is not metric in the strict sense since it can not be ad-invariant because a complex Lie algebra cannot preserve a hermitian inner product, but it satisfies instead an analogous property. Then we will see how we can define a real one from and an orthogonal inner product that is ad-invariant.

The fact that $\fg_\CC = \im \TT$ is a Lie algebra follows again from $\fg$-equivariance of $\TT$.

\begin{lemma}
  The image of $\TT$ is a Lie subalgebra of $\fgl(\VV)$.
\end{lemma}

\begin{proof}
  By definition,
  \begin{align*}
    [ \TT(u,v), \TT(w,s) ] t &= \TT(u,v) [\![w,s,t]\!] - \TT(w,s) [\![u,v,t]\!]\\
    &= [\![u,v, [\![w,s,t]\!]]\!] - [\![w,s,[\![u,v,t]\!]]\!]\\
    &= [\![[\![u,v,w]\!],s,t]\!] - [\![w,[\![v,u,s]\!],t]\!] && \text{by \eqref{eq:FI-C}.}
  \end{align*}
	
  In terms of $\TT$, this equation becomes
  \begin{equation}
    \label{eq:DD-BL4}
    [ \TT(u,v), \TT(w,s) ] t = \TT([\![u,v,w]\!],s) t - \TT(w,[\![v,u,s]\!])t~.
  \end{equation}
	
  Finally, abstracting $t$ we see that the image of $\TT$ closes under the commutator and is hence a Lie subalgebra of $\fgl(V)$.
\end{proof}

$\fg_\CC$ cannot preserve a hermitian inner product, but rather the notion of unitarity for complex Lie algebras says that
\begin{equation*}
  h(x \cdot u,v) + h(u, c(x) \cdot w) = 0~,
\end{equation*}
where $c$ is a \textbf{conjugation} on $\fg$; that is, $c$ is a complex antilinear, involutive ($c^2 = Id$) automorphism of $\fg$. One can define a real Lie algebra from $\fg_\CC$ and define on it an orthogonal inner product with the standard ad-invariance condition.

The properties of $c$ guarantee that its fixed subspace
\begin{equation*}
  \fg := \left\{x \in \fg \middle | c(x) = x\right\} < \fu(V)
\end{equation*}
is a real Lie algebra, said to be a \textbf{real form} of $\fg_\CC$, which then does leave $h$ invariant.

This suggests defining $c: \fg_\CC \to \fg_\CC$ by $c \TT(u,v) = - \TT(v,u)$. From $\TT( i u, v) = i \TT(u,v)$ and $\TT(u, i v) = -i \TT(u,v)$ for all $u,v\in \VV$, it follows that $c(\TT(i u, v)) = c(i \TT(u,v))$, whereas $-\TT(v,i u) = i \TT(v,u) = -i c(\TT(u,v))$, which shows that $c$ can be extended to a complex antilinear map to all of $\fg_\CC$. Moreover, $c$ is involutive. Finally, it remains to show that $c$ is an automorphism. To do this we
compute
\begin{align*}
  [c \TT(u,v),c \TT(w,s)] &= [-\TT(v,u), -\TT(s,w)]\\
  &= [\TT(v,u), \TT(s,w)]\\
  &= \TT(\TT(v,u)s, w) - \TT(s, \TT(u,v)w)\\
  &= c( \TT(\TT(u,v)w, s) - \TT(w, \TT(v,u)s))\\
  &= c[\TT(u,v), \TT(w,s)]~.
\end{align*}

The real form $\fg$ is then spanned by $T(u,v):= \TT(u,v) - \TT(v,u)$ for all $u,v\in \VV$. For example, if $(\be_a)$ is a complex basis for $\VV$, then
$\fg$ is spanned by $\TT(\be_a, \be_b)- \TT(\be_b,be_a)$ and $i(\TT(\be_a,\be_b) + \TT(\be_b,\be_a))$. We will now show that $\fg$ is metric.

\begin{proposition}\label{pr:IPonG}
  The bilinear form on $\fg$ defined by
  \begin{equation*}
    \left(T(u,v), T(w,s)\right) := \Re h(T(u,v)w,s)~.
  \end{equation*}
  is symmetric, nondegenerate and ad-invariant.
\end{proposition}

\begin{proof}
  We prove each property in turn. To prove symmetry we simply calculate:
  \begin{align*}
    h(T(u,v)w, s) &= h(\TT(u,v)w, s)- h(\TT(v,u)w, s)\\
    &= h([\![u,v,w]\!], s) - h([\![v,u,w]\!], s)\\
    &= h(u,[\![s,w,v]\!]) - h(v,[\![s,w,u]\!]) &&\text{using \eqref{eq:symmetry-C}}\\
    &= h(u,\TT(s,w)v) - h(v, \TT(s,w)u)\\
    &= h(\TT(w,s)u, v) - \overline{h(\TT(s,w)u, v)}~,
  \end{align*}
  hence taking real parts we find
  \begin{equation*}
    \Re h(T(u,v)w,s) = \Re h (T(w,s) u,v)~.
  \end{equation*}

  To prove nondegeneracy, let us assume that some linear combination $x = \sum_i T(x_i, y_i)$ is orthogonal to all $T(u,v)$ for $u,v\in V$:
  \begin{equation*}
    \Re h(x \cdot u, v) = 0~.
  \end{equation*}
	
  Now, since $h$ is nondegenerate, so is $\Re h$ because $\Im h(u,v) = \Re h(-i u,v)$, hence this means that $x \cdot u = 0$ for all $u$, showing
  that the endomorphism $x =0$.

  Finally, we show that it is ad-invariant. A simple calculation using $\fg$ equivariance of $\TT$ shows that
  \begin{equation*}
    [T(u,v), T(w,s)] = T(T(u,v)w,s) + T(w, T(u,v)s)~,
  \end{equation*}
  hence
  \begin{align*}
    \left(T(s,t),[T(u,v),T(w,r)]\right) &= \left(T(s,t), T(T(u,v)w, r) + T(w, T(u,v)r)\right)\\
    &= \re h(T(s,t)T(u,v) w, r) + \re h( T(s,t)w, T(u,v) r)\\
    &= \re h(T(s,t)T(u,v) w, r) - \re h( T(u,v) T(s,t)w, r)\\
    &= \re h([T(s,t), T(u,v)] w, r) \\
    &= \left([T(s,t), T(u,v)], T(w,r)\right)~.
  \end{align*}
\end{proof}

We have extracted a metric real Lie algebra $\left(\fg, (-,-) \right)$ and a unitary representation $(\VV,h, \left\langle -,-\right\rangle)$ from a complex metric 3-Leibniz algebra. 

In summary, we have proved the following:

\begin{theorem}
  There is a one-to-one correspondence between isomorphism classes of complex unitary 3-Leibniz algebras $\VV$ and isomorphic classes of pairs $(\fg, \VV)$
  where $\fg$ is a metric real Lie algebra and $\VV$ is a faithful complex unitary representation.
\end{theorem}

\subsubsection{An example}

We shall illustrate this construction with the explicit example in \cite{BL4} and see that the Faulkner Lie algebra $\fg$ is the gauge algebra in the $N=6$ theory that appeared originally in \cite{MaldacenaBL}. 

\begin{example}\label{eg:BL4}
  Let $V$ denote the space of $m \times n$ complex matrices. For all $u,v,w \in \VV$, define
  \begin{equation*}
    [\![u,v,w]\!] := u v^* w - w v^* u~,
  \end{equation*}
  where $v^*$ denotes the hermitian adjoint of $v$. The generators of $\fg$ are $T(u,v) = \TT(u,v) - \TT(v,u)$, where $\TT(u,v) \cdot w = [\![u,v,w]\!] = [w,u;v]$, in the notation of  \cite{BL4}.
  The action of $T(u,v)$ on $\VV$ is given by
  \begin{align*}
    T(u,v) w &= \TT(u,v) w - \TT(v,u) w\\
    &= u v^* w - w v^* u - v u^* w + w u^* v \\
    &= w (u^* v - v^* u) + (u v^* - v u^*) w
  \end{align*}
  hence it consists of a linear combination of left multiplication by the skewhermitian $m\times m$ matrix $u v^*-v u^*$ and right multiplication by
  the skewhermitian $n\times n$ matrix $u^* v-v^* u$. Let us decompose them into traceless + scalar matrices as follows
  \begin{equation*}
    u v^* - v u^* = A(u,v) + \tfrac{i}{m}\alpha(u,v) 1_m \qquad\text{and}\qquad
    u^* v - v^* u = B(u,v) - \tfrac{i}{n}\alpha(u,v) 1_n~, 
  \end{equation*}
  where $A(u,v)$ and $B(u,v)$ are traceless, hence in $\fsu(m)$ and $\fsu(n)$, respectively, and $\alpha(u,v) = 2 \Im\tr u v^*$. Into the action of
  $T(u,v)$ on $w$, we find
  \begin{equation*}
    T(u,v) w = A(u,v) w + w B(u,v) + i \alpha(u,v) (\tfrac1m - \tfrac1n) w~.
  \end{equation*}
	
  The hermitian inner product on $\VV$ is given by $h(u,v) = 2 \tr u v^*$, where the factor of $2$ is for later convenience. The invariant inner product
  on $\fg$ in proposition~\ref{pr:IPonG} is given by polarizing
  \begin{equation*}
    \left(T(u,v),T(u,v)\right) = \tr A(u,v)^2 - \tr B(u,v)^2  - \alpha(u,v)^2  (\tfrac1m - \tfrac1n)~.
  \end{equation*}
	
  Therefore we see that if $m=n$, then $\fg_\CC \cong \fsu(n) \oplus \fsu(n)$ acting on $\VV \cong \CC^n \otimes (\CC^n)^*$, which is the bifundamental
  $(\boldsymbol{n},\overline{\boldsymbol{n}})$. The inner product is $\left(X_L\oplus X_R, Y_L\oplus Y_R\right) = \tr (X_L Y_L) - \tr (X_R
  Y_R)$, for $X_L,Y_L,X_R,Y_R\in\fsu(n)$, with the traces in the fundamental. If $m\neq n$, then $\fg_\CC \cong \fsu(m) \oplus \fsu(n) \oplus \fu(1)$,
  which is the quotient of $\fu(m) \oplus \fu(n)$ by the kernel of its action on $V$. The inner product on the semisimple part is the same as in the
  case $m = n$, whereas the inner product on the centre is positive-definite (resp. negative-definite) accordingly to whether $m<n$
  (resp. $m>n$).
\end{example}

\subsection{The quaternionic case}
\label{sec:FaulkH}
\subsubsection{From the Lie algebra to the 3-algebra}

Let $\left(\fg, [-,-], (-,-)\right)$ be a metric real Lie algebra and $\left(\VV_H, \omega( -,-) \right) = \left(\VV,J, \omega( -,-) \right)$ a faithful quaternionic unitary representation of $\fg$ as defined in \ref{sec:QUrep}. We see this case as a special case of the complex one. The reason being that, in the case of an honest quaternionic representation, the dualising procedure employed here would map to a quaternionification of the Lie algebra, but such an object does not exist; although see \cite{MR1877855} for a possibly related concept.

We can, however, define a quaternionic Faulkner map $\TT_H$ by:

\begin{equation}
  \left(\TT_H(u,v), \XX\right) = \omega(\XX\cdot u, v)
\end{equation}
to build a quaternionic 3-bracket. Recalling that $\TT_H(u,v) = \TT(u,Jv)$, we have that this map goes from $\VV \times \VV \to \fg_\CC$ where $\fg_\CC$ is the complexification of $\fg$. Also, from this relation we deduce that the map $\TT_H$ is complex bilinear. The resulting 3-bracket is:

\begin{equation}
\label{def:quat3br}
\TT_H(u,v) \cdot w := [\![u,v,w]\!]_H,
\end{equation}
$\forall u, v, w \in \VV$.

\begin{proposition}
  The 3-bracket $[\![u,v,w]\!]_H: = \TT_H(u,v) \cdot w$ on the vector space $\VV$ equipped with quaternionic structure $J$  defines a quaternionic metric 3-Leibniz algebra with symplectic form $\omega(u,v) = h(u,Jv)$.
\end{proposition}

\begin{proof}
The new bracket satisfies the fundamental identity \eqref{eq:FI-H}, the symplectic \eqref{eq:symplectic-H} and symmetry \eqref{eq:symmetry-H} conditions inherited from the properties of the complex 3-bracket. 
\end{proof}

\textit{Some remarks}

\begin{itemize}

	\item Since $\TT_H$ is complex bilinear, we have $[\![u,v,w]\!]_H$ that is complex trilinear.
	\item $\TT_H(u,Jv) = \TT_H(v,Ju)$, which implies symmetry under exchange of the first two entries:  $[\![u,v,w]\!]_H =  [\![v,u,w]\!]_H$
	\item $\overline{\TT_H(u,v)} =- \TT_H(Ju,Jv) $
	\item $J \circ \TT_H(u,v) = \overline{\TT_H(u,v)} \circ J$
	\item $[\![-,-,-]\!]_H$ is complex tri-linear. Follows immediately from $\TT$ sesquilinear and $J$ complex anti-linear.
	\item We obtain again the two special cases from the fourth-rank tensor $\eR_H (u,v,w,x) := \left(\TT_H(u, v), \TT_H(w,x)\right) \in S^2 S^2 \VV_H \cong \VV_H^{\yng(2,2)} \oplus S^4 \VV_H$. When $\eR_H \in S^4 \VV_H$, the 3-bracket is totally symmetric. This corresponds to \textbf{quaternionic triple systems}. The other special class corresponds to the case $\eR_H \in \VV_H^{\yng(2,2)}$, or equivalently,
	
\begin{equation}
  \label{eq:quat-aLTS}
  \TT_H(u,v) \cdot w + \TT_H(v,w) \cdot u + \TT_H(w,u) \cdot v = 0,
\end{equation}
or in terms of the 3-bracket
\begin{equation}
\label{eq:quat-aLTS-3b}
  [\![u,v,w]\!]_H + [\![v,w,u]\!]_H + [\![w,u,v]\!]_H = 0.
\end{equation}

Equivalently, taking the symmetry condition into account, we may write the above two conditions as
\begin{equation}
  \TT_H(u,u) \cdot u = 0 \qquad\text{and}\qquad [\![u,u,u]\!]_H = 0,
\end{equation}
respectively, for all $u \in \VV_H$. Either of these conditions defines a \textbf{(quaternionic) anti-Lie triple system}. They are the relevant representations for the $N=4,5$ theories.	
\end{itemize}

\subsubsection{From the 3-algebra to the Lie algebra}

Again, this case is just a special case of the complex one, hence all the theorems are inherited from it. Given a quaternionic unitary 3-Leibniz algebra $(\VV,J, [\![-,-,-]\!]_H, h, \omega)$ we extract first a complex Lie algebra $\fg_\CC = \im \TT(u,Jv)$ and then a \textbf{real} one $\fg$ spanned by $T(u,v):= \TT(u,Jv) - \TT(v,Ju)$. The latter is metric, as the orthogonal inner product:

  \begin{equation}
    \left(T(u,v), T(w,s)\right) := \Re h(T(u,v)w,s)
  \end{equation}
is ad-invariant.

Finally, $(\VV, h)$ is a complex-unitary representation of $\fg$ and, since $\VV$ possesses a quaternionic structure map $J$ which, by definition, preserves the 3-bracket, and a symplectic form  $\omega$ which is ad-invariant, we have that $(\VV,J, \omega)$ is a complex-quaternionic representation of $\fg$.

In summary, combining the last two sections based on results inherited from the complex case, we have proved the following:

\begin{theorem}
  There is a one-to-one correspondence between isomorphism classes of quaternionic unitary 3-Leibniz algebras $\VV_H$ and isomorphic classes of pairs $(\fg, \VV_H)$
  where $\fg$ is a metric real Lie algebra and $\VV_H$ is a faithful quaternionic unitary representation.
\end{theorem}

\section{Lie superalgebra embeddings}
\label{sec:LSAembed}

Of the six extreme cases of 3-Leibniz algebras, three play an active role in dual theories to M2-branes: 3-Lie algebras (3LA), anti Jordan triple systems (aJTS) and quaternionic anti Lie triple systems (aLTS). As it turns out, they share another characteristic: they can be embedded in a Lie superalgebra. This provides yet another way to describe these objects: either as ternary algebras, Lie algebras with a unitary representation or as a Lie superalgebra. 

Moreover, in the case where the 3-Leibniz algebra (or representation of the Lie algebra) $V$ is \textbf{positive-definite}, the notions of simplicity of the 3-algebra, of the Lie superalgebra and irreducibility of the representation are equivalent. However, this notion does not imply simplicity of the corresponding Lie algebra $\fg$. We will see in chapter 5 that this is very important for highly supersymmetric SCCSM and, indeed, in none of the cases that give rise to $N>4$ supersymmetry is the Lie algebra simple. This fact might explain in part why these theories took a relatively long time to be discovered. In most theories physicists have encountered in the past, one could, without loss of generality, require the gauge algebra to be simple, because if it were semisimple the theory would decompose in separate theories in each of the simple factors. This assumption is wrong for Bagger-Lambert theories. Restricting to the case of simple Lie algebras only, it is impossible to find these theories with $N>4$ supersymmetry, because there aren't any with simple gauge Lie algebra.

The other three special cases (Lie triple systems, Jordan triple systems and quaternionic triple systems) can be embedded in graded Lie algebras. For example, it is well-known in the mathematical literature that Lie triple systems can be embedded in a 2-graded metric Lie algebra and Jordan triple systems in a 3-graded one. The situation is summarised in table \ref{tab:special}. In that table and in the rest of the section we use the same notation used in \ref{sec:relat-betw-real} and let $U$, $V$ and $W$ stand for a real, complex or quaternionic representation, respectively. With the distinction, however, that for us quaternionic representations are always complex representations in the image of $r'$.

\begin{table}[ht!]
  \centering
  \begin{tabular}{|c|>{$}c<{$}|>{$}c<{$}|l|}
    \hline
    Class & \text{type} & \eR & \multicolumn{1}{|c|}{Embedding Lie (super)algebra}\\\hline
    LTS & \RR & U^{\yng(2,2)} & 2-graded metric Lie algebra\\
    JTS & \CC & S^2V \otimes S^2\Vbar & 3-graded complex metric Lie algebra\\
    QTS & \HH & S^4W & 3-graded complex metric Lie algebra\\
    aLTS & \HH & W^{\yng(2,2)} & complex metric Lie superalgebra\\
    aJTS & \CC & \Lambda^2V \otimes \Lambda^2\Vbar & 3-graded complex metric Lie superalgebra\\
    3LA & \RR & \Lambda^4 U & 3-graded metric Lie superalgebra\\
    \hline
  \end{tabular}
  \vspace{8pt}
  \caption{Lie-embeddable unitary representations of metric Lie algebras}
  \label{tab:special}
\end{table}

Before diving into the Lie embeddings, we want to point out that some of these extreme cases are related to each other. In section \ref{sec:some-relat-betw} we review these relations. Then, we focus on the three extreme cases that are of physical importance for Bagger-Lambert theories. In section \ref{sec:extremeLSA}, we recover the construction of Lie superalgebras that appeared in detail in \cite{Lie3Algs} for the case of aJTS and aLTS representations (see theorem~22 for the aJTS case and the discussion around equation (45) for the aLTS case). We also state for completeness simplicity results from \cite{Palmkvist} and Figueroa-O'Farrill's work appeared in \cite{JMFSimplicity}. In section \ref{sec:extremegLA} we review the well-known results for the remaining three exceptional cases on how they can be embedded in graded Lie algebras.

\subsection{Some relations between extreme cases}
\label{sec:some-relat-betw}

Some of the special cases are related to each other via the maps in section~\ref{sec:relat-betw-real} consistent with the requirements of supersymmetry of the corresponding Chern-Simons theory. Let $\Dar(\fg,\KK)_{\text{C}}$ denote the unitary representations of $\fg$ of type $\KK$ and class $\text{C}$, where $\KK = \RR, \CC, \HH$ and $\text{C}$ can be either 3LA, LTS, aJTS, JTS, aLTS or QTS.

\begin{proposition}
\label{prop:relations}
  The following relations hold:
  \begin{enumerate}
  \item $V \in \Dar(\fg,\CC)_{\text{aJTS}} \iff \Vbar \in \Dar(\fg,\CC)_{\text{aJTS}}$
  \item $V \in \Dar(\fg,\CC)_{\text{JTS}} \iff \Vbar \in  \Dar(\fg,\CC)_{\text{JTS}}$
  \item $V \in \Dar(\fg,\CC)_{\text{aJTS}} \Leftarrow \rf{V} \in \Dar(\fg,\RR)_{\text{3LA}}$
  \item $V \in \Dar(\fg,\CC)_{\text{JTS}} \iff \rf{V} \in \Dar(\fg,\RR)_{\text{LTS}}$
  \item $U_\CC \in \Dar(\fg,\CC)_{\text{JTS}} \implies U$ is trivial
  \item $U \in \Dar(\fg,\RR)_{\text{3LA}} \iff U_\CC \in \Dar(\fg,\CC)_{\text{aJTS}}$
  \item $W \in \Dar(\fg,\HH)_{\text{QTS}} \iff \rh{W} \in \Dar(\fg,\CC)_{\text{JTS}}$
  \item $\rh{W} \in \Dar(\fg,\CC)_{\text{aJTS}} \implies W$ is trivial
  \item $V \in \Dar(\fg,\CC)_{\text{aJTS}} \iff V_\HH \in \Dar(\fg,\HH)_{\text{aLTS}}$
  \end{enumerate}
\end{proposition}

\begin{proof}
  First of all (1) and (2) follow because in both cases the fourth-rank tensor $\eR$ lives in a self-conjugate representation: $\Lambda^2V \otimes \Lambda^2\Vbar$ or $S^2V \otimes S^2\Vbar$, respectively.

  To prove (3), let $V \in \Dar(\fg,\CC)$ and let $U = \rf{V} \in \Dar(\fg,\RR)_{\text{3LA}}$. The relation between $\TT$ on $V$ and $T$ on $\rf{V}$ is given by lemma~\ref{le:T-map-C}. It follows that
  \begin{align*}
    \TT(u,v)\cdot w &= T(u,v) \cdot w + T(u,Iv)\cdot Iw \\
    &= T(u,v) \cdot w + I T(u,Iv)\cdot w &&\text{since $I$ is $\fg$-invariant}\\
    &= - T(w,v) \cdot u - I T(w,Iv)\cdot u && \text{since $r(V) \in \Dar(\fg,\RR)_{\text{3LA}}$}\\
    &= - T(w,v) \cdot u - T(w,Iv)\cdot I u && \text{since $I$ is $\fg$-invariant}\\
    &= - \TT(w,v) \cdot u,
  \end{align*}
  whence $V \in \Dar(\fg,\CC)_{\text{aJTS}}$.

  We prove (4) along similar lines. In one direction, let $\rf{V} \in \Dar(\fg,\RR)_{\text{LTS}}$ and use lemma~\ref{le:T-map-C} and the $\fg$-invariance of $I$ to calculate
  \begin{equation*}
    \TT(u,v)\cdot w - \TT(w,v)\cdot u = T(u,v)\cdot w + I T(u,Iv)\cdot w - T(w,v)\cdot u - I T(w,Iv)\cdot u.
  \end{equation*}
  The first and third terms and the second and fourth terms combine, using equation \eqref{eq:real-LTS}, to produce
  \begin{equation*}
    \TT(u,v)\cdot w - \TT(w,v)\cdot u = - T(w,u) \cdot v - I T(w,u) \cdot Iv,
  \end{equation*}
  which vanishes due to the $\fg$-invariance of the complex structure $I$. In the other direction, from lemma~\ref{le:T-map-C} we see that $T(u,v)$ is the real part of $\TT(u,v)$:
  \begin{equation*}
    T(u,v) = \half \left(\TT(u,v) - \TT(v,u)\right).
  \end{equation*}

Writing down the LTS condition \eqref{eq:real-LTS} in full, we find
  \begin{equation*}
		\begin{aligned}[m]
    T(u,v) \cdot w  &+ T(v,w) \cdot u +   T(w,u) \cdot v = \half \left(\TT(u,v) - \TT(v,u) \right) \cdot w  \\
    &+ \half \left(\TT(v,w) - \TT(w,v) \right) \cdot u + \half \left(\TT(w,u) - \TT(u,w) \right) \cdot v,
		\end{aligned}
 \end{equation*}
  which cancels pairwise using the JTS condition \eqref{eq:T-HSS}.
  
  To prove (5), let $V = U_\CC \in \Dar(\fg,\CC)$. Then $V$ admits a real structure $R$ compatible with the hermitian structure $h(Ru,Rv) = h(v,u)$, from where it follows that the map $\TT: V \times V \to \fg_\CC$ obeys $\TT(Ru,Rv) = \overline{\TT(u,v)}$. Under the action of $R$, $V$ decomposes as $V = U \oplus i U$, where $U$ and $iU$ are the real eigenspaces of $R$ with eigenvalues $\pm1$, respectively. It follows that if $u,v\in U$ then $\TT(u,v)$ is real and, since $\overline{\TT(u,v)} = - \TT(v,u)$, it is skewsymmetric. Hence it defines a real alternating map $U \times U \to \fg$. This map is seen to be the map $T$ in \eqref{eq:T-map-R}, since compatibility of $R$ with $h$ says that $h$ on $U$ is real, so that it agrees with the inner product $\left<-,-\right>$ on $U$. Now let $w \in U$ and consider $\TT(u,v)\cdot w$. If $V \in \Dar(\fg,\CC)_{\text{JTS}}$, then in particular $\TT(u,v)\cdot w = + \TT(w,v)\cdot u$, but since $\TT(u,v) \cdot w = - \TT(v,u)\cdot w$ we see that $\TT(u,v) \cdot w=0$. 
  
  If, on the contrary, $V = U_\CC \in \Dar(\fg,\CC)_{\text{aJTS}}$, then $\TT(u,v) \cdot w = - \TT(w,v)\cdot u$, hence it is totally skewsymmetric and $U \in \Dar(\fg,\RR)_{\text{3LA}}$, proving the reverse implication in (6). To finish proving (6), notice that if $U \in \Dar(\fg,\RR)_{\text{3LA}}$, then $\TT(u,v) \cdot w = - \TT(w,v) \cdot u$ for all $u,v,w \in U$. Now use the sesquibilinearity of $\TT(u,v) \cdot w$ to show that this is satisfied for all $u,v,w\in V$.
  
  To prove (7) simply notice that if $\rh{W} \in \Dar(\fg,\CC)_{\text{JTS}}$, so that the map $\TT$ satisfies the JTS condition \eqref{eq:T-HSS}, then in particular
  \begin{equation*}
    \TT(u,Jv) \cdot w = + \TT(w,Jv) \cdot u,
  \end{equation*}
  which together with the symmetry condition $\TT(u,Jv) = \TT(v,Ju)$ says that $\TT(u,Jv)\cdot w$ is totally symmetric, whence $W \in \Dar(\fg,\HH)_{\text{QTS}}$. The argument is clearly reversible, so we get both implications.
  
  If instead $\rh{W} \in \Dar(\fg,\CC)_{\text{aJTS}}$, then $\TT(u,Jv)\cdot w$ is symmetric in $u\leftrightarrow v$ but skewsymmetric in $u\leftrightarrow w$, whence it has to vanish, which says that $\rh{W}$ and hence $W$ is a trivial representation. This proves (8).
  
  Finally, let us prove (9). Let $W = V_\HH$. Recall that we do not work with $W$ but with its image $\rh{W}$ under $r'$, which from proposition~\ref{prop:diamond}(9) is given by $\rh{V_\HH} \cong V \oplus \Vbar$. We will denote vectors in $\Vbar$ by $\overline v$, for $v \in V$. Then the quaternionic structure $J$ on $V \oplus \Vbar$ is defined by
  \begin{equation*}
    J v = \overline v \qquad\text{and}\qquad J \overline v = - v \qquad \text{for all $v \in V$}.
  \end{equation*}
  The hermitian structure on $V \oplus \Vbar$ is given by the hermitian structures on $V$ and $\Vbar$ and declaring the direct sum to be orthogonal. The complex symplectic structure on $V \oplus \Vbar$ is such that $V$ and $\Vbar$ are Lagrangian submodules\footnote{In this case being Lagrangian just means $\omega(V, V) =0$ and $\omega(\overline V, \overline V) =0$, since $V$ and $\overline V$ already have half the dimension of the total space.} and
  \begin{equation*}
    \omega(u, \overline v) = - h(u,v).
  \end{equation*}
  The only nonzero components of the map $\TT$ are
  \begin{equation}
    \label{eq:Propi}
    \TT(u,J\overline v) = - \TT(u,v).
  \end{equation}
  The aLTS condition \eqref{eq:quat-aLTS} is satisfied if and only if
  \begin{equation*}
    \TT(u,J\overline v) \cdot w +   \TT(\overline v, Jw) \cdot u +   \TT(w, Ju) \cdot \overline v = 0.
  \end{equation*}
  The last term vanishes since $V$ is a Lagrangian submodule, hence the aLTS condition is equivalent to
  \begin{equation*}
    \TT(u,J\overline v) \cdot w +   \TT(\overline v, Jw) \cdot u = 0,
  \end{equation*}
  which using equation \eqref{eq:Propi} is equivalent to
  \begin{equation*}
    \TT(u,v) \cdot w + \TT(w, v) \cdot u = 0,
  \end{equation*}
  which is equivalent to the aJTS condition \eqref{eq:T-BL4} on $V$.
\end{proof}

\subsection{Extreme cases embeddable in Lie superalgebras}
\label{sec:extremeLSA}
Lets first recall the definition of a Lie superalgebra.

\begin{definition}
A \textbf{Lie superalgebra} is a $Z_2$-graded algebra $L= L_0 \oplus L_1$ together with a product $[-, -]$ called the Lie superbracket or \textbf{supercommutator} that satisfies:
	\begin{itemize}
	\item grading $[L_i,L_j]\subseteq L_{i+j  (mod 2)}$, 
	\item super skew-symmetry $[x,y]=-(-1)^{|x| |y|}[y,x]$ and
	\item the super Jacobi identity 
	$ (-1)^{|z| |x|}[x,[y,z]]+(-1)^{|x| |y|}[y,[z,x]]+(-1)^{|y| |z|}[z,[x,y]]=0$ 
	$\forall x, y, z \in L$. 
	\end{itemize}
		Where $|x|$ denotes the degree of $x$ (either 0 or 1) and the degree of [x,y] is the sum of degree of x and y modulo 2. $L_0$ is called \textit{even part} of the algebra and $L_1$ \textit{odd part}.
\end{definition}

The key idea to embed triple systems in Lie superalgebras is to take as the even part the Faulkner Lie algebra $\fg$ and as the odd part the representation $V$. The Lie bracket on $\fg$ defines the Lie superbracket on the even part and the action of $\fg$ on $V$ defines the Lie superbracket $[L_0,L_1] \subseteq L_1$. Then one can use the Faulkner map $T$ to define a bracket that takes $V \times V \to \fg$, hence respecting the grading.

A Lie superalgebra can also be graded, as long as the new grading is consistent with the pre-existing  $Z_2$-grading. In the following we will encounter 3-graded Lie superalgebras. That means that $L= W_{-1}\oplus W_0 \oplus W_1$ and $[W_i,W_j]\subseteq W_{i+j}$. To be consistent with the $Z_2$-grading it is enough, for example, that $L_0 = W_0$ and $L_1 = W_{-1}\oplus W_1$.

\subsubsection{aJTS embedded in 3-graded metric Lie superalgbras}
Let $V \in \Dar(\fg,\CC)_{\text{aJTS}}$. Then on the 3-graded vector space $V \oplus \fg_\CC \oplus \Vbar$ we define the structure of a complex 3-graded Lie superalgebra using the Lie algebra structure on $\fg_\CC$, the action of $\fg_\CC$ on $V$ and $\Vbar$ and the map $\TT$ defined in equation \eqref{eq:T-map-C}, but thought of as a complex bilinear map $\TT: V \times \Vbar \to \fg_\CC$. The identity \eqref{eq:T-BL4} that defines the aJTS corresponds now to the one component of the Jacobi identity which is not already trivially satisfied by the construction. The resulting complex 3-graded Lie superalgebra is metric relative to the inner product on $\fg_\CC$ and the symplectic structure on $V \oplus \Vbar$ defined by declaring $V$ and $\Vbar$ to be Lagrangian subspaces and $\left(u,\bar v\right) = h(u,v)$. This complex Lie superalgebra is the complexification of a metric Lie superalgebra with underlying vector space $\fg \oplus \rf{V}$ and with inner product defined by the one on $\fg$ together with the imaginary part of the hermitian inner product on $V$, which is a symplectic structure on $\rf{V}$. This construction appeared already in \cite{Lie3Algs} and was considered further in \cite{Palmkvist} and \cite{JMFSimplicity}. 

\subsubsection{aLTS embedded in metric Lie superalgbras}

Similarly, let $W\in \Dar(\fg,\HH)_{\text{aLTS}}$. Consider the 2-graded complex vector space $\fg_\CC \oplus W$, with $\fg_\CC$ in degree 0 and $W$ in degree 1. We define the Lie bracket by extending the one on $\fg_\CC$ and the action of $\fg_\CC$ on $W$ by $[u,v] = \TT(u,Jv)$ for $u,v \in W$. Then the identity \eqref{eq:quat-aLTS} that defines the aLTS is the one component of the Jacobi identity for a complex Lie superalgebra which is not automatically satisfied in the case of any $W \in \Dar(\fg,\HH)$. This Lie superalgebra is metric relative to the inner product on $\fg_\CC$ and to the complex symplectic form $\omega$ on $W$. This construction appeared in \cite{Lie3Algs} already and was considered further in \cite{JMFSimplicity}.

\subsubsection{3LA embedded in 3-graded metric Lie superalgbras}
Finally, we discuss the case of 3LA representations. It follows from proposition~\ref{prop:relations}(6) that if $U \in \Dar(\fg,\RR)_{\text{3LA}}$, then its complexification $V = U_\CC \in \Dar(\fg,\CC)_{\text{aJTS}}$. By theorem~22 in \cite{Lie3Algs} and using that $\Vbar \cong V$ in this case, we may define a 3-graded metric Lie superalgebra structure on $V \oplus \fg_\CC \oplus V$. Furthermore, since $V$ here is the complexification of a real representation, this complex Lie superalgebra is the complexification of a metric Lie superalgebra which, unlike in the general case of aJTS representations, is also 3-graded. We can see this explicitly as follows. Consider the 3-graded real vector space $U_{-1} \oplus \fg_0 \oplus U_1$, where the subscripts reflect the degree. For every $u\in U$, we will write $u_1$ and $u_2$, respectively, the corresponding vectors in $U_1$ and $U_{-1}$. We write $u_a$ generically, where $a=1,2$. Then we define the following Lie brackets in addition to the ones of $\fg$:
\begin{equation}
  [X, u_a] := (X\cdot u)_a \qquad\text{and}\qquad [u_a, v_b] := \epsilon_{ab} T(u,v),
\end{equation}
with $\epsilon_{ab}$ the Levi-Civita symbol with $\epsilon_{12} = 1$, say. The inner product is defined to be the one on $\fg$ extended by
\begin{equation}
  \left(u_a, v_b\right) = \epsilon_{ab} \left<u,v\right>.
\end{equation}

The Jacobi identity \eqref{eq:Jacobi} is then satisfied and the resulting inner product is ad-invariant.

Conversely, given any 3-graded metric Lie superalgebra $U_{-1} \oplus \fg_0 \oplus U_1$ with $U_1$ and $U_{-1}$ both isomorphic to an orthogonal representation $U$ of $\fg$, then the 3-bracket $[u,v,w]$ on $U$ defined by
\begin{equation}
  [[u_a,v_b],w_c] = \epsilon_{ab} [u,v,w]_c,
\end{equation}
defines a metric 3-Lie algebra structure on $U$.

In summary, we have the following characterisation of metric 3-Lie algebras.

\begin{theorem}
  \label{th:3LAchar}
  Metric 3-Lie algebras $\left(U,[-,-,-],\left<-,-\right>\right)$ are in one-to-one correspondence with metric 3-graded Lie superalgebras $U_{-1} \oplus \fg_0 \oplus U_1$, where $U_1$ and $U_{-1}$ are both isomorphic to $U$, a faithful orthogonal representation of $\fg$.
\end{theorem}

\begin{example}
  \label{ex:S4embedding}
  As shown by Ling \cite{LingSimple} there is a unique complex simple 3-Lie algebra. There is a unique real form of this 3-Lie algebra which is metric relative to a positive-definite inner product. The corresponding vector space is $\RR^4$ with the standard Euclidean inner product and $\fg  = \fso(4)$ the Lie algebra of skewsymmetric endomorphisms, with inner product  given under the isomorphism $\fso(4) = \fsu(2) \oplus \fsu(2)$ by the Killing form on on the first $\fso(3)$ and the negative of the Killing form on the second. The corresponding 3-graded Lie superalgebra is a ``compact'' real form of $A(1,1)$ in the Kac classification \cite{KacSuperSketch}. Notice that the Killing form of $A(1,1)$ vanishes identically, but here we see that it does nevertheless have a non-degenerate inner product.
\end{example}

\subsubsection{Simplicity}
\label{sec:simplicity}

We have seen above that to every \textit{Lie-embeddable representation}\footnote{We call Lie-embeddable representation to the 6 extreme cases of representations for a metric Lie algebra 3LA, LTS, JTS, aJTS, QTS and aLTS.} of a metric Lie algebra one can attach a triple system and a Lie (super)algebra. In principle, there are three separate notions of simplicity or irreducibility one can consider: irreducibility of the representation, simplicity of the embedding Lie (super)algebra and simplicity of the triple system - this latter one being defined as the non-existence of proper ideals in the triple system, ideals being defined as kernels of homomorphisms. 

For the case of \emph{positive-definite} aJTS representations, this has been discussed recently in \cite{Palmkvist} and from the present point of view in Figueroa-O'Farrill's work \cite{JMFSimplicity}, where LTS and aLTS representations are also treated. The following theorems are proved in \cite{JMFSimplicity}; except for the first which was already known. In that paper anti-Jordan triple systems had not yet been identified for what they were, and they are referred to as $N=6$ triple systems instead. 

We find useful to include here Figueroa-O'Farrill's results for their relevance when identifying indecomposable Bagger-Lambert theories, as we will see in chapter \ref{sec:3dscfts}.

\begin{theorem}
  \label{thm:LTS-simplicity}
  Let $\fg$ be a metric Lie algebra, $U \in \Dar(\fg,\RR)_{\text{LTS}}$ faithful and positive-definite and let $\fk = \fg \oplus U$ denote its embedding 2-graded Lie algebra. The following are equivalent:
  \begin{enumerate}
  \item $U \in \Irr(\fg,\RR)_{\text{LTS}}$,
  \item $U$ is a simple Lie triple system,
  \item $\fk$ is a simple Lie algebra or else $U \cong \fg$ (as representations) and $\fg$ is a simple Lie algebra.
  \end{enumerate}
\end{theorem}

\begin{theorem}
  \label{thm:BL4-simplicity}
  Let $\fg$ be a metric Lie algebra, $V \in \Dar(\fg,\CC)_{\text{aJTS}}$ faithful and positive-definite and let $\fk = V \oplus \fg_\CC \oplus \Vbar$ denote its embedding 3-graded Lie superalgebra. The following are equivalent:
  \begin{enumerate}
  \item $V \in \Irr(\fg,\CC)_{\text{aJTS}}$,
  \item $V$ is a simple anti-Jordan triple system,
  \item $\fk$ is a simple Lie superalgebra.
  \end{enumerate}
\end{theorem}

\begin{theorem}
  \label{thm:aLTS-simplicity}
  Let $\fg$ be a metric Lie algebra, $W \in \Dar(\fg,\HH)_{\text{aLTS}}$ positive-definite and let $\fk = \fg_\CC \oplus W$ denote its embedding Lie superalgebra. The following are equivalent:
  \begin{enumerate}
  \item $W \in \Irr(\fg,\HH)_{\text{aLTS}}$,
  \item $W$ is a simple quaternionic anti-Lie triple system,
  \item $\fk$ is a simple Lie superalgebra.
  \end{enumerate}
\end{theorem}

Similar results can be proved also for the Jordan triple systems, but as they do not play such an important role in the study of superconformal Chern-Simons theories, we will not mention them here. It was shown in corollary (6) of \cite{SCCS3Algs} that there are no positive-definite QTS representations, hence this question does not arise in this case. There do exist, however, indefinite QTS associated to  hyperkäler manifolds which have been classified \cite{MR1913815}. Finally, example \ref{ex:S4embedding} shows that the same result holds for the unique positive-definite $U \in \Irr(\fg,\RR)_{\text{3LA}}$, whose associated triple system is the unique positive-definite nonabelian simple 3-Lie algebra in \cite{Filippov} and which embeds in the simple Lie superalgebra $A(1,1)$.

The above results allow a classification of positive-definite irreducible representations which are  Lie-embeddable. This is summarised in table \ref{tab:irreducible-LE} of chapter \ref{sec:3dscfts} for those of classes 3LA, aJTS and aLTS, as these are the ones relevant for the study of three-dimensional superconformal Chern-Simons-matter theories. The other positive-definite Lie-embeddable classes are associated with the Riemannian and hermitian symmetric spaces and that classification is classical and can be found, for example, in \cite{Helgason}.

\subsection{Extreme cases embeddable in graded Lie algebras}
\label{sec:extremegLA}

As mentioned above Lie triple systems, Jordan triple systems and quaternionic triple systems can be embedded in graded Lie algebras. Lets first recall what they are.

\begin{definition}
A \textbf{graded Lie algebra} is a Lie algebra endowed with a gradation which is compatible with the Lie bracket. In other words, it is a vector space $\fl$ with a gradation $\fl=\bigoplus_{i\in{\mathbb Z}} {\fl}_i $ together with a Lie bracket that satisfies the standard axioms of a Lie algebra and respects this gradation:
\begin{equation}
  [{\fl}_i,{\fl}_j]\subseteq {\fl}_{i+j}
\label{eq:LieBrgrad}
\end{equation}
  \end{definition}
	
\subsubsection[LTS embedded in 2-graded metric Lie algbras]{Lie triple systems embedded in 2-graded metric Lie algbras}

Let $U$ define a LTS or equivalently $U \in \Dar(\fg,\RR)_{\text{LTS}}$, then on the 2-graded vector space $\fg \oplus U$, with $\fg$ having degree 0 and $U$ having degree 1, one can define the structure of a graded metric Lie algebra in the following way. The Lie bracket is given by the Lie bracket of $\fg$, the action of $\fg$ on $U$ and the Faulkner map $T$ defined by equation \eqref{eq:T-map-R}. Then identity \eqref{eq:real-LTS} is the one component of the Jacobi identity which is not implicit in the construction. The inner product consisting of the one on $\fg$ and the one on $U$, with both spaces being mutually perpendicular, is invariant under the adjoint action. Conversely, given any 2-graded metric Lie algebra, the degree-1 subspace as a representation of the degree-0 Lie subalgebra is an LTS representation.

\subsubsection[JTS embedded in 3-graded metric Lie algbras]{Jordan triple systems embedded in 3-graded metric Lie algbras}

Now let $V \in \Dar(\fg,\CC)_{\text{JTS}}$. In this case we can define on the 3-graded vector space $V \oplus \fg_\CC \oplus \Vbar$ - with degrees $-1,0,1$, respectively - the structure of a Lie algebra by adding to the Lie bracket on $\fg_\CC$ and the action of $\fg_\CC$ on $V$ and $\Vbar$, the Faulkner map $\TT$ defined in equation \eqref{eq:T-map-C}, but viewed here as a complex bilinear map $V \times \Vbar \to \fg_\CC$. Then the defining condition \eqref{eq:T-HSS} for a JTS representation implies the two components of the Jacobi identity which are not already trivially satisfied. The 3-graded Lie algebra $V \oplus \fg_\CC \oplus \Vbar$ is metric relative to the complex inner product defined by the one on $\fg_\CC$ and by $h$, thought of as a complex bilinear inner product $V \times \Vbar \to \CC$. Relative to this inner product, the subspaces $V$ and $\Vbar$ are isotropic abelian Lie subalgebras. These representations are in one-to-one correspondence with hermitian symmetric spaces. Indeed, the 3-graded Lie algebra $V \oplus \fg_\CC \oplus \Vbar$ is the complexification of a 2-graded real metric Lie algebra $\fg \oplus \rf{V}$, and the inner product is given by the one on $\fg$ together with the real part of the hermitian inner product on $V$.

\subsubsection[QTS embedded in 3-graded metric Lie algbras]{Quaternionic triple systems embedded in 3-graded metric Lie algbras}

Finally, if $W$ is a QTS representation of $\fg$, then by proposition~\ref{prop:relations}(7), $\rh{W}$ gives rise to a JTS representation, hence it admits an embedding Lie algebra with a 3-grading, but with both the subspaces of degree $\pm 1$ isomorphic to $\rh{W}$.

\chapter[Metric 3-Lie algebras]{Maximal supersymmetry and metric 3-Lie algebras}
\label{sec:3LieAlg}

In this chapter we study in detail one of the special classes of 3-Leibniz algebras: metric 3-Lie algebras, where the 3-bracket is totally antisymmetric. These ternary algebras are of particular importance because, in the recent history of M2-branes, they were the first ternary algebras to appear and also because they are the ones that give rise to maximal supersymmetry in Bagger-Lambert theories. 

Extensive study on the structure of 3-Lie algebras (and more generally n-Lie algebras) was done by Ling in his PhD thesis \cite{LingSimple} following the work of Filippov \cite{Filippov} and Kasymov \cite{kasymov1987theory}. In particular, he was able to classify all simple 3-Lie algebras. In this chapter we will reproduce some of their notation and results and take them as a starting point for our structure theorems.

To the best of our knowledge, 3-Lie algebras have not been classified. If we restrict ourselves to the class of \textit{metric} 3-Lie algebras a general classification still does not exist but we will prove a structure theorem that says how they are constructed and we do classify low index ones (index less than 2). Also, we are able to provide a full classification of a subclass of the metric ones: those admitting a maximally isotropic centre.

It is remarkable that the situation is very analogous to Lie algebras. They have not been classified either, not even in the metric case. However for the metric case there exists a structure theorem by Medina and Revoy \cite{MedinaRevoy} (see also work of Stanciu and Figueroa-O'Farrill \cite{FSalgebra}) which is analogous to the one we will prove for 3-Lie algebras. 

Also, metric Lie algebras admitting a maximally isotropic centre were classified by Kath and Olbrich \cite{KathOlbrich2p}. As it turns out, even though we were not aware of this result at the time, the method we use to classify 3-Lie algebras with a maximally isotropic centre is similar to theirs. Nevertheless, we will see that the 3-Lie algebra case is more involved and requires new ideas.

The content of this chapter is based on \cite{Lor3Lie, 2p3Lie, 2pBL}. On paper \cite{Lor3Lie}, we classified Lorentzian 3-Lie algebras on  \cite{Lor3Lie}, then we did the same with those of index 2 in \cite{2p3Lie} and in that paper we also proved the structure theorem. Carrying this classification to higher index proved to be an un-tamed problem. However, we realised that we could actually classify the class of 3-Lie algebras of physical interest (those with a maximally isotropic centre) following a completely different approach, that is not an increasing index strategy. The physical motivation for studying this class of 3-Lie algebras will be made clear in section \ref{sec:N8unitarity}, where we will argue that this is the necessary condition for the corresponding maximally supersymmetric theory to be \textit{unitarisable}.

This chapter is organised as follows. In the first section we prove a structure theorem on 3-Lie algebras that says that they are build out of one-dimensional and simple ones under the iteration of two operations that will be defined. We then use this to classify 3-Lie algebras of index less than 2. Finally, in section \ref{sec:max-iso-cent} we classify 3-Lie algebras with a maximally isotropic centre.

\section{Structure of metric 3-Lie algebras}

In order to classify metric 3-Lie algebras, it is enough to classify indecomposable ones since, by definition, any other algebra is a direct sum of those. To do this we first introduce some basic 3-algebraic concepts by analogy with the theory of Lie algebras in chapter \ref{ch:liealgebras}. Most of these concepts can be found in the foundational paper of Filippov \cite{Filippov}.

Recall that a (finite-dimensional, real) \textbf{3-Lie algebra} consists of a finite-dimensional, real vector space $V$ together with a 3-bracket, which is totally antisymmetric and satisfies the fundamental identity \eqref{def:FI}. We specify the totally skewsymmetric 3-bracket either by $[-,-,-]$ or by a map $\Phi: \Lambda^3 V \to V$.

 It is called \textbf{metric} if it admits a nondegenerate symmetric bilinear form $\left<-,-\right> :S^2 V \to \RR$,
 which is ad-invariant. 

Given two metric 3-Lie algebras $(V_1,\Phi_1,b_1)$ and $(V_2,\Phi_2,b_2)$, we may form their \textbf{orthogonal direct sum}
$(V_1\oplus V_2,\Phi_1\oplus \Phi_2, b_1 \oplus b_2)$, by declaring that
\begin{align*}
  [x_1,x_2,y] = 0 \qquad\text{and}\qquad
  \left<x_1,x_2\right> = 0~,
\end{align*}
for all $x_i\in V_i$ and all $y\in V_1 \oplus V_2$. The resulting object is again a metric 3-Lie algebra. 

A metric 3-Lie algebra is then said to be \textbf{indecomposable} if it is not isomorphic to an orthogonal direct sum of metric 3-Lie algebras $(V_1\oplus V_2, \Phi_1\oplus \Phi_2, b_1\oplus b_2)$ with $\dim V_i > 0$.

 It is clear that one dimensional 3-Lie algebras are indecomposable. Like for Lie algebras, we will see that a class of indecomposable 3-Lie algebras is that of simple ones. In fact, once again, the only other kind of indecomposable metric 3-Lie algebras are those defined as ``double extensions''. 

\subsection{Simple 3-Lie algebras}
\label{sec:simple3LA}

From now on let $(V,[-,-,-])$ be a 3-Lie algebra. Given subspaces $W_i\subset V$, we will let $[W_1 W_2 W_3]$ denote the subspace of $V$ spanned by elements of the form $[w_1,w_2,w_3] \in V$, where $w_i \in W_i$.

To define the concept of simplicity, we need to define first that of an ideal. Notice that we have a choice on how to extend the notion of ideal from that defined for Lie algebras \eqref{def:LAIdeal}. A subspace $I \subset V$ could be called an ideal, if $[I,V,V]\subset I$ or $[I,I,V]\subset I$. Both notions exist in the literature, but only the former matches with the concept of ideal in the theory of Lie algebras, in the sense that using that definition there exists a theorem analogous to the one for Lie algebras that says that there is a one-to-one correspondence between ideals and kernels of homomorphisms.

In summary, a subspace $I \subset V$ is an \textbf{ideal}, written $I \lhd V$, if $[I,V,V]\subset I$.

In particular, ideals are 3-Lie subalgebras, where a \textbf{subalgebra} is a subspace $W$ satisfying $[W,W,W] \subset W$. Again, this ensures that $W$ itself satisfies all the axioms to be a 3-Lie algebra.

The image of $\Phi: \Lambda^3 V \to V$,  $V' = [V,V,V] \subset V$, is an ideal called the \textbf{derived ideal} of $V$. Another ideal is provided by the \textbf{centre} $Z$, defined by
\begin{equation*}
  Z = \left\{z \in V\middle | [z,x,y]=0,~\forall x,y\in V \right\}~.
\end{equation*}
In other words, $[Z,V,V]=0$. More generally the \textbf{centraliser}
$Z(W)$ of a subspace $W \subset V$ is defined by
\begin{equation*}
  Z(W) = \left\{z \in V\middle | [z,w,y]=0,~\forall w\in W,y\in V
  \right\}~,
\end{equation*}
or equivalently $[Z(W),W,V]=0$ (thus $Z(V)=Z$). It follows from the fundamental identity
\eqref{def:FI} that $Z(W)$ is a subalgebra.

An ideal $I \lhd V$ is \textbf{minimal} if any other ideal $J \lhd V$ contained in $I$ is either $0$ or $I$. Conversely, an ideal $I \lhd V$ is \textbf{maximal} if any other ideal $J \lhd V$ containing $I$ is either $I$ or $V$.

A 3-Lie algebra is \textbf{simple} if it is not one-dimensional and every ideal $I\lhd V$ is either $0$ or $V$ (that is, it has no proper ideals) and \textbf{semisimple} if it is the direct sum of simple 3-Lie algebras. 

Ling studied simple n-Lie algebras in his thesis \cite{LingSimple} by studying their corresponding Faulkner Lie algebra, that is the Lie algebra defined by the image of the adjoint maps or, as he called it, the Lie algebra of inner derivations. He showed that any simple n-Lie algebra is indecomposable and also classified simple n-Lie algebras (both over the reals and over the complex).

In particular he showed that every real simple 3-Lie algebra is isomorphic to either one of the four-dimensional 3-Lie algebras defined, relative to a basis $\be_i$, by
  \begin{equation}
    \label{eq:simple-3-Lie}
    [\be_i,\be_j,\be_k] = \sum_{\ell=1}^4 \varepsilon_{ijk\ell}
    \lambda_\ell \be_\ell~,
  \end{equation}
  for some $\lambda_\ell$, all nonzero; or to $S_{4,4}$, which has real dimension 8 and relative to a basis ${\be_i, \be_{i+4}}$ for $i = 1,...,4$ has brackets:

\begin{equation}
\label{eq:simple-3-Lie-8D}
  \begin{aligned}[m]
		[\be_i,\be_j,\be_k] &= \sum_{\ell=1}^4 \varepsilon_{ijk\ell} \be_\ell\\
		[\be_{i+4},\be_{j+4},\be_{k+4}] &= \sum_{\ell=1}^4 \varepsilon_{ijk\ell} \be_{\ell+4}\\
		[\be_i,\be_j,\be_{k+4}] &= \sum_{\ell=1}^4 \varepsilon_{ijk\ell} \be_{\ell+4}\\\\
		[\be_i,\be_{j+4},\be_{k+4}] &= - \sum_{\ell=1}^4 \varepsilon_{ijk\ell} \be_\ell
		~,
	\end{aligned}
\end{equation}
where $1 \leq i,j,k,\ell \leq 4$. Which is the \textit{realification} of the single complex simple 3-Lie algebra. 

Let us now introduce an inner product, $b$, so that $(V,\Phi,b)$ is a
metric 3-Lie algebra. The concepts of orthogonal complement, nondegenerate subspace, isotropic and coisotropic subspaces are exactly the same as those defined for Lie algebras.

The four-dimensional simple 3-Lie algebras admit invariant metrics of any signature: Euclidean, Lorentzian or split as they leave invariant the diagonal metric with entries $(1/\lambda_1, 1/\lambda_2, 1/\lambda_3,
1/\lambda_4)$. One can further change to a basis where the $\lambda_i$ are signs.  

In particular this shows that up to homotethy (i.e. a rescaling of the inner product) there are three unique four-dimensional simple metric 3-Lie algebras with Euclidean, Lorentzian and split signatures, corresponding to choosing $\lambda_i$ to be $(1,1,1,1)$, $(-1,1,1,1)$ and $(-1,-1,1,1)$, respectively and that we denote $S_{0,4} := S_4, S_{1,3}$ and $S_{2,2}$. 

As explained for example in \cite{jin2009real} and up to homotethy and change of basis, there is one single eight-dimensional simple 3-Lie algebra, $S_{4,4}$ which has split signature $(4,4)$. 

In summary, there are, up to homotethy and change of basis, four real metric simple 3-Lie algebras: $S_{0,4}, S_{3,1}, S_{1,3}, $ and $S_{2,2}$, which are four-dimensional and have signatures Euclidean, Lorentzian and split; and $S_{4,4}$, which is eight-dimensional and has split signature. We want to point out here that this last case has escaped recent literature, including \cite{2p3Lie,Lor3Lie,2pBL}. This does not change the well-known statement in the context of M2-branes that there is a unique simple 3-Lie algebra which is positive-definite, since $S_{4,4}$ has split signature.

\subsection{A structure theorem on metric 3-Lie algebras}
\label{sec:structure}

We now investigate the structure of metric 3-Lie algebras. By definition, they are the direct sum of indecomposable ones. We know that one-dimensional and simple 3-Lie algebras are indecomposable, it only remains to see whether there are any other indecomposable ones. Such algebras, if they exist, posses at least one proper ideal (otherwise they would be simple) and hence also a minimal ideal $I$. The key idea is to exploit the properties of that ideal to learn more about the indecomposable 3-Lie algebra. We summarise here the steps to follow:

\begin{itemize}
\label{it:structure}
	\item We will see that $I$ is contained in its orthogonal complement $I^\perp$. This is only possible if the the inner product is not positive-definite. Hence the first conclusion we draw is that there are no other indecomposable Euclidean 3-Lie algebras except for the one-dimensional and simple ones. 
	\item We have then that $I \subset  I^\perp \subset V$. We will call $U$ the part of $V$ which is not in $I^\perp$ and $W$ the part of  $I^\perp$ which is not in $I$ so that $V = U \oplus W \oplus I$.
	\item We will  then see that $U$ has the same dimension as $I$ and is a simple or one dimensional 3-Lie algebra and that $W$ is also a metric 3-Lie algebra. The key to give these subspaces such structures is to identify them with certain quotients of 3-Lie algebras by their ideals, which have the structure of 3-Lie algebras.
	\item Finally we will write the possible 3-brackets on $V = U \oplus W \oplus I$ and argue that $V$ is the double extension of the metric 3-Lie algebra $W$ by a one-dimensional or simple 3-Lie algebra $U$. Where a double extension is defined as follows.
\end{itemize}

\begin{definition}\label{def:doublext}
  Let $W$ be a metric 3-Lie algebra and let $U$ be a 3-Lie algebra. Then by the \textbf{double extension of $W$ by $U$} we
  mean the metric 3-Lie algebra on the vector space $W \oplus U \oplus U^*$ with the following nonzero 3-brackets:
  \begin{itemize}
  \item $[UUU] \subset U$ being the bracket of the 3-Lie algebra $U$;
  \item $[UUU^*]\subset U^*$ being the 	action of $\ad U$ on $U^*$, the dual space to $U$;
  \item $[UUW] \subset W$ being the action of $\ad U$ on $W$;
  \item $[UWW] \subset W \oplus U^*$, where the $U^*$ component is related to the previous bracket by
    \begin{equation*}
      \left<[u_1,w_1,w_2],u_2\right>  =
      \left<[u_1,u_2,w_1],w_2\right>~.
    \end{equation*}
  \item $[WWW] \subset W \oplus U^*$, where the $W$ component is the bracket of the 3-Lie algebra $W$ and the $U^*$ component is     related to the $W$ component of the previous bracket by
    \begin{equation*}
      \left<[w_1,w_2,w_3],u_1\right>  =
      \left<u_1,w_1,w_2],w_3\right>~.
    \end{equation*}
  \end{itemize}
	
	  These brackets are subject to the fundamental identity. Two of these identities can be interpreted as saying that the bracket $[UUW]
  \subset W$ defines a Lie algebra homomorphism $\ad U \to \Der^0W$, where $\Der^0W$ is the Lie algebra of skewsymmetric derivations of the 3-Lie algebra $W$, whereas the map $\Lambda^3W \to U^*$ defining the $U^*$ component of the $[WWW]$ bracket is $\ad
  U$-equivariant. We have not found similarly transparent interpretations for the other Jacobi identities. The above
  3-brackets leave invariant the inner product on $V$ with components 
\begin{itemize} \item $\left<W,W\right>$, being the inner product on the metric 3-Lie algebra $W$;
  \item $\left<U,U^*\right>$, being the natural dual pairing; and
  \item $\left<U,U\right>$, being any $\ad U$-invariant symmetric bilinear form.
  \end{itemize}
\end{definition}

Furthermore, we will see that actually $I$ is abelian (or equivalently $[I\, I\, I]=0$) and, because $I$ is isotropic, from ad-invariance of the metric we have that $[I\, I\, I^\perp]=0$ also; so combining the two  $[I\, I\, V]=0$. 

The theorem that we are going to prove is as follows.

\begin{theorem}\label{th:indecmetric3lie}
  Every indecomposable metric 3-Lie algebra $V$ is either one-dimensional, simple or else it is the double extension of a
  metric 3-Lie algebra $W$ by a one-dimensional or simple 3-Lie algebra $U$.
\end{theorem}

Any metric 3-Lie algebra will be an orthogonal direct sum of indecomposables, each one being either one-dimensional, simple or
a double extension of a metric 3-Lie algebra, which itself is an orthogonal direct sum of indecomposables of strictly lower dimension.
Continuing in this way, we arrive at the following characterisation.

\begin{corollary}\label{co:metric3lie}
  The class of metric 3-Lie algebras is generated by the simple and one-dimensional 3-Lie algebras under the operations of orthogonal
  direct sum and double extension.
\end{corollary}

We proceed to prove this theorem and corollary by following the steps listed in \ref{it:structure}.

\subsubsection{Properties of the minimal ideal}

Let $V$ be an indecomposable metric 3-Lie algebra. Then $V$ is either simple, one-dimensional or it possesses a proper ideal $I$. Given that $V$ is finite-dimensional, the following lemma guarantees that there is always a minimal and a maximal proper ideal. 

\begin{lemma}
  If $I,J$ are ideals of $V$ then so are their intersection $I\cap J$ and their linear span $I + J$, defined as the smallest vector subspace containing their union $I \cup J$.
\end{lemma}

\begin{proof}
  Since $I\cap J \subset I$, $[I\cap J, V, V] \subset I$ and since $I\cap J \subset J$, $[I\cap J, V, V] \subset J$, hence $[I\cap J, V, V] \subset I \cap J$. Similarly, $[I+J,V,V] \subset [I,V,V] + [J, V, V] \subset I + J$.
\end{proof}

Lets take into account now that we are concerned with \textsl{metric} 3-Lie algebras. Then, for every ideal $I$ we can define its orthogonal complement $I^\perp$ and we have the following further properties.

\begin{proposition}\label{pr:ideals}
  Let $V$ be a metric 3-Lie algebra and $I \lhd V$ be an ideal. Then
	\begin{enumerate}
  \item $I^\perp \lhd V$ is also an ideal;
  \item $I^\perp\lhd Z(I)$; and
  \item if $I$ is minimal then $I^\perp$ is maximal.
  \end{enumerate}
\end{proposition}

\begin{proof}
  \begin{enumerate}
  \item For all $x,y\in V$, $u\in I$ and $v\in I^\perp$,
    $\left<[v,x,y],u\right> = - \left<[x,y,u],v\right> = 0$, since
    $[x,y,u]\in I$. Therefore $[v,x,y] \in I^\perp$.
  \item For all $u \in I^\perp$, $v\in I$ and $x,y\in V$, consider
    $\left<[u,v,x],y\right> = - \left<[x,y,v],u\right> =0$ since $I$
    is an ideal, which means that $[u,v,x]=0$, hence
    $[I,I^\perp,V]=0$.
  \item Let $J \supset I^\perp$ be an ideal. Taking perpendiculars,
    $J^\perp \subset I$. Since $I$ is minimal, $J^\perp = 0$ or
    $J^\perp = I$, hence $J = V$ or $J= I^\perp$ and $I^\perp$ is
    maximal.
  \end{enumerate}
\end{proof}

From now on, we call $I$ the \textit{minimal} ideal in $V$. If $I$ is nondegenerate, we have that $I \cap I^\perp = 0$ and hence $V$ is indecomposable. So, for indecomposability, $I$ must be either isotropic or coisotropic. Since $I$ is minimal it must be isotropic, $I \subset I^\perp$. Similarly, from proposition \eqref{pr:ideals} we know that $I^\perp$ is maximal and hence we have that $I^\perp$ is co-isotropic. 

Furthermore, also from proposition \eqref{pr:ideals} we have that $I^\perp\lhd Z(I)$. However, since the centralizer of $I$, $Z(I)$, is an ideal and $I^\perp$ is maximal $Z(I)$ must be all of $V$. This means that $I$ is abelian or equivalently $[I,I,I]=0$. 

Also, from ad-invariance of the metric and the definition of orthogonal complement it follows

\begin{equation}
\left\langle [I,I,I^\perp],I\right\rangle = -\left\langle [I,I,I],I^\perp\right\rangle = 0
\end{equation}

Since the inner product is nondegenerate, this implies $[I,I,I^\perp] = 0$, which combined with $I$ abelian implies $[I,I,V]=0$. 

To summarise so far, if $V$ is indecomposable it is either one-dimensional, simple or it posses a proper minimal ideal which is isotropic, abelian, satisfies $[I,I,V]=0$ and whose orthogonal complement, $I^\perp$ , is a maximal coisotropic ideal in $V$.

\subsubsection{Decomposition of V}

We define $U$ and $W$ as the part of $V$ which is not in $I^\perp$ and the part of  $I^\perp$ which is not in $I$ respectively, so that $V = U \oplus W \oplus I$. We will prove that $U$ can be identified with $\Ubar := V/I^\perp$ and $W$ with $\Wbar :=I^\perp/I$. This will be useful because quotients of 3-Lie algebras by their ideals have many known properties that we discuss now.

The main property of ideals, which in fact motivates their definition, is that the quotient of a 3-Lie algebra by one of its ideals is again a 3-Lie algebra. The proof is part of a bigger theorem, which is the 3-algebraic version of one of the three isomorphism theorems. We will only need a simplified version of the first one that appeared for the first time in \cite{Filippov}.

\begin{lemma}
\label{le:idker}
  There is a one-to-one correspondence between ideals and kernels of  homomorphisms.
\end{lemma}

\begin{proof}
  It follows immediately from its definition that the kernel of a homomorphism is an ideal. Conversely, if $I \lhd V$, then
  $V/I$ is a 3-Lie algebra with bracket
  \begin{equation*}
    [x+I, y+I, z+I] = [x,y,z] + I~,
  \end{equation*}
  and the canonical projection $V \to V/I$ is a homomorphism with
  kernel $I$.
\end{proof}

This tells us that both $V/I^\perp$ and $I^\perp/I$ are 3-Lie algebras. Furthermore, when the ideal we are quotienting by is maximal, the resulting algebra can only be simple or one-dimensional.

\begin{lemma}\label{le:simplequot}
  If $K\lhd V$ is a maximal ideal, then $V/K$ is simple or
  one-dimensional.
\end{lemma}

\begin{proof}
  Let $\pi: V \to V/K$ denote the natural surjection, suppose that
  $J\subset V/K$ is an ideal and let let $\pi^{-1}J = \left\{x \in
    V \middle | \pi(x) \in J\right\}$. Then $\pi^{-1}J$ is an
  ideal of $V$: $\pi[\pi^{-1}J,V,V] = [J, V/K, V/K] \subset
  J$, hence $[\pi^{-1}J,V,V] \subset \pi^{-1}J$. Since $K = \ker
  \pi$, $K$ is contained in $\pi^{-1}J$, but since $K$ is maximal
  $\pi^{-1}J=K$ or $\pi^{-1}J=V$. In the former case, $J =
  \pi\pi^{-1}J = \pi K = 0$ and in the latter $J = \pi\pi^{-1}J = \pi
  V = V/K$. Hence $V/K$ has no proper ideals.
\end{proof}

If the ideal is coisotropic, then the algebra obtained by quotienting it by its orthogonal complement is also metric.

\begin{lemma}\label{le:coisoquot}
  Let $J\lhd V$ be a coisotropic ideal of a metric 3-Lie algebra.
  Then $J/J^\perp$ is a metric 3-Lie algebra.
\end{lemma}

 \begin{proof}
   Since $J$ is coisotropic, $J^\perp \lhd J$, hence $J/J^\perp$
   becomes a 3-algebra. The induced inner product is nondegenerate since we are
   quotienting precisely by $J^\perp$, which is the radical of the
   restriction of the inner product to $J$.
 \end{proof}

Applying these lemmas to the maximal and coisotropic ideal $I^\perp$, we have that:

\begin{itemize}
	\item $\Ubar := V/I^\perp$ is simple or one dimensional and metric.
	\item $\Wbar :=I^\perp/I$ is a metric 3-Lie algebra
\end{itemize}

The next step is to identify $I$ with $U^*$ and the algebras $\Ubar$ and $\Wbar$ with the spaces $U$ and $W$ in the decomposition $V = U \oplus W \oplus I$.

\subsubsection{Identifying quotients and subspaces}

The inner product on $V$ induces a nondegenerate pairing $g: \Ubar \otimes I \to \RR$. Indeed,
let $[u] = u + I^\perp \in \Ubar$ and $v\in I$. Then we define
$g([u],v) = \left<u,v\right>$, which is independent of the
coset representative for $[u]$. In particular, $I \cong \Ubar^*$ is
either one-, four- or eight-dimensional. If the signature of the inner product
of $\Wbar$ is $(p,q)$, that of $V$ is $(p+r,q+r)$ where $r = \dim I =
\dim \Ubar$.

Let $u_i \in V$ be $r = dim \Ubar$ elements of $V$  such that $u_i \not\in I^\perp$, hence their image in $\Ubar$ generate it. Notice that, by definition of $U$, the $u_i$ also span it and therefore we have that $U \cong \Ubar$. 

Since $I \cong  \Ubar^*$ is induced by the inner product, there are $r$ elements $v_i \in I$ such that
$\left<u_i,v_j\right> = \delta_{i,j}$. The subspace spanned by the $u_i$ and the $v_i$ is
$U \oplus I$ and is therefore nondegenerate. Hence, as a vector space, we have an
orthogonal decomposition $V = U \oplus I \oplus W$, where $W$ is the
perpendicular complement of $U \oplus I = \RR(u_i,v_i)$. 

Notice that $W \subset I^\perp$, and that $I^\perp = I \oplus W$ as a vector space. Indeed, the projection $I^\perp \to \Wbar$ maps $W$ isomorphically onto $\Wbar$.

\subsubsection{$V$ is a double extension}

Let us now write the possible 3-brackets for $V = W \oplus I \oplus U$. First of all, by proposition~\ref{pr:ideals} (2), $[V,I^\perp,I]=0$. Since $U<V$, $[U U U]\subset U$ and since $I$ is an ideal,
$[U U I]\subset I$. Similarly, since $W \subset I^\perp$ and $I^\perp \lhd V$ is an ideal, $[W W W]\subset W \oplus I$. We write this as
\begin{equation*}
  [w_1,w_2,w_3] := [w_1,w_2,w_3]_W + \varphi(w_1,w_2,w_3)~,
\end{equation*}
where $[w_1,w_2,w_3]_W$ defines a 3-bracket on $W$, which is isomorphic to the Lie 3-bracket of $\Wbar = I^\perp/I$, and $\varphi:\Lambda^3W \to I$ is to be understood as an abelian extension. It remains to understand $[U W W]$ and $[U U W]$. We notice that because $W \subset I^\perp$, which is an ideal, \emph{a priori} $[U W W] \subset W \oplus I$ and $[U U W] \subset
W \oplus I$.  However,
\begin{equation*}
  \left<[U U W],U\right> = - \left<[U U U], W\right> = 0
\end{equation*}
hence the component of $[U U W]$ along $I$ vanishes, so that $[U U W]\subset W$. Furthermore, the fundamental identity makes $W$ into an $\ad U$-representation and $\varphi$ into an $\ad U$-equivariant map.

Similarly,
\begin{equation*}
  \left<[U W W],W\right> = - \left<[W W W], U\right>~,
\end{equation*}
hence the $W$ component of $[U W W]$ is determined by the map
$\varphi$ defined above; whereas the $I$ component
\begin{equation*}
  \left<[U W W],U\right> =  \left<[U U W], W\right>
\end{equation*}
is thus determined by the action of $\ad U$ on $W$.

In summary, we have the following nonzero 3-brackets
\begin{equation*}
  [U U U] \subset U \qquad
  [U U I] \subset I \qquad 
  [U U W] \subset W \qquad
  [U W W] \subset W \oplus I \qquad
  [W W W] \subset W \oplus I~,
\end{equation*}
which we proceed to explain. The first bracket is simply the fact that $U<V$ is a subalgebra, whereas the second makes $I$ into a representation of $U$. In fact, $I\cong U^*$ as a representation has a name which is the coadjoint representation. The third bracket defines an action of $\ad U$ on $W$ and this also determines the $I$-component of
the fourth bracket. The $W$-component of the fourth bracket is determined by the $I$-component of the last bracket. The last bracket defines a 3-Lie algebra structure on $W \oplus I$, which is an abelian extension of the 3-Lie algebra structure on $\Wbar$ by a ``cocycle'' $\varphi: \Lambda^3 W \to I$. The inner product is such that $\left<W,W\right>$ and $\left<U,I\right>$ are nondegenerate and the only other nonzero inner product is $\left<U,U\right>$ which can be \emph{any} $\ad U$-invariant symmetric bilinear form on $U$, not necessarily nondegenerate.

Remembering that $I \cong U^*$ we have that $V$ is the \emph{double extension} \eqref{def:doublext} of the metric 3-Lie algebra $W$ by the simple or one dimensional 3-Lie algebra $U$ and we have finally proved our main structure theorem and corollary for metric 3-Lie algebras \eqref{th:indecmetric3lie}.

\begin{remark}
  It can be shown that if $U$ is simple and $[U U W]=0$ then the   resulting double extension is decomposable. Indeed, if $[U U W]=0$,   then by the fundamental identity the $U^*$ component in $[W W W]$ would have to be invariant under $\ad U$. If $U$ is simple, then this means that this component is absent, hence $W$ would be a subalgebra and indeed an ideal since the $U^*$ component in $[U W W]$ is also absent. But $W$ is nondegenerate, hence it decomposes $V$.
\end{remark}

\subsubsection{A special case: $U$ is one-dimensional}

This case is of special relevance for low index 3-Lie algebras, because the index of $V$ is given by the index of $W$ plus the dimension of $U$. When $dim U = 1$, the structure is simplified since $[U,U,V]= [U,U,U]=0$ and the $U^*$ component of the $[UWW]$ also vanishes. In particular we have that the only nonzero 3-brackets take the form 

\begin{equation}
  \label{eq:1dim-3-brackets}
  \begin{aligned}[m]
    [u,w_1,w_2] &= [w_1,w_2]\\
    [w_1,w_2,w_3] &= - \left<[w_1,w_2],w_3\right> v +
    [w_1,w_2,w_3]_W~,
  \end{aligned}
\end{equation}
which defines $[w_1,w_2]$ and $[w_1,w_2,w_3]_W$ and where $w_i\in W$.
The fundamental identity is equivalent to the following two conditions:
\begin{enumerate}
\item $[w_1,w_2]$ defines a Lie algebra structure on $W$, which   leaves the inner product invariant due to the skewsymmetry of $\left<[w_1,w_2],w_3\right>$; and
\item $[w_1,w_2,w_3]_W$ defines a metric 3-Lie algebra structure on $W$ which is invariant under the Lie algebra structure.
\end{enumerate}

Hence $W$ has a\textit{ double personality }as a 3-Lie algebra and as a Lie algebra. Following the notation we use in this thesis, we denote its Lie algebra personality as $\fw$ and we write $V= V(W)$ to say that $V$ is the double extension of a metric 3-Lie algebra $W$ by a one dimensional 3-Lie algebra $U$.

\subsection{Classification of low index 3-Lie algebras}

\subsubsection{Euclidean 3-Lie algebras}
\label{sec:claseucl}
It follows that if $V$ is positive-definite (or index 0) and indecomposable, since all ideals are nondegenerate and there cannot be isotropic ideals, it is either one-dimensional or simple, hence isomorphic to $S_{0,4}$. This result, originally due to Nagy \cite{NagykLie} (see also \cite{GP3Lie,GG3Lie}), was conjectured in \cite{FOPPluecker}. 

In other words, Euclidean metric 3-Lie algebras are \textbf{reductive}, this is of the form $A \oplus S$ where $A$ is abelian and $S$ semisimple. Furthermore, since $V$ is positive-definite the only simple factor possible is the unique Euclidean simple 3-Lie algebra $S_{0,4}$ so $S$ is a direct sum of zero or more such factors.

\subsubsection{Classification of metric 3-Lie algebras with signature $(1,p)$}
\label{sec:Lorm3LA}

A Lorentzian (or index 1) 3-Lie algebra decomposes into one Lorentzian indecomposable factor and zero or more indecomposable Euclidean
factors. As discussed above, the indecomposable Euclidean 3-Lie algebras are either one-dimensional or simple ($S_{0,4}$). On the other hand, an
indecomposable Lorentzian 3-Lie algebra is either one-dimensional, simple or is a double extension $V(\fw)$ of a Lie algebra $\fw$ by a one-dimensional 3-Lie algebra. 

Notice that a Lorentzian 3-Lie algebra can never be a double extension by a simple one because simple 3-Lie algebras have dimension either 4 or 8 and the index of the double extension is the index of $W$ plus the dimension of $U$, hence this construction would only be possible for a resulting algebra of index at least 4.

For Lorentzian 3-Lie algebras which are double extensions of $W$ by a one-dimensional one, $W$ must be Euclidean and we prove now that for $V$ to be indecomposable $[-,-,-,]_W$  necessarily vanishes, hence $W$ is actually abelian as a 3-Lie algebra (but not as a Lie algebra).

The main idea of the proof is to exploit the double personality of $W$ as a Lie algebra and a 3-Lie algebra. Since it is Euclidean, it must be reductive in both cases. In particular, as a 3-Lie algebra, it is of the form $W = A \oplus S$ where $A$ is abelian and $S$ semi-simple. Now, $S$ can only be of the form $S=  S_1 \oplus
\dots \oplus S_m$, where each $S_i$, $i=1,\dots,m$ is isomorphic to $S_{0,4}$. The first step of the proof is to show that the Lie algebra personality of $W$, that we call $\fw$, also has the same decomposition as a vector space and each of those four-dimensional simple factors must be, simultaneously, a reductive Lie algebra. By exploiting the properties of four-dimensional reductive Lie algebras, we will prove that $V$ can be decomposed into $S$ and its orthogonal complement, contradicting its indecomposability. 

Notice that the \emph{Lie algebra} structure on $W$ is such that its adjoint maps $\ad_x(-):=[x,-]$ preserve both the 3-brackets and the inner product,
hence $\ad W$ is contained in $\fso(A) \oplus \fso(S_1) \oplus \dots \oplus \fso(S_m)$. 

Indeed, for any $x\in W$, $\ad_x$ preserves the Lie 3-bracket, hence also the ``volume'' forms on each
of the simple factors. In turn this means that $\ad_x$ preserves the subspaces $S$ themselves. To see this, let
$(\be_1,\be_2,\be_3,\be_4)$ be a basis for one of the simple factors, say $S_1$, and let $\be_1\wedge \be_2 \wedge \be_3 \wedge \be_4$ be
the corresponding volume form. Invariance under $\ad_x$ means 

\begin{equation*}
  \begin{aligned}[m]
  &[x,\be_1] \wedge \be_2 \wedge \be_3 \wedge \be_4 + 
  \be_1 \wedge [x,\be_2] \wedge \be_3 \wedge \be_4 \\
	&+ 
  \be_1 \wedge \be_2 \wedge [x,\be_3] \wedge \be_4 + 
  \be_1 \wedge \be_2 \wedge \be_3 \wedge [x,\be_4] = 0~.
  \end{aligned}
\end{equation*}

Now by invariance of the inner product, $[x,\be_i] \perp \be_i$, hence we may write it as $[x,\be_i] = y_i + z_i$, where $y_i \in S_1 \cap \be_i^\perp$ and $z_i \in S_1^\perp$. Inserting it back into the above equation,
\begin{equation*}
  z_1 \wedge \be_2 \wedge \be_3 \wedge \be_4 + 
  \be_1 \wedge z_2 \wedge \be_3 \wedge \be_4 + 
  \be_1 \wedge \be_2 \wedge z_3 \wedge \be_4 + 
  \be_1 \wedge \be_2 \wedge \be_3 \wedge z_4 = 0~.	
\end{equation*}
The above four terms are linearly independent, hence $z_i=0$ and $\ad_x$ indeed preserves $S_1$. This means that each simple factor is a submodule of the adjoint representation and, hence so
is their direct sum. Finally, by invariance of the inner product, so is its perpendicular complement $A$. In other words, the adjoint representation is contained in $\fso(A) \oplus \fso(S_1) \oplus \dots \oplus \fso(S_m)$.

This decomposition of the adjoint representation now implies a decomposition of the Lie algebra itself as $W = \fg \oplus \fh_1 \oplus \dots \oplus \fh_m$, where $\fg$ is a Euclidean Lie algebra (i.e., with $\ad \fg < \fso(A)$) and each $\fh_i$ is a four-dimensional, real, Euclidean Lie algebra (i.e., $\ad\fh_i <\fso(S)$). Indeed, if $x$ and $y$ belong to different orthogonal summands of the vector space $W$, then $[x,y]$ belongs to the same summand as $y$ when understood as $\ad_x(y)$ and to the same summand as $x$ when understood as $\ad_y(x)$. Since these summands are
orthogonal, $[x,y]=0$.

Now Euclidean Lie algebras are reductive; that is, a direct sum of a compact semisimple Lie algebra and an abelian Lie algebra. By inspection there are precisely two isomorphism classes of four-dimensional Euclidean Lie algebras: the abelian $4$-dimensional Lie algebra $\RR^4$ and $\fso(3) \oplus \RR$. Hence $\fso(S)$ has to be isomorphic to one of those.

We will now show that any $S$ summand in $W$ factorises $V$, contradicting the assumption that $V$ is indecomposable.

Consider one such $S$ summand, say $S_1$. The corresponding Lie algebra $\fh_1$ is either abelian or isomorphic to $\fso(3) \oplus \RR$. If $\fh_1$ is abelian, so that the structure constants vanish, then for any $x \in S_1$, $[u,x,V]=0$ and $[x,y,V]=0$ for any $y \in W$ perpendicular to $S_1$ . Hence $S_1\lhd V$ is a nondegenerate ideal, contradicting the indecomposability of $V$.

If $\fh_1 \cong \fso(3) \oplus \RR$, its adjoint algebra $\ad\fh_1$ is an $\fso(3)$ subalgebra of $\fso(S_1) \cong \fso(4)$, which therefore leaves a line $\ell \subset S_1$ invariant. The Lie
algebra structure on $\fh_1$ thus coincides with that given by the Lie bracket $[-,-]_x = [x,-,-]_W$, for some $x\in \ell$, induced from the 3-Lie algebra structure on $S_1$. In other words, $[u,y,z] =
[y,z]_x = [x,y,z]_W$ for all $y,z\in W$. This allows us to ``twist'' $S_1$ into a nondegenerate ideal of $V$.  Indeed, define now 
\begin{equation}
  \label{eq:newcoords}
  u' = u - x - \half |x|^2 v \qquad\text{and}\qquad
  y' = y + \left<y,x\right> v~,
\end{equation}
for all $y \in S_1$. Then $[u',y',z']= 0$ for all $y,z\in S_1$,
and, using that $v$ is central,
\begin{align*}
  [y',z',w'] = [y,z,w] &= - \left<[y,z],w\right> v + [y,z,w]_W\\
  &= - \left<[x,y,z]_W,w\right> v + [y,z,w]_W \\
  &=  \left<[y,z,w]_W,x\right> v + [y,z,w]_W\\
  &= [y,z,w]'_W~.
\end{align*}
Moreover, for every $y\in S_1$,
\begin{equation*}
  \left<u', y'\right> = \left<u - x- \half |x|^2 v, y +
    \left<x,y\right> v\right> = \left<x,y\right> \left<u,v\right> -
  \left<x,y\right> = 0~,
\end{equation*}
and finally
\begin{equation*}
  \left<u',u'\right> = \left<u-x-\half |x|^2 v, u-x-\half |x|^2
    v\right> = - |x|^2 \left<u,v\right> + \left<x,x\right> = 0~.
\end{equation*}
In other words, the subspace of $V$ spanned by the $y'$ for
$y\in S_1$ is a nondegenerate ideal of $V$, contradicting again the
fact that $V$ is indecomposable.

Consequently there can be no $S$'s in $W$, hence as a 3-Lie algebra, $W$ is abelian. As a Lie algebra it is Euclidean, hence reductive. However the abelian summand commutes with $u$, hence it is
central in $V$, again contradicting the fact that it is indecomposable. Therefore as a Lie algebra $W$ is compact and semisimple. In summary, we have proved the following

\begin{theorem}\label{th:lorentzian}
  Let $(V,[-,-,-],b)$ be an indecomposable Lorentzian 3-Lie algebra.
  Then it is either one-dimensional, simple, or else there is a Witt
  basis $(u,v,x_a)$, with $u,v$ complementary null directions, such
  that the nonzero 3-brackets take the form
  \begin{equation*}
    [u,x_a,x_b] = f_{ab}{}^c x_c \qquad\text{and}\qquad
    [x_a,x_b,x_c] = - f_{abc} v~,
  \end{equation*}
  where $[x_a,x_b] = f_{ab}{}^c x_c$ makes the span of the $(x_a)$
  into a compact semisimple Lie algebra $\fg$ and $f_{abc} =
  \left<[x_a,x_b],x_c\right>$.
\end{theorem}

We denote these latter 3-Lie algebras by $V=V(\fg)$. They have been discovered independently in
\cite{GMRBL,BRGTV,HIM-M2toD2rev}, although in some cases in a slightly
different form. 

Paraphrasing the theorem, the class of indecomposable Lorentzian 3-Lie algebras are in one-to-one correspondence with the class of Euclidean metric semisimple Lie algebras, by which we mean a compact semisimple Lie algebra \emph{and} a choice of invariant inner product. This choice involves a choice of scale for each simple factor.

A final remark is that the classification of indecomposable Lorentzian 3-Lie algebras is analogous to the classification of indecomposable Lorentzian Lie algebras, which as shown in \cite{MedinaLorentzian} (see also \cite[Â§2.3]{FSPL}) are either one-dimensional, simple, or obtained as a double extension \cite{MedinaRevoy,FSsug,FSalgebra} of an abelian Euclidean Lie algebra $\fg$ by a one-dimensional Lie algebra acting on $\fg$ via a skew-symmetric endomorphism. In the 3-Lie algebra case, we have an analogous result, with the action of the endomorphism being replaced by a semisimple Lie algebra.

\subsubsection{Classification of metric 3-Lie algebras with signature $(2,p)$}

In \cite{2p3Lie} we classified metric 3-Lie algebras of index $2$ with the ultimate goal of extracting the subset of them of physical interest, namely those with a maximally isotropic centre. However, later in \cite{2pBL} we found a much shorter and direct way of classifying metric 3-Lie algebras with a maximally isotropic of any index, theorem that we discuss at full length in the following section. We refer the reader interested in index 2 3-Lie algebras to \cite{2p3Lie} for a detailed account of their classification.

\section{A classification of metric 3-Lie algebras with maximally isotropic centre}
\label{sec:max-iso-cent}

In section \ref{sec:metric-lie-algebras} we reviewed  Kath and Olbrich's \cite{KathOlbrich2p} classification of metric Lie algebras with a maximally isotropic centre. Here we present an analogous theorem for 3-Lie algebras. Although the spirit of the proof is the same, this case is a bit more involved and, as we will see, requires novel ideas. 

Let us first define metric 3-Lie algebras with a maximally isotropic centre. The centre $Z$ of a 3-Lie algebra $V$ is a subspace $Z \in V$ that satisfies $[Z,V,V]=0$. It is called isotropic if it is contained in its orthogonal complement $Z \in Z^\perp$, which in particular implies $\left\langle Z,Z\right\rangle=0$. This is only possible if the metric of the 3-Lie algebra is not positive definite and hence has index $r<dim(V)$. The dimension of an isotropic space is then bounded above by the index of the 3-Lie algebra. Hence a maximally isotropic centre is a centre which is isotropic and has the maximal dimension allowed, namely the index of the 3-Lie algebra $r$. In other words, a \textbf{3-Lie algebra with a maximally isotropic centre} is an algebra such that its centre $Z$ ($[Z,V,V] = 0$) has dimension equal to its index $r$ and satisfies $\left\langle Z,Z\right\rangle=0$. We summarise now the main steps in the proof.

\begin{enumerate}
\item We start by choosing an appropriate basis for the total vector space $V$ constructed by expanding a basis for the centre $Z$ in a way consistent with the fact that it is
 maximal and isotropic. This basis decomposes $V$ as a vector space in $V = Z \oplus Z^* \oplus W$, where $Z$ and $Z^*$ are   nondegenerately paired and $W$ is positive-definite. 

\item Then, we write down the most general brackets of a metric 3-Lie algebra on $V$ with respect to this basis.
\item Imposing ad-invariance of the metric will relate some of the structure constants.
\item  Imposing the fundamental identity will constrain most of the structure constants and we will be able to interpret them as several structures on the Euclidean space $W$.
\item First we will find the structure of a metric 3-Lie algebra on $W$. Furthermore, for $V$ to be indecomposable, it must be abelian.
\item Then, we will find that other structure constants define a family of reductive Lie algebras on $W$and show that they are all proportional to a reductive Lie algebra structure $\fg \oplus \fz$ on $W$, where $\fg$ is semisimple and $\fz$ is abelian.
\item Another set of structure constants will turn out to define a family $J_{ij}$ of commuting endomorphisms spanning an abelian Lie subalgebra $\fa < \fso(\fz)$. Under the action of $\fa$, $\fz$ breaks up into a direct sum of   irreducible 2-planes $E_\pi$ and a Euclidean vector space $E_0$ on which the $J_{ij}$ act trivially.
\item Finally, the remaining structure constants will  define elements $K_{ijk} \in E_0$ which are subject to   a quadratic equation.
\end{enumerate}

\subsection{Preliminary form of the 3-algebra}
\label{sec:preliminary-form-3}

Let $V$ be a finite-dimensional metric 3-Lie algebra with index $r>0$ and admitting a maximally isotropic centre. Let $v_i$, $i=1,\dots,r$,
denote a basis for the centre. Since the centre is (maximally)
isotropic, $\left<v_i,v_j\right>=0$, and since the inner product on
$V$ is nondegenerate, there exist $u_i$, $i=1,\dots,r$ satisfying
$\left<u_i,v_j\right> = \delta_{ij}$. Furthermore, it is possible to
choose the $u_i$ such that $\left<u_i,u_j\right>=0$. The space spanned by the $u$'s is the dual space to the centre $Z^*$. The
perpendicular complement, $W$, to the $2r$-dimensional subspace $Z \oplus Z^*$ is therefore positive-definite. In other
words, $V$ admits a vector space decomposition:
\begin{equation}
  V = \bigoplus_{i=1}^r \left( \RR u_i \oplus \RR v_i\right) \oplus W = Z \oplus Z^*\oplus W
\end{equation}
Since the $v$'s are central, similarly to the Lie algebra case, metricity of $V$ implies that the $u$'s
cannot appear in the right-hand side of any 3-bracket. If ${e_a}$ is a basis for $W$, the most general brackets we can have are:

\begin{equation}
  \label{eq:pre-3-brackets}
  \begin{aligned}[m]
    [u_i,u_j, u_k] &= F_{ijk}{}^a e_a + \sum_{l=1}^r F_{ijk}{}^l v_l\\
    [u_i,u_j,e_a] &=  F_{ija}{}^b e_b + \sum_{l=1}^r F_{ija}{}^l v_l\\
		[u_i,e_a,e_b] &= F_{iab}{}^c e_c + \sum_{l=1}^r F_{iab}{}^l v_l\\
    [e_a,e_b, e_c] &= F_{abc}{}^d e_d + \sum_{l=1}^r F_{abc}{}^l v_l,
  \end{aligned}
\end{equation}
where indices $a, b, c$ indicate a component in $W$, downstairs $i, j, k, l$ refer to $u_i, u_j, u_k, u_l$ components respectively whereas upstairs ones refer to a $v_i, v_j, v_k, v_l$ components. Again, this can be done because the $v$'s don't appear inside the brackets and the $u$'s don't appear on the right-hand side.

Just like with Lie algebras, ad-invariance of the metric relates several structure constants, for example:

\begin{equation}
\left\langle  [u_i, u_j,e_a] , u_k\right\rangle = F_{ijak} = - \left\langle  e_a , [u_i,u_j, u_k]\right\rangle = - \left\langle  e_a , F_{ijk}{}^b e_b\right\rangle
\end{equation}

Hence, $F_{ijak} = - F_{ijk}{}^b  \left\langle  e_a , e_b\right\rangle$. The most general Lie brackets on $V$ are now:

\begin{equation}
  \label{eq:3-brackets}
  \begin{aligned}[m]
    [u_i,u_j,u_k] &= K_{ijk} + \sum_{\ell=1}^r L_{ijk\ell} v_\ell\\
    [u_i,u_j,x] &= J_{ij} x - \sum_{k=1}^r \left<K_{ijk},x\right> v_k\\
    [u_i,x,y] &= [x,y]_i - \sum_{j=1}^r \left<x,J_{ij} y\right> v_j\\
    [x,y,z] &= [x,y,z]_W - \sum_{i=1}^r \left<[x,y]_i,z\right> v_i,
  \end{aligned}
\end{equation}
where we have changed slightly the notation by using $x,y,z \in W$ rather than the basis on $W$. Also, in light of the properties that the fundamental identity will provide to these structure constants, we have re-named them. For instance, we will see that the structure constants  $F_{iab}{}^c$ define alternating bilinear maps from $W \to W$ for each $i= 1, \cdots, r$ and satisfy the Jacobi identity \eqref{eq:Jacobi}, so we have renamed them $F_{iab}{}^c := [e_a, e_b]_i$. In addition, they obey
\begin{equation}
  \label{eq:metricity-i}
  \left<[x,y]_i,z\right> = \left<x,[y,z]_i\right>.
\end{equation} 

Similarly,  $F_{ija}{}^b$ define endomorphisms from $W \to W$ that preserve the metric for each $i,j$, so we have renamed them $F_{ija}{}^b= J_{ij} \in \fso(W)$. $K_{ijk} \in W$ and $L_{ijk\ell} \in \RR$ are skewsymmetric in their indices and $[-,-,-]_W : W \times W \times W \to W$ is an alternating trilinear map which obeys
\begin{equation}
  \label{eq:metricity-W}
  \left<[x,y,z]_W,w\right> = - \left<[x,y,w]_W,z\right>.
\end{equation}

The following lemma is the result of a straightforward, if somewhat
lengthy, calculation.

\begin{lemma}
  The fundamental identity \eqref{def:FI} of the 3-Lie algebra $V$
  defined by \eqref{eq:3-brackets} and ad-invariance of the metric are equivalent to the following
  conditions, for all $t,w,x,y,z\in W$:
  \begin{subequations}\label{eq:FI-V-pre}
    \begin{align}
      [t,w,[x,y,z]_W]_W &= [[t,w,x]_W,y,z]_W +[x,[t,w,y]_W,z]_W + [x,y,[t,w,z]_W]_W \label{eq:W-3la}\\
      [w,[x,y,z]_W]_i &= [[w,x]_i,y,z]_W + [x,[w,y]_i,z]_W + [x,y,[w,z]_i]_W \label{eq:ad-i-der-W}\\
      [x,y,[z,t]_i]_W &= [z,t,[x,y]_i]_W + [[x,y,z]_W,t]_i + [z,[x,y,t]_W]_i \label{eq:C9-W}\\
      J_{ij}[x,y,z]_W &=  [J_{ij}x,y,z]_W + [x,J_{ij}y,z]_W + [x,y,J_{ij}z]_W \label{eq:J-der-W}\\
      J_{ij}[x,y,z]_W - [x,y,J_{ij}z]_W  &= [[x,y]_i,z]_j - [[x,y]_j,z]_i \label{eq:C3-W}\\
      [x,y,K_{ijk}]_W &= J_{jk} [x,y]_i + J_{ki}[x,y]_j + J_{ij}[x,y]_k \label{eq:C6-W}\\
      [J_{ij}x,y,z]_W &= [[x,y]_i,z]_j + [[y,z]_j,x]_i + [[z,x]_i,y]_j \label{eq:C2-W}\\
      J_{ij}[x,y,z]_W &= [z,[x,y]_j]_i + [x,[y,z]_j]_i + [y,[z,x]_j]_i \label{eq:C2-W-bis}\\
      [x,y,K_{ijk}]_W &= J_{ij} [x,y]_k - [J_{ij} x,y ]_k - [x, J_{ij} y ]_k \label{eq:C1-W}\\
      J_{ik}[x,y]_j - J_{ij}[x,y]_k &= [J_{jk}x,y]_i + [x,J_{jk}y]_i \label{eq:J-der-2-1}\\
      [x,J_{jk}y]_i &= [J_{ij}x,y]_k + [J_{ki}x,y]_j + J_{jk}[x,y]_i \label{eq:J-der-2-2}\\
      [K_{ijk},x]_\ell &= [K_{\ell ij},x]_k + [K_{\ell jk},x]_i + [K_{\ell ki},x]_j \label{eq:K-ad-x}\\
      [K_{ijk},x]_\ell - [K_{ij\ell},x]_k &= \left(J_{ij}J_{k\ell} - J_{k\ell}J_{ij}\right) x  \label{eq:C4-W}\\
      [x,K_{jk\ell}]_i &= \left(J_{jk}J_{i\ell} + J_{k\ell}J_{ij} + J_{j\ell}J_{ki}\right) x  \label{eq:C5-W}\\
      J_{im}K_{jk\ell} &= J_{ij} K_{k\ell m} + J_{ik} K_{\ell mj} + J_{i\ell} K_{jkm} \label{eq:C7-bis}\\
      J_{ij}K_{k\ell m} &= J_{\ell m}K_{ijk}  + J_{mk}K_{ij\ell} + J_{k\ell}K_{ijm}  \label{eq:C7-W}\\
      \left<K_{ijm},K_{nk\ell}\right> + \left<K_{ijk},K_{\ell mn}\right>  &= \left<K_{ijn},K_{k\ell m}\right> + \left<K_{ij\ell},K_{mnk}\right> . \label{eq:C8-W}
    \end{align}
  \end{subequations}
\end{lemma}

Not all of these equations are independent, but we will not
attempt to select a minimal set here, since we will be able to
dispense with some of the equations easily.

\subsection{$W$ is abelian}
\label{sec:w-abelian}

Equation \eqref{eq:W-3la} says that $W$ becomes a 3-Lie algebra under
$[-,-,-]_W$, which is also metric by \eqref{eq:metricity-W}. Since $W$ is
positive-definite, as we saw in section \ref{sec:claseucl} it is reductive, hence isomorphic to an
orthogonal direct sum $W = S \oplus A$, where $S$ is semisimple and
$A$ is abelian. Furthermore, $S$ is an orthogonal direct sum of
several copies of the unique positive-definite simple 3-Lie algebra
$S_4$ \cite{Filippov,LingSimple}. 

We will show that as metric 3-Lie algebras $V = S \oplus S^\perp$. hence, if $V$ is indecomposable, then $S=0$ and $W=A$ is abelian as a 3-Lie algebra. This is an extension of the result in section \ref{sec:Lorm3LA} for indecomposable Lorentzian 3-Lie algebras by which semisimple factors
decompose a one-dimensional double extension, and we will, in fact, follow a similar method to the one in that section by which we perform an isometry on $V$ which manifestly exhibits a nondegenerate
ideal isomorphic to $S$ as a 3-Lie algebra.

Consider then the isometry $\varphi: V \to V$, defined by
\begin{equation}
  \label{eq:isometry}
    \varphi(v_i) = v_i \qquad \varphi(u_i) = u_i - s_i -\half
    \sum_{j=1}^r \left<s_i,s_j\right> v_j \qquad  \varphi(x) = x +
    \sum_{i=1}^r \left<s_i,x\right> v_i,
\end{equation}
for $x\in W$ and for some $s_i \in W$. This is obtained by
extending the linear map $v_i \to v_i$ and $u_i \mapsto u_i - s_i$ to
an isometry of $V$. Under $\varphi$ the 3-brackets
\eqref{eq:3-brackets} take the following form
\begin{equation}
  \label{eq:3-brackets-phi}
  \begin{aligned}[m]
    [\varphi(u_i),\varphi(u_j),\varphi(u_k)] &= \varphi(K^\varphi_{ijk}) + \sum_{\ell=1}^r L^\varphi_{ijk\ell} v_\ell\\ 
    [\varphi(u_i),\varphi(u_j),\varphi(x)] &= \varphi(J^\varphi_{ij} x) - \sum_{k=1}^r \left<K^\varphi_{ijk},x\right> v_k\\
    [\varphi(u_i),\varphi(x),\varphi(y)] &= \varphi([x,y]^\varphi_i) - \sum_{j=1}^r \left<x,J^\varphi_{ij} y\right> v_j\\
    [\varphi(x),\varphi(y),\varphi(z)] &= \varphi([x,y,z]_W) - \sum_{i=1}^r \left<[x,y]^\varphi_i,z\right> v_i,
  \end{aligned}
\end{equation}
where
\begin{equation}
  \label{eq:isometry-changes}
  \begin{aligned}[m]
    [x,y]^\varphi_i &= [x,y]_i + [s_i,x,y]_W\\
    J^\varphi_{ij} x &= J_{ij} x + [s_i,x]_j - [s_j,x]_i + [s_i,s_j,x]_W\\
    K^\varphi_{ijk} & = K_{ijk} - J_{ij} s_k - J_{jk} s_i - J_{ki} s_j
    + [s_i,s_j]_k + [s_j,s_k]_i + [s_k,s_i]_j - [s_i,s_j,s_k]_W\\
    L^\varphi_{ijk\ell} &= L_{ijk\ell} + \left<K_{jk\ell},s_i\right> -
    \left<K_{k\ell i},s_j\right> + \left<K_{\ell ij},s_k\right> -
    \left<K_{ijk},s_\ell\right>\\
    &\quad - \left<s_i,J_{k\ell}s_j\right> -
    \left<s_k,J_{j\ell}s_i\right> - \left<s_j,J_{i\ell}s_k\right> +
    \left<s_\ell,J_{jk}s_i\right> + \left<s_\ell,J_{ki}s_j\right> +
    \left<s_\ell,J_{ij}s_k\right>\\
    &\quad + \left<[s_i,s_j]_\ell, s_k\right> - \left<[s_i,s_j]_k,
      s_\ell\right> - \left<[s_k,s_i]_j, s_\ell\right> -
    \left<[s_j,s_k]_i, s_\ell\right> +
    \left<[s_i,s_j,s_k]_W,s_\ell\right>.
  \end{aligned}
\end{equation}

\begin{lemma}
  There exists $s_i\in S$ such that the following conditions are met
  for all $x \in S$:
  \begin{equation}
    [x,-]^\varphi_i = 0 \qquad J^\varphi_{ij} x = 0 \qquad
    \left<K^\varphi_{ijk},x\right> = 0.
  \end{equation}
\end{lemma}

Assuming for a moment that this is the case, the only nonzero
3-brackets involving elements in $\varphi(S)$ are
\begin{equation}
  [\varphi(x),\varphi(y),\varphi(z)] = \varphi([x,y,z]_W),
\end{equation}
and this means that $\varphi(S)$ is a nondegenerate ideal of $V$,
hence $V = \varphi(S) \oplus \varphi(S)^\perp$. But this violates
the indecomposability of $V$, unless $S=0$.

\begin{proof}[Proof of the lemma]
  To show the existence of the $s_i$, let us decompose $S = S_4^{(1)} \oplus \dots \oplus S_4^{(m)}$ into $m$ copies of the unique simple positive-definite 3-Lie algebra $S_4$. As shown in \cite[§3.2]{Lor3Lie}, since $J_{ij}$ and $[x,-]_i$ define skewsymmetric derivations of $W$, they preserve the decomposition of $W$ into $S\oplus A$ and that of $S$ into its simple factors. One consequence of this fact is that $J_{ij}x \in S$ for all $x \in S$ and $[x,y]_i \in S$ for all $x,y\in S$, and
  similarly if we substitute $S$ for any of its simple factors in the previous statement. Notice in addition that putting $i=j$ in equation \eqref{eq:C2-W}, $[-,-]_i$ obeys the Jacobi identity \eqref{eq:Jacobi}. Hence on any one of the simple factors of $S$ - let's call it generically $S_4$ - the bracket $[-,-]_i$ defines the structure of a four-dimensional Lie algebra. This Lie algebra is
  metric by equation \eqref{eq:metricity-i} and positive-definite. There are (up to isomorphism) precisely two four-dimensional positive-definite metric Lie algebras: the abelian Lie algebra and $\fso(3) \oplus \RR$. In either case, as shown in \cite[§3.2]{Lor3Lie}, there exists a unique $s_i \in S_4$ such that $[s_i,x,y]_W = [x,y]_i$ for $x,y\in S_4$, (in the former
  case, $s_i=0$). Since this is true for all simple factors, we conclude that there exists $s_i \in S$ such that $[s_i,x,y]_W = [x,y]_i$ for $x,y\in S$ and for all $i$.

  Now equation \eqref{eq:C2-W} says that for all $x,y,z\in S$,
  \begin{align*}
    [J_{ij}x,y,z]_W &= [[x,y]_i,z]_j + [[y,z]_j,x]_i + [[z,x]_i,y]_j\\
    &= [s_j,[s_i, x,y]_W,z]_W + [s_i,[s_j,y,z]_W,x]_W + [s_j,[s_i, z,x]_W,y]_W\\
    &= [[s_i,s_j,x]_W,y,x]_W, &&\text{using \eqref{eq:W-3la}}
  \end{align*}
  which implies that $J_{ij}x - [s_i,s_j,x]_W$ centralises $S$, and thus is in $A$. However, for $x \in S$, both $J_{ij}x \in S$
  and $[s_i,s_j,x]_W \in S$, so that $J_{ij}x = [s_i,s_j,x]_W$. Similarly, equation \eqref{eq:C1-W} says that for all $x,y \in S$,
  \begin{align*}
    [x,y,K_{ijk}]_W &= J_{ij} [x,y]_k - [J_{ij} x,y ]_k - [x, J_{ij} y]_k\\
    &= [s_i,s_j, [s_k,x,y]_W]_W - [s_k,[s_i,s_j,x]_W,y ]_W - [s_k, x, [s_i,s_j y]_W]_W\\
    &= [[s_i,s_j,s_k]_W,x,y]_W, && \text{using \eqref{eq:W-3la}}
  \end{align*}
  which implies that $K_{ijk}-[s_i,s_j,s_k]_W$ centralises $S$, hence $K_{ijk} -
  [s_i,s_j,s_k]_W  = K^A_{ijk} \in A$. Finally, using the explicit formulae for
  $J^\varphi_{ij}$ and $K^\varphi_{ijk}$ in equation
  \eqref{eq:isometry-changes}, we see that for all all $x \in S$,
  \begin{align*}
    J^\varphi_{ij} x &= J_{ij} x + [s_i,x]_j - [s_j,x]_i + [s_i,s_j,x]_W\\
    &= [s_i,s_j,x]_W + [s_j,s_i,x]_W - [s_i,s_j,x]_W + [s_i,s_j,x]_W = 0\\
    \intertext{and}
    K^\varphi_{ijk} &= K_{ijk} - J_{ij} s_k - J_{jk} s_i - J_{ki} s_j +
    [s_i,s_j]_k + [s_j,s_k]_i + [s_k,s_i]_j - [s_i,s_j,s_k]_W\\
    &= K^A_{ijk} + [s_i,s_j,s_k]_W - [s_i,s_j,s_k]_W - [s_j,s_k,s_i]_W -
    [s_k,s_i,s_j]_W\\
    & \quad +  [s_k, s_i,s_j]_W + [s_i,s_j,s_k]_W +
    [s_j,s_k,s_i]_W - [s_i,s_j,s_k]_W= K^A_{ijk},
  \end{align*}
  hence $\left<K^\varphi_{ijk},x\right> = 0$ for all $x \in S$.
\end{proof}

We may summarise the above discussion as follows.

\begin{lemma}
  Let $V$ be a finite-dimensional indecomposable metric 3-Lie algebra of index $r>0$ with a maximally isotropic centre. Then as a vector space
  \begin{equation}
    V = \bigoplus_{i=1}^r \left( \RR u_i \oplus \RR v_i\right) \oplus
    W,
  \end{equation}
  where $W$ is positive-definite, $u_i,v_i \perp W$,
  $\left<u_i,u_j\right> = 0$, $\left<v_i,v_j\right> = 0$ and
  $\left<u_i,v_j\right>= \delta_{ij}$. The $v_i$ span the maximally
  isotropic centre. The nonzero 3-brackets are given by
  \begin{equation}
    \label{eq:V3b}
    \begin{aligned}[m]
      [u_i,u_j,u_k] &= K_{ijk} + \sum_{\ell=1}^r L_{ijk\ell} v_\ell\\ 
      [u_i,u_j,x] &= J_{ij} x - \sum_{k=1}^r \left<K_{ijk},x\right> v_k\\
      [u_i,x,y] &= [x,y]_i - \sum_{j=1}^r \left<x,J_{ij} y\right> v_j\\
      [x,y,z] &= - \sum_{i=1}^r \left<[x,y]_i,z\right> v_i,
    \end{aligned}
  \end{equation}
  for all $x,y,z\in W$ and for some $L_{ijk\ell} \in \RR$, $K_{ijk} \in W$, $J_{ij} \in \fso(W)$, all of which are totally skewsymmetric in their indices, and bilinear alternating brackets $[-,-]_i : W \times W \to W$ satisfying equation \eqref{eq:metricity-i}. Furthermore, the fundamental identity of the 3-brackets \eqref{eq:V3b} is equivalent to the following conditions on
  $K_{ijk}$, $J_{ij}$ and $[-,-]_i$:
  \begin{subequations}\label{eq:FI-V3b}
    \begin{align}
      [x, [y,z]_i]_j &=  [[x,y]_j,z]_i + [y,[x,z]_j]_i \label{eq:C2}\\
      [[x,y]_i,z]_j &=  [[x,y]_j,z]_i  \label{eq:C3}\\
      J_{ij} [x,y]_k &= [J_{ij} x,y ]_k + [x, J_{ij} y ]_k \label{eq:C1}\\
      0 &= J_{j\ell} [x,y]_i + J_{\ell i}[x,y]_j + J_{ij}[x,y]_\ell \label{eq:C6}\\
      [K_{ijk},x]_\ell - [K_{ij\ell},x]_k &= \left(J_{ij}J_{k\ell} - J_{k\ell}J_{ij}\right) x  \label{eq:C4}\\
      [x,K_{jk\ell}]_i &= \left(J_{jk}J_{i\ell} + J_{k\ell}J_{ij} + J_{j\ell}J_{ki}\right) x  \label{eq:C5}\\
      J_{ij}K_{k\ell m} &= J_{\ell m}K_{ijk}  + J_{mk}K_{ij\ell} + J_{k\ell}K_{ijm}  \label{eq:C7}\\
      0 &= \left<K_{ijn},K_{k\ell m}\right> + \left<K_{ij\ell},K_{mnk}\right> - \left<K_{ijm},K_{nk\ell}\right> - \left<K_{ijk},K_{\ell mn}\right> . \label{eq:C8}
    \end{align}
  \end{subequations}
\end{lemma}

There are less equations in \eqref{eq:FI-V3b} than are obtained from \eqref{eq:FI-V-pre} by simply making $W$ abelian. It is not hard to show that the equations in \eqref{eq:FI-V3b} imply the rest. The study of equations \eqref{eq:FI-V3b} will take us until the end of this section. The analysis of these conditions breaks naturally into several steps. In the first step we solve equations \eqref{eq:C2} and \eqref{eq:C3} for the $[-,-]_i$. We then solve equations \eqref{eq:C1} and \eqref{eq:C6}, which in turn allow us to solve equations \eqref{eq:C4} and \eqref{eq:C5} for the $J_{ij}$. Finally we solve equation \eqref{eq:C7}. We not solve equation \eqref{eq:C8}. In fact, this equation defines an algebraic variety (an intersection of conics) which parametrises these 3-algebras.

\subsection{Solving for the $[-,-]_i$}
\label{sec:solving-brackets}

Condition \eqref{eq:C2} for $i=j$ says that $[-,-]_i$ defines a Lie algebra structure on $W$, denoted $\fg_i$. By equation \eqref{eq:metricity-i}, $\fg_i$ is a metric Lie algebra. Since the inner product on $W$ is positive-definite, $\fg_i$ is reductive, hence $\fg_i = [\fg_i,\fg_i] \oplus \fz_i$, where $\fs_i := [\fg_i,\fg_i]$ is the semisimple derived ideal of $\fg_i$ and $\fz_i$
is the centre of $\fg_i$. The following lemma will prove useful.

\begin{lemma}
  \label{lem:g1-simple}
  Let $\fg_i$, $i=1,\dots,r$, be a family of reductive Lie algebras sharing the same underlying vector space $W$ and let $[-,-]_i$ denote the Lie bracket of $\fg_i$. Suppose that they satisfy
  equations \eqref{eq:C2} and \eqref{eq:C3} and in addition that one of these Lie algebras, $\fg_1$ say, is simple. Then for all $x,y\in
  W$,
  \begin{equation}
    [x,y]_i = \kappa_i [x,y]_1,
  \end{equation}
  where $\kappa_i \in \RR$.
\end{lemma}

\begin{proof}
  Equation \eqref{eq:C2} says that for all $x \in W$, $\ad_i x :=[x,-]_i$ is a derivation of $\fg_j$, for all $i,j$. In particular, $\ad_1 x$ is a derivation of $\fg_i$. Since derivations preserve the centre, $\ad_1 x : \fz_i \to \fz_i$, hence the subspace $\fz_i$ is an ideal of $\fg_1$. Since by hypothesis, $\fg_1$ is simple, we must have that either $\fz_i = W$, in which case $\fg_i$ is abelian and the lemma holds with $\kappa_i=0$, or else $\fz_i=0$, in which case $\fg_i$ is semisimple. It remains therefore to study this case.

  Equation \eqref{eq:C2} again says that $\ad_i x$ is a derivation of $\fg_1$. Since all derivations of $\fg_1$ are inner, this means that there is some element $y$ such that $\ad_i x = \ad_1 y$. This element is moreover unique because $\ad_1$ has trivial kernel. In other words, this defines a linear map
  \begin{equation}
    \label{eq:psi-i}
    \psi_i: \fg_i \to \fg_1 \qquad\text{by}\qquad \ad_i x = \ad_1
    \psi_i x \qquad \forall x \in W.
  \end{equation}
  This linear map is a vector space isomorphism since $\ker \psi_i \subset \ker \ad_i = 0$, for $\fg_i$ semisimple. Now suppose that $I \lhd \fg_i$ is an ideal, hence $\ad_i(x) I \subset I$ for all $x
  \in \fg_i$. This means that $\ad_1(y) I \subset I$ for all $y \in \fg_1$, hence $I$ is also an ideal of $\fg_1$. Since $\fg_1$ is simple, this means that $I=0$ or else $I=W$; in other words, $\fg_i$ is simple.

  Now for all $x,y,z \in W$, we have
  \begin{align*}
    [\psi_i[x,y]_i,z]_1 &= [[x,y]_i,z]_i && \text{by equation \eqref{eq:psi-i}}\\
    &= [x,[y,z]_i]_i - [y,[x,z]_i]_i && \text{by the Jacobi identity of $\fg_i$}\\
    &= [\psi_ix,[\psi_iy,z]_1]_1 - [\psi_iy,[\psi_ix,z]_1]_1 && \text{by  equation \eqref{eq:psi-i}}\\
    &= [[\psi_i x, \psi_i y]_1, z]_1 && \text{by the Jacobi identity of $\fg_1$}
  \end{align*}
  and since $\fg_1$ has trivial centre, we conclude that
  \begin{equation*}
    \psi_i[x,y]_i = [\psi_i x,\psi_i y]_1,
  \end{equation*}
  hence $\psi_i: \fg_i \to \fg_1$ is a Lie algebra isomorphism.

  Next, condition \eqref{eq:C3} says that $\ad_1 [x,y]_i = \ad_i [x,y]_1$, hence using equation \eqref{eq:psi-i}, we find that $\ad_1 [x,y]_i = \ad_1 \psi_i [x,y]_1$, and since $\ad_1$ has
  trivial kernel,  $[x,y]_i = \psi_i [x,y]_1$. We may rewrite this equation as $\ad_i x = \psi_i \ad_1 x$ for all $x$, which again by virtue of \eqref{eq:psi-i}, becomes $\ad_1 \psi_i x = \psi_i \ad_1
  x$, hence $\psi_i$ commutes with the adjoint representation of $\fg_1$. Since $\fg_1$ is simple, Schur's lemma says that $\psi_i$ must be a multiple, $\kappa_i$ say, of the identity. In other words,  $\ad_i x = \kappa_i \ad_1 x$, which proves the lemma.
\end{proof}

Let us now consider the general case when none of the $\fg_i$ are simple. Let us focus on two reductive Lie algebras, $\fg_i = \fz_i\oplus \fs_i$, for $i=1,2$ say, sharing the same underlying vector space $W$. We will further decompose $\fs_i$ into its simple ideals
\begin{equation}
  \fs_i = \bigoplus_{\alpha =1}^{N_i} \fs_i^\alpha.
\end{equation}
For every $x \in W$, $\ad_1 x$ is a derivation of $\fg_2$, hence it
preserves the centre $\fz_2$ and each simple ideal $\fs_2^\beta$.
This means that $\fz_2$ and $\fs_2^\beta$ are themselves ideals of
$\fg_1$, hence
\begin{equation}
  \fz_2 = E_0 \oplus \bigoplus_{\alpha \in I_0} \fs_1^\alpha
  \qquad\text{and}\qquad
  \fs_2^\beta = E_\beta \oplus \bigoplus_{\alpha \in I_\beta}
  \fs_1^\alpha \qquad \forall \beta\in\left\{1,2,\dots,N_2\right\},
\end{equation}
and where the index sets $I_0,I_1,\dots,I_{N_2}$ define a partition of $\left\{1,\dots,N_1\right\}$, and
\begin{equation}
  \fz_1 = E_0 \oplus E_1 \oplus \cdots \oplus E_{N_2}
\end{equation}
is an orthogonal decomposition of $\fz_1$. But now notice that the restriction of $\fg_1$ to $E_\beta \oplus \bigoplus_{\alpha \in I_\beta} \fs_1^\alpha$ is reductive, hence we may apply lemma~\ref{lem:g1-simple} to each simple $\fs_2^\beta$ in turn. This allows us to conclude that for each $\beta$, either $\fs_2^\beta =E_\beta$ or else $\fs_2^\beta = \fs_1^\alpha$, for some $\alpha \in \left\{1,2,\dots, N_1\right\}$ which depends on $\beta$, and in this latter case, $[x,y]_{\fs_2^\beta} = \kappa [x,y]_{\fs_1^{\alpha}}$, for some nonzero constant $\kappa$.

This means that, given any one Lie algebra $\fg_i$, any other Lie algebra $\fg_j$ in the same family is obtained by multiplying its simple factors by some constants (which may be different in each factor and may also be zero) and maybe promoting part of its centre to be semisimple.

The metric Lie algebras $\fg_i$ induce the following orthogonal decomposition of the underlying vector space $W$. We let $W_0 = \bigcap_{i=1}^r \fz_i$ be the intersection of all the centres of the
reductive Lie algebras $\fg_i$. Then we have the following orthogonal direct sum $W = W_0 \oplus \bigoplus_{\alpha =1}^N W_\alpha$, where restricted to each $W_{\alpha>0}$ at least one of the Lie algebras, $\fg_i$ say, is simple and hence all other Lie algebras $\fg_{j\neq i}$ are such that for all $x,y\in W_\alpha$,
\begin{equation}
  [x,y]_j = \kappa_{ij}^\alpha [x,y]_i \qquad \exists
  \kappa_{ij}^\alpha \in \RR.
\end{equation}

To simplify the notation, we define a semisimple Lie algebra structure $\fg$ on the perpendicular complement of $W_0$, whose Lie bracket $[-,-]$ is defined in such a way that for all $x,y\in
W_\alpha$, $[x,y] := [x,y]_i$, where $i\in\{1,2,\dots,r\}$ is the smallest such integer for which the restriction of $\fg_i$ to $W_\alpha$ is simple. That such an integer $i$ exists follows from
the definition of $W_0$ and of the $W_\alpha$. It then follows that the restriction to $W_\alpha$ of every other $\fg_{j\neq i}$ is a
(possibly zero) multiple of $\fg$.

We summarise this discussion in the following lemma, which summarises the solution of equations \eqref{eq:C2} and \eqref{eq:C3}.

\begin{lemma}
  \label{lem:C2C3}
  Let $\fg_i$, $i=1,\dots,r$, be a family of metric Lie algebras sharing the same underlying Euclidean vector space $W$ and let $[-,-]_i$ denote the Lie bracket of $\fg_i$. Suppose that they
  satisfy equations \eqref{eq:C2} and \eqref{eq:C3}. Then there is an orthogonal decomposition
  \begin{equation}
    \label{eq:W-decomp}
    W = W_0 \oplus \bigoplus_{\alpha =1}^N W_\alpha,
  \end{equation}
  where
  \begin{equation}
    \label{eq:g-brackets}
    [x,y]_i =
    \begin{cases}
      0 & \text{if $x,y\in W_0$;}\\
      \kappa_i^\alpha [x,y] & \text{if $x,y \in W_\alpha$,}
    \end{cases}
  \end{equation}
  for some $\kappa_i^\alpha \in \RR$ and where $[-,-]$ are the Lie  brackets of a semisimple Lie algebra $\fg$ with underlying vector space $\bigoplus_{\alpha =1}^N W_\alpha$.
\end{lemma}

\subsection{Solving for the $J_{ij}$}
\label{sec:solving-Js}

Next we study the equations \eqref{eq:C1} and \eqref{eq:C6}, which
involve only $J_{ij}$. Equation \eqref{eq:C1} says that each $J_{ij}$
is a derivation over the $\fg_k$ for all $i,j,k$. Since derivations
preserve the centre, every $J_{ij}$ preserves the centre of every
$\fg_k$ and hence it preserves their intersection $W_0$. Since
$J_{ij}$ preserves the inner product, it also preserves the
perpendicular complement of $W_0$ in $W$, which is the underlying
vector space of the semisimple Lie algebra $\fg$ of the previous
lemma. Equation \eqref{eq:C1} does not constrain the component of
$J_{ij}$ acting on $W_0$ since all the $[-,-]_k$ vanish there, but it
does constrain the components of $J_{ij}$ acting on $\bigoplus_{\alpha
  =1}^N W_\alpha$. Fix some $\alpha$ and let $x,y\in W_\alpha$. Then
by virtue of equation \eqref{eq:g-brackets}, equation \eqref{eq:C1}
says that
\begin{equation}
  \kappa_k^\alpha \left( J_{ij} [x,y] - [J_{ij}x,y] - [x,J_{ij}y]
  \right) = 0.
\end{equation}

Given any $\alpha$ there will be at least some $k$ for which $\kappa_k^\alpha \neq 0$ implying that $J_{ij}$ is a derivation of $\fg$. Since $\fg$ is semisimple, this derivation is inner and
there exists a unique $z_{ij} \in \fg$, such that $J_{ij} y = [z_{ij},y]$ for all $y \in \fg$. Since the simple ideals of $\fg$ are submodules under the adjoint representation, $J_{ij}$ preserves each of the simple ideals and hence it preserves the decomposition \eqref{eq:W-decomp}. Let $z_{ij}^\alpha$ denote the component of $z_{ij}$ along $W_\alpha$. Equation \eqref{eq:C6} can now be rewritten for $x,y\in W_\alpha$ as
\begin{equation}
  \kappa_i^\alpha [z_{j\ell}^\alpha, [x,y]] + \kappa_j^\alpha
  [z^\alpha_{\ell i},[x,y]] + \kappa_\ell^\alpha [z^\alpha_{ij},[x,y]]
  = 0.
\end{equation}
The centre of $\fg$ being trivial is equivalent to
\begin{equation}
  \label{eq:kwedgez=0}
  \kappa_i^\alpha z_{j\ell}^\alpha + \kappa_j^\alpha z^\alpha_{\ell i}
  + \kappa_\ell^\alpha z^\alpha_{ij} = 0,
\end{equation}
which can be written more suggestively as $\kappa^\alpha \wedge
z^\alpha = 0$, where $\kappa^\alpha \in \RR^r$ and $z^\alpha \in
\Lambda^2\RR^r \otimes W_\alpha$. This equation has as unique
solution $z^\alpha = \kappa^\alpha \wedge s^\alpha$, for some $s^\alpha
\in \RR^r\otimes W_\alpha$, or in indices
\begin{equation}
  \label{eq:zijalpha}
  z^\alpha_{ij} = \kappa_i^\alpha s_j^\alpha - \kappa_j^\alpha
  s_i^\alpha \qquad \exists s_i^\alpha \in W_\alpha.
\end{equation}
Let $s_i = \sum_\alpha s_i^\alpha \in \fg$ and consider now the
isometry $\varphi: V \to V$ defined by
\begin{equation}
  \begin{aligned}[m]
    \varphi(v_i) &= v_i\\
    \varphi(z) &= z\\
    \varphi(u_i) &= u_i - s_i - \half \sum_j \left<s_i,s_j\right> v_j\\
    \varphi(x) &= x + \sum_i \left<s_i,x\right> v_i,
  \end{aligned}
\end{equation}
for all $z\in W_0$ and all $x\in \bigoplus_{\alpha =1}^N W_\alpha$.
The effect of such a transformation on the 3-brackets \eqref{eq:V3b}
is an uninteresting modification of $K_{ijk}$ and $L_{ijk\ell}$
and the more interesting disappearance of $J_{ij}$ from the 3-brackets
involving elements in $W_\alpha$. Indeed, for all $x \in W_\alpha$,
we have
\begin{align*}
  [\varphi(u_i),\varphi(u_j),\varphi(x)] &= [u_i-s_i,u_j-s_j,x]\\
  &= [u_i,u_j,x] + [u_j,s_i,x] - [u_i,s_j,x] + [s_i,s_j,x]\\
  &= J_{ij} x + [s_i,x]_j - [s_j,x]_i + \text{central terms}\\
  &= [z_{ij}^\alpha, x] + \kappa_j^\alpha [s_i^\alpha,x] -
  \kappa_i^\alpha [s_j^\alpha,x] + \text{central terms}\\
  &= [z_{ij}^\alpha + \kappa_j^\alpha s_i^\alpha - \kappa_i^\alpha
  s_j^\alpha , x] + \text{central terms}\\
  &= 0 + \text{central terms},
\end{align*}
where we have used equation \eqref{eq:zijalpha}.

This means that, without loss of generality, we may assume that $J_{ij}
x =0$ for all $x \in W_\alpha$ for any $\alpha$. Now, consider
equation \eqref{eq:C5} for $x \in \bigoplus_{\alpha =1}^N W_\alpha$.
The right-hand side vanishes, hence $[K_{ijk},x]_\ell=0$. Also, if $x
\in W_0$, then $[K_{ijk},x]_\ell=0$ because $x$ is central with
respect to all $\fg_\ell$. Therefore we see that $K_{ijk}$ is central
with respect to all $\fg_\ell$, and hence $K_{ijk} \in W_0$.

In other words, we have proved the following

\begin{lemma}
  \label{lem:C1C6C5}
  In the notation of lemma~\ref{lem:C2C3}, the nonzero 3-brackets for
  $V$ may be brought to the form
  \begin{equation}
    \label{eq:V3b2}
    \begin{aligned}[m]
      [u_i,u_j,u_k] &= K_{ijk} + \sum_{\ell=1}^r L_{ijk\ell} v_\ell\\
      [u_i,u_j,x_0] &= J_{ij} x_0 - \sum_{k=1}^r \left<K_{ijk},x_0\right> v_k\\
      [u_i,x_0,y_0] &= - \sum_{j=1}^r \left<x_0,J_{ij} y_0\right> v_j\\
      [u_i,x_\alpha,y_\alpha] &= \kappa_i^\alpha [x,y]\\
      [x_\alpha,y_\alpha,z_\alpha] &= -
      \left<[x_\alpha,y_\alpha],z_\alpha\right> \sum_{i=1}^r
      \kappa_i^\alpha v_i,
    \end{aligned}
  \end{equation}
  for all $x_\alpha,y_\alpha,z_\alpha\in W_\alpha$, $x_0,y_0 \in W_0$
  and for some $L_{ijk\ell} \in \RR$, $K_{ijk} \in W_0$ and
  $J_{ij} \in \fso(W_0)$, all of which are totally skewsymmetric in
  their indices.
\end{lemma}

Since their left-hand sides vanish, equations \eqref{eq:C4} and
\eqref{eq:C5} become conditions on $J_{ij} \in \fso(W_0)$:
\begin{align}
  J_{ij}J_{k\ell} - J_{k\ell}J_{ij} &= 0, \label{eq:JJ1}\\
  J_{jk}J_{i\ell} + J_{k\ell}J_{ij} + J_{j\ell}J_{ki} &= 0. \label{eq:JJ2}
\end{align}
The first condition says that the $J_{ij}$ commute, hence the inner product on $W_0$ being positive-definite, they must belong
to the same Cartan subalgebra $\fh \subset \fso(W_0)$. Let $H_\pi$, for $\pi=1,\dots,\lfloor\frac{\dim W_0}{2}\rfloor$, denote a
basis for $\fh$, with each $H_\pi$ corresponding to the generator of infinitesimal rotations in mutually orthogonal 2-planes in
$W_0$. In particular, this means that $H_\pi H_\varrho = 0$ for $\pi \neq \varrho$ and that $H_\pi^2 = - \Pi_\pi$, with $\Pi_\pi$
the orthogonal projector onto the 2-plane labelled by $\pi$. We write $J_{ij}^\pi \in \RR$ for the component of $J_{ij}$ along
$H_\pi$. Fixing $\pi$ we may think of $J_{ij}^\pi$ as the components of $J^\pi \in \Lambda^2 \RR^r$. Using the relations obeyed
by the $H_\pi$, equation \eqref{eq:JJ2} separates into $\lfloor\frac{\dim W_0}{2}\rfloor$ equations, one for each value of $\pi$,
which in terms of $J^\pi$ can be written simply as $J^\pi \wedge J^\pi = 0$. This is a special case of a Plücker relation and
says that $J^\pi$ is decomposable; that is, $J^\pi = \eta^\pi \wedge \zeta^\pi$ for some $\eta^\pi, \zeta^\pi \in \RR^r$. In
other words, the solution of equations \eqref{eq:JJ1} and \eqref{eq:JJ2} is
\begin{equation}
  J_{ij} = \sum_\pi \left(\eta_i^\pi \zeta^\pi_j - \eta_j^\pi \zeta^\pi_i\right)
  H_\pi
\end{equation}
living in a Cartan subalgebra $\fh \subset \fso(W_0)$.

\subsection{Solving for the $K_{ijk}$}
\label{sec:solving-K}

It remains to solve equations \eqref{eq:C7} and \eqref{eq:C8} for
$K_{ijk}$. The linear equation \eqref{eq:C7} on $K \in \Lambda^3\RR^r
\otimes W_0$ and says that it is in the kernel of a linear map
\begin{equation}
  \label{eq:cocycle}
  \begin{CD}
    \Lambda^3\RR^r \otimes W_0 @>>> \Lambda^2\RR^r \otimes
    \Lambda^3\RR^r \otimes W_0
  \end{CD}
\end{equation}
defined by
\begin{equation}
\label{eq:cocycle-eqn}
  K_{ijk} \mapsto J_{ij} K_{k\ell m} - J_{\ell m}K_{ijk}  - J_{mk}K_{ij\ell} - J_{k\ell}K_{ijm}.
\end{equation}
The expression in the right-hand side is manifestly skewsymmetric in
$ij$ and $k\ell m$ separately, hence it belongs to $\Lambda^2\RR^r
\otimes \Lambda^3\RR^r \otimes W_0$ as stated above. For generic $r$
(here $r\geq 5$) we may decompose
\begin{equation}
  \Lambda^2\RR^r \otimes \Lambda^3\RR^r = Y^{\yng(2,2,1)}\RR^r \oplus
  Y^{\yng(2,1,1,1)}\RR^r \oplus \Lambda^5 \RR^r,
\end{equation}
where $Y^{\text{Young tableau}}$ denotes the corresponding Young
symmetriser representation. Then one can see that the right-hand side
of \eqref{eq:cocycle-eqn} has no component in the first of the above
summands and hence lives in the remaining two summands, which are
isomorphic to $\RR^r \otimes \Lambda^4 \RR^r$.

We now observe that via an isometry of $V$ of the form
\begin{equation}
  \begin{aligned}[m]
    \varphi(v_i) &= v_i\\
    \varphi(x_\alpha) &= x_\alpha\\
    \varphi(u_i) &= u_i + t_i - \half \sum_j \left<t_i,t_j\right> v_j\\
    \varphi(x_0) &= x_0 - \sum_i \left<x_0, t_i\right> v_i,
  \end{aligned}
\end{equation}
for $t_i \in W_0$, the form of the 3-brackets \eqref{eq:V3b2} remains
invariant, but with $K_{ijk}$ and $L_{ijk\ell}$ transforming by
\begin{align}
  K_{ijk} &\mapsto K_{ijk} + J_{ij} t_k + J_{jk} t_i + J_{ki} t_j,\\
\intertext{and}
  \begin{split}
    L_{ijk\ell} &\mapsto L_{ijk\ell} +
    \left<K_{ijk},t_\ell\right> - \left<K_{\ell ij},t_k\right> +
    \left<K_{k\ell i},t_j\right> - \left<K_{jk\ell},t_i\right>\\
    &\qquad + \left<J_{ij} t_k,t_\ell\right> + \left<J_{ki}
      t_j,t_\ell\right> + \left<J_{jk} t_i,t_\ell\right> +
    \left<J_{i\ell} t_j,t_k\right> + \left<J_{j\ell} t_k,t_i\right> +
    \left<J_{k\ell} t_i,t_j\right>,
  \end{split}
\end{align}
respectively. In particular, this means that there is an ambiguity in
$K_{ijk}$, which can be thought of as shifting it by the image of the
linear map
\begin{equation}
  \label{eq:coboundary}
  \begin{CD}
    \RR^r \otimes W_0 @>>> \Lambda^3\RR^r \otimes W_0
  \end{CD}
\end{equation}
defined by
\begin{equation}
  t_i \mapsto J_{ij} t_k + J_{jk} t_i + J_{ki} t_j.
\end{equation}
The two maps \eqref{eq:cocycle} and \eqref{eq:coboundary} fit together
in a complex
\begin{equation}
  \label{eq:complex}
  \begin{CD}
    \RR^r \otimes W_0 @>>> \Lambda^3\RR^r \otimes W_0 @>>> \RR^r \otimes
    \Lambda^4\RR^r \otimes W_0,
  \end{CD}
\end{equation}
where the composition vanishes \emph{precisely} by virtue of equations
\eqref{eq:JJ1} and \eqref{eq:JJ2}. We will show that this complex is
acyclic away from the kernel of $J$, which means that without loss
of generality we can take $K_{ijk}$ in the kernel of $J$ subject to
the final quadratic equation \eqref{eq:C8}.

Let us decompose $W_0$ into an orthogonal direct sum
\begin{equation}
  W_0 =
  \begin{cases}
     \bigoplus\limits_{\pi=1}^{(\dim W_0)/2} E_\pi, & \text{if $\dim W_0$ is even, and}\\[18pt]
     \RR w \oplus \bigoplus\limits_{\pi=1}^{(\dim W_0-1)/2} E_\pi, & \text{if $\dim W_0$ is odd,}
  \end{cases}
\end{equation}
where $E_\pi$ are mutually orthogonal 2-planes and, in the second case,
$w$ is a vector perpendicular to all of them. On $E_\pi$ the Cartan
generator $H_\pi$ acts as a complex structure, and hence we may identify
each $E_\pi$ with a complex one-dimensional vector space and $H_\pi$ with
multiplication by $i$. This decomposition of $W_\pi$ allows us to
decompose $K_{ijk} = K^w_{ijk} + \sum_\pi K^\pi_{ijk}$, where the first
term is there only in the odd-dimensional situation and the
$K^\pi_{ijk}$ are complex numbers. The complex \eqref{eq:complex}
breaks up into $\lfloor\frac{\dim W_0}{2}\rfloor$ complexes, one for
each value of $\pi$. If $J^\pi = 0$ then $K_{ijk}^\pi$ is not constrained
there, but if $J^\pi = \eta^\pi \wedge \zeta^\pi \neq 0$ the complex turns
out to have no homology, as we now show.

Without loss of generality we may choose the vectors $\eta^\pi$ and
$\zeta^\pi$ to be the elementary vectors $e_1$ and $e_2$ in $\RR^r$, so
that $J^\pi$ has a $J^\pi_{12}=1$ and all other $J^\pi_{ij}=0$. Take $i=1$
and $j=2$ in the cocycle condition \eqref{eq:cocycle}, to obtain
\begin{equation}
  K^\pi_{k\ell m} = J^\pi_{\ell m} K^\pi_{12k} + J^\pi_{mk} K^\pi_{12\ell} +
  J^\pi_{k\ell} K^\pi_{12m}.
\end{equation}
It follows that if any two of $k,\ell,m >2$, then $K^\pi_{k\ell m} = 0$.
In particular $K^\pi_{1ij} = K^\pi_{2ij} = 0$ for all $i,j>2$, hence only
$K^\pi_{12k}$ for $k>2$ can be nonzero. However for $k>2$, $K^\pi_{12k} =
J^\pi_{12} e_k$, with $e_k$ the $k$th elementary vector in $\RR^r$, and
hence $K^\pi_{12k}$ is in the image of the map \eqref{eq:coboundary};
that is, a coboundary. This shows that we may assume without loss of
generality that $K^\pi_{ijk} = 0$. In summary, the only components of
$K_{ijk}$ which survive are those in the kernel of all the $J_{ij}$.
It is therefore convenient to split $W_0$ into an orthogonal direct
sum
\begin{equation}
  W_0 = E_0 \oplus \bigoplus_\pi E_\pi,
\end{equation}
where on each 2-plane $E_\pi$, $J^\pi = \eta^\pi \wedge \zeta^\pi \neq 0$,
whereas $J_{ij} x = 0$ for all $x \in E_0$. Then we can take $K_{ijk}
\in E_0$.

Finally it remains to study the quadratic equation \eqref{eq:C8}.
First of all we mention that this equation is automatically satisfied
for $r\leq 4$. To see this notice that the equation is skewsymmetric
in $k,\ell, m,n$, hence if $r<4$ it is automatically zero. When $r=4$,
we have to take $k,\ell,m,n$ all different and hence the equation
becomes
\begin{equation*}
  \left<K_{ij1},K_{234}\right> - \left<K_{ij2},K_{341}\right> +
  \left<K_{ij3},K_{412}\right> - \left<K_{ij4},K_{123}\right> = 0,
\end{equation*}
which is skewsymmetric in $i,j$. There are six possible choices for
$i,j$ but by symmetry any choice is equal to any other up to
relabelling, so without loss of generality let us take $i=1$ and $j=2$,
hence the first two terms are identically zero and the two remaining
terms satisfy
\begin{equation*}
  \left<K_{123},K_{412}\right> - \left<K_{124},K_{123}\right> = 0,
\end{equation*}
which is identically true. This means that the cases of index $3$
and $4$ are classifiable using our results. By contrast, the cases of
index $5$ and above seem not to be tame, as we illustrate with an example. Let us take the case of $r=5$ and $\dim E_0 = 1$, so that
the $K_{ijk}$ can be taken to be real numbers. The solutions to
\eqref{eq:C8} now describe the intersection of five quadrics in
$\RR^{10}$:
\begin{gather*}
K_{125} K_{134} - K_{124} K_{135} + K_{123} K_{145} = 0\\
K_{125} K_{234} - K_{124} K_{235} + K_{123} K_{245} = 0\\
K_{135} K_{234} - K_{134} K_{235} + K_{123} K_{345} = 0\\
K_{145} K_{234} - K_{134} K_{245} + K_{124} K_{345} = 0\\
K_{145} K_{235} - K_{135} K_{245} + K_{125} K_{345} = 0,
\end{gather*}
hence the solutions define an algebraic variety. One possible branch
is given by setting $K_{1ij}=0$ for all $i,j$, which leaves
undetermined $K_{234}$, $K_{235}$, $K_{245}$ and $K_{345}$. There are
other branches which are linearly related to this one: for
instance, setting $K_{2ij}=0$, etc., but there are also other
branches which are not linearly related to it.

\subsection{Summary and main theorem}
\label{sec:summary-conclusion}

Let us summarise the above results in the following structure
theorem.

\begin{theorem}
  \label{thm:main}
  Let $V$ be a finite-dimensional indecomposable metric 3-Lie algebra
  of index $r>0$ with a maximally isotropic centre. Then $V$ admits a
  vector space decomposition into $r+M+N+1$ orthogonal subspaces
  \begin{equation}
    \label{eq:decomp-V}
    V = \bigoplus_{i=1}^r \left(\RR u_i \oplus \RR v_i\right) \oplus
    \bigoplus_{\alpha =1}^N W_\alpha \oplus \bigoplus_{\pi=1}^M E_\pi
    \oplus E_0,
  \end{equation}
  where $W_\alpha$, $E_\pi$ and $E_0$ are positive-definite subspaces
  with the $E_\pi$ being two-dimensional, and where
  $\left<u_i,u_j\right> = \left<v_i,v_j\right> = 0$ and
  $\left<u_i,v_j\right> = \delta_{ij}$. The 3-Lie algebra is defined
  in terms of the following data:
  \begin{itemize}
  \item $0 \neq \eta^\pi \wedge \zeta^\pi \in \Lambda^2 \RR^r$ for each $\pi=1,\dots,M$,
  \item $0 \neq \kappa^\alpha \in \RR^r$ for each $\alpha =1, \dots, N$,
  \item a metric simple Lie algebra structure $\fg_\alpha$ on each $W_\alpha$,
  \item $L \in \Lambda^4\RR^r$, and
  \item $K \in \Lambda^3\RR^r \otimes E_0$ subject to the equation
    \begin{equation*}
     \left<K_{ijn},K_{k\ell m}\right> + \left<K_{ij\ell},K_{mnk}\right> - \left<K_{ijm},K_{nk\ell}\right> - \left<K_{ijk},K_{\ell mn}\right> = 0,
    \end{equation*}
  \end{itemize}
  by the following 3-brackets, \footnote{We understand that if a 3-bracket is not listed here it vanishes. Also every
    summation is written explicitly, so the summation convention is not in force. In particular, there is no sum over $\pi$ in
    the third and fourth brackets.}
  \begin{equation}
    \label{eq:V3b-main}
    \begin{aligned}[m]
      [u_i,u_j,u_k] &= K_{ijk} + \sum_{\ell=1}^r L_{ijk\ell} v_\ell\\
      [u_i,u_j,x_0] &= - \sum_{k=1}^r \left<K_{ijk},x_0\right> v_k\\
      [u_i,u_j,x_\pi] &= J^\pi_{ij} H_\pi x_\pi\\
      [u_i,x_\pi,y_\pi] &= - \sum_{j=1}^r \left<x_\pi,J^\pi_{ij} H_\pi y_\pi\right> v_j\\
      [u_i,x_\alpha,y_\alpha] &= \kappa_i^\alpha [x_\alpha,y_\alpha]\\
      [x_\alpha,y_\alpha,z_\alpha] &= - \left<[x_\alpha,y_\alpha],z_\alpha\right> \sum_{i=1}^r \kappa_i^\alpha v_i,
    \end{aligned}
  \end{equation}
  for all $x_0 \in E_0$, $x_\pi,y_\pi \in E_\pi$ and $x_\alpha, y_\alpha, z_\alpha \in W_\alpha$, and where $J^\pi_{ij} =
  \eta^\pi_i \zeta^\pi_j - \eta^\pi_j \zeta^\pi_i$ and $H_\pi$ a complex structure on each 2-plane $E_\pi$. The resulting 3-Lie
  algebra is indecomposable provided that there is no $x_0 \in E_0$ which is perpendicular to all the $K_{ijk}$, hence in
  particular $\dim E_0 \leq \binom{r}{3}$.
\end{theorem}

\subsection{Examples for low index}
\label{sec:examples}

Let us now show how to recover the known classifications in index
$\leq 2$ from theorem~\ref{thm:main}.

Let us consider the case of minimal positive index $r=1$. In that
case, the indices $i,j,k,l$ in theorem~\ref{thm:main} can only take
the value $1$ and therefore $J_{ij}$, $K_{ijk}$ and $L_{ijkl}$ are not
present. Indecomposability of $V$ forces $E_0=0$ and $E_\pi=0$,
hence letting $u:=u_1$ and $v:=v_1$, we have $V = \RR u \oplus \RR v
\oplus \bigoplus_{\alpha =1}^N W_\alpha$ as a vector space, with
$\left<u,u\right> = \left<v,v\right> = 0$, $\left<u,v\right> = 1$ and
$\bigoplus_{\alpha =1}^N W_\alpha$ Euclidean. The 3-brackets are:
\begin{equation}
\label{eq:Wg3b}
  \begin{aligned}[m]
    [u,x_\alpha,y_\alpha] &= [x_\alpha,y_\alpha]\\
    [x_\alpha,y_\alpha,z_\alpha] &= - \left<[x_\alpha,y_\alpha],z_\alpha\right>  v,
  \end{aligned}
\end{equation}
for all $x_\alpha, y_\alpha, z_\alpha \in W_\alpha$ and where we have
redefined $ \kappa^\alpha [x_\alpha,y_\alpha] \to
[x_\alpha,y_\alpha]$, which is a simple Lie algebra on each
$W_\alpha$. This agrees with the classification of Lorentzian $3$-Lie
algebras in section \ref{sec:Lorm3LA}. We conclude that all indecomposable Lorentzian 3-Lie algebras which are of the form of a double extension admit a maximally isotropic centre. 

Let us now consider $r=2$. According to theorem~\ref{thm:main}, those
with a maximally isotropic centre may now have a nonvanishing
$J_{12}$ while $K_{ijk}$ and $L_{ijkl}$ are still absent.
Indecomposability of $V$ forces $E_0=0$. Therefore $W_0 = \bigoplus_{\pi=1}^M E_\pi$
and, as a vector space, $V = \RR u_1 \oplus \RR v_1 \oplus \RR u_2
\oplus \RR v_2 \oplus W_0 \oplus \bigoplus_{\alpha =1}^N W_\alpha$ with
$\left<u_i,u_j\right> = \left<v_i,v_j\right> = 0$,
$\left<u_i,v_j\right> = \delta_{ij}$, $\forall i,j = 1,2$ and $W_0
\oplus \bigoplus_{\alpha =1}^N W_\alpha$ is Euclidean. The $3$-brackets are
now:
\begin{equation} \label{eq:2pIndec}
  \begin{aligned}[m]
    [u_1,u_2,x_\pi] &= J x_\pi\\
    [u_1,x_\pi,y_\pi] &= - \left<x_\pi,J y_\pi\right> v_2\\
    [u_2,x_\pi,y_\pi] &=  \left<x_\pi,J y_\pi\right> v_1\\
    [u_1,x_\alpha,y_\alpha] &= \kappa_1^\alpha [x_\alpha,y_\alpha]\\
    [u_2,x_\alpha,y_\alpha] &= \kappa_2^\alpha [x_\alpha,y_\alpha]\\
    [x_\alpha,y_\alpha,z_\alpha] &= - \left<[x_\alpha,y_\alpha],z_\alpha\right>  \kappa_1^\alpha v_1 %
    - \left<[x_\alpha,y_\alpha],z_\alpha\right> \kappa_2^\alpha v_2,
  \end{aligned}
\end{equation}
for all $x_\pi,y_\pi \in E_\pi$ and $x_\alpha, y_\alpha, z_\alpha \in
W_\alpha$. This agrees with the classification in \cite{2p3Lie} of
finite-dimensional indecomposable $3$-Lie algebras of index 2 whose
centre contains a maximally isotropic plane. In that paper such
algebras were denoted $V_{\text{IIIb}}(E,J,\fl,\fh,\fg,\psi)$ with
underlying vector space $\RR(u,v) \oplus \RR(\be_+,\be_-) \oplus E
\oplus\fl \oplus\fh\oplus \fg$ with $\left<u,u\right> =
\left<v,v\right> = \left<\be_\pm,\be_\pm\right> = 0$,
$\left<u,v\right>=1=\left<\be_+,\be_-\right>$ and all $\oplus$
orthogonal. The nonzero Lie 3-brackets are given by
\begin{equation}
  \label{eq:type-IIIb}
  \begin{aligned}[m]
    [u,\be_-,x] &= J x\\
    [u,x,y] &= \left<J x,y\right>\be_+\\
    [\be_-,x,y] &= - \left<Jx,y\right> v\\
    [\be_-,h_1,h_2] &= [h_1,h_2]_{\fh}\\
    [h_1,h_2,h_3] &= -\left<[h_1,h_2]_{\fh},h_3\right> \be_+
  \end{aligned}\qquad
  \begin{aligned}[m]
    [u,g_1,g_2] &= [\psi g_1,g_2]_{\fg}\\
    [\be_-,g_1,g_2] &= [g_1,g_2]_{\fg}\\
    [g_1,g_2,g_3] &= - \left<[g_1,g_2]_{\fg},g_3\right> \be_+ - \left<[\psi g_1,g_2]_{\fg}, g_3\right> v\\
    [u,\ell_1,\ell_2] &= [\ell_1,\ell_2]_{\fl}\\
    [\ell_1,\ell_2,\ell_3] &= -\left<[\ell_1,\ell_2]_{\fl},\ell_3\right> v,
  \end{aligned}
\end{equation}
where $x,y\in E$, $h,h_i\in\fh$, $g_i\in\fg$ and $\ell_i\in\fl$.

To see that this family of $3$-algebras is of the type \eqref{eq:2pIndec} it is enough to identify
\begin{equation}
  u_1 \leftrightarrow u  \qquad v_1 \leftrightarrow v \qquad u_2 \leftrightarrow e_- \qquad v_2 \leftrightarrow e_+
\end{equation}
as well as
\begin{equation}
  W_0 \leftrightarrow E \qquad\text{and}\qquad \bigoplus_{\alpha =1}^N  W_\alpha  \leftrightarrow  \fl \oplus\fh\oplus \fg,
\end{equation}
where the last identification is not only as vector spaces but also as Lie algebras, and set
\begin{equation} \label{eq:2pkappas}
  \begin{aligned}[m]
  \kappa_1|_{\fh} &= 0\\
  \kappa_1|_{\fl} &= 1 \\
  \kappa_1|_{\fg_{\alpha}} &= \psi_{\alpha}
  \end{aligned}\qquad\qquad
  \begin{aligned}[m]
    \kappa_2|_{\fh} &= 1\\
  \kappa_2|_{\fl} &= 0\\
  \kappa_2|_{\fg_{\alpha}} &= 1,
  \end{aligned}
\end{equation}
to obtain the map between the two families. As shown in \cite{2p3Lie}
there are 9 different types of such 3-Lie algebras, 
depending on which of the four ingredients $(E,J)$, $\fl$, $\fh$ or
$(\fg,\psi)$ are present. 

The next case is that of index $r=3$, where there are up to three nonvanishing $J_{ij}$ and one $K_{123} := K$, while $L_{ijkl}$ is still not present. Indecomposability of $V$ forces $\dim E_0 \leq 1$.
As a vector space, $V$ splits up as
\begin{equation}
  V = \bigoplus_{i=1}^3 \left(\RR u_i \oplus \RR v_i\right) \oplus
  \bigoplus_{\alpha =1}^N W_\alpha \oplus \bigoplus_{\pi=1}^M E_\pi
  \oplus E_0,
\end{equation}
where all $\oplus$ are orthogonal except the second one, $W_\alpha$,
$E_0$ and $E_\pi$ are positive-definite subspaces with $\dim E_0 \leq
1$, $E_\pi$ being two-dimensional, and where $\left<u_i,u_j\right> =
\left<v_i,v_j\right> = 0$ and $\left<u_i,v_j\right> = \delta_{ij}$.
The 3-brackets are given by
\begin{equation}\label{eq:3pIndec}
  \begin{aligned}[m]
    [u_1,u_2,u_3] &= K \\
    [u_i,u_j,x_0] &= - \sum_{k=1}^r \left<K_{ijk},x_0\right> v_k\\
    [u_i,u_j,x_\pi] &= J^\pi_{ij} H_\pi x_\pi\\
    [u_i,x_\pi,y_\pi] &= - \sum_{j=1}^r \left<x_\pi,J^\pi_{ij} H_\pi y_\pi\right> v_j\\
    [u_i,x_\alpha,y_\alpha] &= \kappa_i^\alpha [x_\alpha,y_\alpha]\\
    [x_\alpha,y_\alpha,z_\alpha] &= - \left<[x_\alpha,y_\alpha],z_\alpha\right> \sum_{i=1}^r \kappa_i^\alpha v_i,
  \end{aligned}
\end{equation}
for all $x_0 \in E_0$, $x_\pi,y_\pi \in E_\pi$ and $x_\alpha, y_\alpha, z_\alpha \in W_\alpha$, and where $J^\pi_{ij} = \eta^\pi_i
\zeta^\pi_j - \eta^\pi_j \zeta^\pi_i$ and $H_\pi$ a complex structure on each 2-plane $E_\pi$.

Finally, let us remark that the family of admissible 3-Lie algebras found in \cite{Ho:2009nk} are included in theorem~\ref{thm:main}. In that paper, a family of solutions to equations \eqref{eq:FI-V-pre} was found by setting each of the Lie algebra structures $[-,-]_i$ to be nonzero in orthogonal subspaces of $W$. This corresponds, in our language, to the particular case of allowing precisely one $\kappa_i^\alpha$ to be nonvanishing in each $W_{\alpha}$.

Notice that, as shown in \eqref{eq:2pkappas}, already in \cite{2p3Lie} there are examples of admissible 3-Lie algebras of index
$2$ which are not of this form as both $\kappa_1$ and $\kappa_2$ might be nonvanishing in the $\fg_{\alpha}$ factors.

To solve the rest of the equations, two ansätze are proposed in \cite{Ho:2009nk}:
\begin{itemize}
\item the trivial solution with nonvanishing $J$,
  i.e. $\kappa_i^\alpha = 0$, $K_{ijk}=0$ for all $i,j,k = 1,...,r$ and
  for all $\alpha$; and
\item precisely one $\kappa_i^\alpha = 1$ for each $\alpha$ (and
  include those $W_{\alpha}$'s where all $\kappa$'s are zero in $W_0$)
  and one $J_{ij} := J \neq 0$ assumed to be an outer
  derivation of the reference Lie algebra defined on $W$.
\end{itemize}

As pointed out in that paper, $L_{ijkl}$ is not constrained by the fundamental identity, so it can in principle take any value,
whereas the ansatz provided for $K_{ijk}$ is given in terms of solutions of an equation equivalent to \eqref{eq:C8}. In the
Lagrangians they considered, both $L_{ijkl}$ and $K_{ijk}$ are set to zero.

One thing to notice is that in all these theories there is certain redundancy concerning the index of the 3-Lie algebra. If the
indices in the nonvanishing structures $\kappa_i^\alpha$, $J_{ij}$, $K_{ijk}$ and $L_{ijkl}$ involve only numbers from 1 to
$r_0$, then any 3-Lie algebra with such nonvanishing structures and index $r \geq r_0$ gives rise to the equivalent theories.

In this light, in the first ansatz considered, one can always define the non-vanishing $J$ to be $J_{12}$ and then the
corresponding theory will be equivalent to one associated to the index-2 3-Lie algebras in \cite{2p3Lie}.

In the second case, the fact that $J$ is an outer derivation implies that it must live on the abelian part of $W$ as a Lie algebra because the semisimple part does not possess outer derivations. This coincides with what was shown above, i.e., that $J|_{W_{\alpha}} = 0$ for each $\alpha$. Notice that each Lie algebra $[-,-]_i$ identically vanishes in $W_0$, therefore the
structure constants of the 3-Lie algebra do not mix $J$ and $[-,-]_i$. The theories in \cite{Ho:2009nk} corresponding to this ansatz also have $K_{ijk}=0$, hence they are again equivalent to the theory corresponding to the index-2 3-Lie algebra which was
denoted $V(E,J,\fh)$ in \cite{2p3Lie}.

\chapter[3D superconformal Chern-Simons-matter]{Three-dimensional superconformal Chern-Simons-matter theories}
\label{sec:3dscfts}

In this chapter we review superconformal Chern-Simons theories in 3-dimensions coupled to matter (SCCSM) with supersymmetry from $N=1$ to $N=8$. This class of conformal gauge theories (or a subset of them) are conjectured to be dual to configurations of M2-branes in different $AdS_4$ supergravity backgrounds. 

Such Lagrangians have been very difficult to find for over a decade for $N \geq 4$ supersymmetry. One reason for this might be, as we will see, a novel feature of these theories: the amount of supersymmetry preserved is related to the Lie algebra for the gauge theory and the particular representation the fields live in. In particular, to obtain $N \geq 4$ this algebra must not be simple. Hence, if one restricts to simple Lie algebras (assuming for example that it is $\fu(n)$) thinking that theories based on semi-simple ones will decompose in independent theories on each factor, it is impossible to construct such theories with $N \geq 4$. 

The situation changed in 2007 when Bagger and Lambert proposed a Lagrangian for an $N=8$ superconformal Chern-Simons-matter (SCCSM) theory in 3 dimensions \cite{BL1,BL2}. The novel feature of this theory was that the gauge symmetry was based in a 3-Lie algebra instead of a Lie algebra. After this, many other theories were constructed. Some were also based in 3-algebras but perhaps not of Lie type \cite{CherSaem}, some others in more standard Lie algebras \cite{GaiottoWitten} and some were written in both languages \cite{MaldacenaBL,BL4}. In this chapter we provide a uniform framework in which to understand all these theories and illustrate how the Faulkner construction we discussed in chapter \ref{sec:3algebras} implies that all these theories can be equivalently written in terms of 3-Leibniz algebras or Lie algebras and their representations.

We start with the most general $N=1$ SCCSM Lagrangian. Such a Lagrangian has three main parts: a kinetic part for the matter fields, a Chern-Simons part and a superpotential. Then, we show how by choosing a particular form of the superpotential, one can obtain the most general $N=2$ SCCSM Lagrangian. The only prerequisite for this to be possible will be that the matter fields live in a complex unitary representation of the gauge Lie algebra. Similarly, we show how another choice of the $N=1$ superpotential enhances supersymmetry to $N=3$. This choice of superpotential will turn out to be rigid, not allowing any room for continuing this process to $N>3$ supersymmetry. We will work through all these Lagrangians mainly in $N=1$ superspace, both on and off-shell. However from $N=3$ on-wards we will be forced to work on-shell only because an off-shell formalism would require to use projective or harmonic space techniques adding a complexity to the discussion that will not be necessary for our analysis. 

Once we prove that the $N=3$ superpotential is rigid, the only way to achieve $N>3$ supersymmetry is by considering special cases of the $N<4$ theories. We will then discuss how choosing the representation of the Lie algebra to be of the some of the extreme cases embeddable in a Lie superalgebra discussed in section \ref{sec:extremeLSA} provides enhancement to $N>3$. A key feature in this discussion will be R-symmetry of the superpotential. It is a necessary condition for a Lagrangian to be invariant under $N$-extended supersymmetry that it is invariant under $\fso(N)$ R-symmetry. However, since we will be working in $N=1$ superspace, the choice of $N=1$ superspace parameter breaks this R-symmetry and the necessary condition for an enhanced $N$-extended superconformal theory to be derivable from an $N=1$ superpotential is that the superpotential be invariant under those R-symmetries preserving the choice of $N=1$ superspace parameter, which is an $\fso(N-1)$ subalgebra of the $\fso(N)$ R-symmetry. It will turn out that in all cases considered the theories resulting from imposing $\fso(N-1)$ invariance of the superpotential will indeed be $N$-supersymmetric, rendering this condition necessary and sufficient.

We want to remark that we consider the class of SCCSM theories in 3-dimensions from a merely classically field theoretic point of view. That is, we fix the symmetries we would like the Lagrangians to have and then construct the most general Lagrangians whose integrals are invariant under those symmetries. We do not enter into matters of quantisation or interpretation of these theories in the context of M-theory by looking at their moduli spaces or other techniques. However, the discussion in this chapter provides a solid and uniform framework in which to understand all known 3-dimensional SCCSM theories establishing what representation-theoretic conditions must be obeyed in order to realise a particular amount of superconformal symmetry. 

In section \ref{sec:N8unitarity} we review our results found in \cite{Lor3Lie,2pBL} on theories with maximal supersymmetry, $N=8$. Such theories are based in a 3-Lie algebra, which is initially taken to be positive-definite to ensure unitarity. However, the drought of Euclidean 3-Lie algebras and the special interest of the maximally supersymmetric case were motivations to explore more general cases where the 3-Lie algebra has an inner product of arbitrary index. The results on metric 3-Lie algebras that we discussed on chapter \ref{sec:3LieAlg} allowed us to generalise to arbitrary index $r$ the methodology developed in \cite{GMRBL,BRGTV,HIM-M2toD2rev} to render theories based in Lorentzian 3-Lie algebras unitary.

This chapter is organised as follows. In section \ref{sec:SCalgebra} we describe the superconformal algebras for SCCSM theories in 3 dimensions and discuss the consequences such algebras have on the possible matter content of these theories. In section \ref{sec:N1susy} we discuss the $N=1$ theory from which all the others will follow by two supersymmetry enhancing mechanisms. In section \ref{sec:from1to3} we discuss the first mechanism: the choice of a particular form of the $N=1$ superpotential provides supersymmetry enhancement to $N=2,3$. After establishing that the $N=3$ superpotential is rigid, we explore in section \ref{sec:Ngt3} an alternative way to achieve $N>3$ based on imposing $\fso(N-1)$-invariance of the superpotential by choosing a specific class of 3-algebra where the matter fields are valued. This way we recover all the known theories in the literature with $N>3$ and prove that in this context $N=7$ supersymmetry implies $N=8$. We dedicate section \ref{sec:N8unitarity} to the case of maximal supersymmetry, $N=8$, in an attempt to overcome the scarcity of such theories consequence of the fact that there is a unique positive-definite 3-Lie algebra. We close this chapter and this thesis with a summary \ref{sec:summary} of the results we will find throughout and an outlook on future research in the field. The content of this chapter is based on \cite{2pBL,Lie3Algs, SCCS3Algs}.

\section{Superconformal algebra and its consequences}
\label{sec:SCalgebra}

In this section we review the general constraints that representation theory imposes in the content of 3-dimensional superconformal Chern-Simons-matter theories. First, we specify the superconformal algebra under which such theories must be invariant, and then we analyse the consequences this has for the matter fields allowed. This overview of the representation theory involved will facilitate the construction of the different N-supersymmetric Lagrangians in sections to follow. We close this section with a note on indecomposability of these theories, as we aim to classify only indecomposable ones and avoid redundancies.

\subsection{Matter content of 3D superconformal theories}
\label{sec:matter-reps}

\subsubsection{Superconformal algebra}

Superconformal field theories in three-dimensional Minkowski spacetime $\RR^{1,2}$ are invariant under a conformal superalgebra. These algebras were classified by Nahm in the 70's. Those in $\RR^{1,2}$ they are denoted type VII in \cite[proposition~2.2]{Nahm} and are indexed by a positive integer $N$ that coincides with the amount of supersymmetry preserved. The even Lie algebra is isomorphic to $\fso(N) \oplus \fso(2,3)$, where $\fso(N)$ is the R-symmetry of the field theory and $\fso(2,3)$ is the three-dimensional conformal algebra. The odd subspace is in the tensor product representation of the vector of $\fso(N)$ and the spinor of $\fso(2,3)$. The spinor representation of $\fso(2,3)$ defines an isomorphism $\fso(2,3) \cong \fsp(4,\RR)$, which means that the spin representation is real, four-dimensional and symplectic. In other words, the conformal superalgebra is isomorphic to the orthosymplectic Lie superalgebra $\fosp(N|4)$. 

This Lie superalgebra can be understood as endomorphisms of the vector superspace $\RR^{N|4}$ of $N$ even and $4$ odd dimensions preserving a Euclidean structure on the even subspace and a symplectic structure on the odd subspace. The supercharges in the superalgebra are the odd endomorphisms which map the even and odd subspaces to each other. Standard arguments restricting unitary interacting field theories to admit at most 32 supercharges impose an upper bound $N\leq 8$ for the interesting theories. For this reason we restrict our attention to $N\leq 8$. Having said that, such arguments must be taken with a pinch of salt for theories in less than 4 dimensions which do not admit a dimensional oxidation to four dimensions, as is the case with the theories we are concerned with here. See for example \cite{Bergshoeff:2008ta} for a topologically massive Chern-Simons theory admitting $N>8$.

\subsubsection{Field content}

The field content of these theories, ignoring auxiliary fields, is comprised of a gauge field $A_\mu$ valued in a metric Lie algebra $\fg$, and two matter fields: a bosonic scalar $X$ and a fermionic Majorana spinor $\Psi$ on $\RR^{1,2}$. These matter fields must realise the $\fso(N)$ R-symmetry of the theory and live in a representation of the Lie algebra $\fg$. In order to build a Lagrangian, this representation must be unitary, that is possess an ad-invariant inner product. The data necessary to define this matter content is then a metric Lie algebra and a unitary representation of it. As we know from the Faulkner construction, this data is equivalent to a metric 3-Leibniz algebra.

\begin{table}[ht!]
  \centering
  \begin{tabular}{|>{$}c<{$}>{$}c<{$}>{$}c<{$}|}
    \hline
    \text{Metric 3-Leibniz algebra} & \Leftrightarrow & \text{Metric Lie algebra and unitary representation}\\
    \left(\fM, [-,-,-], \left\langle -,-\right\rangle \right) &  \Leftrightarrow & \left(\fg,[-,-], (-,-)\right) \text{and} \left(\fM,  \left\langle -,-\right\rangle \right) \\\hline    
  \end{tabular}
  \vspace{8pt}
  \caption{Necessary data to define field content in 3D SCCSM}
  \label{tab:datafc}
\end{table}

Lets denote schematically by $\fB \otimes \fM_1$ and $\fF\otimes \fM_2$, the representations where the matter fields $X$ and $\Psi$ live in respectively, where $\fB$ and $\fF$ are the bosonic and fermionic R-symmetry representations and $\fM_1,\fM_2$ are representations of $\fg$. The supersymmetry transformations take the generic form \begin{equation}
  \label{eq:susy-sketch}
  \delta_\epsilon X = \overline{\epsilon}\Psi \qquad\text{and}\qquad \delta_\epsilon \Psi = dX \cdot \epsilon + \cdots,
\end{equation}
where $\cdot$ is the Clifford action and $\epsilon$ is the supersymmetry parameter which is a vector under the R-symmetry, but inert under $\fg$, reflecting the fact that for a rigidly supersymmetric theory, supersymmetry and gauge transformations commute. It then follows that $\fM_1 = \fM_2$, hence we will drop the subscript, and focusing on the R-symmetry, we see that letting $\fV$ denote the vector representation of the R-symmetry,
\begin{equation*}
  \fB \subset \fV \otimes \fF \qquad\text{and}\qquad   \fF \subset \fV \otimes \fB.
\end{equation*}
This suggests taking $\fB$ and $\fF$ to be spinor representations in such a way that the above inclusions are induced from the Clifford actions $\fV \otimes \fF \to \fB$ and $\fV \otimes \fB \to \fF$, respectively. We will do so. This means that when $N$ is odd, bosons and fermions will be in the same representation, whereas if $N$ is even, since Clifford multiplication by vectors reverses chirality, the fermionic representation will be obtained from the bosonic one by changing the chirality of the spinor representations.

\begin{table}[ht!]
  \centering
  \begin{tabular}{|>{$}c<{$}|>{$}c<{$}|>{$}c<{$}|}
    \hline
    N & \fso(N) & \text{spinor irreps}\\\hline
    2 &  \fu(1) & \CC\\
    3 &  \fsp(1) & \HH\\
    4 &  \fsp(1) \oplus \fsp(1) & \HH \oplus \HH\\
    5 &  \fsp(2) & \HH^2\\
    6 &  \fsu(4) & \CC^4\\
    7 &  \fso(7) & \RR^8\\
    8 &  \fso(8) & \RR^8 \oplus \RR^8\\\hline
  \end{tabular}
  \vspace{8pt}
  \caption{Spinor representations of $\fso(N)$ for $N\leq 8$}
  \label{tab:spinors}
\end{table}

Table~\ref{tab:spinors} summarises the spinor representations for $N\leq 8$. It lists the exceptional low-dimensional isomorphisms which are induced by the spinor representations and lists the types of representation with their dimension. For $N$ odd there is a unique irreducible spinor representation (up to isomorphism) which is real for $N\equiv \pm 1 \pmod{8}$ and quaternionic for $N\equiv \pm 3 \pmod{8}$. For $N$ even there are two, distinguished by chirality. They are complex for $N\equiv\pm 2 \pmod{8}$, with opposite chiralities being related by complex conjugation, real for $N\equiv 0 \pmod 8$ and quaternionic for $N\equiv 4 \pmod 8$. It will be convenient to introduce the following notation for the spinor representations: for $N$ odd, we let $\Delta^{(N)}$ denote the unique irreducible spinor representation of $\fso(N)$, whereas for $N$ even, we let $\Delta^{(N)}_\pm$ denote the unique irreducible spinor representation of $\fso(N)$ with positive/negative chirality, with the understanding that for $N=2,6$, $\overline{\Delta^{(N)}_\pm} = \Delta^{(N)}_\mp$.

The degrees of freedom described by the matter fields are fundamentally real and hence this fact determines the type of the representation $\fM$ in terms of the type of the relevant spinor representation. This means that if the spinor representation is real then so must $\fM$, whereas if the spinor representation is quaternionic then so must $\fM$, but we are then supposed to take the fields to be in the underlying real representation of the tensor product of the two quaternionic representations. In practical terms, this means imposing a reality condition on the fields which involves the symplectic structure of both the spinor representation and $\fM$. Finally, if the spinor representation is complex, we can take $\fM$ to be complex without loss of generality, with the understanding that we may think of both real and quaternionic representations as special types of complex representations. In this case, the matter fields take values in the real representation given by their real and imaginary parts. In conclusion, for $N=1,7,8$ the representations $\fM$ are real, for $N=3,4,5$ quaternionic and for $N=2,6$ complex.

Summarising so far, for odd $N$ the bosonic and fermionic matter fields both take values in the representation $\Delta^{(N)} \otimes \fM$, with the condition that for $N=3,5$, when $\Delta^{(N)}$ is quaternionic, fields must obey the natural reality condition. For even $N$ the bosonic matter fields can take values in the representation $\Delta^{(N)}_+ \otimes \fM_1 \oplus \Delta^{(N)}_- \otimes \fM_2$, whereas the fermionic matter fields take values in $\Delta^{(N)}_- \otimes \fM_1 \oplus \Delta^{(N)}_+ \otimes \fM_2$, where a priori both representations $\fM_1$ and $\fM_2$ can be different. Again, if $N=4$, then all representations are quaternionic, so that we must impose the natural symplectic reality condition on the fields. If $N=2,6$ then all representations are complex and we must consider the representation made up of by the real and imaginary parts of the fields or, said differently, to consider both the fields and their complex conjugates. In this case one may ignore the distinction between $\fM_1$ and $\fM_2$ because taking real and imaginary parts of $\Delta^{(N)}_+ \otimes \fM_1 \oplus \Delta^{(N)}_- \otimes \fM_2$ is the same as taking real and imaginary parts of $\Delta^{(N)}_+ \otimes (\fM_1\oplus\overline\fM_2)$, so that we can always take the matter fields to be in (the underlying real form of) a particular chiral spinor representation of $\fso(N)$. For $N=4$ the symplectic reality condition for the matter fields does not eliminate the distinction and we will see in later sections how this can lead to the notion of ``twisted'' and ``untwisted'' $N=4$ hypermultiplets according to the relative chiralities of the spinor representation of $\fso(4)$ they transform under. Similarly for $N=8$ the matter representations are real and one might expect to be able to distinguish between different types of matter in $\Delta^{(8)}_\pm$. However, this case of maximal supersymmetry will turn out to be rather special in that we will find one can obtain any $N=8$ supermultiplet from an $N=7$ one where both $\Delta^{(8)}_\pm$ are identified with $\Delta^{(7)}$ under the embedding $\fso(7) \into \fso(8)$ thus eliminating the apparent distinction between the two possible types of matter.

Table \eqref{tab:pre-matter-reps} displays this discussion. We use in it the more compact notation of denoting $U, V$ and $W$, real, complex and quaternionic representations of a Lie algebra. In other words, we keep $\fM$ for a generic representation but rename it $U, V, W$ when it belongs to $\Dar(\fg,\RR)$, $\Dar(\fg,\CC)$ and $\Dar(\fg,\HH)$ respectively. Note that the spinor representation $\Delta^{(1)}$ can be dropped because it is trivial.

\begin{table}[ht!]
  \centering
  \begin{tabular}{|>{$}c<{$}|>{$}c<{$}|>{$}c<{$}|>{$}c<{$}|}
    \hline
    N & \text{Field for }\fM & \text{Representation for X} & \text{Representation for } \Psi \\\hline
    8 & \RR & \Delta^{(8)}_+ \otimes U & \Delta^{(8)}_+ \otimes U \\
    7 & \RR & \Delta^{(7)}\otimes U & \Delta^{(7)}\otimes U\\
    6 & \CC & \Delta_+^{(6)} \otimes V \oplus \Delta_-^{(6)} \otimes \Vbar & \Delta_-^{(6)} \otimes V \oplus \Delta_+^{(6)} \otimes \Vbar\\
    5 & \HH & \Delta^{(5)} \otimes W & \Delta^{(5)} \otimes W \\
    4 & \HH & \Delta_+^{(4)} \otimes W_1 \oplus \Delta_-^{(4)} \otimes W_2 & \Delta_-^{(4)} \otimes W_1 \oplus \Delta_+^{(4)} \otimes W_2 \\
    3 & \HH & \Delta^{(3)} \otimes W & \Delta^{(3)} \otimes W \\
		2 & \CC & \Delta_+^{(2)} \otimes V \oplus \Delta_-^{(2)} \otimes \Vbar & \Delta_-^{(2)} \otimes V \oplus \Delta_+^{(2)} \otimes \Vbar\\
		1 & \RR & \Delta^{(1)}\otimes U \cong U & \Delta^{(1)}\otimes U \cong U \\\hline
  \end{tabular}
  \vspace{8pt}
  \caption{Representations for the matter fields in 3D SCCSM}
  \label{tab:pre-matter-reps}
\end{table}

\subsubsection{Constraints from supersymmetry}
\label{sec:constr-from-supersym}

We will see that for the $N=1,2,3$ theories one can take the matter to be in any real, complex or quaternionic unitary representations, respectively, whereas for $N>3$ the allowed representations are subject to further restrictions. We will start with the most general $N=1$ Lagrangian and work our way up to increasing amounts of supersymmetry by investigating the conditions for supersymmetry enhancement. We will find that, by selecting a particular $N=1$ superpotential, one can achieve $N=2$ and $N=3$ supersymmetry but not more because the $N=3$ superpotential will turn out to be rigid.

Theories with $N>3$ are then forced to be special cases of those theories with $N\leq 3$. To find the conditions on them that lead to supersymmetry enhancement we exploit R-symmetry of the superpotential. It is a necessary condition for a theory to be invariant under N-extended supersymmetry, that its Lagrangian be invariant under $\fso(N)$ R-symmetry. However, since we will be using $N=1$ superspace formalism, this translates into the condition for the $N=1$ superpotential that it must be invariant under those R-symmetries preserving the choice of $N=1$ superspace parameter, which is an $\fso(N-1)$ subalgebra of the $\fso(N)$ R-symmetry. Such $\fso(N-1)$-invariance of the superpotential can be achieved by expressing the matter fields, initially in spinor representations of $\fso(N)$, in terms of spinor representations of $\fso(N-1)$.

This way representation theory also helps to explain the conditions for supersymmetry enhancement. Table \ref{tab:spinor-decomp} summarises how the spinor representations decompose as a result of the embedding of the R-symmetry Lie algebras $\fso(N-1) \into \fso(N)$. The notation $\rf{V}$, introduced in section \ref{sec:relat-betw-real}, means the real representation obtained from the complex representation $V$ by restricting scalars to $\RR$.

\begin{table}[ht!]
  \centering
  \begin{tabular}{|>{$}c<{$}|>{$}r<{$}>{$}c<{$}>{$}l<{$}|}
    \hline
    N & \fso(N) & \supset & \fso(N-1)\\\hline
    8 & \Delta^{(8)}_\pm &\cong & \Delta^{(7)}\\
    7 & \Delta^{(7)} &\cong & \rf{\Delta_+^{(6)}}\\
    6 & \Delta_\pm^{(6)} &\cong & \Delta^{(5)}\\
    5 & \Delta^{(5)} &\cong & \Delta_+^{(4)} \oplus \Delta_-^{(4)}\\
    4 & \Delta_\pm^{(4)} &\cong & \Delta^{(3)}\\
    3 & \Delta^{(3)} &\cong & \Delta_+^{(2)} \oplus \Delta_-^{(2)}\\\hline
  \end{tabular}
  \vspace{8pt}
  \caption{Spinor representations under $\fso(N-1) \into \fso(N)$}
  \label{tab:spinor-decomp}
\end{table}

This then implies the decomposition of the matter representations from $N$- to ($N-1$)-extended supersymmetry which is summarised in table~\ref{tab:susy-enhancement}. Again,  we use notation introduced in section \ref{sec:relat-betw-real}. In particular, $U_\CC$ is the complexification of a real representation $U$, whereas $V_\HH$ is the quaternionification of a complex representation $V$ and $\rh{W}$ is a complex representation obtained from a quaternionic representation $W$ by forgetting the quaternionic structure. As usual, square brackets denote the underlying real representation, so that if $V$ is a complex representation with a real structure, then $[V]_\CC \cong V$.

\begin{table}[ht!]
  \centering
  \begin{tabular}{|>{$}c<{$}|>{$}c<{$}|>{$}c<{$}|}
    \hline
    N & N-\text{matter representation} & (N-1)-\text{matter representation}\\\hline
    8 & \Delta^{(8)}_+ \otimes U & \Delta^{(7)} \otimes U\\
    7 & \Delta^{(7)}\otimes U & [ (\Delta_+^{(6)} \oplus \Delta_-^{(6)}) \otimes U_\CC] \\
    6 & \Delta_+^{(6)} \otimes V \oplus \Delta_-^{(6)} \otimes \Vbar & \Delta^{(5)} \otimes V_\HH\\
    5 & \Delta^{(5)} \otimes W & \Delta_+^{(4)} \otimes W \oplus \Delta_-^{(4)} \otimes W\\
    4 & \Delta_+^{(4)} \otimes W_1 \oplus \Delta_-^{(4)} \otimes W_2 & \Delta^{(3)} \otimes (W_1 \oplus W_2)\\
    3 & \Delta^{(3)} \otimes W & (\Delta_+^{(2)} \oplus \Delta_-^{(2)})\otimes \rh{W}\\\hline
  \end{tabular}
  \vspace{8pt}
  \caption{Decomposition of the matter representations in 3D SCCS}
  \label{tab:susy-enhancement}
\end{table}

One may understand the following supersymmetry enhancements by looking at the $N$-extended matter representation in terms of the ($N-1$)-extended representation and then comparing with the generic ($N-1$)-extended representation. In practice one finds the $N$-extended matter representation in the second column of table~\ref{tab:susy-enhancement}, then moves over to the third column which shows this representation in terms of ($N-1$)-extended supersymmetry and then moves back to the second column but one row below to compare with the generic ($N-1$)-extended representations. 

In the following diagram the starting point is $N=4$ supersymmetry. We will find in following sections that $N=4$ supersymmetric theories arise from $N=3$ supersymmetric ones when $W_1,W_2$ are not only quaternionic representations, but they are of anti-Lie triple system type. For enhancement to $N=6$ $V$ must be of anti-Jordan triple system type and finally for $N=7,8$ $U$ must be a 3-Lie algebra. Then one can understand the enhancements $N=4\to N=5$, $N=5 \to N=6$ and $N=6 \to N>6$, as follows.
\begin{itemize}
\item In $N=4$, $W_1,W_2 \in \Dar(\fg,\HH)_{\text{aLTS}}$ and the enhancement to $N=5$ occurs precisely when $W_1 = W_2$:
  \begin{equation}
    \label{eq:4to5}
    \xymatrix{\Delta^{(5)} \otimes W \ar@{.>}[r] & \Delta_+^{(4)} \otimes W \oplus \Delta_-^{(4)} \otimes W \ar@{.>}[dl]\\
      \Delta_+^{(4)} \otimes W_1 \oplus \Delta_-^{(4)} \otimes W_2 & \\}
  \end{equation}
\item In $N=5$, $W \in \Irr(\fg,\HH)_{\text{aLTS}}$ and the enhancement to $N=6$ occurs when $W = V_\HH$, for $V \in \Irr(\fg,\CC)_{\text{aJTS}}$:
  \begin{equation}
    \label{eq:5to6}
    \xymatrix{\Delta_+^{(6)} \otimes V \oplus \Delta_-^{(6)} \otimes \Vbar \ar@{.>}[r] &  \Delta^{(5)} \otimes V_\HH\ar@{.>}[dl]\\
      \Delta^{(5)} \otimes W & \\}
  \end{equation}
\item Finally, in $N=6$, $V \in \Irr(\fg,\CC)_{\text{aJTS}}$ and enhancement to $N=7$ occurs when $V=U_\CC$ for $U \in \Irr(\fg,\RR)_{\text{3LA}}$:
  \begin{equation}
    \label{eq:6to7}
    \xymatrix{\Delta_+^{(7)} \otimes U \ar@{.>}[r] &  [\![\Delta_+^{(6)} \otimes U_\CC]\!] \ar@{.>}[dl]\\
      [\![\Delta_+^{(6)} \otimes V]\!] & \\}
  \end{equation}
\end{itemize}

We also see from table~\ref{tab:susy-enhancement} that enhancement from $N=7$ to $N=8$ does not constrain the representation further. This suggests that $N=7$ implies $N=8$ and we will show in section~\ref{sec:nequal8} that this is indeed the case.

\subsection{Notion of indecomposability}

Given two $N$-extended superconformal Chern-Simons theories with matter with data $(\fg_1,\fM_1)$ and $(\fg_2,\fM_2)$ one can add their Lagrangians to obtain a theory with the same amount of supersymmetry and with data $(\fg_1 \oplus \fg_2, (\fM_1\otimes \KK) \oplus (\KK \otimes \fM_2))$, where $\KK = \RR,\CC$ denotes the relevant trivial one-dimensional representation. In other words, superconformal Chern-Simons theories admit direct sums and hence there is a notion of indecomposability; namely, an indecomposable theory is one which cannot be decoupled as a direct sum of two nontrivial theories.

For $N<4$ indecomposability places very weak constraints on the allowed representations. For example, if the Chern-Simons Lie algebra $\fg$ is simple, then any direct sum of nontrivial irreducible unitary representations of the right type will give rise to an indecomposable theory, the Chern-Simons terms acting as the ``glue'' binding the matter together. However, for $N>4$ we will see that indecomposability coincides with irreducibility of the matter representation (or of the corresponding 3-algebra) which in turns corresponds to simplicity of the embedding Lie superalgebra in which these triple systems can be embedded, as we discussed in section \ref{sec:simplicity}. Recalling that the embedding Lie superalgebra has the Lie algebra as its even part, this puts severe constrains on the gauge algebra that can be used in a theory with $N>3$ supersymmetry. For $N=4$ we will see that the situation is a bit more complicated but their indecomposability is also (partially) related to irreducibility of the representation and again restricts the possible Lie algebras.

\section{$N=1$ Supersymmetry}
\label{sec:N1susy}
We are now ready to start building Lagrangians for superconformal Chern-Simons-matter theories in 3 dimensions. Our starting point is the generic off-shell $N=1$ theory. The ingredients required to define such theory at a classical level are:

\begin{itemize}
	\item A metric Lie algebra $\left(\fg,(-,-)\right)$,
	\item a unitary representation of it $\left(\fM, \left\langle -,-\right\rangle\right)$,
	\item a real superpotential $\sW$, which is a $\fg$-invariant function on $\fM$ that must be quartic for the theory to be conformal at classical level.
\end{itemize}

Once more, thanks to the Faulkner construction, the first two ingredients are equivalent to a metric 3-Leibniz algebra. Given these ingredients, the $N=1$ superspace formalism provides the most general Chern-Simons-matter Lagrangian which is invariant under $N=1$ supersymmetry transformations. The field content is fixed by the $N=1$ supermultiplets in three dimensions and the Lagrangian contains a kinetic term for the matter fields, a Chern-Simons term for the gauge field and the quartic superpotential. We review now this formalism. 

\subsection{$N=1$ superspace formalism}

\subsubsection{Conventions and spinors}

First of all we need to fix some conventions. We take the \textbf{Minkowski metric} in three dimensions to have mostly plus signature and denote it by $\eta_{\mu\nu}$, where $\mu, \nu = 0,1,2$. We choose the orientation tensor $\varepsilon_{\mu\nu\rho}$ such that $\varepsilon_{012} = 1$. 

The \textbf{Clifford algebra} $\Cl(1,2)$ has two inequivalent representations, called Majorana spinors, both of which are real and two-dimensional. Having chosen one of these representations, the Clifford algebra acts via $2\times2$ real matrices $\gamma_\mu$ which obey $\gamma_\mu \gamma_\nu + \gamma_\nu \gamma_\mu = 2 \eta_{\mu\nu} 1$. A suitable choice is for example $\gamma_0 = i \sigma_2$, $\gamma_1 = \sigma_1$ and $\gamma_2 = \sigma_3$, where $\sigma_\mu$ are the Pauli matrices. 

They act on spinors $\xi$ with two real components. We define ${\bar \xi} := \xi^t \gamma^0$, which implies ${\bar \chi} \xi = {\bar \xi} \chi$ and ${\bar \chi} \gamma_\mu \xi = - {\bar \xi} \gamma_\mu \chi$ for any fermionic\footnote{The fact that $\chi = (\chi_1, \chi_2)$ and $\xi= (\xi_1, \xi_2)$ are fermionic means that their components anticommute, for example $\chi_1\xi_2 = - \xi_2\chi_1$} Majorana spinors $\chi$ and $\xi$. Some useful identities are:

\begin{itemize}
	\item $\gamma_{\mu\nu} = \varepsilon_{\mu\nu\rho} \gamma^\rho$,
	\item $\gamma_{\mu\nu\rho} = \varepsilon_{\mu\nu\rho} 1$ and
	\item the \textbf{Fierz identity} $\xi {\bar \chi} = -\half \left[ ( {\bar \chi} \xi ) 1 + ( {\bar \chi} \gamma^\mu \xi ) \gamma_\mu \right]$.
\end{itemize}

\subsubsection{$N=1$ supermultiplets in three dimensions}
\label{sec:N1supermult}

There are two kinds of $N=1$ supermultiplets in three dimensions: {\emph{gauge}} and {\emph{matter}}. A \textbf{gauge supermultiplet} consists of a bosonic gauge field $A_\mu$ and a fermionic Majorana spinor $\chi$. We will assume that both fields take values in a Lie algebra $\fg$ that is equipped with an ad-invariant inner product $(-,-)$. It has been sometimes assumed in the past that the Lie algebra is semisimple and that these fields are valued in its adjoint representation. This provides a natural inner product on $\fg$, the Killing form, which is ensured to be positive-definite if the Lie algebra is semisimple. However, since the gauge fields in a Chern-Simons theory are non-propagating, it is not necessary for the inner-product to be positive-definite. In fact in several important examples the inner product $(-,-)$ on $\fg$ has split signature and it is not the Killing form.

A \textbf{matter supermultiplet} consists of a bosonic scalar field $X$ and an auxiliary field $C$ plus a fermionic Majorana spinor $\Psi$. We will assume that all the matter fields take values in a faithful unitary representation $\left(\fM, \left\langle -,-\right\rangle \right)$ of $\fg$. As discussed in \ref{sec:matter-reps}, the R-symmetry is trivially satisfied in this case. They can be collected into a matter superfield $\Xi = X + {\bar \theta} \Psi + \half {\bar \theta} \theta C$ where the superspace coordinate $\theta$ is a fermionic Majorana spinor.

Requiring the representation to be faithful is done for convenience. Recall that a representation $\rho: \fg \to \fgl(\fM)$ is faithful if $\rho$ has trivial kernel. Normally this assumption can be made without any loss of generality: the kernel $\fk$ of $\rho$ is an ideal, hence the quotient $\fg/\fk$ is also a Lie algebra. Therefore, if $\rho$ is not a faithful representation of $\fg$, it is enough to consider this quotient instead and then $\rho$ is a faithful representation of it. 

However here one has to be careful when replacing the Lie algebra $\fg$ with the quotient $\fg/\fk$ because the gauge fields take values in it. In order to do this replacement without consequences for the resulting theory, one needs to show that the fields taking values in the kernel of $\rho$, $\fk$, somehow decouple. The gauge fields $A$ enter the matter Lagrangian via covariant derivatives of the form $d + \rho(A)$, hence any gauge field in $\fk$ appears only in the Chern-Simons term. The precise way in which this happens depends on the choice of ad-invariant inner product on $\fg$. If $\fk$ is a nondegenerate ideal, so that $\fg = \fk \oplus \fk^\perp$, then it is not hard to see that since both $\fk$ and $\fk^\perp$ are orthogonal ideals, hence the theory decouples into a Chern-Simons term for $\fk$ and a Chern-Simons term for $\fk^\perp$. In the case when $\fk$ is not nondegenerate, preliminary results with abelian quiver theories suggest that one ends up with a Chern-Simons-matter theory for $\fk^\perp/(\fk \cap\fk^\perp)$ which is also a metric Lie algebra of which the representation is faithful. Hence also in this case one can take $\fg$ to be faithful without loss of generality. 

\subsubsection{Supersymmetry transformations}

The coupling of gauge and matter supermultiplets is achieved using the action $\cdot$ of the Lie algebra $\fg$ on its representation $\fM$. Making use of this fact, the $N=1$ supersymmetry transformations for the matter and gauge fields are

\begin{equation}
  \label{eq:susy}
  \begin{aligned}[m]
    \delta X &= {\bar \epsilon} \Psi\\
    \delta \Psi &= - (D_\mu X) \gamma^\mu \epsilon + C \epsilon\\
    \delta C &= - {\bar \epsilon} \gamma^\mu ( D_\mu \Psi ) - {\bar \epsilon} \chi \cdot X\\
    \delta A_\mu &= {\bar \epsilon} \gamma_\mu \chi\\
    \delta \chi &= \half F_{\mu\nu} \gamma^{\mu\nu} \epsilon,
  \end{aligned}
\end{equation}
where the parameter $\epsilon$ is a fermionic Majorana spinor and $D_\mu \phi = \partial_\mu \phi + A_\mu \cdot \phi$ for any field $\phi$ valued in $\fM$. 

The derivative $D_\mu$ is covariant with respect to the gauge transformations $\delta \phi = - \Lambda \cdot \phi$ and $\delta A_\mu = \partial_\mu \Lambda + [ A_\mu , \Lambda]$, for any gauge parameter $\Lambda$ valued in $\fg$ and where $[-,-]$ denotes the Lie bracket on $\fg$. The curvature of this covariant derivative is $\fg$-valued and defined by $F_{\mu\nu} = [ D_\mu , D_\nu ] = \partial_\mu A_\nu - \partial_\nu A_\mu + [ A_\mu , A_\nu ]$. The commutator of two supersymmetry transformations in \eqref{eq:susy} closes off-shell giving a translation on $\RR^{1,2}$ plus a gauge transformation.  

It is worth emphasising that up to this point there is no need to define inner products for the Lie algebra $\fg$ or its representation $\fM$. It is in order to construct a Lagrangian that this extra data is needed.

\subsection{Off-shell $N=1$ Lagrangian}
\label{N1_offLagr}
There are three distinct contributions making up the most general Chern-Simons-matter Lagrangian that is invariant under \eqref{eq:susy}: the supersymmetric Chern-Simons term $\eL_{CS}$, the supersymmetric matter term $\eL_{M}$ and the superpotential $\sW$.

Given a gauge supermultiplet $(A_\mu , \chi)$ valued in the Lie algebra $\fg$ with ad-invariant inner product $(-,-)$, the canonical Chern-Simons term is: 

\begin{equation}
\label{eq:susy-cs}
  \eL_{CS} = -\varepsilon^{\mu\nu\rho} \left( A_\mu , \partial_\nu A_\rho + \tfrac{1}{3} [ A_\nu , A_\rho ] \right)  - ( {\bar \chi} , \chi ),
\end{equation}
whose integral is invariant under the last two supersymmetry transformations in \eqref{eq:susy}. It is also manifestly gauge-invariant as a consequence of the ad-invariance of $(-,-)$.

Given a matter supermultiplet $(X,\Psi ,C)$ valued in a real\footnote{We take the representation to be real as a \textit{minimal} requirement. In other words, a complex or quaternionic representation would also be valid here, since they can be thought of as real representations with extra structure on them.} representation $\fM$ of $\fg$ with an invariant positive-definite symmetric inner product $\left< -,- \right>$, that is coupled to the gauge supermultiplet $(A_\mu , \chi )$, the canonical supersymmetric kinetic terms for its component fields are:

\begin{equation}
\label{eq:susy-m}
  \eL_{M} = -\half \left< D_\mu X , D^\mu X \right> + \half \left< {\bar \Psi} , \gamma^\mu D_\mu \Psi \right> + \half \left< C , C \right> - \left< X ,  {\bar \chi} \cdot \Psi \right>.
\end{equation}

Replacing covariant with partial derivatives in the first two terms and dropping the fourth term would describe the supersymmetric Lagrangian for the matter fields in the ungauged theory. As it is, the integral of \eqref{eq:susy-m} is invariant under \eqref{eq:susy} with the fourth term describing an additional gauge-matter coupling that is required to cancel the supersymmetry variation of the first three terms in the gauged theory.

Notice that the integral of $\eL_{CS} + \eL_{M}$ from \eqref{eq:susy-cs} and \eqref{eq:susy-m} is classically scale-invariant with respect to the fields $(X,\Psi ,C,A_\mu , \chi )$ being assigned weights $( \half ,1, \tfrac{3}{2} ,1, \tfrac{3}{2} )$. 

Finally, we need to add non-kinetic terms for the matter fields. One could add a mass term of the form $\half \int d^2 \theta \left< \Xi , \Xi \right> = \left< X , C \right> - \half \left< {\bar \Psi} , \Psi \right>$ which is manifestly supersymmetric, but this would break the classical scale invariance. More generally, one could consider a superpotential

\begin{equation}
\label{eq:superpotential}
  \sW = \int d^2 \theta \; \fW (\Xi) = C^a \frac{\partial}{\partial X^a} \, \fW (X) - \tfrac{1}{2} {\bar \Psi}^a \Psi^b \,
  \frac{\partial^2}{\partial X^a \partial X^b} \, \fW (X),
\end{equation}
where $\fW$ is an arbitrary polynomial function on $\fM$ and the matter superfield components have been written relative to a basis $\{e_a\}$ for $\fM$. The function $\fW$ must be:

\begin{itemize}
	\item \textbf{Quartic} for scale-invariance and
	\item \textbf{$\fg$-invariant} for supersymmetry.
\end{itemize}

Given a $\fg$-invariant inner product on $\fM$, $\left\langle -,-\right\rangle$, we have several ingredients at our disposal to build such a potential. One is the $\fg$-invariant inner product itself which allows us to construct terms proportional to $\left< \Xi , \Xi \right>^2$. However, these can be thought of as arising as marginal deformations of an existing theory by the square of the $\fg$-invariant operator $\left< \Xi , \Xi \right>$. Such operators are unprotected from quantum corrections given only $N=1$ supersymmetry and in what follows we will not consider superpotentials which take the form of such deformations.

Other ingredients at our disposal, remembering that $\Xi$ is valued in a unitary representation of $\fg$, are the representation map $\fg \mapsto \fso(\fM)$ and the ad-invariant inner product on the Lie algebra $\fg$, $(-,-)$. Rather, one can use the transpose of this map, the Faulkner map (that we defined in section \ref{sec:unitrep} and called $T: \fM \times \fM \to \fg$ in the real case and $\TT$, $\TT_H$ in the complex and quaternionic cases). Then, one can build terms proportional to
  
\begin{equation}
  \label{superpotTerms}
  \begin{aligned}[m]
		& \left(T(-,-), T(-,-)\right),\\
		& \left(\TT(-,-), \TT(-,-)\right) \text{ or} \\
    & \left(\TT_H(-,-), \TT_H(-,-)\right).\\
  \end{aligned}
\end{equation}
where the second term can only be built if the real representation is equipped with a complex structure (hence it is a complex or quaternionic representation) and the third if it has both a complex and a quaternionic structure (hence it is a quaternionic representation). 

In the next sections we will see how a choice of coefficient in the second term enhances supersymmetry to $N=2$. Also we will see how a particular combination of the second and third terms, when $\fM$ is quaternionic,  gives enhancement to $N=3$ and that such superpotential is rigid.

\subsection{On-shell $N=1$ Lagrangian}

Before discussing the superpotentials leading to increased amounts of supersymmetry, let us conclude this subsection by noting the on-shell form of the generic $N=1$ supersymmetric Chern-Simons-matter Lagrangian $\eL^{N=1} = \eL_{CS} + \eL_{M} + \sW$, after integrating out the auxiliary fields $\chi$ and $C$. Their respective equations of motion are $\chi = \half T (X, \Psi )$ and $\left< C,- \right> = - d \fW (X)$, the terms in the equation for $C$ being thought of as $\fM^*$-valued. Substituting these expressions into the Lagrangian gives
\begin{equation}
  \label{eq:susy-lag-on-shell}
  \begin{aligned}[m]
    \eL_{on}^{N=1} &= \eL_{M} + \eL_{CS} + \sW =   -\half \left< D_\mu X , D^\mu X \right>   + \half \left< {\bar \Psi} , \gamma^\mu D_\mu \Psi \right>\\
    &		-\half \left< d \fW (X) , d \fW(X) \right> %
		-\varepsilon^{\mu\nu\rho} \left( A_\mu , \partial_\nu A_\rho + \tfrac{1}{3} [ A_\nu , A_\rho ] \right)\\
		&- \half {\bar \Psi}^a \Psi^b \partial_a \partial_b \fW (X) + \tfrac{1}{4} ( T (X, {\bar \Psi} ) , T (X, \Psi )).
  \end{aligned}
\end{equation}

In a slight abuse of notation, the expression $-\half \left< d \fW (X) , d \fW(X) \right>$ for the scalar potential is shorthand for $-\half g^{ab} \partial_a \fW(X) \partial_b \fW(X)$ with $g^{ab}$ denoting components of the matrix inverse of $\left< e_a , e_b \right>$ on $\fM$. Notice that the effect of integrating out the auxiliary fields has been to generate a sextic potential for the scalar fields and various scalar-fermion Yukawa couplings.

This on-shell Lagrangian \eqref{eq:susy-lag-on-shell} is invariant under the $N=1$ supersymmetry transformations \eqref{eq:susy}, upon substituting into their expressions the field equations for the auxiliary fields $\chi$ and $C$. These supersymmetry transformations are
\begin{equation}
  \label{eq:susy-onshell}
  \begin{aligned}[m]
    \delta X &= {\bar \epsilon} \Psi\\
    \delta \Psi &= - (D_\mu X) \gamma^\mu \epsilon - d \fW (X) \epsilon\\
    \delta A_\mu &= \half T \left( X , {\bar \epsilon} \gamma_\mu \Psi \right) ,
  \end{aligned}
\end{equation}
which close up to a translation on $\RR^{1,2}$ plus a gauge transformation, using the equations of motion from \eqref{eq:susy-lag-on-shell}.

\section{From $N=1$ to $N=3$}
\label{sec:from1to3}

In this section we explore the minimal superpotential that needs to be added to the $N=1$ Lagrangian $\eL_{CS} + \eL_{M}$ to enhance supersymmetry. We will see that this procedure can not be pushed beyond $N=3$ because the superpotential is then rigid, therefore further enhancement needs to be achieved by other methods that we discuss in section \ref{sec:Ngt3}. 

\subsection{N=2 supersymmetry for $\fM$ complex}
\label{sec:nequal2}

We first review the most general $N=2$ superconformal Chern-Simons-matter Lagrangian using $N=2$ superspace formalism and then we explain how it can be obtained from the previous $N=1$ Lagrangian $\eL^{N=1}$, by choosing an appropriate superpotential.

\subsubsection{$N=2$ supermultiplets and supersymmetry transformations}
\label{N2superspace}

A complex notation for the $N=2$ case is more convenient, since the matter fields live in the tensor product of the spinor representation of the R-symmetry $\fso(2)$, which is $\CC$, with a complex representation of $\fg$, as discussed in section \ref{sec:matter-reps}. This can be done because the two representations of the Clifford algebra $\Cl(1,2)$ are real and two-dimensional, and one can identify $\CC$ with $\RR^2$.

In $N=2$ superspace formalism there is a \textbf{matter chiral superfield} $\Xi_{\CC} = X + {\bar \theta_{\CC}^*} \Psi + \half {\bar \theta_{\CC}^*} \theta_{\CC}^* F$, where $\theta_{\CC}$ is a complex superspace coordinate, and a \textbf{gauge vector superfield} ${\sf V} = {\bar \theta_{\CC}^*} \gamma^\mu \theta_{\CC} D_\mu + i {\bar \theta_{\CC}^*} \theta_{\CC} \sigma - \tfrac{i}{4} ( {\bar \theta_{\CC}^*} \theta_{\CC}^* )( {\bar \theta_{\CC}} \theta_{\CC} ) D + \half ( {\bar \theta_{\CC}^*} \theta_{\CC}^* ) {\bar \theta_{\CC}} \chi_{\CC} - \half ( {\bar \theta_{\CC}} \theta_{\CC} ) {\bar \theta_{\CC}^*} \chi_{\CC}^*$, which is pure imaginary. Again, we assume that the gauge fields take values in a metric Lie algebra $\fg$, and the matter fields in a faithful unitary representation of it $\fM$, which this time is complex. Associated with this enhanced $N=2$ supersymmetry there is a $\fso(2) \cong \fu (1)$ R-symmetry under which the superspace coordinate $\theta_{\CC}$ has $\fu (1)$ R-charge -1 and so $\Xi_{\CC}$ has R-charge $\half$ while ${\sf V}$ is neutral.

However, we will work in $N=1$ superspace formalism for theories realising more than $N=2$ supersymmetry. To ensure continuity in the discussion, we find it convenient to discuss theories with $N=2$ supersymmetry in $N=1$ superspace formalism, hence we will use the gauge supermultiplet $( A_\mu , \chi_{\CC} , \sigma , D)$ and the matter supermultiplet $(X, \Psi, F)$ instead of the $N=2$ superfields $\Xi_{\CC}$ and $\sf V$.

\begin{table}[ht!]
  \centering
  \begin{tabular}{|c|c|c|}
    \hline
    $N$ & Gauge & Matter\\\hline
    1 &  $( A_\mu , \chi )$ & $(X,\Psi ,C)$\\
    2 &  $( A_\mu , \chi_{\CC} , \sigma , D)$ & $(X,\Psi ,F)$\\\hline
  \end{tabular}
  \vspace{8pt}
  \caption{$N=1$ and $N=2$ supermultiplets}
  \label{tab:12supermult}
\end{table}

The $N=2$ supersymmetry transformations for the gauge supermultiplet are
\begin{equation}
  \label{eq:susy2g}
  \begin{aligned}[m]
    \delta A_\mu &= {\mathrm{Re}} \left( {\bar \epsilon_{\CC}}^* \gamma_\mu \chi_{\CC} \right) \\
    \delta \chi_{\CC} &= \half F_{\mu\nu} \gamma^{\mu\nu} \epsilon_{\CC} + i ( D_\mu \sigma ) \gamma^\mu \epsilon_{\CC} - iD\, \epsilon_{\CC} \\
    \delta \sigma &= - {\mathrm{Im}} \left( {\bar \epsilon_{\CC}}^* \chi_{\CC} \right) \\
    \delta D &= {\mathrm{Im}} \left( {\bar \epsilon_{\CC}}^* \left( \gamma^\mu D_\mu \chi_{\CC} +i [ \sigma , \chi_{\CC} ] \right) \right) ,
  \end{aligned}
\end{equation}
where the parameter $\epsilon_{\CC}$ is a complex spinor on $\RR^{1,2}$. $\chi_{\CC}$ and $\epsilon_{\CC}$ have R-charge -1 (their complex conjugates having charge +1) while $A_\mu$, $\sigma$ and $D$ are uncharged.

These are the $N=2$ supersymmetry transformations for the matter supermultiplet
\begin{equation}
  \label{eq:susy2m}
  \begin{aligned}[m]
    \delta X &= {\bar \epsilon_{\CC}}^* \Psi \\
    \delta \Psi &= - ( D_\mu X ) \gamma^\mu \epsilon_{\CC} + F\, \epsilon_{\CC}^* -i\sigma \cdot X \, \epsilon_{\CC} \\
    \delta F &= - {\bar \epsilon_{\CC}} \left(  \gamma^\mu D_\mu \Psi + \chi_{\CC} \cdot X - i\sigma \cdot \Psi \right) .
  \end{aligned}
\end{equation}

The matter fields $(X,\Psi ,F)$ have charges $(\half , -\half , -\tfrac{3}{2} )$ under the $\fu (1)$ R-symmetry. Notice that, as expected, one can recover the $N=1$ transformations \eqref{eq:susy} as a subalgebra of these by taking $\epsilon_\CC = \epsilon$ to be real and identifying $F = C + i\sigma \cdot X$.

\subsubsection{Off-shell $N=2$ Lagrangian}

The most general Chern-Simons-matter $N=2$ Lagrangian has again three parts: a kinetic term for the matter fields, $\eL_{M}^{N=2}$, a Chern-Simons term for the gauge fields, $\eL_{CS}^{N=2}$, and a superpotential which is of the so-called F-type, $\sW_F$. We discuss them now in turn. 

The $N=2$ supersymmetric Chern-Simons Lagrangian whose integral is invariant under \eqref{eq:susy2g} is

\begin{equation}\label{eq:susy-cs2}
  \eL_{CS}^{N=2} = -\varepsilon^{\mu\nu\rho} \left( A_\mu , \partial_\nu A_\rho + \tfrac{1}{3} [ A_\nu , A_\rho ] \right)  - ( {\bar \chi} , \chi ) - ( {\bar {\hat \chi}} , {\hat \chi} ) +2\, (\sigma ,D) ,
\end{equation}
and the standard off-shell gauged $N=2$ supersymmetric matter Lagrangian is
\begin{equation}\label{eq:susy-m2a}
  \begin{aligned}[m]
  \eL_{M}^{N=2} =&-\half \left< D_\mu X , D^\mu X \right> + \half \left< {\bar \Psi} , \gamma^\mu D_\mu \Psi \right> + \half \left< F , F \right> - \left< X , {\bar \chi_{\CC}}^* \cdot \Psi \right> \\
  &+\half \left< X , i D \cdot X \right> - \half \left< {\bar \Psi} , i \sigma \cdot \Psi \right> -\half \left< \sigma \cdot X , \sigma \cdot X \right>,
  \end{aligned}
\end{equation}
whose integral is invariant under the $N=2$ supersymmetry transformations \eqref{eq:susy2m} and \eqref{eq:susy2g}. In this Lagrangian and in the rest of this chapter the inner product $\left<-,- \right>$ represents the real part of the hermitian inner product $h(-,-)$ when the representation of the Lie algebra is complex or quaternionic.

To the $N=2$ supersymmetric Chern-Simons-matter Lagrangian $\eL_{CS}^{N=2} + \eL_{M}^{N=2}$, one can add an F-term superpotential. We turn back to $N=2$ superspace formalism for a moment because it provides an automatically $N=2$ supersymmetric superpotential
\begin{equation}\label{eq:ftermsuperpotential}
  \sW_F = \int d^2 \theta_{\CC} \; \fW_F ( \Xi_{\CC} ) + \int d^2 \theta_{\CC}^* \; \fW_F ( \Xi_{\CC} )^* ,
\end{equation}
provided that $\fW_F$ is a $\fg$-invariant holomorphic function of the matter fields. Scale-invariance again requires $\fW_F$ to be a quartic function. Notice that \eqref{eq:ftermsuperpotential} does not require an inner product on $\fM$ but demands it is complex. Also the chiral superspace measure guarantees that $\sW_F$ is invariant under the $\fu (1)$ R-symmetry of the $N=2$ superalgebra. 

\subsubsection{$N=2$ Lagrangian from the $N=1$ Lagrangian}

This $N=2$ Lagrangian $\eL^{N=2} = \eL_{CS}^{N=2} + \eL_{M}^{N=2} + \sW_F$ can be obtained from the generic $N=1$ Chern-Simons-matter Lagrangian $\eL^{N=1} = \eL_{CS} + \eL_{M} + \sW$, by a particular choice of the $N=1$ superpotential $\sW$. We will see this in two parts. First, we describe how one can obtain $\eL_{CS}^{N=2} + \eL_{M}^{N=2}$ when the representation $\fM$ is complex (or quaternionic) and then we look at the F-term superpotential $\sW_F$.

As mentioned briefly in section \ref{N1_offLagr}, when the matter representation is complex $\fM \in \Dar(\fg,\CC)$, there is a new map available to construct a superpotential, the complex Faulkner map $\TT$, and one can add a superpotential to the $N=1$ Lagrangian proportional to $(\TT (\Xi ,\Xi ),\TT (\Xi ,\Xi ) )$. In particular, if the superpotential $\sW$ is chosen to be precisely
 
\begin{equation}\label{eq:superpotentialcomplex}
  \sW_{\CC} = \tfrac{1}{16}\int d^2 \theta \; (\TT (\Xi ,\Xi ),\TT (\Xi ,\Xi ) ) ,
\end{equation}
it gives rise to an enhanced $N=2$ supersymmetry when added to the $N=1$ Lagrangian $\eL_{CS} + \eL_{M}$. That is, it provides precisely the additional gauge-matter couplings that are required for $N=2$ supersymmetry and is equivalent to $\eL_{CS}^{N=2} + \eL_{M}^{N=2}$.

Notice that, by proposition \eqref{pr:T-map-C}, $\TT (\Xi ,\Xi )$ is pure imaginary (indeed $\TT (\Xi ,\Xi ) = i T(\Xi , I \Xi )$ from lemma~\ref{le:T-map-C}), hence \eqref{eq:superpotentialcomplex} is real. Notice also that $T(\Xi , \Xi ) \equiv 0$ and so there is no possibility to build an alternative superpotential based on the real part $T$ of $\TT$ here. 

To understand the equivalence with $\eL_{CS}^{N=2} + \eL_{M}^{N=2}$, we need to re-write this Lagrangian in a way that includes the extra component fields needed to extend $N=1$ supermultiplets into the $N=2$ supermultiplets. The way to do this is to note that one can obtain \eqref{eq:superpotentialcomplex} via integrating out an auxiliary matter supermultiplet $\Pi = \sigma - {\bar \theta} {\hat \chi} + \half {\bar \theta} \theta D$, which is just an $N=1$ matter superfield but this time valued in $\fg$ instead of $\fM$\footnote{The supersymmetry transformations for these fields just follow from \eqref{eq:susy} by taking the action $\cdot$ of $\fg$ to be the adjoint action of $\fg$ on itself.}. The real superpotential
\begin{equation}\label{eq:superpotentialcomplexaux}
  \int d^2 \theta \; ( \Pi , \Pi ) + \tfrac{i}{2} ( \Pi , \TT (\Xi ,\Xi ) ) ,
\end{equation}
then gives precisely $\sW_{\CC}$ \eqref{eq:superpotentialcomplex} after integrating out $\Pi$. Classical scale invariance here follows from the auxiliary components $(\sigma ,{\hat \chi},D)$ being assigned weights $( 1, \tfrac{3}{2} ,2)$.

It is no coincidence that the names of the component fields in the auxiliary matter superfield match the extra field components needed to construct the $N=2$ supermultiplets defined in section \ref{N2superspace} from the $N=1$ supermultiplets. It is enough to redefine $F := C + i\sigma \cdot X$ and assemble the fermions into a complex spinor $\chi_{\CC} = \chi + i {\hat \chi}$ to obtain all the component fields of the $N=2$ superfields. Note that $( {\bar \chi} , \chi ) + ( {\bar {\hat \chi}} , {\hat \chi} ) = ( {\bar \chi_{\CC}}^* , \chi_{\CC} )$, with $\chi_{\CC}^* = \chi - i {\hat \chi}$ denoting the complex conjugate of $\chi_{\CC}$. 

Written in this way, it is straight forward to see that the superpotential \eqref{eq:superpotentialcomplexaux} provides exactly the terms needed to complement $\eL_{CS} + \eL_{M}$ into $\eL_{CS}^{N=2} + \eL_{M}^{N=2}$. Indeed, adding the first term in \eqref{eq:superpotentialcomplexaux} to the supersymmetric Chern-Simons term \eqref{eq:susy-cs} gives
\begin{equation}
  \eL_{CS} + \int d^2 \theta \; ( \Pi , \Pi ) = -\varepsilon^{\mu\nu\rho} \left( A_\mu , \partial_\nu A_\rho + \tfrac{1}{3} [ A_\nu , A_\rho ] \right)  - ( {\bar \chi} , \chi ) - ( {\bar {\hat \chi}} , {\hat \chi} ) +2\, (\sigma ,D) 
\end{equation}
which is precisely the $N=2$ Chern-Simons Lagrangian $\eL_{CS}^{N=2}$ in \eqref{eq:susy-cs2}. The supersymmetry enhancement here can be seen to arise from the choice of taking either $\chi$ or ${\hat \chi}$ to describe the superpartner of the gauge field $A_\mu$ in the $N=1$ gauge supermultiplet. 

On the other hand, combining the remaining term in \eqref{eq:superpotentialcomplexaux} with the $N=1$ matter term \eqref{eq:susy-m} gives
\begin{equation}\label{eq:susy-m2}
  \begin{aligned}[m]
  \eL_{M} +\half \int d^2 \theta \, \left< \Xi , i \Pi \cdot \Xi \right> =& -\half \left< D_\mu X , D^\mu X \right> + \half \left< {\bar \Psi} , \gamma^\mu D_\mu \Psi \right> + \half \left< C , C \right> - \left< X , {\bar \chi_{\CC}}^* \cdot \Psi \right> \\
  &+\half \left< X , i D \cdot X \right> - \half \left< {\bar \Psi} , i \sigma \cdot \Psi \right> + \left< X , i \sigma \cdot C \right>.
  \end{aligned}
\end{equation}
which, using $F := C + i\sigma \cdot X$ becomes precisely $\eL_{M}^{N=2}$.

Lets look now at the $N=2$ F-term superpotential in \eqref{eq:ftermsuperpotential}. Notice that one can generally obtain it off-shell from a particular $N=1$ superpotential of the form $\int d^2 \theta \; 2\, \Re \, \fW_F (\Xi )$ when $\fM$ is of complex type, where $\Xi = X + {\bar \theta} \Psi + \half {\bar \theta} \theta F$ is the $N=1$ superfield on which the the chiral $N=2$ superfield $\Xi_{\CC}$ is constructed. The extra data here being precisely the quartic $\fg$-invariant holomorphic function $\fW_F$. It is worth pointing out that one recovers precisely the same F-term superpotential from the aforementioned $N=1$ superpotential based on the $N=1$ superfield $\Xi = X + {\bar \theta} \Psi + \half {\bar \theta} \theta C$ we had been using before. This follows from the fact that $F - C = i\sigma \cdot X$ and so the potential discrepancy between the resulting superpotentials is proportional to $\left< i\sigma \cdot X , \partial \fW_F (X) \right>$ which vanishes identically as a consequence of $\fW_F$ being $\fg$-invariant.

In summary, we have seen that, when $\fM \in \Dar(\fg,\CC)$, one can obtain the general $N=2$ Chern-Simons-matter Lagrangian $\eL_{CS}^{N=2} + \eL_{M}^{N=2} + \sW_F$ from the choice of $N=1$ superpotential
\begin{equation}\label{eq:superpotential2}
  \sW_2 = \sW_{\CC} + \sW_F = \int d^2 \theta \; \left(\tfrac{1}{16} (\TT (\Xi ,\Xi ),\TT (\Xi ,\Xi ) ) + 2\, \Re \, \fW_F (\Xi )\right).
\end{equation}

\subsubsection{R-symmetry of the $N=2$ superpotential}
\label{sec:der-nequal2superpotential}

Looking at the different components of a generic quartic $N=1$ superpotential we can prove that $\sW_2$ is in fact the unique choice of $N=1$ superpotential giving rise to an on-shell Lagrangian which is invariant under the $\fu(1)$ R-symmetry that is necessary for $N=2$ supersymmetry. 

Whenever $\fM \in \Dar(\fg,\CC)$, one can decompose a generic quartic $N=1$ superpotential into its $(4,0)+(3,1)+(2,2)+(1,3)+(0,4)$ components, with respect to the complex structure on $\fM$. We have found that enhancement to $N=2$ supersymmetry for the $N=1$ Chern-Simons-matter Lagrangian is guaranteed provided the $(3,1)+(1,3)$ component is absent and the $(2,2)$ component is $\sW_\CC$. We will see now that this is in fact the only component which is $\fu(1)$-invariant.

We assign the bosonic matter field $X$ $\fu(1)$ R-charge $\half$ and to the fermionic matter field $\Psi$ $-\half$, with their complex conjugates having the opposite R-charges. This assignment implies that all the terms in \eqref{eq:susy-lag-on-shell} are automatically $\fu(1)$-invariant (as a consequence of the hermitian inner product $h$ on $\fM$ being complex-sesquilinear) except for the scalar-fermion Yukawa couplings.

It is convenient to break up the Yukawa couplings into the contributions transforming with different overall $\fu(1)$ charges. To this end, let us first decompose the quartic superpotential $\fW = \fW_{4,0} + \fW_{3,1} + \fW_{2,2} + \fW_{1,3} + \fW_{0,4}$ with respect to the complex structure on $\fM$, where $\fW_{p,4-p} = {\overline{\fW_{4-p,p}}}$ for all $p=0,1,2,3,4$ since $\fW$ is real. Each component $\fW_{p,4-p}$ is a quartic tensor which is taken to be complex-linear in its first $p$ arguments and complex-antilinear in its remaining $4-p$ arguments. Thus $\fW_{2,2} (X)$ is real and uncharged, $\fW_{3,1} (X)$ has charge $1$ and $\fW_{4,0} (X)$ has charge $2$. The other contribution to the Yukawa couplings involves $\TT (X, \Psi )$ which has charge $1$ (and its complex conjugate $-\TT (\Psi ,X)$ with charge $-1$). Assembling these contributions to the Yukawa couplings, we see that there are separate contributions from terms with overall charges $0$, $\pm 1$ and $\pm 2$. The uncharged contributions are therefore unconstrained and the $\fu(1)$ R-symmetry can only be realised if the complex terms with charges $-1$ and $-2$ (and their complex conjugates with charges $1$ and $2$) vanish identically.

The component $\fW_{4,0}$ (and its complex conjugate) only appear in a term with no overall $\fu(1)$ charge. This unconstrained component is to be identified with the F-term superpotential in the $N=2$ theory. There is only one contribution to the charge $-1$ term involving the component $\fW_{3,1}$ and it is straightforward to check that this term can vanish only if $\fW_{3,1} =0$. The remaining contributions to the charge $-2$ terms involve $\fW_{2,2}$ and it is easily checked that they vanish only if $\fW_{2,2} (X) = \tfrac{1}{16} ( \TT (X,X) , \TT (X,X) )$. Thus we have established that $\sW_\CC + \sW_F$ in \eqref{eq:superpotential2} is the most general $N=1$ superpotential which can realise the aforementioned $\fu(1)$ R-symmetry in the on-shell Lagrangian. The fact that we have already established that this theory is invariant under the $N=2$ superconformal algebra thus means that the realisation of this $\fu(1)$ R-symmetry is in fact necessary and sufficient in this instance for $N=2$ enhancement.

\subsubsection{On-shell N=2 Lagrangian}
\label{sec:on-shell-3dsusy-complex}

Before going on to look at further types of supersymmetry enhancing superpotentials which exist when $\fM$ is quaternionic, let us conclude this subsection by noting the on-shell form of the generic $N=2$ supersymmetric Chern-Simons-matter Lagrangian $\eL_{on}^{N=2} =\eL_{CS}^{N=2} + \eL_{M}^{N=2} + \sW_F$, after integrating out the auxiliary fields $\chi_{\CC}$, $D$ and $F$. Their equations of motion are respectively $\chi_{\CC}^* = \half \, \TT (X, \Psi )$, $\sigma = - \tfrac{i}{4} \, \TT (X,X)$ and $\left< F,- \right> = - \partial \fW_F (X)$. Substituting these expressions into the Lagrangian gives
\begin{equation}
  \label{eq:susy-lag-on-shell-complex}
  \begin{aligned}[m]
    \eL_{on}^{N=2} &= \eL_{M}^{N=2} + \eL_{CS}^{N=2} + \sW_F =   -\half \left< D_\mu X , D^\mu X \right> + \half \left< {\bar \Psi} , \gamma^\mu D_\mu \Psi \right>  \\
    & -\varepsilon^{\mu\nu\rho} \left( A_\mu , \partial_\nu A_\rho + \tfrac{1}{3} [ A_\nu , A_\rho ] \right)- \fV_D (X) - \fV_F (X) \\
		& - \half {\bar \Psi}^a \Psi^b \partial_a \partial_b \fW_F (X) - \half {\bar \Psi}^{\bar a} \Psi^{\bar b} \partial_{\bar a} \partial_{\bar b} \fW_F (X)^* \\
    &- \tfrac{1}{4} ( \TT (X, {\bar \Psi} ) , \TT (\Psi , X )) -\tfrac{1}{8} ( \TT (X,X) , \TT ( {\bar \Psi} , \Psi )).
  \end{aligned}
\end{equation}
where we have introduced the positive-definite D-term and F-term sextic scalar potentials $\fV_D (X) = \tfrac{1}{32} \left< \TT (X,X) \cdot X , \TT (X,X) \cdot X \right>$ and $\fV_F (X) = \half \left< \partial \fW_F (X) , \partial \fW_F (X) \right>$ and the indices are with respect to a complex basis $\{ \be_a \}$ on $\fM$.

Invariance of \eqref{eq:susy-lag-on-shell-complex} under the $N=2$ supersymmetry transformations \eqref{eq:susy2g} and \eqref{eq:susy2m} can be established after substituting the equations of motion for $\chi_{\CC}$, $D$ and $F$. These supersymmetry transformations are
\begin{equation}
  \label{eq:susy2-onshell}
  \begin{aligned}[m]
    \delta X &= {\bar \epsilon_{\CC}}^* \Psi \\
    \delta \Psi &= - ( D_\mu X ) \gamma^\mu \epsilon_{\CC} - \partial \fW_F (X) \, \epsilon_{\CC}^* - \tfrac{1}{4} \TT (X,X) \cdot X \, \epsilon_{\CC} \\
    \delta A_\mu &= \half T \left( X , {\bar \epsilon_{\CC}}^* \gamma_\mu \Psi \right) ,
  \end{aligned}
\end{equation}
and close up to a translation on $\RR^{1,2}$ plus a gauge transformation, using the equations of motion from \eqref{eq:susy-lag-on-shell-complex}.

\subsection{N=3 supersymmetry for $\fM$ quaternionic}
\label{sec:nequal3}

From this point onwards we use only $N=1$ superspace formalism. This forces us to work mostly on-shell, because for the matter fields to realise more than $N=2$ supersymmetry off-shell would require the use of rather elaborate harmonic or projective superspace techniques that are unnecessary for our present analysis.

Again, we review first the most general $N=3$ superconformal Chern-Simons-matter Lagrangian and then we explain how it can be obtained from the $N=1$ Lagrangian $\eL^{N=1}$, by choosing an appropriate superpotential.

As it turns out, most of the terms in the $N=3$ Chern-Simons-matter Lagrangian are actually invariant under $N=4$ supersymmetry transformations and only the Chern-Simons part breaks that symmetry to $N=3$. For that reason we find it convenient to work with $N=4$ supermultiplets, that we review next.

\subsubsection{$N=4$ supermultiplets and supersymmetry transformations}

An $N=2$ gauge superfield $(A_\mu , \chi_\CC , \sigma , D)$ and an $N=2$ matter superfield $( \tau_\CC , \zeta_\CC , E_\CC )$ together comprise an $N=4$ vector gauge supermultiplet in three dimensions. We are only concerned with the on-shell $N=4$ matter supermultiplet, comprised by the matter fields $X$ and $\Psi$.

The off-shell $N=4$ supersymmetry transformations for the gauge supermultiplet are
\begin{equation}
  \label{eq:susy4g}
  \begin{aligned}[m]
    \delta A_\mu &= {\mathrm{Re}} \left( {\bar \epsilon_{\CC}}^* \gamma_\mu \chi_\CC + {\bar \eta_{\CC}}^* \gamma_\mu \zeta_\CC \right) \\
    \delta \chi_{\CC} &= \half F_{\mu\nu} \gamma^{\mu\nu} \epsilon_{\CC} + i ( D_\mu \sigma ) \gamma^\mu \epsilon_{\CC} - iD\, \epsilon_{\CC} \\
    &\quad + ( D_\mu \tau_\CC^* ) \gamma^\mu \eta_\CC^* - i [ \sigma , \tau_\CC^* ] \eta_\CC^* - E_\CC \, \eta_\CC \\
    \delta \zeta_{\CC} &= \half F_{\mu\nu} \gamma^{\mu\nu} \eta_{\CC} + i ( D_\mu \sigma ) \gamma^\mu \eta_{\CC} + \left( iD - [ \tau_\CC , \tau_\CC^* ] \right) \eta_{\CC} \\
    &\quad - ( D_\mu \tau_\CC^* ) \gamma^\mu \epsilon_\CC^* + i [ \sigma , \tau_\CC^* ] \epsilon_\CC^* + E_\CC^* \, \epsilon_\CC \\
    \delta \sigma &= - {\mathrm{Im}} \left( {\bar \epsilon_{\CC}}^* \chi_{\CC} + {\bar \eta_{\CC}}^* \zeta_{\CC} \right) \\
    \delta \tau_\CC^* &= {\bar \epsilon_{\CC}} \zeta_\CC - {\bar \eta_{\CC}} \chi_\CC \\
    \delta D &= {\mathrm{Im}} \left( {\bar \epsilon_{\CC}}^* \left( \gamma^\mu D_\mu \chi_{\CC} +i [ \sigma , \chi_{\CC} ] \right) - {\bar \eta_{\CC}}^* \left( \gamma^\mu D_\mu \zeta_{\CC} +i [ \sigma , \zeta_{\CC} ] \right) -2\, [ {\bar \eta_{\CC}}^* \chi_\CC^* , \tau_\CC^* ] \right) \\
    \delta E_\CC &= - {\bar \epsilon_{\CC}} \left( \gamma^\mu D_\mu \zeta_\CC^* -i [\sigma , \zeta_\CC^* ] + [ \chi_\CC , \tau_\CC ] \right) + {\bar \eta_{\CC}}^* \left( \gamma^\mu D_\mu \chi_\CC +i [\sigma , \chi_\CC ] - [ \zeta_\CC^* , \tau_\CC^* ] \right) ,
  \end{aligned}
\end{equation}
where the parameters $\epsilon_\CC$ and $\eta_\CC$ are complex spinors on $\RR^{1,2}$, with $\fu(1)$ R-charges $-1$ and $0$ respectively. These transformations close off-shell for any $\epsilon_\CC$ and $\eta_\CC$, thus generating an $N=4$ superconformal algebra. 

Upon setting $\eta_\CC =0$ in \eqref{eq:susy4g} one recovers the $N=2$ supersymmetry transformations in \eqref{eq:susy2g} for $(A_\mu , \chi_\CC , \sigma , D)$ and those in \eqref{eq:susy2m} for the components of the auxiliary superfield $\Pi_\CC = \tau_\CC + {\bar \theta_{\CC}^*} \zeta_{\CC}^* + \half {\bar \theta_{\CC}^*} \theta_{\CC}^* E_\CC$. 

The on-shell $N=4$ supersymmetry transformations for on-shell $N=4$ hypermultiplet formed by the matter fields $X$ and $\Psi$ are
\begin{equation}
  \label{eq:susy4m}
  \begin{aligned}[m]
    \delta X &= ( {\bar \epsilon_{\CC}}^* + {\bar \eta_{\CC}} J ) \Psi \\
    \delta \Psi &= - \gamma^\mu ( \epsilon_\CC - \eta_\CC J ) D_\mu X - ( \tau_\CC^* \cdot JX ) \epsilon_\CC^* - ( \tau_\CC^* \cdot X ) \eta_\CC^* - i( \epsilon_\CC - \eta_\CC J ) \sigma \cdot X.
  \end{aligned}
\end{equation}

Notice that one recovers precisely the $N=2$ supersymmetry transformations in \eqref{eq:susy2m} for the matter fields upon setting $\eta_\CC = 0$ and imposing the equation of motion $F = - \tau_\CC^* \cdot JX$ for the auxiliary matter field $F$. Similarly one can obtain the full $N=4$ transformations in \eqref{eq:susy4m} from two sets of $N=2$ transformations in \eqref{eq:susy2m}, one with matter fields $(X, \Psi )$ and parameter $\epsilon_\CC$ and the other with matter fields $(-JX, \Psi )$ and parameter $\eta_\CC$.

These transformations are gauged with respect to the off-shell $N=4$ vector supermultiplet described above. It can be checked that the $N=4$ supersymmetry transformations \eqref{eq:susy4m} combined with \eqref{eq:susy4g} close up to the equation of motion
\begin{equation}\label{eq:susy4mpsieom}
   \gamma^\mu D_\mu \Psi  -i\sigma \cdot \Psi + \chi_\CC \cdot X - \tau_\CC^* \cdot J\Psi - \zeta_\CC \cdot JX=0,
\end{equation}
for the fermionic field $\Psi$. 

\subsubsection{$N=3$ superymmetric Lagrangian}

The $N=3$ off-shell supersymmetric Chern-Simons Lagrangian is given by:

\begin{equation}\label{eq:susy-cs3}
  \begin{aligned}[m]
  \eL_{CS}^{N=3}=& -\varepsilon^{\mu\nu\rho} \left( A_\mu , \partial_\nu A_\rho + \tfrac{1}{3} [ A_\nu , A_\rho ] \right) +2\, (\sigma ,D) - ( \tau_\CC , E_\CC ) - ( \tau_\CC^* , E_\CC^* ) \\
  & - ( {\bar \chi_{\CC}}^* , \chi_{\CC} ) +\half ( {\bar \zeta_{\CC}} , \zeta_{\CC} ) +\half ( {\bar \zeta_{\CC}}^* , \zeta_{\CC}^* ),
  \end{aligned}
\end{equation}

Indeed, the integral of this Lagrangian is only invariant the subset of $N=3$ supersymmetry transformations generated by the parameters $\epsilon_\CC$ and ${\mathrm{Im}} \, \eta_\CC$ in the $N=4$ transformations \eqref{eq:susy4g}. In other words, invariance under \eqref{eq:susy4g} is only possible provided ${\mathrm{Re}} \, \eta_\CC = 0$. 

The on-shell $N=4$ supersymmetric matter Lagrangian is

\begin{equation}\label{eq:susy-m3a}
  \begin{aligned}[m]
  \eL_{M}^{N=4} =& -\half \left< D_\mu X , D^\mu X \right> + \half \left< X , i D \cdot X \right> -\half \left< \sigma \cdot X , \sigma \cdot X \right> - \half \left< E_\CC \cdot X , JX \right> \\
	&-\half \left< \tau_\CC \cdot X , \tau_\CC \cdot X \right> + \half \left< {\bar \Psi} , \gamma^\mu D_\mu \Psi \right>  - \half \left< {\bar \Psi} , i \sigma \cdot \Psi \right> \\
	&+ \left< {\bar \Psi} , \chi_{\CC} \cdot X \right> - \half \left< {\bar \Psi} , \tau_\CC^* \cdot J \Psi \right> - \left< {\bar \Psi} , \zeta_\CC \cdot J X \right> ,
  \end{aligned}
\end{equation}
whose integral is indeed invariant under the on-shell $N=4$ supersymmetry transformations \eqref{eq:susy4m}.

The $N=3$ Chern-Simons-matter Lagrangian is given by $\eL_{CS}^{N=3} + \eL_{M}^{N=4}$ and, as we will see, there is no option for adding any extra superpotential to this combination without breaking supersymmetry. 

As we discussed in section \ref{sec:matter-reps}, both for $N=3$ and $N=4$ supersymmetry, the matter fields must be valued in the tensor product of the corresponding $\fso(N)$ representation with a \textsl{quaternionic} representation of the Lie algebra. The $\fso(4)$ R-symmetry is not manifest in the way we wrote the Lagrangian in this section. In following sections we will re-write this Lagrangian in a way that the $\fso(4)$ symmetry is explicit and also see how the $\fso(3)$-symmetry embeds in it so that one can obtain an $N=3$ Lagrangian with manifest $\fso(3)$-symmetry. In a similar fashion as we did for the $N=2$ case in section \ref{sec:der-nequal2superpotential}, we will see that the superpotential that satisfies the necessary condition for $N=3$ supersymmetry of possessing $\fso(3)$ R-symmetry, automatically renders the matter part of the Lagrangian plus the superpotential invariant under the on-shell $N=4$ supersymmetry transformations described above. In other words, one gets $N=4$ supersymmetry for free, except in the Chern-Simons part of the Lagrangian. Before moving on to this, lets see how one can obtain the $N=3$ Chern-Simons-matter Lagrangian $\eL_{CS}^{N=3} + \eL_{M}^{N=4}$ from a particular choice of $N=1$ superpotential on $\eL^{N=1}$.

\subsubsection{$N=3$ Lagrangian from the $N=1$ Lagrangian}

For $N=3,4$ supersymmetry the unitary representation of the Lie algebra in which the matter fields are valued must be quaternionic $\fM = W \in \Dar(\fg,\HH)$\footnote{Recall that we view quaternionic representations as complex representations with a quaternionic structure map $J$. In the language of \ref{sec:relat-betw-real}, we work not with $W$ but with $\rh{W}$; although by a slight abuse of notation we will say that fields take values in $W$ when in fact they take values in $\rh{W}$.}. In that case, there exists a new Faulkner map at our disposal $\TT_H (-,-)=\TT (-,J-)$, to build new terms for the superpotential. One can choose for the $N=2$ Lagrangian $\eL^{N=2}$ an F-term superpotential of the form
\begin{equation}\label{eq:nequal3ftermsuperpotential}
   \fW_F ( \Xi_{\CC} ) = \tfrac{1}{32} ( \TT ( \Xi_\CC , J \Xi_\CC ) , \TT ( \Xi_\CC , J \Xi_\CC ) ),
\end{equation}
where $\Xi_\CC$ is a chiral $N=2$ matter superfield, just as in the previous section but $W$-valued this time. This superpotential is $\fg$-invariant and the coefficient is again fixed by the requirement that it gives rise to an enhanced $N=3$ supersymmetry, when added to $\eL_{CS}^{N=2} + \eL_{M}^{N=2}$. Equivalently, following the results of the previous section, this F-term superpotential also arises from choosing $N=1$ superpotential $\sW_2$ in \eqref{eq:superpotential2} to be
\begin{equation}
\label{eq:superpotential3}
\begin{aligned}[m]
  \sW_{\HH} &= \tfrac{1}{16} \int d^2 \theta \; \left[ ( \TT (\Xi ,\Xi ), \TT (\Xi ,\Xi ) ) + \Re\, ( \TT (\Xi , J \Xi ), \TT (\Xi , J \Xi ) )  \right] \\
  &= \tfrac{1}{16} \int d^2 \theta \; \left[ - ( T (\Xi , I \Xi ), T (\Xi , I \Xi ) ) + ( T (\Xi , J \Xi ) , T (\Xi , J \Xi ) ) \right.\\
	&\left.- ( T (\Xi , I J \Xi ) , T (\Xi , I J \Xi ) )  \right] ,
\end{aligned}
\end{equation}
where $\Xi$ is a $W$-valued $N=1$ matter superfield.

Again, the introduction of the extra field components to see the enhanced $N=3$ supersymmetry can be achieved by using a similar method to that which was employed in the previous subsection for understanding enhancement from $N=1$ to $N=2$ supersymmetry. In this case we note that one can obtain the $N=2$ F-term superpotential based on \eqref{eq:nequal3ftermsuperpotential} via integrating out an auxiliary chiral $N=2$ matter superfield $\Pi_\CC = \tau_\CC + {\bar \theta_{\CC}^*} \zeta_{\CC}^* + \half {\bar \theta_{\CC}^*} \theta_{\CC}^* E_\CC$ that is valued in $\fg_\CC$ rather than $W$. Similarly, the $N=2$ supersymmetry transformations for these fields just follow from \eqref{eq:susy2m} by taking the action $\cdot$ of $\fg$ to be the adjoint action of $\fg_\CC$ on itself. The F-term superpotential resulting from
\begin{equation}\label{eq:superpotentialquaternionaux}
  -\half ( \Pi_\CC , \Pi_\CC ) - \tfrac{1}{4} ( \Pi_\CC , \TT (\Xi_\CC , J \Xi_\CC ) ),
\end{equation}
then gives precisely \eqref{eq:nequal3ftermsuperpotential} after integrating out $\Pi_\CC$. Classical scale invariance here follows from the auxiliary components $(\tau_\CC , \zeta_\CC , E_\CC )$ being assigned weights $( 1, \tfrac{3}{2} ,2)$ while their $\fu(1)$ R-charges are $(1,0,-1)$.

Written this way, we can now see that the $N=2$ Lagrangian with this choice of F-term superpotential gives precisely the $N=3$ Lagrangian. Indeed,  adding the first term in \eqref{eq:superpotentialquaternionaux} to the $N=2$ supersymmetric Chern-Simons term \eqref{eq:susy-cs2} gives precisely the Lagrangian $\eL_{CS}^{N=3}$ for $N=3$ supersymmetric Chern-Simons theory \eqref{eq:susy-cs3}. 

\begin{equation}
  \begin{aligned}[m]
  \eL_{CS}^{N=3}=  \eL_{CS}^{N=2} &-\half  \int d^2 \theta_\CC \; ( \Pi_\CC , \Pi_\CC ) -\half  \int d^2 \theta_\CC^* \; ( \Pi_\CC^* , \Pi_\CC^* ),
  \end{aligned}
\end{equation}

Now, adding the second term in \eqref{eq:superpotentialquaternionaux} to the $N=2$ supersymmetric matter Lagrangian \eqref{eq:susy-m2a} gives 

\begin{equation}\label{eq:susy-m3}
  \begin{aligned}[m]
  \eL_{M}^{N=2} &-\half  \int d^2 \theta_\CC \; ( \Pi_\CC , \TT (\Xi_\CC , J \Xi_\CC ) ) +\half  \int d^2 \theta_\CC^* \; ( \Pi_\CC^* , \TT ( J \Xi_\CC ,\Xi_\CC ) ) \\
  =& -\half \left< D_\mu X , D^\mu X \right> + \half \left< X , i D \cdot X \right> -\half \left< \sigma \cdot X , \sigma \cdot X \right> + \half \left< F , F \right> \\
  &+ \half \left< {\bar \Psi} , \gamma^\mu D_\mu \Psi \right>  - \half \left< {\bar \Psi} , i \sigma \cdot \Psi \right> + \left< {\bar \Psi} , \chi_{\CC} \cdot X \right> \\
  &- \half \left< {\bar \Psi} , \tau_\CC^* \cdot J \Psi \right> - \left< {\bar \Psi} , \zeta_\CC \cdot J X \right>  - \half \left< E_\CC \cdot X , JX \right> + \left< F , \tau_\CC^* \cdot JX \right>,
  \end{aligned}
\end{equation}
and upon integrating out the auxiliary matter field $F$, thus fixing $F = - \tau_\CC^* \cdot JX$, one obtains the on-shell $N=4$ supersymmetric matter Lagrangian \eqref{eq:susy-m3a} and the equation of motion for the fermionic field $\Psi$ \eqref{eq:susy4mpsieom}, necessary for the $N=4$ supersymmetry transformations \eqref{eq:susy4m} and \eqref{eq:susy4g} to close.

We prove next that $\sW_\HH$ in \eqref{eq:superpotential3} is the \emph{unique} choice of $N=1$ superpotential giving rise to an on-shell Lagrangian which is invariant under the $\fso(3)$ R-symmetry that is necessary for $N=3$ supersymmetry. However, before doing that we discuss how to re-write this Lagrangian in such a way that the R-symmetry is manifest.

\subsubsection{R-symmetry and uniqueness of $N=3$ superpotential}
\label{sec:nequal3rsymmetry}

An $N=3$ supersymmetric Lagrangian is expected to have $\fso(3) \cong \fsu(2)$ R-symmetry. However, since we wrote this Lagrangian using $N=4$ supermultiplets and most of the action is in fact $N=4$ supersymmetric, it is more natural to write the Lagrangian in a way that is covariant under $\fso(4) \cong \fso(3) \oplus \fso(3) \cong \fsu(2) \oplus \fsu(2)$. Then, the $\fso(3)$ R-symmetry can be recovered by embedding the $\fsu(2)$ diagonal subalgebra of $\fso(4)$. 

To write down the supersymmetry transformations and the Lagrangian in a manifestly $\fso(4)$-invariant way, we need to arrange the component fields into different representations of it forming $\fso(4)$ tensors. Then, as long as all the $\fso(4)$ indices are contracted, the result will be automatically $\fso(4)$-invariant. We recall that $\fso(3) \cong \fsu(2) \cong \fusp(2) \cong \fu(2) \cap \fsp(2, \CC)$. Some representations of $\fsu(2)$ are easily defined via representations of $\fsp(2, \CC)$. 

First, we fix some notation. The vector space for the defining representation of $\fu(2)$ is $\CC^2$. Relative to a basis $\{ \be_\alpha \}$ on $\CC^2$, we denote by $v^\alpha$ the components of a complex vector $v$ which transforms in the defining representation of $\fu(2)$ acting on $\CC^2$. Identifying the complex conjugate with the dual of this vector, the components of the complex conjugate vector $u^*$ are written $u_\alpha$ with the index downstairs (whereby $u_\alpha v^\alpha$ is $\fu(2)$-invariant with repeated indices summed). With respect to this basis, $\varepsilon_{\alpha\beta} = - \varepsilon_{\beta\alpha}$ denotes the component of the ${\mathfrak{sp}}(2,\CC )$-invariant holomorphic 2-form $\varepsilon = \be^1 \wedge \be^2$ on $\CC^2$. This invariant tensor is the one we will use to raise and lower $\alpha, \beta, \gamma$... indices. 

The way the different representations can be identified is as follows. Lets first identify representations of $\fusp(2) \cong \fsu(2)$ and then of the direct product of two of them (which gives a representation of $\fso(4)$). A tensor $w$ in the adjoint representation ${\bf 3}$ of $\fusp(2) = \fu(2) \cap {\mathfrak{sp}}(2,\CC )$ can be taken to have complex components $w^{\alpha\beta} = w^{\beta\alpha}$ obeying the reality condition $w_{\alpha\beta} = \varepsilon_{\alpha\gamma} \varepsilon_{\beta\delta} w^{\gamma\delta}$. Consequently, the tensor $w^\alpha{}_\beta := \varepsilon_{\beta\gamma} w^{\alpha\gamma}$ can be thought of as a skew-hermitian $2\times2$ matrix (in the sense that $w^\alpha{}_\beta = - w_\beta{}^\alpha$) as corresponds to the adjoint representation of $\fusp(2)$. A vector $v \in W$ in the defining representation of $\fu(2)$ is in the fundamental representation ${\bf 2}$ of $\fusp(2)$ if it obeys the pseudo-reality condition $J v^\alpha = \varepsilon_{\alpha\beta} v^\beta$. This implies $h( u^\alpha , v^\alpha ) = - \varepsilon_{\alpha\beta} \, \omega ( u^\alpha , v^\beta )$ for any $u$ and $v$ in the fundamental representation.

We assign the singlet representation to the gauge superfield $A_\mu$ (that is, it has no $\fso(4)$ index) and we combine the auxiliary fields $( \chi_\CC , \zeta_\CC , \sigma , \tau_\CC , D , E_\CC )$ into the following $\fusp(2)$ tensors 
\begin{equation}\label{eq:nequal3quaterniondata}
\begin{aligned}[m]
\left(\chi_{\alpha\beta} \right) &= \begin{pmatrix} \chi_\CC & \zeta_\CC^* \\ - \zeta_\CC & \chi_\CC^* \end{pmatrix}, \quad %
\left(\Sigma_\alpha{}^\beta \right) = \begin{pmatrix} i \sigma & \tau_\CC^* \\ - \tau_\CC & -i\sigma \end{pmatrix}, \quad \\%
\left(\Delta^\alpha{}_\beta \right) &= \begin{pmatrix} iD - \half [ \tau_\CC , \tau_\CC^* ] & E_\CC^* \\ - E_\CC & -iD + \half [ \tau_\CC , \tau_\CC^* ] \end{pmatrix}.
\end{aligned}
\end{equation}

The $N=4$ supersymmetry parameter can also be arranged into a $\fusp(2)$ tensor $\left( \epsilon_{\alpha\beta} \right) = \begin{pmatrix} \epsilon_\CC & \eta_\CC^* \\ - \eta_\CC & \epsilon_\CC^* \end{pmatrix}$ which obeys $\epsilon_{\alpha\beta} = \varepsilon_{\alpha\gamma} \varepsilon_{\beta\delta} \, \epsilon^{\gamma \delta}$ with $\epsilon^{\alpha\beta}$ defined as the complex conjugate of $\epsilon_{\alpha\beta}$. Since they are skew-hermitian, $\Sigma_\alpha{}^\beta$ and $\Delta^\alpha{}_\beta$ are taken to transform in the ${\bf 3}$ of $\fusp(2)$ while the fermionic matrices $\chi_{\alpha\beta}$ and $\epsilon_{\alpha\beta}$ transform in the reducible ${\bf 2} \otimes {\bf 2} = {\bf 3} \oplus {\bf 1}$ representation of $\fusp(2)$. 

Now, to turn them into $\fso(4)$ representations, we denote the matrices above as $\chi_{\alpha{\dot \beta}}$, $\Sigma_{\dot \alpha}{}^{\dot \beta}$, $\Delta^{\alpha}{}_{\beta}$ and $\epsilon_{\alpha{\dot \beta}}$, where indices $\alpha$ and ${\dot \alpha}$ denote the fundamental representations of the left and right $\fsu(2) = \fusp(2)$ factors in $\fso(4)$. Since $\chi_{\alpha\beta}$ and $\epsilon_{\alpha\beta}$ are already in a reducible representation of $\fsu(2)$ which is the product of two, now they are simply in the $({\bf 2},{\bf 2})$ of the $N=4$ R-symmetry algebra $\fso(4) \cong \fsu(2) \oplus \fsu(2)$. On the other hand, $\Sigma_\alpha{}^\beta$ and $\Delta^\alpha{}_\beta$ are in the ${\bf 3}$ of $\fusp(2)$, so it is natural to take the singlet representation ${\bf 1}$ in the other $\fsu(2)$ factor and declare them to be in the $({\bf 1},{\bf 3})$ and $({\bf 3},{\bf 1})$ representations of $\fsu(2) \oplus \fsu(2)$ respectively. 

The $N=3$ structure is then recovered by embedding the R-symmetry $\fusp(2)$ as the diagonal subalgebra of $\fso(4)$. Notice that this selects precisely the one supersymmetry parameter ${\mathrm{Re}} \, \eta_\CC$ which could not preserve the supersymmetric Chern-Simons term in \eqref{eq:susy-cs3} to be the singlet in the decomposition ${\bf 2} \otimes {\bf 2} = {\bf 3} \oplus {\bf 1}$ for the $N=4$ parameter $\epsilon_{\alpha {\dot \beta}}$.

In this notation, the $N=4$ supersymmetry transformations in \eqref{eq:susy4g} take the form
\begin{equation}
  \label{eq:susy4g1}
  \begin{aligned}[m]
    \delta A_\mu &= \half {\mathrm{Re}} \left( {\bar \epsilon^{\alpha {\dot \beta}}} \gamma_\mu \chi_{\alpha {\dot \beta}} \right) \\
    \delta \chi_{\alpha {\dot \beta}} &= \half F_{\mu\nu} \gamma^{\mu\nu} \epsilon_{\alpha {\dot \beta}} - ( \gamma^\mu \epsilon_{\alpha {\dot \gamma}} ) D_\mu \Sigma^{\dot \gamma}{}_{\dot \beta} + \half \epsilon_{\alpha {\dot \gamma}} [ \Sigma^{\dot \gamma}{}_{\dot \delta} , \Sigma^{\dot \delta}{}_{\dot \beta} ] + \Delta_\alpha{}^\gamma \epsilon_{\gamma {\dot \beta}} \\
    \delta \Sigma^{\dot \alpha}{}_{\dot \beta} &= {\bar \epsilon^{\gamma {\dot \alpha}}} \chi_{\gamma {\dot \beta}} - \half \delta^{\dot \alpha}_{\dot \beta} \, {\bar \epsilon^{\gamma {\dot \delta}}} \chi_{\gamma {\dot \delta}} \\
    \delta \Delta_\alpha{}^\beta &= {\bar \epsilon_{\alpha {\dot \gamma}}} \left( \delta^{\dot \gamma}_{\dot \delta} \, \gamma^\mu D_\mu + \Sigma^{\dot \gamma}{}_{\dot \delta} \right) \chi^{\beta {\dot \delta}} - \half \delta_\alpha^\beta \; {\bar \epsilon_{\gamma {\dot \gamma}}} \left( \delta^{\dot \gamma}_{\dot \delta} \, \gamma^\mu D_\mu + \Sigma^{\dot \gamma}{}_{\dot \delta} \right) \chi^{\gamma {\dot \delta}}.
  \end{aligned}
\end{equation}

The $N=3$ supersymmetric Chern-Simons term \eqref{eq:susy-cs3} is
\begin{equation}\label{eq:susy-cs3a}
   \eL_{CS}^{N=3} = -\varepsilon^{\mu\nu\rho} \left( A_\mu , \partial_\nu A_\rho + \tfrac{1}{3} [ A_\nu , A_\rho ] \right) +  (\Sigma_\alpha{}^\beta , \Delta_\beta{}^\alpha ) - \tfrac{1}{6} ( \Sigma_\alpha{}^\beta , [ \Sigma_\beta{}^\gamma , \Sigma_\gamma{}^\alpha ] ) - \half ( {\bar \chi^{\beta\alpha}} , \chi_{\alpha\beta} ).
\end{equation}
It is worth emphasising that the $- \tfrac{1}{6} \Sigma^3$ term is purely to account for the fact the diagonal elements of $\Delta^\alpha{}_\beta$ in \eqref{eq:nequal3quaterniondata} involve the shifted auxiliary field $D + \tfrac{i}{2} [ \tau_\CC , \tau_\CC^* ]$ rather than just $D$. The reason that this shifted definition is useful is that it allows one to obtain from \eqref{eq:susy4g1} precisely the $N=2$ supersymmetry structure with parameter $\epsilon_\CC$ we have found already by simply setting the other parameter $\eta_\CC =0$. Thus, just as in \eqref{eq:susy-cs3}, there are no cubic terms involving any of the auxiliary fields in \eqref{eq:susy-cs3a}. Notice that \eqref{eq:susy-cs3a} cannot be expressed invariantly in terms of the vector supermultiplet fields in the representations of $\fso(4)$ described above. This is another signal of only $N=3$ supersymmetry for the Chern-Simons term.

The on-shell components of the $N=4$ supermultiplet matter fields $X$ and $\Psi$ can be assembled into the vectors
\begin{equation}\label{eq:nequal3quaterniondata2}
 \left( X^\alpha \right) = \begin{pmatrix} X \\ JX \end{pmatrix}, \quad %
\left( \Psi_{\dot \alpha} \right) = \begin{pmatrix} \Psi \\ J\Psi \end{pmatrix},
\end{equation}
which satisfy the pseudo-reality conditions $J X^\alpha = \varepsilon_{\alpha\beta} X^\beta$ and $J \Psi_{\dot \alpha} = \varepsilon^{{\dot \alpha}{\dot \beta}} \Psi_{\dot \beta}$ identically and therefore are in the ${\bf 2}$ representation of $\fusp(2)$. They can both be thought of as inhabiting the graph of $J$ in $W \oplus W$. To turn them into $\fso(4)$ tensors, it is enough to take them to be in the singlet representation of the other $\fsu(2)$ factor so that they are in the representations $({\bf 2},{\bf 1})$ and $({\bf 1},{\bf 2})$ of $\fso(4)$ respectively. 

The on-shell $N=4$ supersymmetry transformations \eqref{eq:susy4m} for \eqref{eq:nequal3quaterniondata2} can now be more compactly expressed as
\begin{equation}
  \label{eq:susy4mH}
  \begin{aligned}[m]
    \delta X^\alpha &= {\bar \epsilon^{\alpha{\dot \beta}}} \Psi_{\dot \beta} \\
    \delta \Psi_{\dot \alpha} &= - ( \gamma^\mu \epsilon_{\beta {\dot \alpha}} ) D_\mu X^\beta - \left( \Sigma_{\dot \alpha}{}^{\dot \beta} \cdot X^\gamma \right) \epsilon_{\gamma {\dot \beta}},
  \end{aligned}
\end{equation}
and the $N=4$ supersymmetric matter Lagrangian \eqref{eq:susy-m3a} becomes
\begin{equation}\label{eq:susy-m3b}
  \begin{aligned}[m]
  \eL_{M}^{N=4} =& -\tfrac{1}{4} \left< D_\mu X^\alpha , D^\mu X^\alpha \right> + \tfrac{1}{4} \left< X^\alpha , \Delta^\alpha{}_\beta \cdot X^\beta \right> -\tfrac{1}{8} \left< \Sigma_{\dot \alpha}{}^{\dot \beta} \cdot X^\gamma , \Sigma_{\dot \alpha}{}^{\dot \beta} \cdot X^\gamma \right>  \\
  &+ \tfrac{1}{4} \left< {\bar \Psi_{\dot \alpha}} , \gamma^\mu D_\mu \Psi_{\dot \alpha} \right>  - \tfrac{1}{4} \left< {\bar \Psi_{\dot \alpha}} , \Sigma_{\dot \alpha}{}^{\dot \beta} \cdot \Psi_{\dot \beta} \right> + \half \left< {\bar \Psi_{\dot \alpha}} , \chi_{\beta{\dot \alpha}} \cdot X^\beta \right>.
  \end{aligned}
\end{equation}
Notice that, for example, $\left< u^\alpha , v^\alpha \right> = 2 \left< u , v \right> = - \varepsilon_{\alpha\beta} \, {\mathrm{Re}} \, \omega ( u^\alpha , v^\beta )$ for any $u,v \in W$, with $u^\alpha = ( u , Ju)$ and $v^\alpha = ( v , Jv)$, and the contraction of $\alpha$ indices here is $\fusp(2)$-invariant as a consequence of $h(-,-)$ being complex sesquilinear (remember that here $\left< -,- \right> = Re h(-,-)$). The same applies for contracted ${\dot \alpha}$ indices with respect to the other $\fusp(2)$ factor in $\fso(4)$. Thus the matter Lagrangian \eqref{eq:susy-m3b} is manifestly $\fso(4)$-invariant as expected from the fact that it is $N=4$ supersymmetric.

The lack of off-shell $N=4$ supersymmetry and $\fso(4)$-invariance for the Chern-Simons term propagates into the form of the equations of motion for some of the auxiliary fields. In particular, the equations of motion for $\chi_{\CC}$, $\zeta_\CC$, $D$ and $E_\CC$ collect into the following $\fusp(2)$ representations
\begin{equation}\label{eq:nequal3quaterniondata3}
\left( \chi_{\alpha\beta} \right) = -\half \, \TT ( \Psi_\alpha , X^\beta ), \quad %
\left( \Sigma_\alpha{}^\beta \right) = \tfrac{1}{4} \, \TT ( X^\beta , X^\alpha ),
\end{equation}
with the indices matching as a consequence of $\TT$ being a complex sesquilinear map. These $\fusp(2)$-invariant equations are not $\fso(4)$-invariant since they would not make sense after sprinkling dots commensurate with the $\fso(4)$ representations that the fields were declared to inhabit above. Notice though that the first equation in \eqref{eq:nequal3quaterniondata3} would have been $\fso(4)$-invariant, with $\chi_{\alpha {\dot \beta}} = -\half \, \TT ( \Psi_{\dot \beta} , X^\alpha )$, if it had been the transposed vector appearing on the right hand side. This seemingly innocuous statement will turn out to be a key feature of realising $N=4$ supersymmetry to be described in the next section.

Let us now close by noting the manifestly $\fusp(2)$-invariant form of the on-shell $N=3$ Lagrangian \eqref{eq:susy-lag-on-shell-quaternionic} given by
\begin{equation}
  \label{eq:susy-lag-on-shell-quaternionic2}
  \begin{aligned}[m]
    \eL^{N=3} =&  -\varepsilon^{\mu\nu\rho} \left( A_\mu , \partial_\nu A_\rho + \tfrac{1}{3} [ A_\nu , A_\rho ] \right) -\tfrac{1}{4} \left< D_\mu X^\alpha , D^\mu X^\alpha \right> + \tfrac{1}{4} \left< {\bar \Psi}_\alpha , \gamma^\mu D_\mu \Psi_\alpha \right>  \\
    &- \tfrac{1}{8} ( \TT ( X^\alpha , {\bar \Psi}_\beta ) , \TT ( \Psi_\alpha , X^\beta )) -\tfrac{1}{16} ( \TT ( X^\alpha , X^\beta ) , \TT ( {\bar \Psi}_\alpha , \Psi_\beta )) \\
    &+\tfrac{1}{384} \left( [ \TT ( X^\alpha , X^\beta ) , \TT ( X^\beta , X^\gamma ) ] , \TT ( X^\gamma , X^\alpha ) \right) \\
		&- \tfrac{1}{128} \left< \TT ( X^\alpha , X^\beta ) \cdot X^\gamma , \TT ( X^\alpha , X^\beta ) \cdot X^\gamma \right>,
  \end{aligned}
\end{equation}
and the $N=3$ supersymmetry transformations in \eqref{eq:susy3-onshell} which become 
\begin{equation}
  \label{eq:susy3-onshell2}
  \begin{aligned}[m]
    \delta X^\alpha &= {\bar \epsilon}^{\alpha\beta} \Psi_\beta \\
    \delta \Psi_{\alpha} &= - ( \gamma^\mu \epsilon_{\alpha\beta} ) D_\mu X^\beta - \tfrac{1}{4} \TT ( X^\beta , X^\alpha ) \cdot X^\gamma \, \epsilon_{\beta\gamma} \\
    \delta A_\mu &= \tfrac{1}{4} T \left( X^\alpha , {\bar \epsilon}^{\alpha\beta} \gamma_\mu \Psi_\beta \right) .
  \end{aligned}
\end{equation}

\subsection{Rigidity of the $N=3$ superpotential}
\label{sec:der-nequal3superpotential}

We proof rigidity of the $N=3$ superpotential by imposing the necessary condition for an $N=1$ superpotential to realise $N=3$ supersymmetry, that it must have a $\fso(3)$ R-symmetry. In this case it will turn out that the only option is in fact the $N=3$ superpotential described above which does realise $N=3$ supersymmetry, rendering this condition also sufficient in this case.

Since a theory with enhanced $N=3$ superconformal symmetry can be thought of as a special kind of $N=2$ superconformal Chern-Simons-matter theory, we can use the results on the on-shell $N=2$ Lagrangian \eqref{eq:susy-lag-on-shell-complex} which we know already has $\fso(2)$ R-symmetry and deduce the constraints which an enhanced $\fusp(2) > \fu(1)$ R-symmetry puts on the generic holomorphic $N=2$ F-term superpotential $\fW_F$.

With respect to this desired enhancement, as discussed in the previous section, we can collect the $W$-valued bosonic and fermionic matter fields $X$ and $\Psi$ into the fields $X^\alpha = (X,JX)$ and $\Psi_\alpha = ( \Psi , J \Psi )$, obeying identically the pseudo-reality conditions $J X^\alpha = \varepsilon_{\alpha\beta} X^\beta$ and $J \Psi_\alpha = \varepsilon^{\alpha\beta} \Psi_\beta$, corresponding to the fundamental representation of $\fusp(2)$. It follows that the kinetic terms $\left< D_\mu X , D^\mu X \right> = \half \left< D_\mu X^\alpha , D^\mu X^\alpha \right>$ and $\left< {\bar \Psi} , \gamma^\mu D_\mu \Psi \right> = \half \left< {\bar \Psi}_\alpha , \gamma^\mu D_\mu \Psi_\alpha \right>$ in \eqref{eq:susy-lag-on-shell-complex} are automatically $\fusp(2)$-invariant, but the same cannot be said of the Yukawa couplings and scalar potential. The extra conditions for $\fusp(2)$ R-symmetry in these terms can be deduced most easily by focusing initially on the Yukawa couplings. We will see that the condition that they be $\fusp(2)$-invariant then guarantees that the scalar potential is too.

Consider the Yukawa couplings in \eqref{eq:susy-lag-on-shell-complex} which do not involve the F-term superpotential. Since $\TT ( X^\alpha , X^\alpha )$ vanishes identically, it turns out that there are not many options for writing down manifestly $\fusp(2)$-invariant terms which recover these Yukawa couplings. In particular, one finds that only from $-\tfrac{1}{16} ( \TT ( X^\alpha , X^\beta ) , \TT ( {\bar \Psi}_\alpha , \Psi_\beta ) )$ can one recover the second term while the first term must come from $- \tfrac{1}{8} ( \TT ( X^\alpha , {\bar \Psi}_\beta ) , \TT ( \Psi_\alpha , X^\beta ) )$ (other possible $\fusp(2)$-invariant permutations of indices can be rewritten in terms of these using the pseudo-reality condition for the matter fields). These contributions do not give just the corresponding terms of \eqref{sec:on-shell-3dsusy-complex} or else it would already be $\fusp(2)$-invariant. Indeed, the point of doing this is to isolate the terms responsible for obstructing the enhanced $\fusp(2)$ R-symmetry. Combining these extra terms with the contribution from the F-term superpotential (which one finds also cannot be $\fusp(2)$-invariant on its own) to the Yukawa couplings gives an overall obstruction term which vanishes only if $\fW_F (X) = \tfrac{1}{32} ( \TT (X,JX), \TT (X,JX) )$. 

This is precisely the F-term superpotential \eqref{eq:nequal3ftermsuperpotential} which gives rise to the $N=1$ superpotential $\sW_\HH$ \eqref{eq:superpotential3}. The fact that we have already established this theory to be invariant under the $N=3$ superconformal algebra means that the realisation of this $\fusp(2)$ R-symmetry in the Yukawa couplings is in fact necessary and sufficient for $N=3$ enhancement here. 

Recalling that $N=3$ theories should describe M2-brane near-horizon geometries of the form $\AdS_4 \times X_7$, where $X_7$ is a 3-Sasakian 7-manifold, a nice consistency check is that there exists a corresponding infinitesimal rigidity theorem \cite[theorem~13.3.24]{MR2382957} for such geometries.

In summary, we have shown that for $\fM = W \in \Dar(\fg,\HH)$, the $N=1$ superpotential $\sW_{\HH}$ in \eqref{eq:superpotential3} added to the $N=1$ Chern-Simons-matter Lagrangian $\eL_{CS} + \eL_M$ (or equivalently the F-term superpotential in \eqref{eq:nequal3ftermsuperpotential} added to the $N=2$ Chern-Simons-matter Lagrangian $\eL_{CS}^{N=2} + \eL_{M}^{N=2}$) gives rise to precisely the $N=3$ Chern-Simons-matter Lagrangian $\eL_{CS}^{N=3} + \eL_{M}^{N=4}$. Moreover, $\sW_\HH$ in \eqref{eq:superpotential3} is the \emph{unique} choice of $N=1$ superpotential giving rise to an on-shell Lagrangian which is invariant under the $\fusp(2)$ R-symmetry that is necessary for $N=3$ supersymmetry. Thus one establishes that the class of $N=3$ superconformal Chern-Simons-matter theories is rigid in the sense that the superpotential function $\sW_\HH$ is fixed uniquely by the requirement of $N=3$ supersymmetry.

It is worth stressing that, were it not for the Chern-Simons term, the quaternionic structure of $W$ would have allowed an even greater enhancement to $N=4$ supersymmetry here. The obvious question is therefore whether there are special kinds of quaternionic unitary representations for which the realisation of (at least) $N=4$ superconformal symmetry is possible? This is indeed the case and the resulting theories will be detailed in the next section. 

Before moving on to this, we describe the on-shell form of the $N=3$ Chern-Simons-matter Lagrangian and supersymmetry transformations. 

\subsubsection{On-shell N=3 supersymmetric Lagrangian}
\label{sec:on-shell-3dsusy-quaternionic}

Having already integrated out $F$ in order to obtain the matter Lagrangian \eqref{eq:susy-m3a}, it remains to impose the equations of motion $\chi_\CC^* = \half \TT ( X , \Psi )$, $\zeta_\CC^* = -\half \TT ( X , J \Psi )$, $\sigma = -\tfrac{i}{4} \TT (X,X)$ and $\tau_\CC = -\tfrac{1}{4} \TT (X, J X)$ for the respective auxiliary fields $\chi_{\CC}$, $\zeta_\CC$, $D$ and $E_\CC$ in the $N=4$ vector supermultiplet. Substituting these expressions into the Lagrangian gives
\begin{equation}
  \label{eq:susy-lag-on-shell-quaternionic}
  \begin{aligned}[m]
    \eL_{on}^{N=3}&=\eL_{CS}^{N=3} + \eL_{M}^{N=4} =  -\varepsilon^{\mu\nu\rho} \left( A_\mu , \partial_\nu A_\rho + \tfrac{1}{3} [ A_\nu , A_\rho ] \right) -\half \left< D_\mu X , D^\mu X \right>  \\
    &- \fV_D (X) - \fV_F (X) + \half \left< {\bar \Psi} , \gamma^\mu D_\mu \Psi \right> - \tfrac{1}{4} ( \TT (X, {\bar \Psi} ) , \TT (\Psi , X ))  \\
    &-\tfrac{1}{8} ( \TT (X,X) , \TT ( {\bar \Psi} , \Psi ))- \tfrac{1}{8} ( \TT (X, J {\bar \Psi} ) , \TT ( X , J \Psi )) -\tfrac{1}{16} ( \TT (X,JX) , \TT ( {\bar \Psi} , J \Psi )) \\
    &- \tfrac{1}{8} ( \TT (JX, {\bar \Psi} ) , \TT ( J X , \Psi )) -\tfrac{1}{16} ( \TT (JX,X) , \TT ( J {\bar \Psi} , \Psi )),
  \end{aligned}
\end{equation}
where the F-term scalar potential here is $\fV_F (X) = \tfrac{1}{32} \left< \TT (X,JX) \cdot X , \TT (X,JX) \cdot X \right>$ while $\fV_D (X)$ is just as in \eqref{eq:susy-lag-on-shell-complex}.

Invariance of \eqref{eq:susy-lag-on-shell-quaternionic} under the $N=3$ supersymmetry transformations \eqref{eq:susy4g} and \eqref{eq:susy4m} can be established after substituting the equations of motion for $\chi_{\CC}$, $\zeta_\CC$, $D$, $E_\CC$ and $F$. These supersymmetry transformations are
\begin{equation}
  \label{eq:susy3-onshell}
  \begin{aligned}[m]
    \delta X &= ( {\bar \epsilon_{\CC}}^* + {\bar \eta_{\CC}} J ) \Psi \\
    \delta \Psi &= - \gamma^\mu ( \epsilon_\CC - \eta_\CC J ) D_\mu X - \tfrac{1}{4} \TT (JX,X) \cdot ( \epsilon_\CC^* J + \eta_\CC^* ) X - \tfrac{1}{4} ( \epsilon_\CC + \eta_\CC J ) \TT(X,X) \cdot X \\
    \delta A_\mu &= \half T \left( X , ( {\bar \epsilon_{\CC}}^* - {\bar \eta_{\CC}}^* J ) \gamma_\mu \Psi \right) ,
  \end{aligned}
\end{equation}
and close up to a translation on $\RR^{1,2}$ plus a gauge transformation, using the equations of motion from \eqref{eq:susy-lag-on-shell-quaternionic}.

\section{$N > 3$ Supersymmetry}
\label{sec:Ngt3}

Having proved that the $N=3$ superpotential is rigid, the only way to achieve supersymmetry enhancement is by choosing special cases of the $N<4$ theories. In order to analyse how to do that, we find it is crucial to examine one requirement that the superpotential of a theory with $N$-extended supersymmetry must satisfy: the fact that it realises an enhanced $\fso(N-1)$ global symmetry. For a given value of $N>3$, this property needs not guarantee that the resulting on-shell Chern-Simons-matter theory realises an enhanced $N$-extended superconformal symmetry. However, for such an on-shell theory to be derivable from an $N=1$ superpotential requires that superpotential to be invariant under those R-symmetries preserving the choice of $N=1$ superspace parameter, which is an $\fso(N-1)$ subalgebra of the $\fso(N)$ R-symmetry. Nevertheless in all the cases we consider we will find that this property does give rise to superconformal symmetry enhancement and thereby this construction will recover all the known examples of $N \geq 4$ superconformal Chern-Simons-matter theories. 

In this section we discuss first $N=4$ theories. We will find that a generic $N=3$ theory, where the representation of the Lie algebra to be used (or equivalently the 3-algebra to be used) is of quaternionic type, is enhanced to $N=4$ theory when $W$ is not only quaternionic but is the direct sum of one or more anti-Lie triple systems. This condition recovers all known $N=4$ supersymmetric Chern-Simons-matter theories discovered first in \cite{GaiottoWitten, pre3Lee}.

Then we move on to discuss further enhancement of supersymmetry by imposing invariance of the $N=1$ superpotential under $\fso(N-1)$  with the help of representation theory as discussed in \ref{sec:matter-reps}, where we showed how one can express an $N$-extended matter representation in terms of an $(N-1)$-extended representation. In each case, from $N=5$ to $N=8$, we discuss first the new form of the superpotential $\sW_\HH$ after choosing the appropriate representation. Then we display the on-shell form of the corresponding $N$-supersymmetric Lagrangian and the supersymmetry representations. Finally, results from section \ref{sec:LSAembed} will allow us to establish all the possible gauge Lie algebras allowed for each irreducible $N$-supersymmetric theory. 

With this procedure we recover all the known 3-dimensional Chern-Simons-matter theories for $N > 3$: the $N=4$ theories of \cite{GaiottoWitten, pre3Lee}, the $N=5$ in \cite{3Lee,BHRSS}, the $N=6$ ABJM theory \cite{MaldacenaBL} and finally the Bagger-Lambert theory \cite{BL1,BL2}. Also, we prove that in this context $N=7$ supersymmetry implies $N=8$.

\subsection{N=4 supersymmetry}
\label{sec:nequal4}

There are two families of $N=4$ Chern-Simons-matter theories in 3 dimensions known in the literature. The first one was obtained by  Gaiotto and Witten in \cite{GaiottoWitten} by considering the $N=3$ theories in section \ref{sec:nequal3} for one of the extremal cases of quaternionic representations of a Lie algebra: the one that defines an anti-Lie triple system. This is the case where the tensor $\eR_H$ lives in $W^{\yng(2,2)}$ or equivalently the 3-bracket in $W$ satisfies the cyclicity condition \eqref{eq:quat-aLTS-3b}. In our notation, we say that the matter representation is $\fM = W \in \Dar(\fg,\HH)_{\text{aLTS}}$. We will see in this section how this condition is equivalent to precisely making the change in the $N=3$ theory that we needed for it to be $N=4$ supersymmetric. Namely, it is equivalent to replacing the $\fusp(2)$-invariant expression $- \half ( {\bar \chi^{\beta \alpha}} , \chi_{\alpha \beta} )$ for the auxiliary fermions in the off-shell $N=3$ Chern-Simons Lagrangian \eqref{eq:susy-cs3a} with the $\fso(4)$-invariant term $- \half ( {\bar \chi^{\alpha {\dot \beta}}} , \chi_{\alpha {\dot \beta}} )$. 

The other known class of $N=4$ theories was found by Hosomichi, Lee, Lee, Lee and Park in \cite{pre3Lee} by the coupling of the so-called twisted hypermultiplets. It was shown in section 3.1.2 in \cite{SCCS3Algs} that this class is equivalent to the one obtained by taking $W$ to be a quaternionic representation formed by the direct sum of two anti-Lie triple systems, $W = W_1 \oplus W_2$ with $W_1,W_2 \in\Dar(\fg,\HH)_{\text{aLTS}}$, which is not necessarily of aLTS class itself. This links with the discussion in \ref{sec:matter-reps}, where we concluded that the matter fields in $N=4$ theories must transform under (the appropriate real form of) the representation $\Delta^{(4)}_+ \otimes W_1 \oplus \Delta^{(4)}_- \otimes W_2$ of $\fso(4) \oplus \fg$. Where we know now that it is not sufficient that $W_1, W_2$ are quaternionic representations, but they must also be of aLTS type. The Gaiotto-Witten class is then a particular case where $W_1 = W_2$, hence the fields transform under $\Delta^{(4)}_+ \otimes W$.

In this section we review these two classes of theories obtained when the representations of the Lie algebra $W_1$ and $W_2$ are taken to be of aLTS type and the constrains on the gauge algebra this condition imposes.

\subsubsection{Gaiotto-Witten $N=4$ theories}
\label{sec:antiLTS-GW}

We begin by considering matter representations valued in a single aLTS $\fM = W \in \Dar(\fg,\HH)_{\text{aLTS}}$. Moreover we demand $W$ to be an irreducible aLTS, i.e. not a direct sum of two aLTS. This class of representation was shown by Gaiotto and Witten in \cite{GaiottoWitten} to give rise to an enhanced $N=4$ superconformal symmetry. 

Such representations $W$ are characterised by the existence of a Lie superalgebra structure on $\fg_\CC \oplus W$, a fact which was first appreciated in \cite{GaiottoWitten}. Notice that the aLTS cyclicity condition \eqref{eq:quat-aLTS} is quite restrictive. In particular it means that $\TT (X,JX) \cdot X = 0$ identically for any $W$-valued field $X$ in this special case. This implies that the F-term superpotential in \eqref{eq:nequal3ftermsuperpotential} vanishes identically. Thus the superpotential $\sW_\HH$ in \eqref{eq:superpotential3} is equal to the minimal superpotential $\sW_{\CC}$ in \eqref{eq:superpotentialcomplex} for $W\in \Dar(\fg,\HH)_{\text{aLTS}}$.

Some crucial identities implied by the aLTS cyclicity condition \eqref{eq:quat-aLTS} are   
\begin{equation}
  \label{eq:antiLTSid1}
  ( T (X,IX) , T (X,IX) ) = ( T (X,JX) , T (X,JX) ) = ( T (X,IJX) , T (X,IJX) ),
\end{equation}
and
\begin{equation}
  \label{eq:antiLTSid2}
  ( T (X,IX) , T (X,JX) ) = ( T (X,IX) , T (X,IJX) ) = ( T (X,JX) , T (X,IJX) ) =0,
\end{equation}
for any $W$-valued field $X$. Remarkably these algebraic conditions imply that the minimal superpotential $\sW_\HH = \sW_\CC$ for $W$ is precisely that which was used by Gaiotto and Witten \cite{GaiottoWitten}, thus allowing the realisation of $N=4$ superconformal symmetry. Also, one can use these identities to express this superpotential as
\begin{equation}\label{eq:superpotential4}
  \sW_{\HH} = -\tfrac{1}{16} \int d^2 \theta \; ( T (\Xi ,J \Xi ), T (\Xi ,J \Xi ) ).
\end{equation}

Writing the $W$-valued $N=1$ superfield $\Xi$ appearing above as $\left( \Xi^\alpha \right) = ( \Xi , J \Xi )$ in terms of a $\fusp(2)$-doublet then the superpotential \eqref{eq:superpotential4} can be re-expressed as
\begin{equation}\label{eq:superpotential4a}
  \sW_{\HH} = -\tfrac{1}{48} \int d^2 \theta \; \half \, \varepsilon_{\alpha\beta} \, \varepsilon_{\gamma\delta} \, ( \TT (\Xi^\alpha ,J \Xi^\gamma ), \TT (\Xi^\beta ,J \Xi^\delta ) ).
\end{equation}

On its own, this is not $\fso(4)$-invariant because the constituent matter fields $X^\alpha$ and $\Psi_{\dot \alpha}$ naturally transform in the fundamental representations of the two different $\fusp(2)$ factors of $\fso(4)$. 

The on-shell form of the Lagrangian $\eL_{GW}^{N=4} = \eL_{CS} + \eL_M + \sW_{\HH}$, using the superpotential \eqref{eq:superpotential4} is given by
\begin{equation}
\label{eq:susy-lag-csm4}
  \begin{aligned}[m]
   \eL_{GW}^{N=4} =& -\varepsilon^{\mu\nu\rho} \left( A_\mu , \partial_\nu A_\rho + \tfrac{1}{3} [ A_\nu , A_\rho ] \right) -\tfrac{1}{2} \left< D_\mu X , D^\mu X \right> + \tfrac{1}{2} \left< {\bar \Psi} , \gamma^\mu D_\mu \Psi \right>  \\
   &+ \tfrac{1}{8} ( T ( {\bar \Psi} , J \Psi ) , T (X,JX) ) + \tfrac{1}{4} ( T ( X ,J {\bar \Psi} ) , T (X,J \Psi ) ) + \tfrac{1}{4} ( T ( X , {\bar \Psi} ) , T ( X , \Psi ) ) \\
   &-\tfrac{1}{32} \left< T (X,JX) \cdot X , T (X,JX) \cdot X \right> ,
  \end{aligned}
\end{equation}
which can be given a manifestly $\fso(4)$-invariant expression as
\begin{equation}
\label{eq:susy-lag-csm4a}
  \begin{aligned}[m]
   \eL_{GW}^{N=4} =& -\varepsilon^{\mu\nu\rho} \left( A_\mu , \partial_\nu A_\rho + \tfrac{1}{3} [ A_\nu , A_\rho ] \right) -\tfrac{1}{4} \left< D_\mu X^\alpha , D^\mu X^\alpha \right> + \tfrac{1}{4} \left< {\bar \Psi}_{\dot \alpha} , \gamma^\mu D_\mu \Psi_{\dot \alpha} \right>  \\
   &+ \tfrac{1}{16} \, \varepsilon_{\alpha\beta} \, \varepsilon^{{\dot \gamma}{\dot \delta}} \, ( \TT ( X^\alpha , J {\bar \Psi}_{\dot \gamma} ) , \TT ( X^\beta , J \Psi_{\dot \delta} ) ) \\
   &-\tfrac{1}{768} \, ( [ \TT( X^\alpha , X^\beta ) , \TT( X^\beta , X^\gamma ) ] , \TT( X^\gamma , X^\alpha ) ).
  \end{aligned}
\end{equation}

This is indeed the $N=4$ Gaiotto-Witten Lagrangian in \cite{GaiottoWitten}. The precise value of the coefficient in the superpotential \eqref{eq:superpotential4} is what has allowed the various Yukawa couplings to assemble themselves in an $\fso(4)$-invariant manner in the second line above and has fixed the overall coefficient for this term.

The on-shell $N=4$ supersymmetry transformations under which the integral of this Lagrangian is invariant are 
\begin{equation}
  \label{eq:susy4-onshell}
  \begin{aligned}[m]
    \delta X &= ( {\bar \epsilon_{\CC}}^* + {\bar \eta_{\CC}} J ) \Psi \\
    \delta \Psi &= - \gamma^\mu ( \epsilon_\CC - \eta_\CC J ) D_\mu X - \tfrac{1}{4} ( \epsilon_\CC - \eta_\CC J ) \TT(X,X) \cdot X \\
    \delta A_\mu &= \half T \left( X , ( {\bar \epsilon_{\CC}}^* + {\bar \eta_{\CC}} J ) \gamma_\mu \Psi \right) ,
  \end{aligned}
\end{equation}
which close up to a translation on $\RR^{1,2}$ plus a gauge transformation, using the equations of motion from \eqref{eq:susy-lag-csm4}. Their $\fso(4)$-covariant expressions are
\begin{equation}
  \label{eq:susy4-onshella}
  \begin{aligned}[m]
    \delta X^\alpha &= {\bar \epsilon}^{\alpha {\dot \beta}} \Psi_{\dot \beta} \\
    \delta \Psi_{\dot \alpha} &= - \left[ \gamma^\mu D_\mu X^\beta + \tfrac{1}{12} \, \TT( X^\beta , X^\gamma ) \cdot X^\gamma \right] \epsilon_{\beta {\dot \alpha}} \\
    \delta A_\mu &= \tfrac{1}{4} \TT \left( X^\alpha , {\bar \epsilon^{\alpha {\dot \beta}}} \gamma_\mu \Psi_{\dot \beta} \right) .
  \end{aligned}
\end{equation}

Comparing the on-shell $N=4$ transformations in \eqref{eq:susy4-onshell} with the $N=3$ ones in \eqref{eq:susy3-onshell}, we note the following differences. First, the second term in the transformation of $\Psi$ in \eqref{eq:susy3-onshell} (that arose from imposing the equation of motion $\tau_\CC = - \tfrac{1}{4} \TT (X,JX)$) is absent in \eqref{eq:susy4-onshell}, which is consistent with the fact that we have no F-term superpotential here. Second, the sign of the parameter $\eta_\CC$ has changed in the second term in the variation of $\Psi$ in \eqref{eq:susy4-onshell} relative to \eqref{eq:susy3-onshell}. Finally, in the on-shell transformation of $A_\mu$ above we have effectively replaced the parameter $\eta_\CC^*$ in \eqref{eq:susy3-onshell} with $- \eta_\CC$ in \eqref{eq:susy4-onshell}. These changes are all necessary for the realisation of $N=4$ supersymmetry on-shell and the final change has a natural interpretation based on the remark at the end of the penultimate paragraph in section~\ref{sec:nequal3rsymmetry}. Namely, it is precisely the on-shell $N=4$ transformation for $A_\mu$ in \eqref{eq:susy4-onshella} that would have resulted from substituting the $\fso(4)$-covariant equation $\chi_{\alpha {\dot \beta}} = -\half \, \TT ( \Psi_{\dot \beta} , X^\alpha )$ for the auxiliary field into the first line of \eqref{eq:susy4g1}. Moreover, this equation of motion could be obtained by replacing the $\fusp(2)$-invariant expression $- \half ( {\bar \chi^{\beta \alpha}} , \chi_{\alpha \beta} )$ for the auxiliary fermions in the off-shell $N=3$ Chern-Simons Lagrangian \eqref{eq:susy-cs3a} with the $\fso(4)$-invariant term $- \half ( {\bar \chi^{\alpha {\dot \beta}}} , \chi_{\alpha {\dot \beta}} )$. One way to think of this change is from Wick rotating one of the four auxiliary fermions changing the ${\mathfrak{so}}(1,3)$-invariant inner product in the $N=3$ case to an $\fso(4)$-invariant one in the $N=4$ case.

\subsubsection{$N=4$ theories for reducible $W$ }
\label{sec:twisthyp}

Lets consider now the case where $W$ is the direct sum of two aLTS, which may or may not be irreducible, $W = W_1 \oplus W_2$ with $W_1,W_2 \in \Dar(\fg,\HH)_{\text{aLTS}}$. The quaternionic hermitian structure on $W$ is defined in the obvious way such that $(W,h ,J) = ( W_1 \oplus W_2 , h_1 \oplus h_2 , J_1 \oplus J_2 )$ in terms of the corresponding structures on $W_1$ and $W_2$. The action of $\fg$ on $W$ is defined by $X \cdot ( v_1 , v_2 ) = ( X \cdot v_1 , X \cdot v_2 )$ for any $X \in \fg$ and the map $\TT = \TT_1 \oplus \TT_2$ decomposes orthogonally in terms of its restrictions to $W_1$ and $W_2$; in other words, $\TT(w_1,w_2) = 0$ for all $w_1\in W_1$ and $w_2 \in W_2$.

Demanding that $W$ be of aLTS itself (and faithful) is too strong and the resulting theory decouples into a Gaiotto-Witten theory for $W_1$ and another for $W_2$. Indeed, consider the aLTS cyclicity condition \eqref{eq:quat-aLTS}
\begin{equation}
  \TT (u,Jv) \cdot w + \TT (v,Jw) \cdot u + \TT (w,Ju) \cdot v =0
\end{equation}
for all $u,v,w \in W$. Decomposing this equation on $W_1 \oplus W_2$ shows that it is identically satisfied on the individual components $W_1$ and $W_2$, since they are both aLTS. However, the contributions from the mixed components imply that $\TT_1 ( u_1 , v_1 ) \cdot w_2 = 0$ and $\TT_2 ( u_2 , v_2 ) \cdot w_1 = 0$, for all $u_1 , v_1 , w_1 \in W_1$ and $u_2 , v_2 , w_2 \in W_2$. Let $\fg_1 = \TT(W_1,W_1)$ and $\fg_2=\TT(W_2,W_2)$. Then $\fg_1$ acts trivially on $W_2$ and $\fg_2$ acts trivially on $W_1$. Since $W$ is faithful, $\fg_1 \cap \fg_2 = 0$ and since $\TT(W_1,W_2)=0$, we see that $\fg_\CC = \fg_1 \oplus \fg_2$. Furthermore the direct sum is orthogonal with respect to the inner product $(-,-)$. Consequently the superconformal Chern-Simons-matter theories based on such data would effectively decouple in terms of distinct Gaiotto-Witten $N=4$ theories on $W_1$ and $W_2$. We will therefore exclude the possibility that $W$ itself be aLTS. Notice that in particular this shows that for the $N=4$ Gaiotto-Witten theories to be irreducible, the aLTS $W$ must be irreducible.

Let us now examine the superpotential $\sW_\HH$ in \eqref{eq:superpotential3} for an $N=1$ matter superfield $\Xi$ valued in $W = W_1 \oplus W_2$. If one decomposes $\sW_\HH$ on $W$ into its component parts on $W_1$ and $W_2$ one obtains a sum of three distinct contributions. Two of these are simply the decoupled superpotentials on the individual components $W_1$ and $W_2$. Since we have assumed that both these components are aLTS then these contributions each agree with the expression \eqref{eq:superpotential4} for the Gaiotto-Witten superpotentials associated with $W_1$ and $W_2$ individually. In addition one has a contribution from the mixture of components on $W_1$ and $W_2$. It is this mixed term which can be thought of as providing a non-trivial F-term superpotential contribution in the resulting theory.

The next question is whether this theory can realise $N=4$ supersymmetry. Recall that the superpotential $\sW_\HH$ in \eqref{eq:superpotential3} gives rise to an on-shell $N=3$ superconformal theory with manifest $\fusp(2)$ R-symmetry. However, on its own, $\sW_\HH$ is generically only invariant under the $\fu(1)$ R-symmetry subalgebra arising from the the $N=2$ supersymmetric framework from whence it came. A crucial indicator of the enhanced $N=4$ supersymmetry in the Gaiotto-Witten theory is that its superpotential $\sW_\HH$ enjoys the larger global symmetry $\fusp(2) > \fu(1)$.

With this in mind we note that if one expresses the $N=1$ matter superfield $\Xi = ( \Xi_1 , \Xi_2 )$ valued in $W = W_1 \oplus W_2$ as a $\fusp(2)$-doublet $\left(\Xi_1^\alpha \right)= ( \Xi_1 , J_1 \Xi_1 )$ and $\left( \Xi_2^\alpha \right)= ( \Xi_2 , J_2 \Xi_2 )$, then the superpotential $\sW_\HH$ is not $\fusp(2)$-invariant. The trick is to instead write the $N=1$ matter superfield as $\Xi = ( \Xi_1 , J_2 \Xi_2 )$ when evaluating $\sW_\HH$ on $W = W_1 \oplus W_2$. This modification allows the superpotential to be expressed in a manifestly $\fusp(2)$-invariant way as

\begin{equation}\label{eq:superpotential4t}
  \sW_{\HH} \mid_{W_1 \oplus W_2} = \sW_{\HH} \mid_{W_1} + \sW_{\HH} \mid_{W_2} + \sW_{\text{\tiny{mixed}}},
\end{equation}

In section 3.1.2 of \cite{SCCS3Algs} it was shown that each of these terms can be written in an $\fso(4)$-invariant way and in fact one obtains precisely the Lagrangian found in \cite{pre3Lee}. The key step in this proof was to declare the matter fields $X_1$, $X_2$, $\Psi_1$ and $\Psi_2$ to transform in the following representations of $\fso(4)$:
\begin{equation}\label{eq:Vredtwistedreps}
 \left(X^\alpha \right)= \begin{pmatrix} X_1 \\ J_1 X_1 \end{pmatrix}, \quad \left( \Psi_{\dot \alpha} \right) = \begin{pmatrix} \Psi_1 \\ J_1 \Psi_1 \end{pmatrix}, \quad \left( {\tilde X}_{\dot \alpha} \right) = \begin{pmatrix} X_2 \\ J_2 X_2 \end{pmatrix}, \quad \left( {\tilde \Psi}^{\alpha} \right) = \begin{pmatrix} \Psi_2 \\ J_2 \Psi_2 \end{pmatrix},
\end{equation}
where $J_1 X^\alpha =  \varepsilon_{\alpha\beta} X^\beta$, $J_1 \Psi_{\dot \alpha} =  \varepsilon^{{\dot \alpha}{\dot \beta}} \Psi_{\dot \beta}$, $J_2 {\tilde X}_{\dot \alpha} =  \varepsilon^{{\dot \alpha}{\dot \beta}} {\tilde X}_{\dot \beta}$ and $J_2 {\tilde \Psi}^\alpha =  \varepsilon_{\alpha\beta} {\tilde \Psi}^{\beta}$ identically. The representations $({\bf 2},{\bf 1})$ and $({\bf 1},{\bf 2})$ for the bosonic and fermionic matter fields in $W_1$ is just as we would expect from \eqref{eq:nequal3quaterniondata2} if they are to comprise an $N=4$ hypermultiplet. The fact that we have defined the matter fields in $W_2$ to transform in the opposite representations of $\fso(4)$ follows from the fact that one requires the matter superfield in $W_2$ to be $J_2 \Xi_2$ rather than $\Xi_2$ in order to obtain the $\fusp(2)$-invariant superpotential \eqref{eq:superpotential4t}. This is still isomorphic to an $N=4$ hypermultiplet representation on $W_2$ though it is useful to distinguish between these two types of $\fso(4)$ representations and the latter is often referred to as a {\emph{twisted}} $N=4$ hypermultiplet in the literature. The $N=4$ supersymmetry transformations for a twisted hypermultiplet follow by acting with the quaternionic structure $J$ on the untwisted $N=4$ hypermultiplet transformations in \eqref{eq:susy4-onshell} and then absorbing the factor of $J$ into the definition of the twisted hypermultiplet matter fields ${\tilde X}$ and ${\tilde \Psi}$. In terms of the subsequent $\fso(4)$-covariant forms of the $N=4$ supersymmetry transformations, the corresponding prescription for going from an untwisted to a twisted hypermultiplet consists of switching all upstairs/downstairs and dotted/undotted indices followed by relabelling all the fields with tildes.

Given these fields definitions, the superpotential \eqref{eq:superpotential4t} was written in an $\fso(4)$-invariant way in section 3.1.2 of \cite{SCCS3Algs} and the resulting Lagrangian was shown to coincide with that in \cite{pre3Lee}, via the coupling of a twisted and untwisted hypermultiplet and which is indeed $N=4$ supersymmetric.

\subsubsection{Gauge Lie algebras for $N=4$ theories}

In summary, we have established that the superpotential ${\sf W}_\HH$ in \eqref{eq:superpotential3}, which generically guarantees $N=3$ supersymmetry when $\fM = W$ is quaternionic unitary, also describes the on-shell theories found in \cite{GaiottoWitten,pre3Lee} with enhanced $N=4$ superconformal symmetry when $W$ is the direct sum of one or more anti-Lie triple systems. It is worth stressing that the guiding principle that has led us to the $N=4$ theories in \cite{GaiottoWitten,pre3Lee} has simply been to look for special cases of quaternionic unitary representations for which the superpotential ${\sf W}_\HH$ can be written in a $\fusp(2)$-invariant way.

Let us conclude by detailing the possible gauge-theoretic structures on which such $N=4$ theories can be based. In the case of the $N=4$ Gaiotto-Witten theory, the fundamental ingredient describing indecomposable $N=4$ Gaiotto-Witten theories is an irreducible aLTS. By theorem~\ref{thm:aLTS-simplicity}, \textit{irreducible} anti-Lie triple systems $W$ are in one-to-one correspondence with metric complex \textit{simple} Lie superalgebras $\fg_\CC \oplus W$. These have been classified and thus one has an indecomposable $N=4$ Gaiotto-Witten theory for each of the classical complex simple Lie superalgebras whose odd component admits a quaternionic structure: namely, $A(m,n)$, $B(m,n)$, $C(n+1)$, $D(m,n)$, $F(4)$, $G(3)$ and $D(2,1;\alpha)$. The complexification of the corresponding Lie algebra is given by the even part of these superalgebras.

For the more general $N=4$ superconformal Chern-Simons-matter theories of \cite{pre3Lee}, let us begin by considering the simplest nontrivial setup where $W$ is the direct sum of \textsl{two} irreducible aLTS $W = W_1 \oplus W_2$ with $W_1,W_2 \in \Irr(\fg,\HH)_{\text{aLTS}}$. Theorem~\ref{thm:aLTS-simplicity} allows us to embed $W_1$ and $W_2$ into complex metric simple Lie superalgebras $G_1$ and $G_2$, respectively. Despite $W = W_1 \oplus W_2$ being a faithful representation of $\fg$, the irreducible aLTS constituents $W_1$ and $W_2$ need not be. Hence the real forms $\fg_1$ and $\fg_2$ of the even parts of $G_1$ and $G_2$ need not be isomorphic to $\fg$ itself (even though $\fg_1$ and $\fg_2$ do collectively span $\fg$). The special case of $W_1 \cong W_2$ where $G_1 \cong G_2$ and thus $\fg_1 \cong \fg_2 \cong \fg$ will be the topic of the next section where it will be shown to give rise to an enhanced $N=5$ superconformal symmetry. In order that the coupling terms between the untwisted and twisted hypermultiplets in the corresponding $N=4$ Lagrangian are non-vanishing, such that we obtain an indecomposable theory, it is necessary for the semisimple Lie algebras $\fg_1$ and $\fg_2$ to have at least one common simple factor. Since both $G_1$ and $G_2$ are simple, they must be one of the classical Lie superalgebras listed at the end of the last paragraph. Hence, from the regular simple Lie superalgebras, one may choose any of the pairs $( G_1 , G_2 )$ = $( A(m,p) , A(n,p) )$, $( B(m,p) , B(n,p) )$, $( B(p,m) , B(p,n) )$, $( B(m,n) , C(n+1) )$, $( B(m,p) , D(n,p) )$, $( C(m+1) , C(m+1) )$, $( D(m,n), C(n+1) )$, $( D(m,p) , D(n,p) )$, $( D(p,m) , D(p,n) )$ and in each case identify the simple Lie algebra factor they have in common. This technique, that we call \textit{domino}, can be continued to incorporate all the exceptional Lie superalgebras too, as well as the additional possibilities which follow from using the various low-dimensional Lie algebra isomorphisms. 

The generalisation of this construction when either of the aLTS constituents $W_1$ or $W_2$ is reducible is straightforward. For example, assume that the untwisted hypermultiplet matter is in an irreducible aLTS $W_1$ but the twisted hypermultiplet matter is taken to be in a reducible aLTS of the form $W_2 \oplus W_3$ where both $W_2$ and $W_3$ are irreducible aLTS representations. Associated with $W_1$, $W_2$ and $W_3$ we have three simple Lie superalgebras $G_1$, $G_2$ and $G_3$ and the construction above can be employed for the two pairs $( G_1 , G_2 )$ and $( G_1 , G_3 )$ such that the ordered triple $( G_3 , G_1 , G_2 )$ is constrained only by the requirement that adjacent simple Lie superalgebras must have identified at least one simple Lie algebra factor in their even components (e.g. $( G_3 , G_1 , G_2 )$ could be $( A(m,p) , A(p,q) , A(q,n) )$ or $( B(m,p) , D(q,p) , D(q,n) )$ to name but two of many possibilities). 

The most general situation can be described such that the faithful reducible representation takes the form $W = \bigoplus_{i=1}^n W_i$ in terms of $n$ irreducible aLTS representations $\{ W_i \mid i=1,...,n \}$ (with $n$ associated simple Lie superalgebras $\{ G_i \mid i=1,...,n \}$) where, for convenience, one can assume there is a relative twist between the hypermultiplet matter in adjacent $W_i$ and $W_{i+1}$ for $i=1,...,n-1$. Hence, one has an indecomposable $N=4$ superconformal Chern-Simons-matter theory for each ordered $n$-tuple of simple Lie superalgebras $( G_1 ,..., G_n )$ such that each adjacent pair have identified at least one simple Lie algebra in their even parts and where there is a relative twist between the hypermultiplet matter in adjacent pairs. Also, when $n$ is even, there is the possibility of identifying a simple Lie algebra factor in the even parts of $G_1$ and $G_n$ at the extremities. There are many possible ways of doing this but most of the generic ones, described first in \cite{pre3Lee}, involve chains of simple Lie superalgebras of the same classical type. To the best of our knowledge, these possibilities comprise all the known examples of $N=4$ superconformal Chern-Simons-matter theories.

\subsection{N=5 supersymmetry} 
\label{sec:nequal5}

We follow now the discussion at the end of section~\ref{sec:constr-from-supersym} and particularly in diagram \eqref{eq:4to5} to obtain enhanced $N=5$ superconformal symmetry from a theory with $N=4$ supersymmetry. We start from the most general $N=4$ theory (that is one where $W = W_1 \oplus W_2$ is the direct sum of two aLTS) but now taking $W_1, W_2 $ to be isomorphic. In other words, the matter content is now taken to transform under two copies of $W\in\Dar(\fg,\HH)_{\text{aLTS}}$. This prescription follows \cite{3Lee} where a new class of $N=5$ superconformal Chern-Simons-matter theories is constructed in this way, as a special case of the class of $N=4$ theories they had found previously in \cite{pre3Lee}.

Employing the notation of the previous section, we identify $W_1 \cong W_2 \cong W$ as quaternionic unitary representations. The untwisted and twisted hypermultiplet matter fields are $X = X_1$, $\Psi = \Psi_1$, ${\tilde X} = X_2$ and ${\tilde \Psi} = \Psi_2$, relative to their counterparts in the previous section, and each field here takes values in $W$. They can be assembled into a pair of $W$-valued $N=1$ matter superfields $\Xi$ and ${\tilde \Xi}$ in the usual way. It is important to stress that, when identifying $W_1 \cong W_2 \cong W$, one is no longer obliged to take $h ( W_1 , W_2 ) =0$ which followed from defining $W_1 \oplus W_2$ as an orthogonal direct sum with respect to the hermitian inner product. What is needed to describe the $N=5$ theory when identifying $W_1 \cong W_2 \cong W$ is to evaluate all the inner products involving the matter fields appearing in section \ref{sec:twisthyp} using the single hermitian inner product $h$ on $W$.

With respect to this structure, the expression in \eqref{eq:superpotential4t} for the superpotential $\sW_\HH$ reads
\begin{equation}
\label{eq:superpotential5}
   \begin{aligned}[m]
   \sW_{\HH} = \tfrac{1}{16} \int d^2 \theta \; &\left[ ( \TT ( \Xi  ,\Xi ) - \TT ( {\tilde \Xi}  , {\tilde \Xi} ) , \TT ( \Xi  ,\Xi ) - \TT ( {\tilde \Xi}  , {\tilde \Xi} ) ) \right. \\
   &\left. -2\, ( \TT ( \Xi  , {\tilde \Xi} ) , \TT ( \Xi  , {\tilde \Xi} ) ) -2\, ( \TT ( {\tilde \Xi}  , \Xi ) , \TT ( {\tilde \Xi}  , \Xi ) ) \right] ,
   \end{aligned}
\end{equation}
where $\TT$ is the Faulkner map associated with $W$. The second line contains the contribution from the F-term superpotential and has been simplified using the identity

\begin{equation}
( \TT (u,Ju) , \TT (Jv,v) ) = 2\, ( \TT (u,v) , \TT (u,v) ) ,
\end{equation}
for any $u,v \in W$, which follows using the aLTS cyclicity condition \eqref{eq:quat-aLTS} for $W$.

Whereas the superpotential $\sW_\HH$ in \eqref{eq:superpotential4t} for $W_1 \oplus W_2$ could only be given a $\fusp(2)$-invariant expression, we now show that the superpotential in \eqref{eq:superpotential5} is actually $\fso(4)$-invariant, as is necessary if the theory is to be $N=5$ supersymmetric. 

Since the enhanced $\fso(4)$-invariance of \eqref{eq:superpotential5} (relative to the manifestly $\fusp(2)$-invariant expression in \eqref{eq:superpotential4t}) is not manifest, it will be enlightening to see explicitly how this works. The trick is to not immediately try to write the hypermultiplet matter fields in terms of the representations of $\fso(4)$ that appeared in \eqref{eq:Vredtwistedreps} for the $N=4$ theory. Instead, one must first define the linear combinations $\Xi_\pm := \tfrac{1}{\sqrt{2}} \left( \Xi \pm {\tilde \Xi} \right)$ of the $W$-valued $N=1$ superfields above (such combinations do not exist on $W_1 \oplus W_2$). One then defines $\Xi_+^\alpha = ( \Xi_+ , J \Xi_+ )$ to transform in the $({\bf 2},{\bf 1})$ representation of $\fso(4)$ while $\Xi_{-\, {\dot \alpha}} = ( \Xi_-, J \Xi_- )$ is defined to transform in the opposite $({\bf 1},{\bf 2})$ representation. In terms of these combinations of $N=1$ matter superfields, the superpotential \eqref{eq:superpotential5} takes the manifestly $\fso(4)$-invariant form
\begin{equation}
\label{eq:superpotential5a}
   \begin{aligned}[m]
   \sW_{\HH} = \tfrac{1}{16} \int d^2 \theta \; &\left[ ( \TT ( \Xi_+^\alpha  , \Xi_{-\, {\dot \beta}} ) , \TT ( \Xi_{-\, {\dot \beta}}  , \Xi_+^\alpha ) ) -\tfrac{1}{6} \, ( \TT ( \Xi_+^\alpha  , \Xi_+^\beta ) , \TT ( \Xi_+^\beta  ,\Xi_+^\alpha ) ) \right. \\
   &\left. -\tfrac{1}{6} \, ( \TT ( \Xi_{-\, {\dot \alpha}}  , \Xi_{-\, {\dot \beta}} ) , \TT ( \Xi_{-\, {\dot \beta}}  , \Xi_{-\, {\dot \alpha}} ) ) \right] .
   \end{aligned}
\end{equation}

\subsubsection{On-shell $N=5$ Lagrangian}

Lets now consider the on-shell form \eqref{eq:susy-lag-on-shell} of the Lagrangian $\eL^{N=5} = \eL_{CS} + \eL_M + \sW_{\HH}$ based on the superpotential in \eqref{eq:superpotential5}. Not surprisingly, this gives precisely the $N=4$ Lagrangian in \cite{pre3Lee} after identifying $W_1 \cong W_2 \cong W$.

We define representations of $\fusp(4)$ via straightforward extension of the way we defined representations of $\fusp(2)$. That is, relative to a basis $\{ \be_A \}$ on $\CC^4$, we denote by $v^A$ the components of a complex vector $v$ transforming in the defining representation of $\fu(4)$ (while components of the complex conjugate vector $v^*$ have a downstairs index). With respect to this basis, we take the $\fsp(4,\CC )$-invariant complex symplectic form to be $\Omega = \be^1 \wedge \be^2 + \be^3 \wedge \be^4$. A vector $v \in W$ in the defining representation of $\fu(4)$ is in the fundamental representation ${\bf 4}$ of $\fusp(4) = \fu(4) \cap \fsp(4,\CC )$ if it obeys $J v^A = \Omega_{AB} v^B$. The embedding of $\fso(4)$ in $\fso(5)$ we consider corresponds to fixing a subalgebra $\fusp(2) \oplus \fusp(2) < \fusp(4)$ which defines a decomposition of $\CC^4 = \CC^2 \oplus \CC^2$ with the two $\fusp(2)$ factors in $\fso(4)$ acting on the respective $\CC^2$ components (for convenience we identify $\{ \be_\alpha \}$ with $\be_1$ and $\be_2$ and $\{ \be^{\dot \alpha} \}$ with $\be_3$ and $\be_4$). In terms of this embedding, we assemble the matter fields into
\begin{equation}\label{eq:Nequals5reps}
 \left( X^A \right) = \begin{pmatrix} X^\alpha \\ {\tilde X}_{\dot \alpha} \end{pmatrix}, \quad \left( \Psi^A \right) = \begin{pmatrix} {\tilde \Psi^\alpha} \\ \Psi_{\dot \alpha} \end{pmatrix},
\end{equation}
with both bosons and fermions transforming in the fundamental representation of $\fusp(4)$. The pseudo-reality conditions $J X^A = \Omega_{AB} X^B$ and $J \Psi^A = \Omega_{AB} \Psi^B$ are then identically satisfied.

In terms of the representations \eqref{eq:Nequals5reps}, the on-shell Lagrangian takes the manifestly $\fusp(4)$-invariant expression
\begin{multline}
\label{eq:susy-lag-csm5}
   \eL^{N=5} = -\varepsilon^{\mu\nu\rho} \left( A_\mu , \partial_\nu A_\rho + \tfrac{1}{3} [ A_\nu , A_\rho ] \right) -\tfrac{1}{4} \left< D_\mu X^A , D^\mu X^A \right> + \tfrac{1}{4} \left< {\bar \Psi}^A , \gamma^\mu D_\mu \Psi^A \right> \\
  \qquad + ( {\bar \nu}^A{}_B , \nu_A{}^B ) + 2\, ( {\bar \nu}^A{}_B , \nu^B{}_A ) +\tfrac{1}{15} \, ( [ \mu^A{}_B , \mu^B{}_C ] , \mu^C{}_A ) - \tfrac{3}{40} \left<  \mu^A{}_B \cdot X^C , \mu^A{}_B \cdot X^C \right>,
\end{multline}
where we have defined $\mu^A{}_B := \tfrac{1}{4} \, \TT ( X^A , X^B )$ and $\nu^A{}_B := \tfrac{1}{4} \TT ( X^A , \Psi^B )$.This is indeed precisely the Lagrangian for the $N=5$ superconformal Chern-Simons-matter theory found in \cite{3Lee} and its integral is invariant under the $N=5$ supersymmetry transformations
\begin{equation}
  \label{eq:susy5}
  \begin{aligned}[m]
    \delta X^A &={\bar \epsilon}^A{}_B \Psi^B \\
    \delta \Psi^A &= - \left[ \gamma^\mu D_\mu X^B + \tfrac{1}{3} \, \mu^B{}_C \cdot X^C \right] \epsilon_B{}^A + \tfrac{2}{3} \, \mu^A{}_B \cdot X^C \, \epsilon_C{}^B \\
    \delta A_\mu &= {\bar \epsilon}_A{}^B \gamma_\mu \nu^A{}_B,
  \end{aligned}
\end{equation}
where the complex $N=5$ supersymmetry parameter $\epsilon_{AB} := \Omega_{BC} \, \epsilon_A{}^C$ is skewsymmetric $\epsilon_{AB} = -\epsilon_{BA}$, symplectic traceless $\Omega^{AB} \epsilon_{AB} =0$ and obeys the reality condition $\epsilon_{AB} = \Omega_{AC} \Omega_{BD} \, \epsilon^{CD}$, hence describing five linearly independent Majorana spinors on $\RR^{1,2}$.

\subsubsection{Gauge Lie algebras for $N=5$ theories}

We conclude this section by summarising the consequences of the (ir)reducibility of $W$ for the $N=5$ theory. If $W \in \Dar(\fg,\HH)_{\text{aLTS}}$ is reducible, so that $W = W_1 \oplus W_2$, then as discussed in section 3.1.2 in \cite{SCCS3Algs} the $N=5$ Lagrangian in \eqref{eq:susy-lag-csm5} decouples into the sum of $N=5$ Lagrangians on the individual irreducible components. The potential mixed terms in the Lagrangian vanish identically as a consequence of the mixed components for the different irreducible factors in the aLTS cyclicity condition for $W$. 

It is therefore necessary to take $W \in \Irr(\fg,\HH)_{\text{aLTS}}$ in order to obtain an indecomposable theory. As stated in proposition~\ref{prop:irreducible}, if $W \in \Irr(\fg,\HH)_{\text{aLTS}}$, the underlying complex representation $\rh{W} \in \Dar(\fg,\CC)$ is irreducible unless $W = V_\HH$, so that $\rh{W} \cong V \oplus \Vbar$ for some $V \in \Irr(\fg,\CC)$. This case will be examined in the next section where it will be found to give rise to an enhanced $N=6$ superconformal symmetry. 

We deduce that the indecomposable Chern-Simons-matter theories with \emph{precisely} $N=5$ superconformal symmetry are in one-to-one correspondence with irreducible anti-Lie triple systems $W$ for which $\rh{W} \in \Irr(\fg,\CC)$, which according to theorem~\ref{thm:aLTS-simplicity}, when the inner product on $W$ is positive-definite, are in turn in one-to-one correspondence with complex simple Lie superalgebras $\fg_\CC \oplus W$. Such Lie superalgebras have been classified and are given by the classical simple Lie superalgebras $A(m,n)$, $B(m,n)$, $C(n+1)$, $D(m,n)$, $F(4)$, $G(3)$ and $D(2,1;\alpha )$. The complexification of the possible gauge Lie algebra is then given by the even part of each of these Lie superalgebras. Accordingly, this limits the Lie algebra to be of one of the types listed in table \eqref{tab:LieAlgN5}. Notice in particular that simplicity of the Lie superalgebra does not imply simplicity of the Lie algebra, and in fact none of the Lie algebras in the list are simple. This list exhausts the examples of $N=5$ superconformal Chern-Simons-matter theories that have been obtained already in \cite{3Lee,BHRSS}.

\begin{table}[ht!]
  \centering
  \begin{tabular}{||>{$}c<{$}|>{$}c<{$}|}
    \hline
    \fg & \text{Lie superalgebra} \\\hline
    \fsu(m+1) \oplus \fsu(n+1) \oplus \fu(1) &  A(m,n), m\neq n\\
    \fsu(n+1) \oplus \fsu(n+1) &  A(n,n) \\
    \fusp(2n) \oplus \fu(1) &  C(n+1) \\\hline
    \fso(2m+1) \oplus \fusp(2n) &  B(m,n) \\
    \fso(2m) \oplus \fusp(2n) &  D(m,n) \\
    \fsu(2) \oplus \fsu(2) \oplus \fsu(2) &  D(2,1;\alpha) \\
    \fsu(2) \oplus \fspin(7) &  F(4) \\
    \fsu(2) \oplus \fg_2 &  G(3) \\\hline
  \end{tabular}
  \vspace{8pt}
  \caption{Lie algebras for irreducible $N=5$ theories}
  \label{tab:LieAlgN5}
\end{table}

\subsection{N=6 supersymmetry} 
\label{sec:nequal6}

Following again the prescription in \cite{3Lee}, and as discussed in section \ref{sec:constr-from-supersym}, one can enhance the supersymmetry of an $N=5$ theory to $N=6$ by considering matter representations $W \in \Irr(\fg,\HH)_{\text{aLTS}}$ which are quaternionifications $W = V_\HH$ of some $V \in \Irr(\fg,\CC)$, so that $\rh{W} \cong V \oplus \Vbar$. As shown in proposition~\ref{prop:relations}(9), this means that in fact $V$ is an irreducible anti-Jordan triple system, $V \in \Irr(\fg,\CC)_{\text{aJTS}}$. We recall that this is one of the extreme cases of the complex 3-Leibniz algebras (or equivalently of complex unitary representations of a Lie algebra) which was first used in this context in \cite{BL4}.

We investigate now the structure of the superpotential $\sW_\HH$ in \eqref{eq:superpotential5} for $W = V \oplus \Vbar$\footnote{The quaternionic unitary structure associated with this representation is defined in the proof of proposition~\ref{prop:relations}(9) in section \ref{sec:some-relat-betw}.}. We write the constituent $N=1$ matter superfields as $\Xi = ( \Xi^1 , {\bar \Xi^2} )$ and ${\tilde \Xi} = ( \Xi^3 , {\bar \Xi^4} )$ in terms of four $N=1$ superfields $\Xi^1 , \Xi^2 , \Xi^3 , \Xi^4 \in V$. In terms of these superfields, the superpotential \eqref{eq:superpotential5} becomes
\begin{equation}
\label{eq:superpotential6}
   \begin{aligned}[m]
   \sW_{\HH} = - \tfrac{1}{8} \int d^2 \theta \; &\left[ ( \TT ( \Xi^1 , \Xi^1 ) ,  \TT ( \Xi^2 , \Xi^2 ) ) + ( \TT ( \Xi^1 , \Xi^1 ) ,  \TT ( \Xi^3 , \Xi^3 ) )  \right. \\
   &\left. - ( \TT ( \Xi^1 , \Xi^1 ) ,  \TT ( \Xi^4 , \Xi^4 ) )-( \TT ( \Xi^2 , \Xi^2 ) ,  \TT ( \Xi^3 , \Xi^3 ) )  \right. \\
	   &\left. + ( \TT ( \Xi^2 , \Xi^2 ) ,  \TT ( \Xi^4 , \Xi^4 ) ) + ( \TT ( \Xi^3 , \Xi^3 ) ,  \TT ( \Xi^4 , \Xi^4 ) )  \right. \\
   &\left.+2\, ( \TT ( \Xi^1  , \Xi^2 ) , \TT ( \Xi^4  , \Xi^3 ) ) +2\, ( \TT ( \Xi^2  , \Xi^1 ) , \TT ( \Xi^3  , \Xi^4 ) ) \right] ,
   \end{aligned}
\end{equation}
where $\TT$ is the complex Faulkner map associated with $V$ and the aJTS skewsymmetry condition $\TT ( x , y ) \cdot z =- \TT ( z , y ) \cdot x$ has been used.

The fourth line in \eqref{eq:superpotential6} represents the contribution from the F-term superpotential which was shown to admit a manifestly $\fso(4)$-invariant expression for the ABJM $N=6$ theory in \cite{MaldacenaBL}. To demonstrate how this works, and mimicking the nomenclature in \cite{MaldacenaBL}, let us collect the four $V$-valued $N=1$ superfields into the two pairs $A = ( \Xi^1 , \Xi^4 )$ and $B = ( \Xi^2 , -\Xi^3 )$. It is convenient to take the pairs $A$ and $B$ to transform separately in the fundamental representation of two different copies of $\fsp(2, \CC )$. With respect to the orthonormal bases $\{ \be_\alpha \}$ and $\{ \be_{\dot \alpha} \}$ associated with these two fundamental representations, we take the respective $\fsp(2,\CC)$-invariant symplectic forms to be $\varepsilon = \be^1 \wedge \be^2$ and ${\tilde \varepsilon} = \be^{\dot 1} \wedge \be^{\dot 2}$. In terms of this structure, the forth line in \eqref{eq:superpotential6} can be written as
\begin{equation}
\label{eq:superpotential6fterm}
\tfrac{1}{8} \int d^2 \theta \; \varepsilon_{\alpha\beta} \, {\tilde \varepsilon}^{{\dot \alpha}{\dot \beta}} \, \Re \, ( \TT ( A^\alpha  , B^{\dot \alpha} ) , \TT ( A^\beta  , B^{\dot \beta} ) ).
\end{equation}
which is manifestly invariant under $\fsp(2,\CC ) \oplus \fsp(2,\CC )$. The addition of the kinetic terms for the matter fields to this F-term superpotential however breaks each $\fsp(2, \CC )$ down to $\fusp(2)$ (since both $h( A^\alpha , A^\alpha )$ and $h( B^{\dot \alpha} , B^{\dot \alpha})$ must also be invariant). Hence the resulting symmetry realised by \eqref{eq:superpotential6fterm} is indeed $\fusp(2) \oplus \fusp(2) \cong \fso(4)$.

However, for an $N=6$ theory written in terms of $N=1$ superfields, we expect that the full superpotential should be invariant under the $\fso(5)$ isotropy subalgebra of the $\fso(6)$ R-symmetry preserving our choice of $N=1$ superspace parameter. This is indeed the case and follows by assembling the superfields into the array $( \Xi^1 , \Xi^2 , \Xi^3 , \Xi^4 )$, which is to be thought of as a $V$-valued element of $\CC^4$ whose components we denote by $\Xi^A$ with respect to a basis $\{ \be_A \}$ on $\CC^4$. This $\CC^4$ is to be equipped with an action of $\fsp(4,\CC)$ such that $\Xi^A$ transforms in the fundamental representation. We define the $\fsp(4,\CC)$ subalgebra as those complex linear transformations which preserve the complex symplectic form $\Omega = \be^1 \wedge \be^3 + \be^2 \wedge \be^4$. The superpotential \eqref{eq:superpotential6} then takes the explicitly $\fsp(4,\CC)$-invariant form
\begin{equation}
\label{eq:superpotential6a}
\sW_{\HH} =  -\tfrac{1}{16} \int d^2 \theta \; \left( \delta_A^C \delta_B^D -  \Omega_{AB} \Omega^{CD} \right) ( \TT ( \Xi^A , \Xi^C ) ,  \TT ( \Xi^B , \Xi^D ) ).
\end{equation}

Again, the addition of the kinetic terms for the matter fields breaks this $\fsp(4, \CC )$ symmetry down to the expected $\fusp(4) \cong \fso(5)$ (since $h ( \Xi^A , \Xi^A )$ must also be invariant).

\subsubsection{On-shell $N=6$ Lagrangian}

Let us now consider the on-shell form \eqref{eq:susy-lag-on-shell} of the $N=6$ Lagrangian $\eL^{N=6} = \eL_{CS} + \eL_M + \sW_{\HH}$ based on the superpotential \eqref{eq:superpotential6}. To do this, one has to re-write \eqref{eq:susy-lag-csm5} for the special case of $W = V \oplus \Vbar$ in a form which is explicitly invariant under the $\fso(6) \cong \fsu(4)$ R-symmetry of the $N=6$ superconformal algebra. To this end, we begin by writing the untwisted and twisted hypermultiplet matter fields from the original $N=4$ theory as $X = ( {\bf X}^1 , {\overline{{\bf X}^2}} )$, ${\tilde X} = ( {\bf X}^3 , {\overline{{\bf X}^4}} )$, $\Psi = ( {\bf \Psi}_4 , - {\overline{{\bf \Psi}_3}} )$ and ${\tilde \Psi} = ( {\bf \Psi}_2 , - {\overline{{\bf \Psi}_1}} )$ on $W = V \oplus \Vbar$ in terms of the four $V$-valued bosons $( {\bf X}^1 , {\bf X}^2 , {\bf X}^3 , {\bf X}^4 )$ and fermions $( {\bf \Psi}_1 , {\bf \Psi}_2 , {\bf \Psi}_3 , {\bf \Psi}_4 )$ whose components we will denote by ${\bf X}^A$ and ${\bf \Psi}_A$ respectively. This parametrisation is convenient because it allows one to assemble the components into the $\fusp(4)$-covariant expressions
\begin{equation}\label{eq:Nequals6reps}
 \left( X^A \right) = ( {\bf X}^A , {\overline{\Omega_{AB} {\bf X}^B}} ), \quad \left( \Psi^A \right) = ( \Omega^{AB} \, {\bf \Psi}_B , - {\overline{{\bf \Psi}_A}} ),
\end{equation}
with the components on the left hand sides being defined just as in \eqref{eq:Nequals5reps}. The pseudo-reality conditions $J X^A = \Omega_{AB} X^B$ and $J \Psi^A = \Omega_{AB} \Psi^B$ are identically satisfied, with no constraint on ${\bf X}^A$ and ${\bf \Psi}_A$, from the definition of $J$ acting on $V \oplus \Vbar$. The expressions in \eqref{eq:Nequals6reps} can be understood as describing the canonical embedding of the fundamental representation of $\fusp(4)$ into the real form of the fundamental representation of $\fsu(4)$, where ${\bf X}^A$ and ${\bf \Psi}_A$ respectively transform in the complex representations corresponding to the fundamental ${\bf 4}$ and antifundamental ${\bf {\bar 4}}$ of $\fsu(4)$.

In terms of these representations, the on-shell Lagrangian \eqref{eq:susy-lag-csm5} can be given the following $\fsu(4)$-invariant expression
\begin{equation}
\label{eq:susy-lag-csm6}
  \begin{aligned}[m]
   \eL^{N=6} =& -\varepsilon^{\mu\nu\rho} \left( A_\mu , \partial_\nu A_\rho + \tfrac{1}{3} [ A_\nu , A_\rho ] \right) -\tfrac{1}{2} \left< D_\mu {\bf X}^A , D^\mu {\bf X}^A \right> + \tfrac{1}{2} \left< {\bar {\bf \Psi}}_A , \gamma^\mu D_\mu {\bf \Psi}_A \right> \\
   &+ 2\, ( {\bar {\boldsymbol{\nu}}}^{AB} , {\boldsymbol{\nu}}_{AB} - 2\, {\boldsymbol{\nu}}_{BA} ) + \varepsilon_{ABCD} ( {\bar {\boldsymbol{\nu}}}^{AC} , {\boldsymbol{\nu}}^{BD} ) + \varepsilon^{ABCD} ( {\bar {\boldsymbol{\nu}}}_{AC} , {\boldsymbol{\nu}}_{BD} ) \\
   &+\tfrac{2}{3}\, ( [ {\boldsymbol{\mu}}^A{}_B , {\boldsymbol{\mu}}^B{}_C ] , {\boldsymbol{\mu}}^C{}_A ) - \tfrac{1}{2} \left< {\boldsymbol{\mu}}^A{}_B \cdot {\bf X}^C , {\boldsymbol{\mu}}^A{}_B \cdot {\bf X}^C \right>,
  \end{aligned}
\end{equation}
where we have defined the moment maps ${\boldsymbol{\mu}}^A{}_B := \tfrac{1}{4} \, \TT ( {\bf X}^A , {\bf X}^B )$ and ${\boldsymbol{\nu}}^{AB} := \tfrac{1}{4} \TT ( {\bf X}^A , {\bf \Psi}_B )$ and $\varepsilon = \be^1 \wedge \be^2 \wedge \be^3 \wedge \be^4$ is the $\fsu(4)$-invariant 4-form with respect an orthonormal basis $\{ \be^A \}$ on the dual of $\CC^4$. Under the embedding of $\fusp(4)$ in $\fsu(4)$ described above, the moment maps here are related to their $N=5$ counterparts defined below \eqref{eq:susy-lag-csm5} such that $\mu^A{}_B = {\boldsymbol{\mu}}^A{}_B - \Omega^{AC} \Omega_{BD} {\boldsymbol{\mu}}^D{}_C$ and $\nu^A{}_B = \Omega_{BC} {\boldsymbol{\nu}}^{AC} - \Omega^{AC} {\boldsymbol{\nu}}_{CB}$ (where ${\boldsymbol{\nu}}_{AB} = - \tfrac{1}{4} \TT ( {\bf \Psi}_B , {\bf X}^A )$ is the complex conjugate of ${\boldsymbol{\nu}}^{AB}$). The components of the $\fsu(4)$-invariant 4-form can be expressed as $\varepsilon_{ABCD} = \Omega_{AB} \Omega_{CD} + \Omega_{AC} \Omega_{DB} + \Omega_{AD} \Omega_{BC}$ in terms of the $\fusp(4)$-invariant symplectic form under the embedding.

Notice that the $N=6$ Lagrangian above has a global $\fu(1)$ symmetry under which the gauge field is uncharged while the bosonic and fermionic matter fields ${\bf X}^A$ and ${\bf \Psi}_A$ both have the same charge. It is to be distinguished from the $\fu(1) < \fsu(4)$ R-symmetry subalgebra that is realised in the description of this theory in terms of $N=2$ superfields under which the bosons and fermions have opposite charges $\half$ and $-\half$. Indeed, this global $\fu(1)$ is a flavour symmetry since it commutes with the $N=6$ superconformal algebra. 

The Lagrangian \eqref{eq:susy-lag-csm6} describes precisely the $N=6$ theory in \cite{MaldacenaBL,BL4,3Lee} and its integral is invariant under the $N=6$ supersymmetry transformations
\begin{equation}
  \label{eq:susy6}
  \begin{aligned}[m]
    \delta {\bf X}^A &={\bar \epsilon}^{AB} {\bf \Psi}_B \\
    \delta {\bf \Psi}_A &= - \left[ \gamma^\mu D_\mu {\bf X}^B + {\boldsymbol{\mu}}^B{}_C \cdot {\bf X}^C \right] \epsilon_{AB} - {\boldsymbol{\mu}}^B{}_A \cdot {\bf X}^C \, \epsilon_{BC} \\
    \delta A_\mu &= - {\bar \epsilon}_{AB} \gamma_\mu {\boldsymbol{\nu}}^{AB} - {\bar \epsilon}^{AB} \gamma_\mu {\boldsymbol{\nu}}_{AB},
  \end{aligned}
\end{equation}
where the complex $N=6$ supersymmetry parameter $\epsilon_{AB}$ here is skewsymmetric $\epsilon_{AB} = -\epsilon_{BA}$ and obeys the reality condition $\epsilon_{AB} = \half \varepsilon_{ABCD} \, \epsilon^{CD}$, hence describing six linearly independent Majorana spinors on $\RR^{1,2}$. The $N=5$ supersymmetry transformations in \eqref{eq:susy5} are recovered following the embedding of $\fusp(4)$ in $\fsu(4)$ described above and then imposing the symplectic tracelessness condition on $\epsilon_{AB}$.

\subsubsection{Gauge Lie algebras for $N=6$ theories}

Given the construction of the $N=6$ theory as an enhanced $N=5$ theory, indecomposability of the $N=6$ theory follows from that of the $N=5$ theory, which required $W \in \Irr(\fg,\HH)_{\text{aLTS}}$. Further enhancement requires $W = V_\HH$ and hence $V \in \Irr(\fg,\CC)_{\text{aJTS}}$. As stated in proposition~\ref{prop:irreducible}, the underlying real representation $\rf{V} \in \Dar(\fg,\RR)$ is still irreducible unless $V = U_\CC$ is the complexification of $U \in \Irr(\fg,\RR)$. As we will see in the next section, in this case  supersymmetry is enhanced to $N=8$. 

We conclude that the Chern-Simons-matter theories with \emph{precisely} $N=6$ superconformal symmetry are in one-to-one correspondence with irreducible anti-Jordan triple systems which are not the complexification of a real representation. According to theorem \ref{thm:BL4-simplicity}, when the inner product on $V$ is positive-definite, such irreducible aJTS are in turn in one-to-one correspondence with complex simple 3-graded Lie superalgebras $V \oplus \fg_\CC \oplus \Vbar$. These have been classified \cite[theorem~4]{KacSuperSketch} and are given by the two classical simple Lie superalgebras $A(m,n)$ and $C(n+1)$. Again, this determines the possible gauge Lie algebras to be used for $N=6$ theories to be of the types listed in table \ref{tab:LieAlgN6}. This conclusion is in accordance with the earlier classification of $N=6$ superconformal Chern-Simons-matter theories in \cite{SchnablTachikawa} and remarks in \cite{GaiottoWitten,3Lee}.

\begin{table}[ht!]
  \centering
  \begin{tabular}{||>{$}c<{$}|>{$}c<{$}|}
    \hline
    \fg & \text{Lie superalgebra} \\\hline
    \fsu(m+1) \oplus \fsu(n+1) \oplus \fu(1) &  A(m,n), m\neq n\\
    \fsu(n+1) \oplus \fsu(n+1) &  A(n,n) \\
    \fusp(2n) \oplus \fu(1) &  C(n+1) \\\hline
  \end{tabular}
  \vspace{8pt}
  \caption{Lie algebras for irreducible $N=6$ theories}
  \label{tab:LieAlgN6}
\end{table}

\subsection{N=8 supersymmetry} 
\label{sec:nequal8}

We describe now how to obtain the theory that was first discovered by Bagger and Lambert \cite{BL1,BL2} and Gustavsson \cite{GustavssonAlgM2} with maximal $N=8$ superconformal symmetry as a special case of the $N=6$ theory encountered in the previous section. 

From the discussion in section \ref{sec:matter-reps}, we know that enhancement to $N=7$ occurs when the representation $V \in \Irr(\fg,\CC)_{\text{aJTS}}$ on which the $N=6$ theory is based is the complexification of a real representation $V = U_\CC$. As shown in proposition~\ref{prop:relations}(6), $V$ being an irreducible\footnote{Notice that $V$ is irreducible as a complex representation but reducible as a real representation.} aJTS means that in fact $U$ is not just any real representation, but an irreducible 3-Lie algebra. The theory obtained by this procedure has initially a manifest global $\fso(7)$ symmetry and is invariant under an $N=7$ superconformal algebra. However, we will see that the Lagrangian for this theory can be rewritten in a manifestly $\fso(8)$-invariant way recovering the maximal $N=8$ superconformal Chern-Simons-matter theory in \cite{BL1,BL2,GustavssonAlgM2}. This proves that for 3-dimensional superconformal Chern-Simons-matter theories, $N=7$ supersymmetry implies $N=8$.

We start by investigating the structure of the superpotential $\sW_\HH$ in \eqref{eq:superpotential6} for the special case when the complex representation is the complexification of a real one $V = U_\CC$. The four $V$-valued $N=1$ matter superfields can be written as $\Xi^1 = \xi^1 + i {\hat \xi}^1$, $\Xi^2 = \xi^2 + i {\hat \xi}^2$, $\Xi^3 = \xi^3 + i {\hat \xi}^3$ and $\Xi^4 = \xi^4 + i {\hat \xi}^4$ in terms of eight $U$-valued $N=1$ superfields $\xi^1 , \xi^2 , \xi^3 , \xi^4 , {\hat \xi}^1 , {\hat \xi}^2 , {\hat \xi}^3 , {\hat \xi}^4$. The superpotential \eqref{eq:superpotential6} then becomes
\begin{equation}
\label{eq:superpotential8}
   \begin{aligned}[m]
   \sW_{\HH} = \tfrac{1}{2} \int &d^2 \theta \; \left[ ( T( \xi^1 , {\hat \xi}^1 ) ,  T( \xi^2 , {\hat \xi}^2 ) ) + ( T( \xi^1 , {\hat \xi}^1 ) ,  T( \xi^3 , {\hat \xi}^3 ) ) - ( T( \xi^2 , {\hat \xi}^2 ) ,  T( \xi^3 , {\hat \xi}^3 ) ) \right. \\
   &\left. + ( T( \xi^2 , {\hat \xi}^2 ) ,  T( \xi^4 , {\hat \xi}^4 ) ) + ( T( \xi^3 , {\hat \xi}^3 ) ,  T( \xi^4 , {\hat \xi}^4 ) ) - ( T( \xi^1 , {\hat \xi}^1 ) ,  T( \xi^4 , {\hat \xi}^4 ) ) \right. \\
   &\left. + ( T( \xi^1 , \xi^2 ) ,  T( {\hat \xi}^3 , {\hat \xi}^4 ) ) + ( T( \xi^3 , \xi^4 ) ,  T( {\hat \xi}^1 , {\hat \xi}^2 ) ) - ( T( \xi^1 , \xi^4 ) ,  T( {\hat \xi}^2 , {\hat \xi}^3 ) ) \right. \\
   &\left. - ( T( \xi^1 , \xi^3 ) ,  T( {\hat \xi}^2 , {\hat \xi}^4 ) ) - ( T( \xi^2 , \xi^4 ) ,  T( {\hat \xi}^1 , {\hat \xi}^3 ) ) - ( T( \xi^2 , \xi^3 ) ,  T( {\hat \xi}^1 , {\hat \xi}^4 ) ) \right. \\
   &\left. +( T( \xi^1 , \xi^2 ) ,  T( \xi^3 , \xi^4 ) ) +( T( {\hat \xi}^1 , {\hat \xi}^2 ) ,  T( {\hat \xi}^3 , {\hat \xi}^4 )) \right] ,
   \end{aligned}
\end{equation}
where $T$ is the real Faulkner map associated with $U$ and the total skewsymmetry of $T(x,y)\cdot z$ has been used. 

For an $N=8$ superconformal Chern-Simons-matter theory written in terms of $N=1$ superfields we expect that the superpotential should be invariant under the $\fso(7)$ subalgebra of the $\fso(8)$ R-symmetry which preserves the $N=1$ superspace parameter. This is indeed the case and follows by assembling the eight real constituent $N=1$ superfields into the array $( \xi^3 , \xi^4 , \xi^1 , \xi^2 , {\hat \xi}^1 , {\hat \xi}^2 , {\hat \xi}^3 , {\hat \xi}^4 )$ which is to be thought of as a $U$-valued element of $\RR^8$. Let us denote its components by $\Xi^I$ with respect to a basis $\{ e_I \}$ on $\RR^8$. This $\RR^8$ is to be equipped with the action of $\fso(7)$ that defines the real spinor representation in seven dimensions. A natural quartic tensor on $\RR^8$ that is preserved by this action of $\fso(7)$ is the Cayley 4-form $\Omega$. If we take $\{ e^I \}$ to define an orthonormal basis on the dual of $\RR^8$ then the Cayley form can be taken to be
\begin{equation}
\label{eq:cayleyform}
\begin{aligned}
\Omega &= e^{1234} +  ( e^{12} - e^{34} ) \wedge ( e^{56} - e^{78} ) +  ( e^{13} + e^{24} ) \wedge ( e^{57} + e^{68} ) \\
&+  ( e^{14} - e^{23} ) \wedge ( e^{58} - e^{67} ) + e^{5678},
   \end{aligned}
\end{equation}
where multiple indices denote wedge products of the corresponding basis elements. In terms of the Cayley form, the superpotential in \eqref{eq:superpotential8} can be written in the manifestly $\fso(7)$-invariant form
\begin{equation}
\label{eq:superpotential8a}
\sW_{\HH} =  \tfrac{1}{48} \int d^2 \theta \; \Omega_{IJKL} (T( \Xi^I , \Xi^J ),T( \Xi^K , \Xi^L ) ).
\end{equation}

This $\fso(7)$-invariant expression for the superpotential that gives rise to the $N=8$ Bagger-Lambert Lagrangian has appeared already in \cite{MauriPetkouBL}. What we have shown above is that it is precisely this superpotential which follows from evaluating $\sW_\HH$ in \eqref{eq:superpotential6} for the special case of $V = U_\CC$.

\subsubsection{On-shell $N=8$ Lagrangian}

We examine now the on-shell form \eqref{eq:susy-lag-on-shell} of the Lagrangian $\eL^{N=8} = \eL_{CS} + \eL_M + \sW_{\HH}$ based on the superpotential \eqref{eq:superpotential8}. There are two ways to evaluate this: one considering it as a special case of the $N=1$ Lagrangian \eqref{eq:susy-lag-on-shell} and the other as a special case of the $N=6$ Lagrangian \eqref{eq:susy-lag-csm6} (which in turn has been constructed as a special case of \eqref{eq:susy-lag-on-shell}). As consistency dictates, both methods give the same answer.

Although the latter is the method that we have followed for the $N=5$ and $N=6$ cases, we find that the first method is the simplest. We integrate out the auxiliary fields in the generic $N=1$ Chern-Simons-matter Lagrangian with the form of the superpotential in \eqref{eq:superpotential8a} based on an $N=1$ matter superfield $\Xi^I$ valued in $\RR^8 \otimes U$. The terms in the matter part of the Lagrangian are evaluated with respect to the natural tensor product inner product involving the unit inner product on $\RR^8$, with components $\delta_{IJ}$, and the inner product $\left< -,- \right>$ on $U$. Doing this one obtains

\begin{equation}
\label{eq:susy-lag-csm8}
  \begin{aligned}[m]
   \eL^{N=8} =& -\varepsilon^{\mu\nu\rho} \left( A_\mu , \partial_\nu A_\rho + \tfrac{1}{3} [ A_\nu , A_\rho ] \right) -\tfrac{1}{2} \left< D_\mu X^I , D^\mu X^I \right> + \tfrac{1}{2} \left< {\bar \Psi}^I , \gamma^\mu D_\mu \Psi^I \right> \\
   &+ \tfrac{1}{8}  \left( 2\, \delta_{IK} \delta_{JL} + \Omega_{IJKL} \right) ( {T ( X^I , \Psi^K )} , T ( X^J , \Psi^L ) ) \\
	&-\tfrac{1}{48} \left< T ( X^I , X^J ) \cdot X^K , T ( X^I , X^J ) \cdot X^K \right>,
  \end{aligned}
\end{equation}
where $X^I$ and $\Psi^I$ are just the bosonic and fermionic components of the $N=1$ superfield $\Xi^I$ and we have used  the identity
\begin{equation}
\label{eq:cayleyid}
\Omega_{IJKL} \Omega^{MNPL} = 6\, \delta_{[I}^M \delta_J^N \delta_{K]}^P - 9\, \delta_{[I}^{[M} \Omega_{JK]}{}^{NP]},
\end{equation}
for the components of the Cayley form to be employed and the contribution of the second term on the right hand side above to the scalar potential can be shown to vanish identically as a consequence of the fundamental identity \eqref{def:FI} for the 3-Lie bracket associated with $T$ on $U$. This form of the Lagrangian is only manifestly $\fso(7)$-invariant but, as we will see, it can be written in an $\fso(8)$-invariant way. 

The second method to obtain \eqref{eq:susy-lag-csm8} is to evaluate the on-shell $N=6$ Lagrangian \eqref{eq:susy-lag-csm6} for $V = U_\CC$. Then, in order that the kinetic terms for the matter fields in the Lagrangians \eqref{eq:susy-lag-csm6} and \eqref{eq:susy-lag-csm8} agree, the identification of $\CC^4$ with $\RR^8$ must be isometric, corresponding to the canonical embedding of $\fsu(4)$ in $\fso(8)$. A convenient way to achieve this for the bosonic fields is to identify the first and last four of the eight real scalars $X^I$ respectively with the real and imaginary parts of the four complex scalars ${\bf X}^A$ in the $N=6$ theory. For the fermionic fields one identifies the first and last four fermions $\Psi^I$ respectively with the imaginary and and real parts of the four complex fermions ${\bf \Psi}_A$ in the $N=6$ theory. This distinction between the way the bosons and fermions are identified is necessary in order to then rewrite the Yukawa couplings for the $N=6$ theory in an $\fso(7)$-invariant way. 

To derive the expression for the Yukawa couplings in the second line of \eqref{eq:susy-lag-csm8} from \eqref{eq:susy-lag-csm6}, one needs the identity $\Omega = \Re\, \varepsilon - \half k \wedge k$ for the Cayley form \eqref{eq:cayleyform} on $\RR^8$ in terms of (the real part of) the holomorphic 4-form $\varepsilon = \be^1 \wedge \be^2 \wedge \be^3 \wedge \be^4$ and the Kähler form $k = \tfrac{i}{2} \be^A \wedge \be_A$ on $\CC^4$ (the latter being a real 2-form and can be expressed as $e^{15} + e^{26} + e^{37} + e^{48}$ on $\RR^8$ under the identification mentioned above). Finally, to derive the scalar potential in \eqref{eq:susy-lag-csm8}, it was useful to first note that the $N=6$ scalar potential in the third line in \eqref{eq:susy-lag-csm6} can be rewritten as $-\tfrac{1}{3} \left( \left< {\boldsymbol{\mu}}^A{}_B \cdot {\bf X}^C , {\boldsymbol{\mu}}^A{}_B \cdot {\bf X}^C \right> - \half \left< {\boldsymbol{\mu}}^A{}_B \cdot {\bf X}^B , {\boldsymbol{\mu}}^A{}_C \cdot {\bf X}^C \right> \right)$.

The integral of the Lagrangian in \eqref{eq:susy-lag-csm8} is invariant under the $N=7$ supersymmetry transformations
\begin{equation}
  \label{eq:susy7}
  \begin{aligned}[m]
    \delta X^I &={\bar \epsilon}^{IJ} \Psi^J \\
    \delta \Psi^I &= - \gamma^\mu D_\mu X^J \, \epsilon^{IJ} + \tfrac{2}{3} \, \mu^{IJ} \cdot X^K \, \epsilon^{JK} + \tfrac{1}{3} \, \Omega^{IJKL} \, \mu^{JK} \cdot X^M \, \epsilon^{LM} \\
    \delta A_\mu &= - 2\, {\bar \epsilon}^{IJ} \gamma_\mu \nu^{IJ},
  \end{aligned}
\end{equation}
where the real $N=7$ supersymmetry parameter $\epsilon_{IJ}$ is skewsymmetric $\epsilon_{IJ} = -\epsilon_{JI}$ and obeys $\epsilon_{IJ} = - \tfrac{1}{6} \Omega_{IJKL} \, \epsilon_{KL}$ which defines the projection onto the seven-dimensional $\fso(7)$-invariant subspace of $\Lambda^2 \RR^8$. Hence, it describes seven linearly independent Majorana spinors on $\RR^{1,2}$. 

Notice that the $N=6$ supersymmetry transformations in \eqref{eq:susy6} can be recovered on $V = U_\CC$ under the identification of $\RR^8$ with $\CC^4$ described above after imposing the the extra condition $k^{IJ} \epsilon_{IJ} =0$ using the Kähler form to eliminate the seventh supersymmetry parameter.

\subsubsection{Gauge Lie algebras for $N=8$ theories}

Before moving on to writing this Lagrangian in a manifestly $\fso(8)$-invariant way, lets investigate the allowed gauge Lie algebras for $N=8$ theories. As it was already discussed in section \ref{sec:simplicity}, there is a unique positive-definite irreducible 3-Lie algebra, which is the simple one $S_4$. In example \ref{eg:A4} it was shown that the corresponding Faulkner Lie algebra is $\fso(4) \cong \fsu(2) \oplus \fsu(2)$ equipped with a split signature inner product.

\subsubsection{$\fso(8)$ R-symmetry}
\label{sec:so8rsymmetry}

To rewrite the Lagrangian \eqref{eq:susy-lag-csm8} in a manifestly $\fso(8)$-invariant way requires a choice of embedding for $\fso(7)$ in $\fso(8)$. There are three real eight-dimensional irreducible representations of $\fso(8)$: the vector representation $\boldsymbol{8}_v$ and the positive and negative chirality spinor representations $\boldsymbol{8}_s$ and $\boldsymbol{8}_c$. Hence, there are three distinct embeddings of $\fso(7)$ in $\fso(8)$ that can be understood as the subalgebras preserving a fixed nonzero element in either $\boldsymbol{8}_v$, $\boldsymbol{8}_s$ or $\boldsymbol{8}_c$. These are distinct only up to the triality symmetry which relates the three representations. In each case, whichever of the three representations the fixed element resides in breaks into the vector and singlet representations ${\bf 7} \oplus {\bf 1}$ of $\fso(7)$ while the remaining two representations both reduce to the spinor representation $\boldsymbol{8}$. 

Both the bosonic and fermionic matter fields $X^I$ and $\Psi^I$ above transform in the $\boldsymbol{8}$ while the supersymmetry parameter $\epsilon_{IJ}$ transforms in the ${\bf 7}$ of $\fso(7)$. We are therefore free to choose any of the three embeddings as long as it is the supersymmetry parameter which embeds into whichever of the three representations of $\fso(8)$ that contains a singlet under the $\fso(7)$ subalgebra. As in \cite{Nahm}, it has been convenient for us so far to assume that the supercharges in an $N$-extended superconformal algebra transform in the vector representation of the $\fso(N)$ R-symmetry. In the case at hand of $N=8$ this would suggest we lift the supercharges into the $\boldsymbol{8}_v$ of $\fso(8)$ with the matter fields lifting to the $\boldsymbol{8}_s$ and $\boldsymbol{8}_c$. However, it will prove more convenient here to lift the supercharges into the $\boldsymbol{8}_s$ of $\fso(8)$ with the matter fields $X^I$ and $\Psi^I$ lifting respectively to the $\boldsymbol{8}_v$ and $\boldsymbol{8}_c$ representations. The main reason for this is that it allows to rewrite things more neatly in terms of elements of the Clifford algebra in eight dimensions and hence recover the well-known form of the $N=8$ Lagrangian originally presented in \cite{BL2}. One can always apply triality to obtain whichever of the three representations of $\fso(8)$ one wants for the supercharges (for example, this has been done explicitly in \cite{Gustavsson:2009pm} in the context of so-called \lq trial BLG' theories).

We adopt the conventions of \cite{AFOS} for the Clifford algebra $\Cl(8)$. An $\fso(8)$-covariant basis for $\Cl(8)$ can be constructed in terms of products of the $8$ real $16\times 16$ skewsymmetric matrices $\Gamma_I$ obeying $\Gamma_I \Gamma_J + \Gamma_J \Gamma_I = -2 \delta_{IJ} \, 1$ \footnote{The index $I$ will be used here to denote the vector representation of $\fso(8)$ and $\delta_{IJ}$ denotes the components of the unit $\fso(8)$-invariant inner product.}. The real 16-dimensional vector space acted upon by the matrices $\Gamma_I$ corresponds to the spinor representation of $\fso(8)$. The chirality matrix is defined by $\Gamma := \Gamma_1 ... \Gamma_8$, which is idempotent and anticommutes with each $\Gamma_I$. The chiral and antichiral representations $\boldsymbol{8}_s$ and $\boldsymbol{8}_c$ correspond respectively to the positive and negative chirality eigenspaces of $\Gamma$. The supersymmetry parameter $\epsilon_{IJ}$ is to be lifted to $\boldsymbol{\epsilon} = \Gamma \boldsymbol{\epsilon}$ in $\boldsymbol{8}_s$ while the fermions $\Psi^I$ are to be lifted to $\boldsymbol{\Psi} = - \Gamma \boldsymbol{\Psi}$ in $\boldsymbol{8}_c$ of $\fso(8)$ (both being also Majorana spinors on $\RR^{1,2}$). The lift of the bosons $X^I$ is more trivial requiring only the reinterpretation of the index $I$ from the $\boldsymbol{8}$ of $\fso(7)$ to the $\boldsymbol{8}_v$ of $\fso(8)$.

We assume the existence of a fixed (commuting) chiral spinor $\vartheta \in \boldsymbol{8}_s$ which we take to be unit normalised such that $\vartheta^t \vartheta = 1$\footnote{Notice that Majorana conjugation is just transposition here since the charge conjugation matrix for $\Cl(8)$ can be taken to be the identity}. This defines the desired embedding of $\fso(7)$ in $\fso(8)$ as the stabiliser of $\vartheta$. In terms of this fixed chiral spinor, one can deduce the precise identifications for the supersymmetry parameter and fermions to be
\begin{equation}\label{eq:Nequals8reps}
 \epsilon_{IJ} = \vartheta^t \Gamma_{IJ} \boldsymbol{\epsilon} , \quad \Psi^I = \vartheta^t \Gamma^I \boldsymbol{\Psi} ,
\end{equation}
where $\Gamma_{IJ} :=  \Gamma_{[I} \Gamma_{J]}$ and the Cayley form $\Omega_{IJKL} = \vartheta^t \Gamma_{IJKL} \vartheta$. Making use of the Fierz identity $\vartheta \vartheta^t = \tfrac{1}{16} \left( 1+ \Gamma + \tfrac{1}{4!} \Omega_{IJKL} \Gamma_{IJKL} \right)$, one can check that the right-hand side of the first equation in \eqref{eq:Nequals8reps} obeys the same projection condition $\epsilon_{IJ} = - \tfrac{1}{6} \Omega_{IJKL} \, \epsilon_{KL}$ satisfied by the left-hand side. Hence the eighth supersymmetry parameter in $\boldsymbol{\epsilon}$ is automatically projected out on the right-hand side.

Substituting this into the Lagrangian \eqref{eq:susy-lag-csm8} gives the sought after $\fso(8)$-invariant expression
\begin{equation}
\label{eq:susy-lag-csm8-bis}
  \begin{aligned}[m]
   \eL^{N=8} =& -\varepsilon^{\mu\nu\rho} \left( A_\mu , \partial_\nu A_\rho + \tfrac{1}{3} [ A_\nu , A_\rho ] \right) -\tfrac{1}{2} \left< D_\mu X^I , D^\mu X^I \right> + \tfrac{1}{2} \left< {\bar {\boldsymbol{\Psi}}}^t , \gamma^\mu D_\mu \boldsymbol{\Psi} \right> \\
   &+\tfrac{1}{8} ( T ( X^I , X^J ) , T( {\bar {\boldsymbol{\Psi}}}^t , \Gamma_{IJ} \boldsymbol{\Psi} )) -\tfrac{1}{48} \left< T ( X^I , X^J ) \cdot X^K , T ( X^I , X^J ) \cdot X^K \right>,
  \end{aligned}
\end{equation}
whose integral is indeed invariant under the $N=8$ supersymmetry transformations
\begin{equation}
  \label{eq:susy8}
  \begin{aligned}[m]
    \delta X^I &= {\bar {\boldsymbol{\epsilon}}}^t \Gamma^I \boldsymbol{\Psi} \\
    \delta \boldsymbol{\Psi} &= - \gamma^\mu D_\mu X^I \, \Gamma_I \boldsymbol{\epsilon} - \tfrac{1}{12} \, \tfrac{1}{3} T ( X^I , X^J ) \cdot X^K \, \Gamma_{IJK} \boldsymbol{\epsilon}  \\
    \delta A_\mu &= - \half \, T ( X^I , {\bar {\boldsymbol{\epsilon}}}^t \gamma_\mu \Gamma_I \boldsymbol{\Psi} ).
  \end{aligned}
\end{equation}

The $N=8$ supersymmetry parameter above can be decomposed as $\boldsymbol{\epsilon} = - \tfrac{1}{8} \epsilon_{IJ} \Gamma_{IJ} \vartheta + \eta \vartheta$ with respect to the $\fso(7)$ subalgebra, where $\eta$ is a single fermionic Majorana spinor on $\RR^{1,2}$. Setting $\eta = 0$ one recovers precisely the $N=7$ supersymmetry transformations in \eqref{eq:susy7}. 

The Lagrangian \eqref{eq:susy-lag-csm8-bis} and supersymmetry transformations \eqref{eq:susy8} indeed agree with those in \cite{BL2}. To be precise, equations (45) and (42) in \cite{BL2} match up with \eqref{eq:susy-lag-csm8-bis} and \eqref{eq:susy8} by identifying their 3-bracket $[-,-,-]$ with $-\half T(-,-)\cdot -$ on $U$ and their Lie algebra inner product with $2 (-,-)$ on $\fg$ here. Notice that the gauge field $A_\mu$ in \cite{BL2} has two indices on the 3-Lie algebra. This corresponds to the fact that $A_\mu$ lives in the image of the Faulkner map $T : V \times V \to \fg$, so it has two indices in $V$ but is actually valued in the Lie algebra $\fg$. The form presented in \cite{BL2} is in terms of projected Majorana spinors of $\Cl(1,10)$ broken to $\Cl(1,2) \otimes \Cl(0,8)$ and one identifies with our expressions above such that the gamma matrices of $\Cl(1,10)$ take the form $\gamma_\mu \otimes \Gamma$ and $1 \otimes i \Gamma_I$.

\section{N=8 and unitarity}
\label{sec:N8unitarity}

\subsection{Motivation and Lorentzian case}

We focus now on the case of maximal supersymmetry, $N=8$. This theory was the first to appear \cite{BL1,BL2} in the recent history of highly supersymmetric Chern-Simons-matter theories in three dimensions. Its interest lies in the fact that a stack of coincident M2-branes in 11-dimensional supergravity has an $AdS_4 \times S^7$ near-horizon geometry and its dual is conjectured to be a conformal field theory in three dimensions with precisely superconformal algebra $\fosp(8|4)$. 

As we have seen, the data needed to achieve $N=8$ supersymmetry is a metric 3-Lie algebra $V$. Up until now we have required the inner product $\left< -,- \right>$ on the 3-algebra to be positive-definite. The reason being that this inner product is used to describe the kinetic terms for the matter fields $X$ and $\Psi$. Therefore, if it is \textsl{not} Euclidean, the associated theory has `wrong' signs for the kinetic terms for those matter fields in the negative-definite directions on $V$, thus carrying negative energy and resulting in a theory which is not unitary as a quantum field theory.

However, it was soon proved in \cite{NagykLie} (see also \cite{GP3Lie,GG3Lie}) that there exists a unique indecomposable Euclidean 3-Lie algebra the simple 3-Lie algebra \cite{Filippov} $S_4$. The corresponding theory gives rise upon quantisation to a family of theories labelled by the integer Chern-Simons level $k$. This theory has been argued to describe at most two M2-branes \cite{LambertTong, DistlerBL} on a certain M-theory orbifold, at least for $k=1,2$. For other values of $k$ its spacetime interpretation is still unclear. This motivated interest in using non-Euclidean 3-Lie algebras instead. However, a mechanism to re-establish unitarity of a theory based in a non positive-definite algebra was needed. 

This was first investigated using a class of 3-Lie algebras of index 1 discovered in \cite{GMRBL,BRGTV,HIM-M2toD2rev}, namely those who are obtained from a double extension of a Euclidean semi-simple Lie algebra $V=V(\fg)$ as defined in theorem \ref{th:lorentzian}. Let $\left\{u,v,e_a\right\}$ be a basis for $V$ such that $u, v$ are null vectors and the vectors $e_a$ area basis for $\fg$ and generate a Euclidean subspace of $V$. What they found was that the matter field components $X^v$ and $\Psi^v$ along one of the two null directions $(u,v)$ in $W(\fg)$ never actually appear in any of the interaction terms in the Bagger-Lambert Lagrangian. The only terms in which they appear are the kinetic terms where they are coupled to $X^u$ and $\Psi^u$. Their corresponding equations of motion force the components $X^u$ and $\Psi^u$ in the other null direction $u$ to take constant values (preservation of maximal supersymmetry in fact requires $\Psi^u =0$). By expanding around the maximally supersymmetric and gauge-invariant vacuum defined by a constant expectation value for $X^u$, one can obtain a unitary quantum field theory. Use of this strategy in \cite{HIM-M2toD2rev} gave the first indication that the resulting theory is nothing but $N=8$ super Yang-Mills theory on $\RR^{1,2}$ with the Euclidean semi-simple gauge algebra $\fg$. The super Yang-Mills theory gauge coupling here being identified with the $SO(8)$-norm of the constant $X^u$. 

This procedure is somewhat reminiscent of the novel Higgs mechanism introduced by Mukhi and Papageorgakis in \cite{MukhiBL} in the context of the Bagger-Lambert theory based on the Euclidean 3-Lie algebra $S_4$. In that case an $N=8$ super Yang-Mills theory with $\fsu(2)$ gauge algebra is obtained, but with an infinite set of higher order corrections suppressed by inverse powers of the gauge coupling. As found in \cite{HIM-M2toD2rev}, the crucial difference is that there are no such corrections present in the Lorentzian case.

We want to point out that one must be wary of naively integrating out the free matter fields $X^v$ and $\Psi^v$ since their absence in any interaction terms in the Bagger-Lambert Lagrangian gives rise to an enhanced global symmetry that is generated by shifting them by constant values. To account for this degeneracy in the action functional, in order to correctly evaluate the partition function, one must gauge the shift symmetry and perform a BRST quantisation of the resulting theory. Fixing this gauged shift symmetry allows one to set $X^v$ and $\Psi^v$ equal to zero while the equations of motion for the new gauge fields sets $X^u$ constant and $\Psi^u = 0$. This more rigorous treatment has been carried out in \cite{BLSNoGhost,GomisSCFT} where the perturbative equivalence between the Bagger-Lambert theory based on $W( \fg )$ and maximally supersymmetric Yang-Mills theory with Euclidean gauge algebra $\fg$ was established (see also \cite{D2toD2}).

The ability to obtain a unitary theory from one based on a Lorentzian 3-Lie algebra prompted several questions. The first one was whether there are any other Lorentzian 3-Lie algebras that can give rise to more interesting unitary Bagger-Lambert theories. More generally, which metric 3-Lie algebras of any index $r$ can be used to obtain unitary theories and what is the physical interpretation of such theories. In our papers \cite{Lor3Lie,2pBL} we addressed these questions and we review the results we obtained in the following sections.

\subsection{Physically admissable 3-Lie algebras}

We consider now the most general metric 3-Lie algebra $\left(V, [-,-,-], \left\langle -,-\right\rangle\right)$ with index $r$. It will be useful to use the expression in \cite{BL2} for the the $N=8$ Lagrangian explicitly in terms of the 3-bracket $[-,-,-]$ and the structure constants $F^A{}_{BCD}$. For the sake of clarity, we focus on just the bosonic contributions since the resulting theories have a canonical maximally supersymmetric completion and none of the manipulations we will perform break any of the supersymmetries of the theories.

Given a basis $\left\{e_A\right\}$ for $V$, the bosonic part of the Bagger-Lambert Lagrangian is given by

\begin{equation}
\label{eq:BLLag}
  \begin{aligned}[m]
  \eL = &-\half \left< D_\mu X_I , D^\mu X_I \right> -\tfrac1{12} \left< [X_I,X_J,X_K],[X_I,X_J,X_K]\right> \\
	&+ \half \left( A^{AB} \wedge d \tilde{A}_{AB} +
  \tfrac23 A^{AB} \wedge \tilde{A}_{AC} \wedge \tilde{A}^C{}_B
\right)~,
  \end{aligned}
\end{equation}
where $(\tilde{A}_\mu)^A{}_B = F^A{}_{BCD} A_\mu^{CD}$, with $F_{ABCD} := \left< [ e_A ,  e_B , e_C ] , e_D \right>$ being the canonical 4-form and $D_\mu \phi^A = \partial_\mu \phi^A + (\tilde{A}_\mu)^A{}_B \phi^B$ for any field $\phi$ valued in $V$. The infinitesimal gauge transformations take the form $\delta \phi^A = - {\tilde \Lambda}^A{}_B \phi^B$ and $\delta (\tilde{A}_\mu)^A{}_B = \partial_\mu {\tilde \Lambda}^A{}_B + (\tilde{A}_\mu)^A{}_C {\tilde \Lambda}^C{}_B - {\tilde \Lambda}^A{}_C (\tilde{A}_\mu)^C{}_B$, where ${\tilde \Lambda}^A{}_B = F^A{}_{BCD} \Lambda^{CD}$ in terms of an arbitrary skewsymmetric parameter $\Lambda^{AB} = - \Lambda^{BA}$.

To obtain this form of the Lagrangian from \ref{eq:susy-lag-csm8} one needs to remove the space-time coordinates and re-instate the coordinates on $V$. Also the inner product on $\fg$ is now divided by 2. If $\left\{e_A\right\}$ is a basis for $V$, this implies:

\begin{equation}
  \begin{aligned}[m]
	&A_\mu = A_\mu^{AB} T(e_A,e_B)\\
	&T(e_A,e_B) \cdot e_C = 2 [e_A,e_B,e_C]\\
	&\left(T(e_A,e_B), T(e_C,e_D)\right) = \left\langle [e_A,e_B,e_C],e_D\right\rangle	\\
	&\left[T(e_A,e_B), T(e_C,e_D)\right] = T\left(T(e_A,e_B)e_C,e_D\right) + T\left(e_C,T(e_A,e_B)e_D\right)
	\end{aligned}
\end{equation}

There always exists a basis for $V$ such that $V = \bigoplus_{i=1}^r ( \RR u_i \oplus \RR v_i ) \oplus W$, where $\left< u_i , u_j \right> = 0 = \left< v_i , v_j \right>$, $\left< u_i , v_j \right> = \delta_{ij}$ and $W$ is a Euclidean vector space. It then follows that, generalising the argument used in the Lorentzian case, one can ensure that none of the $r$ null components $X^{v_i}$ and $\Psi^{v_i}$ of the matter fields appear in any of the interactions in the associated Bagger-Lambert Lagrangian provided that no $v_i$ appear on the left hand side of any of the 3-brackets on $V$. This guarantees an extra shift symmetry for each of these null components suggesting that all the associated negative-norm states in the spectrum of this theory can be consistently decoupled after gauging all the shift symmetries and following BRST quantisation of the gauged theory. 

The condition for the theory to be \textit{unitarisable} is that any 3-bracket containing a $v_i$ vector vanishes or equivalently that all the $v_i$ vectors are in the centre of $V$. This way of formulating the condition however is basis-dependant. A more invariant way of stating it is that $V$ should admit a \emph{maximally isotropic centre}. That is, a subspace $Z\subset V$ of dimension equal to the index of the inner product on $V$, $r$, on which the inner product vanishes identically and which is central, so that $[Z,V,V] = 0$ . The null directions $v_i$ defined above along which we require the extra shift symmetries are thus taken to provide a basis for $Z$. We say that a metric 3-Lie algebra is (physically) \textbf{admissible} if it is indecomposable and admits a maximally isotropic centre. 

In our paper \cite{2pBL} we classified such admissible 3-Lie algebras in a theorem that we reviewed in section \ref{sec:max-iso-cent}. Then, we computed the bosonic contributions to the Bagger-Lambert Lagrangians for such algebras. In that paper, we limited ourselves to expanding the theory around a suitable maximally supersymmetric and gauge-invariant vacuum defined by a constant expectation value for $X^{u_i}$ (with $\Psi^{u_i} = 0$). This is the natural generalisation of the procedure used in \cite{HIM-M2toD2rev} for the Lorentzian theory and coincides with that used in \cite{Ho:2009nk} for more general 3-Lie algebras. However, we did not perform a rigorous analysis of the physical theory in the sense of gauging the shift symmetries and BRST quantisation. 

In the next section we review the general structure of the gauge theories which result from expanding the BLG model based on these physically admissible 3-Lie algebras around a given vacuum expectation value for $X^{u_i}$. Before moving on to that, we note that, as we proved in \cite{Lor3Lie} and in theorem \ref{th:lorentzian} here, indecomposable 3-Lie algebras of index $1$ are either one-dimensional, simple or of the form $V=V(\fg)$. In fact, we saw in section \ref{sec:examples} that only the latter class possesses a maximally isotropic centre, hence this is indeed the only class of Lorentzian 3-Lie algebras that can be \textit{unitarised} via this procedure. 

\subsection{Unitary $N=8$ theories from admissible 3-Lie algebras}

Let us now assume that the indefinite signature metric 3-Lie algebra on which the Lagrangian \eqref{eq:BLLag} is based admits a maximally isotropic centre which we can take to be spanned by the basis elements $v_i$. Then, the 4-form components $F_{v_i
  ABC}$ must all vanish identically.  There are two important physical consequences of this assumption.  The first is that the covariant derivative on a field valued in the $u_i$ directions become just partial derivatives $D_\mu X_I^{u_i} = \partial_\mu X_I^{u_i}$.  The second is that the tensors $F_{ABCD}$ and $F_{ABC}{}^G F_{DEFG} = F_{ABC}{}^g F_{DEFg}$ which govern all the interactions in the Bagger-Lambert Lagrangian contain no legs in the $v_i$ directions.  Therefore the components $A_\mu^{v_i A}$ of the gauge field do not appear at all in the Lagrangian while $X_I^{v_i}$ appear only in the free kinetic term $- D_\mu X_I^{u_i} \partial^\mu X_I^{v_i} = - \partial_\mu X_I^{u_i} \partial^\mu X_I^{v_i}$. Hence, $X_I^{v_i}$ can be integrated out imposing that each $X_I^{u_i}$ be a harmonic function on $\RR^{1,2}$ which must be a constant if the solution is to be nonsingular. We will assume this to be the case, although singular monopole-type solutions may also be worthy of investigation, as in \cite{Verlinde:2008di}. Note that, when considering the full action with the fermions, one can see that in addition to setting $X_I^{u_i}$ constant, one must also set the fermions in all the $u_i$ directions to zero for the preservation of maximal supersymmetry.

The upshot is that, if we denote $\left\{e_a\right\}$ a basis for the Euclidean subspace $W$, we now have $-\half \left< D_\mu X_I , D^\mu X_I \right> = -\half D_\mu X_I^a D^\mu X_I^a$ with contraction over only the Euclidean directions of $V$ and each $X_I^{u_i}$ is taken to be constant. Since both $X_I^{v_i}$ and $A_\mu^{v_i A}$ are now absent, we can define $X_I^i := X_I^{u_i}$ and $A_\mu^{ia} :=A_\mu^{u_i a}$ from here onwards..

Using the structure theorem \ref{sec:max-iso-cent} for admissible 3-Lie algebras, one can calculate the Lagrangian for the BLG model associated with them. Upon expanding this theory around the maximally supersymmetric vacuum defined by constant expectation values $X^{u_i}$ (with all the other fields set to zero) we obtained in \cite{2pBL} standard $N=8$ supersymmetric (but nonconformal) gauge theories. One can think of the vacuum expectation values $X^{u_i}$ as defining a linear map, also denoted $X^{u_i}: \RR^r \to \RR^8$, sending $\xi \mapsto X^\xi := \sum_{i=1}^r \xi_i X^{u_i}$. Then, the physical gauge theory parameters are naturally expressed in terms of components in the image of this map. 

In theorem \ref{thm:main} we saw that the classification of admissible 3-Lie algebras breaks the Euclidean part of $V$, $W$, in several orthogonal Euclidean subspaces $\bigoplus_{\alpha =1}^N W_\alpha \oplus \bigoplus_{\pi=1}^M E_\pi \oplus E_0$ with different structures on them. The resulting Bagger-Lambert Lagrangian turns out to factorise into a sum of decoupled maximally supersymmetric gauge theories on each of the Euclidean components $W_\alpha$ (which we also call $\fg_\alpha$ because they are simple Lie algebras), $E_\pi$ and $E_0$. The physical content and moduli on each component can be summarised as follows:
\begin{itemize}
\item On each $\fg_\alpha$ one has an $N=8$ super Yang-Mills theory. The gauge symmetry is based on the simple Lie algebra
  $\fg_\alpha$. The coupling constant is given by $\| X^{\kappa^\alpha} \|$, which denotes the $SO(8)$-norm of the image of
  $\kappa^\alpha \in \RR^r$ under the linear map $X^{u_i}$. The seven scalar fields take values in the hyperplane $\RR^7 \subset
  \RR^8$ which is orthogonal to the direction defined by $X^{\kappa^\alpha}$. If $X^{\kappa^\alpha}=0$, for a given value of
  $\alpha$, one obtains a degenerate limit corresponding to a maximally superconformal free theory for eight scalar fields and
  eight fermions valued in $\fg_\alpha$.

\item On each plane $E_\pi$ one has a pair of identical free abelian $N=8$ massive vector supermultiplets. The bosonic fields in
  each such supermultiplet comprise a massive vector and six massive scalars. The mass parameter is given by $\| X^{\eta^\pi}
  \wedge X^{\zeta^\pi} \|$, which corresponds to the area of the parallelogram in $\RR^8$ defined by the vectors $X^{\eta^\pi}$
  and $X^{\zeta^\pi}$ in the image of the map $X^{u_i}$. The six scalar fields inhabit the $\RR^6 \subset \RR^8$ which is
  orthogonal to the plane spanned by $X^{\eta^\pi}$ and $X^{\zeta^\pi}$. If $\| X^{\eta^\pi} \wedge X^{\zeta^\pi} \| =0$, for a
  given value of $\pi$, one obtains a degenerate massless limit where the vector is dualised to a scalar, again corresponding to a
  maximally superconformal free theory for eight scalar fields and eight fermions valued in $E_\pi$. Before gauge-fixing, this
  theory can be understood as an $N=8$ super Yang-Mills theory with gauge symmetry based on the four-dimensional Nappi-Witten
  Lie algebra $\fd(E_\pi,\RR)$. Moreover we explain how it can be obtained from a particular truncation of an $N=8$ super
  Yang-Mills theory with gauge symmetry based on any Euclidean semisimple Lie algebra with rank 2, which may provide a more
  natural D-brane interpretation.

\item On $E_0$ one has a decoupled $N=8$ supersymmetric theory involving eight free scalar fields and an abelian Chern-Simons
  term. Since none of the matter fields are charged under the gauge field in this Chern-Simons term then its overall
  contribution is essentially trivial on $\RR^{1,2}$.
\end{itemize}

We refer the reader interested in the technical details of this derivation to \cite{2pBL}.

\section{Summary and outlook}
\label{sec:summary}

In this section we gather our findings of this chapter and discuss the open problems and future directions for research. First, we review the mechanism for supersymmetry enhancement in superconformal Chern-Simons-matter theories in three dimensions. Then, we summarise the initial data needed to build a theory with $N$-supersymmetry and the $N=1$ superpotential that realises such extended supersymmetry. We also discuss the conditions for irreducibility of theories with more than $N=3$ supersymmetry and use them to display a classification of theories with $N>4$. In doing so, we stress the equivalence of the two languages in which these theories can be equivalently expressed: in terms of 3-Leibniz algebras or representation theory of Lie algebras. We then dedicate one paragraph to reviewing our findings for theories with maximal supersymmetry. We finish this section and this thesis with an outlook for future research in the field.

\subsection{Data for superconformal Chern-Simons-matter theories in 3D}

In this chapter we have studied in a systematic way classically superconformal Chern-Simons theories coupled to matter in three dimensions. We have derived from first principles and recovered in a uniform framework all known examples of highly supersymmetric theories in the literature: the $N=4$ theories of \cite{GaiottoWitten, pre3Lee}, the $N=5$ class obtained in \cite{3Lee,BHRSS}, the $N=6$ ABJM theory \cite{MaldacenaBL} (also in the language of \cite{BL4}) and finally the Bagger-Lambert theory \cite{BL1,BL2}. In particular we have established the data that one needs to feed into the $N=1$ Lagrangian \eqref{eq:susy-lag-on-shell} to obtain a theory with $N$-extended supersymmetry.

Our starting point was the $N=1$ theory \eqref{eq:susy-lag-on-shell}. Any such theory is completely determined given a \textbf{real orthogonal 3-Leibniz algebra} as defined in \ref{sec:ro3leib} and a \textbf{real superpotential} $\sW$, which is a $\fg$-invariant quartic function of the matter fields. 

We required the inner product on the 3-algebra to be positive-definite to ensure unitarity of the theory. Recall that, by the Faulkner construction discussed in \ref{sec:FaulkR}, a real orthogonal 3-Leibniz algebra is equivalent to a real orthogonal Lie algebra equipped with a real orthogonal representation. Hence, once the 3-algebra is fixed, this determines uniquely the corresponding real Lie algebra \textit{and} the inner product on it. Alternatively, any pair of a real orthogonal Lie algebra and a unitary representation of it can be used as data for the $N=1$ theory as they together also define a real orthogonal 3-Leibniz algebra.

We found that any $N=2$ theory is completely determined given a \textbf{complex unitary 3-Leibniz algebra} as defined in \ref{sec:cu3leib} and an \textbf{F-term superpotential} $\sW_F$, which is a $\fg$-invariant holomorphic function of the matter fields.

Given this data, the corresponding $N=2$ theory is obtained from the $N=1$ theory \eqref{eq:susy-lag-on-shell} by choosing the $N=1$ superpotential $\sW$ to be \eqref{eq:superpotential2}. Again, thanks to the Faulkner construction described in \ref{sec:FaulkC}, this data is equivalent to a real orthogonal Lie algebra and a complex unitary representation of it. 

Then, we saw how any $N=3$ theory is determined by a \textbf{quaternionic unitary 3-Leibniz algebra} as defined in \ref{sec:qu3leib}. The superpotential is completely fixed by the requirement of $N=3$ supersymmetry. 

Given a quaternionic 3-Leibniz algebra, an $N=3$ theory is obtained from the $N=1$ theory \eqref{eq:susy-lag-on-shell} by choosing the $N=1$ superpotential to be \eqref{eq:superpotential3}. Again, thanks to the Faulkner construction described in \ref{sec:FaulkC}, this data is equivalent to a real orthogonal Lie algebra and a quaternionic unitary representation of it. 

The $N=3$ superpotential being rigid, we found that the only way to achieve further supersymmetry enhancement was by considering special cases of the $N<4$ theories. Our guiding principle to find exactly what particular 3-algebras (or equivalently representations of Lie algebras) would provide an enhancement to N-extended symmetry was the requirement that the $N=1$ superpotential had to be invariant under a global $\fso(N-1)$ symmetry. 

We found that, in order to achieve $N=4$ supersymmetry, the 3-algebra $W$ has to be not only of quaternionic type but in particular it must be a direct sum of one or more anti-Lie triple systems. $N=5$ theories are in one-to-one correspondence with irreducible anti-Lie triple systems. The data required for an $N=6$ theory is an irreducible anti-Jordan triple system and finally, for $N=8$, the only possibility is the single positive-definite 3-Lie algebra. 

The data that needs to be fed to the $N=1$ theory \eqref{eq:susy-lag-on-shell} for subsequent amounts of supersymmetry is summarised in table \ref{tab:input}. In that table every 3-Leibniz algebra is equipped with a \textit{positive-definite} inner product and $\fg$ stands for a real metric Lie algebra (with inner product not necessarily Euclidean).

\begin{table}[ht!]\footnotesize
  \centering
  \begin{tabular}{|>{$}c<{$}|c|c|}
    \hline
    \textbf{N} & \textbf{Input} & \textbf{Alternative input} \\\hline
    1 & $\sW$ and a real orthogonal 3-Leibniz algebra $U$ & $\fg$ and  $U \in \Dar(\fg,\RR)$\\
    2 & $\sW_F$ and a complex unitary 3-Leibniz algebra $V$ & $\fg$ and  $V \in \Dar(\fg,\CC)$\\
    3 & Quaternionic unitary 3-Leibniz algebra $W$ & $\fg$ and  $W \in \Dar(\fg,\HH)$\\
    4 & One or more irreducible anti-Lie triple systems $W = \bigoplus W_i$ & $\fg$ and  $W_i \in \Irr\fg,\HH)_{\text{aLTS}}$\\
    5 & Irreducible anti-Lie triple system & $\fg$ and  $W \in \Irr(\fg,\HH)_{\text{aLTS}}$\\
    6 & Irreducible anti-Jordan triple system $V$ & $\fg$ and  $V \in \Irr(\fg,\CC)_{\text{aJTS}}$\\
    7 & Irreducible 3-Lie algebra $U$ & $\fg$ and  $U \in \Irr(\fg,\RR)_{\text{3LA}}$\\
    8 & Irreducible 3-Lie algebra $U$ & $\fg$ and  $U \in \Irr(\fg,\RR)_{\text{3LA}}$\\
    \hline
  \end{tabular}
  \vspace{8pt}
  \caption{Matter representations for $N$-extended supersymmetry}
  \label{tab:input}
\end{table}

Notice that listing the data in terms of Lie algebras and their representations can be a bit confusing. Take for example the case $N=8$. The input data needed to obtain such a theory is a real metric Lie algebra and a real representation of it of 3-Lie algebra type. The catch is that not any Lie algebra admits a representation of that type. In fact, we know now that only $\fg = \fsu(2) \oplus \fsu(2)$ admits such a representation. When listing the data only in terms of 3-Leibniz algebras, it is clear than the gauge algebra for the theory can not just be any Lie algebra, but only the one given by the Faulkner construction (that is, the one generated by the inner derivations of the 3-algebra) which is totally determined, including the ad-invariant inner product for the Chern-Simons part of the Lagrangian, once the 3-algebra is fixed. 

The Faulkner construction that we specialised to the unitary case in chapter \ref{sec:3algebras} and the work done in this chapter show that, indeed, in the context of 3-dimensional Chern-Simons theories, a formulation in terms of metric 3-Leibniz algebras is completely equivalent to a formulation in terms of the more standard representation theory of Lie algebras. One could argue that this eliminates the need to use 3-algebras to describe these theories, but we believe that the existence of these two descriptions enriches the field and has already proved to be a very useful tool of research. In fact, it was not until a 3-algebra was used in the original Bagger-Lambert proposal \cite{BL1} that a maximal supersymmetric Lagrangian was found at all. The special feature of these theories by which the amount of supersymmetry preserved is directly related to the gauge Lie algebra used is only one example of how counter-intuitive using the wrong language can be. 

\subsection{Superpotentials}
\label{sec:superpotentials}

Superconformal Chern-Simons-matter theories in 3-dimensions with more than $N=2$ supersymmetry are then completely fixed by the 3-algebras in table \ref{tab:input}, that in turn determine uniquely the gauge Lie algebra of the theory and the metric on it. To obtain the on-shell Lagrangian of the corresponding $N$-enhanced theory, all one needs to do is choose the $N=1$ superpotential to be of the form $\sW = \tfrac{1}{16} \int d^2 \theta \, \fW ( \Xi )$ where $\fW ( \Xi )$ is as detailed in table~\ref{tab:superpotentials}. In that table, the only optional piece is the F-term superpotential for $N=2$ theories. For $N\geq 3$, the superpotential is always the same $N=3$ superpotential \eqref{eq:superpotential3}, only expanded out for the particular representation considered. The $N=1$ superfield is in each case valued in an appropriate representation of $\fso(N-1) \oplus \fg$, which are summarised in table \ref{tab:matter-reps}. The tensor $\Omega$ appearing in the $N=6$ row is the $\fso(5) \cong \fusp(4)$-invariant symplectic form on $\Delta^{(5)}$ while in the $N=8$ row it denotes the $\fso(7)$-invariant self-dual Cayley 4-form on $\Delta^{(7)}$. Repeated indices are contracted with respect to the hermitian inner product on $\Delta^{(N-1)}$.

\begin{table}[ht!]\footnotesize
  \centering
  \begin{tabular}{|>{$}c<{$}|>{$}c<{$}|}
    \hline
    N & \fW ( \Xi ) \\\hline   
    2 & ( \TT ( \Xi , \Xi ) , \TT ( \Xi , \Xi ) ) + {\mathrm{Re}} \, \fW_F ( \Xi ) \\
		3 & ( \TT ( \Xi , \Xi ) , \TT ( \Xi , \Xi ) ) + {\mathrm{Re}} \, ( \TT ( \Xi , J \Xi ) , \TT ( \Xi , J \Xi ) )\\		
		4 & \tfrac{1}{6} \, ( \TT_1 ( \Xi^a , \Xi^b ) , \TT_1 ( \Xi^b , \Xi^a ) ) + \tfrac{1}{6} \, ( \TT_2 ( \Xi^a , \Xi^b ) , \TT_2 ( \Xi^b , \Xi^a ) ) - ( \TT_1 ( \Xi^a , \Xi^b ) , \TT_2 ( \Xi^b , \Xi^a ) ) \\
		5 & -\tfrac{1}{6} \, ( \TT ( \Xi^\alpha , \Xi^\beta ) , \TT ( \Xi^\beta , \Xi^\alpha ) ) -\tfrac{1}{6} \, ( \TT ( \Xi^{\dot \alpha} , \Xi^{\dot \beta} ) , \TT ( \Xi^{\dot \beta} , \Xi^{\dot \alpha} ) ) + ( \TT ( \Xi^\alpha , \Xi^{\dot \beta} ) , \TT ( \Xi^{\dot \beta} , \Xi^\alpha ) ) \\
		6 & ( \TT ( \Xi^a , \Xi^b ) , \TT ( \Xi^b , \Xi^a ) ) + \Omega_{ab} \, \Omega^{cd} \, ( \TT ( \Xi^a , \Xi^c ) , \TT ( \Xi^b , \Xi^d ) ) \\
    8 & \tfrac{1}{3} \, \Omega_{abcd} \, ( T ( \Xi^a , \Xi^b ) , T ( \Xi^c , \Xi^d ) ) \\		\hline
  \end{tabular}
  \vspace{8pt}
  \caption{Superpotentials for $N$-extended supersymmetry}
  \label{tab:superpotentials}
\end{table}

\begin{table}[ht!]
  \centering
  \begin{tabular}{|>{$}c<{$}|>{$}c<{$}|>{$}c<{$}|}
    \hline
    N & \text{Matter representation} & \text{Remarks}\\\hline
    1 & U & U \in \Dar(\fg,\RR)\\
    2 & \Delta^{(2)}_\pm \otimes V \oplus  \Delta^{(2)}_\mp \otimes \Vbar & V \in \Dar(\fg,\CC)\\
    3 & \Delta^{(3)} \otimes W & W \in \Dar(\fg,\HH)\\
    4 & \Delta^{(4)}_\pm \otimes W_1 \oplus  \Delta^{(4)}_\mp \otimes W_2 & W_{1,2} \in \Dar(\fg,\HH)_{\text{aLTS}}\\
    5 & \Delta^{(5)} \otimes W & W \in \Irr(\fg,\HH)_{\text{aLTS}}\\
    6 & \Delta^{(6)}_\pm \otimes V \oplus  \Delta^{(6)}_\mp \otimes \Vbar & V \in \Irr(\fg,\CC)_{\text{aJTS}}\\
    7 & \Delta^{(7)} \otimes U & U \in \Irr(\fg,\RR)_{\text{3LA}}\\
    8 & \Delta^{(8)}_\pm \otimes U & U \in \Irr(\fg,\RR)_{\text{3LA}}\\
    \hline
  \end{tabular}
  \vspace{8pt}
  \caption{Matter representations for $N$-extended supersymmetry}
  \label{tab:matter-reps}
\end{table}

A side-effect of this method for supersymmetry enhancement was to prove that in this context $N=7$ supersymmetry implies $N=8$.

\subsection{Indecomposability and irreducibility}
\label{sec:indecomposability}

Throughout this chapter, we have also discussed the conditions for the theories obtained to be indecomposable, that is that they should not decouple into two or more nontrivial theories. We summarise here our findings on the matter.

 For $N<4$ indecomposability places very weak constraints on the allowed representations. For the $N=4$ theories of the type discussed by Gaiotto and Witten \cite{GaiottoWitten}, where the bosonic matter lives in $\Delta^{(4)}_+ \otimes W$, for $W \in \Dar(\fg,\HH)_{\text{aLTS}}$, indecomposability forces $W$ to be irreducible. For the general $N=4$ theories with twisted matter, indecomposability implies the connectedness of the corresponding quiver \cite{pre3Lee}, which imposes conditions - although not irreducibility - on the allowed representations. In particular we saw that the sum of several irreducible aLTS $W = \bigoplus W_i$ must not be of aLTS type itself for the theory to be irreducible. 

For $N>4$ indecomposability turned out to be tantamount to irreducibility of the 3-algebra. The 3-algebras involved in those theories are those that we defined as extreme cases which were embeddable in Lie superalgebras in section \ref{sec:extremeLSA}, namely anti-Lie triple systems, anti-Jordan and 3-Lie algebras. In that section, which is based on \cite{JMFSimplicity},  we discussed how irreducibility of these extreme cases is equivalent to simplicity of the corresponding Lie superalgebra when the inner product on the 3-algebra is positive-definite. This allows for a classification of indecomposable theories with $N>4$-supersymmetry and also provides a list of the allowed gauge Lie algebras for each amount of $N$-extended supersymmetry with $N>4$. The discussion is summarised in table \ref{tab:irreducible-LE}. Notice that simplicity of the embedding Lie superalgebra does not imply the simplicity of the gauge Lie algebra $\fg$ and indeed in none of the cases in the table is $\fg$ allowed to be simple. This fact may explain why these theories took a relatively long time to be discovered.

\begin{table}[ht!]\footnotesize
  \centering
  \begin{tabular}{|>{$}c<{$}|c|>{$}c<{$}|>{$}c<{$}|>{$}c<{$}|}
    \hline
    N & Class & \text{Representation} & \fg & \text{Lie superalgebra} \\\hline
    8 & 3LA & (\boldsymbol{2},\boldsymbol{2}) & \fsu(2) \oplus \fsu(2) &  A(1,1) \\\hline
    5, 6 & aLTS, aJTS & (\boldsymbol{m+1},\overline{\boldsymbol{n+1}})_{m-n} & \fsu(m+1) \oplus \fsu(n+1) \oplus \fu(1) &  A(m,n), m\neq n\\
    5, 6 & aLTS, aJTS & (\boldsymbol{n+1},\overline{\boldsymbol{n+1}}) & \fsu(n+1) \oplus \fsu(n+1) &  A(n,n) \\
    5, 6 & aLTS, aJTS & (\boldsymbol{2n})_{+1} & \fusp(2n) \oplus \fu(1) &  C(n+1) \\\hline
    5 & aLTS & (\boldsymbol{2m+1},\boldsymbol{2n}) & \fso(2m+1) \oplus \fusp(2n) &  B(m,n) \\
    5 & aLTS & (\boldsymbol{2m},\boldsymbol{2n})  & \fso(2m) \oplus \fusp(2n) &  D(m,n) \\
    5 & aLTS & (\boldsymbol{2},\boldsymbol{2},\boldsymbol{2}) & \fsu(2) \oplus \fsu(2) \oplus \fsu(2) &  D(2,1;\alpha) \\
    5 & aLTS & (\boldsymbol{2},\boldsymbol{8}) & \fsu(2) \oplus \fspin(7) &  F(4) \\
    5 & aLTS & (\boldsymbol{2},\boldsymbol{7}) & \fsu(2) \oplus \fg_2 &  G(3) \\\hline
  \end{tabular}
  \vspace{8pt}
  \caption{Irreducible, positive-definite Lie-embeddable representations}
  \label{tab:irreducible-LE}
\end{table}

It must be remarked that all the matter representations for $N>4$ superconformal Chern-Simons theories in our table~\ref{tab:irreducible-LE} had been found already in \cite{BHRSS} via a certain global limit of conformally gauged supergravities in three dimensions. The gauging can be most conveniently described in terms of a so-called \textsl{embedding tensor} and it is the linear constraint imposed on this object by supersymmetry that allows one to identify the different classes of representations in table~\ref{tab:irreducible-LE} with those in table 3 of \cite{BHRSS}. In each case, it is the tensor $\eR$ constructed from the Faulkner maps and defined in section \ref{sec:unitrep} that corresponds to an R-symmetry-singlet of the embedding tensor in the aforementioned global limit. Our classification from first principles establishes that there exist no other indecomposable $N>4$ theories.

\subsection{Maximal supersymmetry and unitarity}

Finally, in section \ref{sec:N8unitarity} we investigated theories with maximal supersymmetry where the inner product on the 3-algebra is not positive-definite. When the 3-Lie algebra is physically admissible (that is, it has a maximally isotropic centre) such theories can be rendered unitary because this algebraic condition ensures that all the negative-norm states in the associated Bagger-Lambert theory can be consistently decoupled from the physical Hilbert space. Unitary theories are obtained by expanding these theories around a suitable vacuum. These typically involve particular combinations of $N=8$ super Yang-Mills and massive vector supermultiplets, discarding an interpretation in terms of M2-branes.

\subsection{Outlook}

The primary motivation to describe these theories was that they are candidates to be dual to configurations of multiple coincident M2-branes in $AdS_4$ supergravity backgrounds. From that point of view, understanding their behaviour upon quantization and finding an interpretation in terms of string/M-theory for all of them is a priority. In order to understand their quantum behaviour one has to take into account monopole operators, whose role has proved crucial for example to match the spectrum of ABJM theories to supergravity and in particular to understand enhancement to $N=8$ supersymmetry \cite{Gustavsson:2009pm,Benna:2009xd,Kwon:2009ar}. From the mathematical point of view, one would like to have a clearer algebraic understanding of these monopole operators in three dimensional SCCSM theories to understand their involvement in the supersymmetry enhancement mechanism. 

Many of these theories have already found interpretation in terms of M-theory. The original $N=8$ theory is thought to describe two coincident M2-branes on a $\ZZ_2$ orbifold when the Chern-Simons level is $k=2$. The ABJM $N=6$ theories in \cite{MaldacenaBL} (see also \cite{KlebanovBL}) describe the low-energy dynamics of multiple coincident M2-brane configurations whose near-horizon geometries are of the form $\AdS_4 \times S^7 / \ZZ_k$, for some positive integer $k$, with maximal $N=8$ supersymmetry recovered only for $k=1,2$ \cite{Lambert:2010ji}. The $N=5$ superconformal field theories in \cite{3Lee,BHRSS} have been argued \cite{ABJ} to describe near-horizon geometries of the form $\AdS_4 \times S^7 / {\hat D}_k$ (${\hat D}_k$ being the binary dihedral group of order $4k$). There is still no interpretation for the $N=5$ SCCSM theories based on the exceptional Lie superalgebras $D(2,1;\alpha)$, $G(3)$ and $F(4)$ and found first in \cite{BergshoeffMM2}. On the other hand, we showed in \cite{deMedeiros:2009pp} that $AdS_4 \times S^7 / \Gamma$ backgrounds where Gamma is one of the $SU(2)$ subgroups $E_6,E_7,E_8$ admit precisely $N=5$ killing spinors, so there is a question on whether these geometries could indeed be their duals. 

A powerful tool to investigate dual geometries to conformal field theories are quivers and \textit{brane tilings}. In \cite{Hanany:2008cd,Hanany:2008fj,Davey:2009et} a method that had been used previously for SCFTs in four dimensions \cite{Franco:2005rj,Franco:2005sm} was generalised to SCFTs in three dimensions to read off the dual geometry from the ingredients of the gauge theory. However these techniques can only be applied to theories whose dual geometry admits a toric description, which in this case is limited to a subclass of $N=2$ supersymmetric theories. 

It would be interesting though to find a uniform framework in which these theories and their duals could be interpreted. A first step towards this goal was presented in \cite{deMedeiros:2009pp} where we classified Freund-Rubin backgrounds of eleven-dimensional supergravity of the form $AdS_4 \times X^7$ which are at least half BPS ($N>3$) and where $X^7$ is a smooth quotient of $S^7$. In the second part of this paper \cite{deMedeiros:2010dn} all backgrounds (not necessarily smooth) with $N>3$ were classified. The next step forward would be to identify the precise relationship between the geometrical parameters that label these theories with the algebraic data that determines 3-dimensional SCCSM with $N>3$, but this task has proved elusive so far.

An interesting generalisation to these theories is to include couplings of the multiple M2-branes to other supergravity background fields. Some work in this direction has been done in \cite{Li:2008eza,Kim:2009nc,Lambert:2009qw}. 

All these new families of Lagrangians have also attracted interest from fields other than strictly M-theory. For instance, they provide a wide range of new examples in which to test the AdS/CFT correspondence. Also, the ABJM model has been conjectured to be integrable \cite{Minahan:2008hf,Gaiotto:2008cg,Gromov:2008qe,Bak:2008cp}, which has led to some significant new results. Three-dimensional SCCSM theories are also of interest to condensed matter physicists because their systems often involve $2+1$ dimensions and Chern-Simons terms are very common. For a good review on the topic see \cite{Hartnoll:2009sz}.

Last, but certainly not least, another highly sought after direction for research would be to find also Lagrangian descriptions for low energy limits to different configurations of stacks of M5-branes, by generalising these theories or by other means. The hope being that then we would be in a much better position to understand M-theory from a much better understanding of its main constituents. Some progress in this direction has been made in \cite{Ho:2008nn,Ho:2008ve,Bandos:2008fr,Gustavsson:2009qd,Benishti:2010jn,Lambert:2010wm} and especially the work of Gaiotto and Maldacena \cite{Gaiotto:2009gz}.

\appendix
\chapter{List of publications}
\label{publications}

The contents of this thesis are the result of the research I have done from September 2007 onwards together with my collaborators Paul de Medeiros, José Figueroa-O'Farrill and Patricia Ritter. The products of this collaboration have previously been published in five papers that we list below. Here, I focused on the results in these papers I had a more personal involvement with.

We also published a sixth paper exploring some aspects of the dual geometries to the conformal field theories studied here.  Although we did not discuss the contents of this last paper in this thesis, we list it here because of its close relation to the subject.

\begin{itemize}
	\item ``\textit{Lorentzian Lie 3-algebras and their Bagger-Lambert moduli space}''\\
Paul de Medeiros, José Figueroa-O’Farrill, Elena Méndez-Escobar.\\
JHEP 07 (2008) 111, e-print:arXiv:0805.4363 [hep-th]
	\item ``\textit{Metric Lie 3-algebras in Bagger-Lambert Theory}"\\
Paul de Medeiros, José Figueroa-O’Farrill, Elena Méndez-Escobar.\\
JHEP 08 (2008) 045, e-print: arXiv:0806.3242 [hep-th] 
\item ``\textit{On the Lie algebraic origin of metric 3-algebras}"\\
Paul de Medeiros, José Figueroa-O’Farrill, Elena Méndez-Escobar, \\
Patricia Ritter. CMP 290 (2009) 871-902, e-print: arXiv:0809.1086 [hep-th] 
\item ``\textit{Metric 3-Lie algebras for unitary Bagger-Lambert theories}"\\
Paul de Medeiros, José Figueroa-O’Farrill, Elena Méndez-Escobar, \\
Patricia Ritter. JHEP 04 (2009) 037, e-print: arXiv:0902.4674 [hep-th] 
\item ``\textit{Superpotentials for superconformal Chern-Simons theories from representation theory}"\\
Paul de Medeiros, José Figueroa-O’Farrill, Elena Méndez-Escobar.\\
J Phys A: Math. Theor. 42 (2009) 485204, e-print: arXiv:0908.2125 [hep-th] 
\item ``\textit{Half-BPS quotients in M-theory: ADE with a twist}"\\
Paul de Medeiros, José Figueroa-O’Farrill, Sunil Gadhia, Elena Méndez-Escobar.\\
JHEP 10 (2009) 038, e-print: arXiv:0909.0163 [hep-th] 
\end{itemize}

\bibliographystyle{utphys}
\bibliography{Thesis}

\end{document}